%% file: jvink_aarv_2col.tex
\documentclass[twocolumn,a4paper]{article}         

\usepackage{graphicx}
\usepackage{mathptmx}  
\usepackage{amsmath}  
\usepackage{amssymb}  
\usepackage{latexsym}
\usepackage{natbib}
\usepackage{natbibspacing}
\usepackage{microtype}
\usepackage[breaklinks=true]{hyperref}
\usepackage{fancyhdr}

\graphicspath{{./Figures/}}

\input{definitions}

\title{
\huge Supernova remnants: the X-ray perspective
}
\author{Jacco Vink}

\begin{document}
\onecolumn

\date{}

\maketitle

\setcounter{tocdepth}{2}

\begin{center}
\begin{minipage}{0.8\textwidth}
\input{jvink_aarv_abstract}

\end{minipage}
\end{center}

\newpage
\tableofcontents
\twocolumn
%
\section{Introduction}
\label{sec:intro}
\input{jvink_aarv_intro}

\section{Supernovae}
\label{sec:supernovae}
\input{jvink_aarv_supernovae}

\input{jvink_aarv_core_collapse}

\input{jvink_aarv_thermonuclear}

\section{The classification of supernova remnants}
\label{sec:snrclasses}
\input{jvink_aarv_snr_classes}

\section{The hydrodynamic structure and evolution of
 supernova remnants}
\label{sec:hydro}
\input{jvink_aarv_hydro}

\section{Collisionless shock heating and particle acceleration}
\label{sec:plasma}
\subsection{Shock heating}
\label{sec:kT}
\input{jvink_aarv_shock_heating}

\subsection{Collisionless shocks}
\label{sec:collisionless}
\input{jvink_aarv_collisionless_shocks}

\subsection{Thermal conduction}
\label{sec:conduction}
\input{jvink_aarv_conduction}

\subsection{Cosmic-ray acceleration by supernova remnant shocks}
\label{sec:cr}
\input{jvink_aarv_cr_acceleration}

\section{X-ray radiation from supernova remnants}
\input{jvink_aarv_nei}

\input{jvink_aarv_lines}

\input{jvink_aarv_radioactivity}

\subsection{Non-thermal emission}
\label{sec:nonthermal}
\input{jvink_aarv_nonthermal}

\section{X-ray spectroscopy with \cxo, \xmm, and \suz}
\label{sec:spectra}
\input{jvink_aarv_xray_instruments}

\section{Type Ia supernova remnants}
\label{sec:typeia}
\subsection{X-ray spectroscopy of Type Ia supernova remnants}
\input{jvink_aarv_typeia_snrs}

\subsection{Evidence for Type Ia progenitor imprints on the
circumstellar medium}
\label{sec:iacsm}
\input{jvink_aarv_ia_csm}

\section{Core collapse supernova remnants}
\subsection{Oxygen-rich supernova remnants}
\label{sec:orich}
\input{jvink_aarv_oxygen_rich}

\subsection{The X-ray emission from SN\,1987A}
\label{sec:sn1987a}
\input{jvink_aarv_sn1987a}

\subsection{Core collapse supernova remnants and neutron stars}
\label{sec:snrs_ns}
\input{jvink_aarv_ns}

\section{Supernova remnants in or approaching the radiative phase}
\label{sec:maturesnrs}

\subsection{From non-radiative to radiative shocks}
\label{sec:radiative}
\input{jvink_aarv_cl_vela}

\subsection{Enhanced metal abundances mature SNRs}
\label{sec:mature_metals}
\input{jvink_aarv_mature_metals}

\subsection{Mixed-morphology supernova remnants}
\label{sec:mixed}
\input{jvink_aarv_mixed}

\section{Shock heating and particle acceleration: observations}
\label{sec:shock_observations}

\subsection{Electron-ion temperature equilibration}
\label{sec:thermaldoppler}
\input{jvink_aarv_thermaldoppler}

\clearpage
\subsection{X-ray synchrotron emitting filaments}
\label{sec:synchrotron}
\input{jvink_aarv_synchrotron}

\subsection{X-ray based evidence for efficient cosmic-ray acceleration}
\label{sec:efficient_crs}
\input{jvink_aarv_efficient_crs}

\subsection{The relation between X-ray synchrotron and \gray\ emission}
\label{sec:xray_gammarays}
\input{jvink_aarv_xray_gamma}

\section{Concluding remarks}
\label{sec:conclusion}
\input{jvink_aarv_conclusion}

\subsection*{acknowledgements}
I am grateful to Eveline Helder and Sjors Broersen for 
carefully reading a draft version of this review. I also appreciate
comments from Joke Claeys on section~\ref{sec:supernovae} and discussion
with her, Alexandros Chiotellis, and Dani Maoz on Type Ia supernovae.
I thank the editor of A\&A Reviews, Thierry Courvoisier, for inviting
me for this review and the patience he showed for the completion 
of the manuscript. 

I want to express my dissappointment that Utrecht University decided to
close its astronomy department, which had a long and fruitful 
tradition of 370~yr.

We acknowledge the permission of the AAS for the reproduction of
the figures \ref{fig:windbubble},\ref{fig:badenes}, \ref{fig:flanagan},
\ref{fig:lopez},\ref{fig:sn1987a},\ref{fig:sn87aspec},\ref{fig:cygnusloop}
and \ref{fig:casaslope}, the permission of the 
Publications of the Astronomical Society of Japan for reproduction of
figures \ref{fig:tychocrmn} and \ref{fig:snr_rrc}, and the permission of
 Astronomy and Astrophysics for the figures \ref{fig:suprathermal}, \ref{fig:stratification}, \ref{fig:chiotellis}, \ref{fig:casadoppler}, \ref{fig:rxj1713}.
Figure~\ref{fig:vela} was taken from \url{http://www.mpe.mpg.de} and appears
with permission from the Max-Planck-Institut für extraterrestrische Physik.


\input{jvink_aarv_arxiv_bib}
\end{document}

%% file: definitions.tex
\newcommand{\asca}{{\it ASCA}}
\newcommand{\astroh}{{\it Astro-H}}
\newcommand{\cgro}{{\it CGRO}}
\newcommand{\comptel}{{\it CGRO-Comptel}}
\newcommand{\egret}{{\it CGRO-EGRET}}
\newcommand{\osse}{{\it CGRO-OSSE}}
\newcommand{\chandra}{{\it Chandra}}
\newcommand{\cxo}{{\it Chandra}}
\newcommand{\einstein}{{\it Einstein}}

\newcommand{\fermi}{{\it Fermi}}
\newcommand{\hegra}{{\it HEGRA}}
\newcommand{\hess}{{\it HESS}}
\def\hst{{\it Hubble Space Telescope}}
\newcommand{\ixo}{{\it IXO}}
\newcommand{\integral}{{\it INTEGRAL}}
\newcommand{\athena}{{\it ATHENA}}
\newcommand{\magic}{{\it MAGIC}}
\newcommand{\next}{{\it Astro-H}}
\newcommand{\rosat}{{\it ROSAT}}
\newcommand{\rxte}{{\it RXTE}}
\newcommand{\sax}{{\it Beppo-SAX}}
\newcommand{\suz}{{\it Suzaku}}
\newcommand{\veritas}{{\it Veritas}}
\newcommand{\xmm}{{\it XMM-Newton}}
\newcommand{\spex}{{\it SPEX}}

\newcommand{\aap}{{A\&A}}
\newcommand{\aaps}{{A\&AS}}
\newcommand{\aapr}{{A\&A Rev.}}

\newcommand{\apj}{{ApJ}}
\newcommand{\aj}{{AJ}}
\newcommand{\apjl}{{ApJ}}
\newcommand{\apjs}{{ApJS}}
\newcommand{\araa}{{ARAA}}
\newcommand{\iaucirc}{{IAU Circ.}}
\newcommand{\mnras}{{MNRAS}}
\newcommand{\nar}{{NewAR}}       
\newcommand{\nat}{{Nat}}

\newcommand{\pasp}{{PASP}}
\newcommand{\pasj}{{PASJ}}
\newcommand{\prc}{{Phys. Rev. C}}
\newcommand{\physrep}{{PhysRep}}
\newcommand{\ssr}{{SSRv}}

\newcommand{\jgr}{{Journal of Geophysical Research}}

\newcommand{\gray}{{$\gamma$-ray}}
\newcommand{\grays}{{$\gamma$-rays}}
\newcommand{\msun}{M$_{\odot}$}
\newcommand{\kms}{{km\,$s^{-1}$}}
\renewcommand{\bar}{\overline}

\newcommand{\arcmin}{{$^{\prime}$}}
\newcommand{\arcsec}{{$^{\prime\prime}$}}
\renewcommand{\deg}{{$^\circ$}}

\newcommand{\net}{{$n_{\rm e}t$}}
\newcommand{\netunit}{{cm$^{-3}$s}}
\newcommand{\kte}{{$kT_{\rm e}$}}

\newcommand{\cc}{{cm$^{-3}$}}

\newcommand{\medfigure}{\columnwidth}

\newcommand{\bigfig}{0.95\textwidth}
\newcommand{\medfig}{\columnwidth}
\newcommand{\twofig}{0.45\textwidth}

\newcommand{\sect}{{Sect.}}

%% file: jvink_aarv_abstract.tex
\begin{abstract}
Supernova remnants are beautiful astronomical objects that are also of high
scientific interest, because they provide insights into supernova explosion mechanisms,
and because they are the likely sources of Galactic cosmic rays.
X-ray observations are an important means to study these objects.
And in particular the advances made in X-ray imaging spectroscopy over the last two decades
has greatly increased our knowledge about supernova remnants.
It has made it possible to map the products of fresh nucleosynthesis, and resulted in the identification of
regions near shock fronts that emit X-ray synchrotron radiation.
Since X-ray synchrotron radiation requires 10-100 TeV electrons, 
which lose their energies rapidly, the study of X-ray synchrotron radiation
has revealed those regions where active and rapid particle acceleration is taking place.

In this text all the relevant aspects of X-ray emission from supernova remnants
are reviewed and put into the context of supernova explosion properties
and the physics and evolution of supernova remnants.
The first half of this review has a more tutorial style and
discusses the basics of supernova remnant physics and
X-ray spectroscopy of the hot plasmas they contain. This includes 
hydrodynamics, shock heating, thermal conduction, radiation processes, 
non-equi\-li\-brium ionization, 
He-like ion triplet lines, and cosmic ray acceleration.
The second half offers a review of the advances
made in field of X-ray spectroscopy of supernova remnants during 
the last 15 year. This period coincides with the availability
of X-ray imaging spectrometers. In addition, I discuss the results of 
high resolution X-ray spectroscopy with the \chandra\ and \xmm\ gratings.
Although these instruments are not ideal for studying extended sources, they
nevertheless provided interesting results for a limited number of remnants.
These results provide a glimpse of what may be achieved with 
future microcalorimeters that will be available on board future X-ray observatories.

In discussing the results of the last fifteen years I have chosen to discuss
a few topics that are of particular interest. These include the
properties of Type Ia supernova remnants, which appear to be regularly shaped
and have stratified ejecta, in contrast  to core collapse supernova remnants,
which have patchy ejecta distributions. For core collapse supernova remnants I discuss
the spatial distribution of fresh nucleosynthesis products, but also their properties in
connection to the neutron stars they contain.

For the mature supernova remnants I focus on the prototypal  supernova
remnants Vela and the Cygnus Loop. And I discuss the
interesting class of mixed-mor\-pho\-lo\-gy remnants. Many of these mature supernova
remnants contain still plasma with enhanced ejecta abundances. Over the
last five years it has also become clear that many mixed-mor\-pho\-lo\-gy remnants 
contain plasma that is overionized. This is in contrast to most other supernova
remnants, which contain underionized plasmas.

This text ends with a review of X-ray synchrotron radiation from shock regions,
which has made it clear that some form of magnetic-field amplification is
operating near shocks, and is an indication of efficient cosmic-ray acceleration.
\end{abstract}

%% file: jvink_aarv_intro.tex
Supernovae play a central role in modern astrophysics. 
They are of prime importance for the chemical evolution
of the Universe, and they are one of the most important sources
of energy for the interstellar medium (ISM). 
Part of that energy is in the form of cosmic rays,
which have an energy density in the Galaxy of
1-2~eV cm$^{-3}$, thus constituting 
about one third of the total energy density of the 
ISM. Finally, supernovae, in particular Type Ia supernovae 
(see \sect~\ref{sec:supernovae}),
play a central role in present day  cosmology as
their brightness allows them to be detected at high redshifts
\citep[up to $z \sim 1.7$,][]{riess07}. Their use has led to the recognition
that the expansion of the Universe is accelerating instead of decelerating
\citep{perlmutter98,garnavich98}.

Supernova remnants (SNRs) are an important means to study supernovae.
Since supernovae are relatively rare (2-3 per century in a typical spiral
galaxy like our own), 
SNRs provide the best way to study the local population of supernovae. In addition, SNRs can reveal details about the explosions that are difficult to obtain
from studying supernovae directly. For example, young SNRs can inform us
about nucleosynthesis yields of individual SNRs and about 
the inherent asymmetries of the explosion itself, as revealed by
the spatial and velocity distribution of heavy elements. Moreover,
SNRs probe the immediate surroundings of supernovae, which are shaped by
their progenitors.

Another aspect of SNRs concerns the interesting physical process that shape their properties. 
SNR shocks provide
the best Galactic examples of high Mach number, collisionless shocks.
The physics of these shocks is not
well understood, as heating of the atoms occurs collisionless, i.e.
shock heating does not operate through
particle-particle (Coulomb) interactions, but 
through electro-magnetic fluctuations and plasma
waves. The SNR shocks are also thought to be the locations 
where part of the explosion energy is converted to
cosmic-ray energy. 
This is supported by observations that indicate that
GeV and TeV particles are present in SNRs through their
radiative signatures:
radio to X-ray synchrotron emission from relativistic electrons,
and \gray\ emission caused by accelerated electrons and ions.

\input{jvink_aarv_fig_tycho}

X-ray observations are one of the most important means to study the many
aspects of SNRs. X-ray spectroscopy is essential to obtain
abundances of the prime nucleosynthesis products of supernovae,
which are the so-called alpha-elements (O, Ne, Mg, Si, S, Ar, Ca)
and iron-group elements (chiefly 
Fe, Ni, and some trace elements with $20 <Z< 28$).
All these elements have prominent emission lines in the 0.5-10 keV band
for temperatures between 0.2-5 keV, which happens to be the
typical electron temperatures of plasma heated by SNR shocks.
The hot plasmas of SNRs are also optically thin, so inferring abundances
is relatively straightforward.
X-ray spectroscopy of SNRs,
therefore, provide us with a record of alpha- and iron-group
element production by supernovae.

X-ray spectroscopy can also be used to study several 
aspects of SNR shock physics.
First of all the line emission provides 
information about the temperature and ionization state of the plasma,
but the absence of lines or weak line emission in young SNRs usually points
toward the contribution of X-ray synchrotron radiation.
Studying X-ray synchrotron radiation offers a powerful diagnostic tool
to study electron cosmic-ray acceleration, 
as the presence of X-ray synchrotron radiation depends sensitively on the shock
acceleration properties. Moreover, the size of the X-ray synchrotron emitting regions
can be used to infer magnetic-field strengths.

One of the major advances in X-ray spectroscopy over the last 15-20 years 
has been the emergence of X-ray imaging spectroscopy. Although presently
the spectral
resolution offered is limited, it provides a direct
way to map the spatial distribution of elements and temperatures
in SNRs. For young SNRs it also helps to
separate thermal from non-thermal (synchrotron) X-ray emission.

The power of X-ray imaging spectroscopy is illustrated in Fig.~\ref{fig:tycho},
which shows Tycho's SNR (the remnant of the historical SN 1572) as observed
by \chandra. The image shows beautifully the distribution of silicon and iron,
but also reveals the synchrotron dominated emission from near the shock front.

Over the last ten years high spectral resolution 
X-ray spectroscopy has gained in importance, due to the availability
of grating spectrometers on board \chandra\ and \xmm.
These spectrometers are also used to study SNRs,
but the contribution of grating
spectrometers to our understanding of SNRs has been limited.
The reason is that the slitless grating spectrometers are not particularly well
suited
for extended sources such as SNRs. Nevertheless, some important
studies of SNRs with properties that make them suitable for 
grating spectroscopy have been published.
With the emergence of high spectral resolution imaging spectroscopy, as
will be possible with the calorimetric spectrometers on board 
\astroh\ \citep{takahashi08} and \ixo/\athena\ \citep{bookbinder10}, 
high resolution X-ray spectroscopy will become much more
prominent.

Here I review the current status of X-ray spectroscopy of SNRs. 
Since there is a lack of good text books or extended reviews on SNRs in
general\footnote{But see \citet{vink06b,reynolds08,badenes10} for recent, 
more topical reviews.}, 
the first part of this review has a tutorial character, describing 
our current understanding 
of the evolution and physics of SNRs. The second half of the review focusses
on the X-ray observations themselves,  with an emphasis on the achievements
obtained with data from the three most recent X-ray observatories:
\chandra, \xmm, and \suz.

%% file: jvink_aarv_fig_tycho.tex
\begin{figure}[b]
\centerline{
\includegraphics[trim=60 40 20 40,clip=true,width=\medfig]{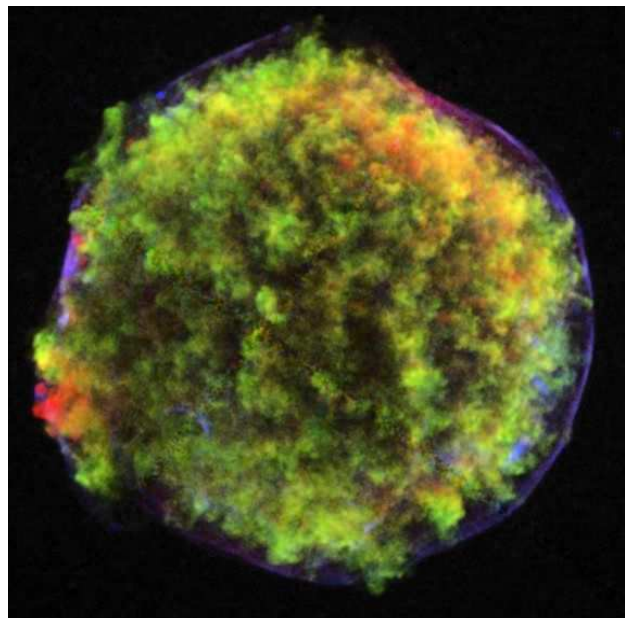}
}
\caption{
Three-color \cxo\ image of Tycho's SNR (SN1572/G120.1+1.4).
The color red shows Fe L-shell emission, green Si XIII, and blue
continuum (4-6 keV). Note the very narrow continuum rims near the shock front
(blueish/purple in this image),
caused by X-ray synchrotron radiation from electrons
with energies up to 100~TeV.
\citep[Image made by the author from \chandra\ data, see also][]{hwang02,warren05}.
\label{fig:tycho}
}
\end{figure}

%% file: jvink_aarv_supernovae.tex
\input{jvink_aarv_fig_sn_types}

Supernovae are divided into two broad categories, reflecting our
understanding of the explosion processes: core collapse supernovae
and thermonuclear supernovae.
In addition an observational classification scheme 
is used (Fig.~\ref{fig:sntypes}) 
that goes back to \citet{minkowski41}, who observed that some supernovae
do not show hydrogen absorption in their spectra (Type I) and others
do (Type II).

Type II supernovae are invariably core collapse supernovae, but
Type I supernovae can be either core collapse
or thermonuclear supernovae. The thermonuclear explosions are associated
with spectroscopic class Type Ia \citep{elias85},
which have Si absorption lines in their spectra. 
Type Ib and Ic supernovae are now understood to be explosions
of stars that have lost their hydrogen-rich envelope as a result
of stellar wind mass loss \citep[e.g.][]{heger03},
or through binary interaction \citep[e.g.][]{podsiadlowski92b}.
For Type Ic the mass loss seems to have removed even the helium-rich
layers of the progenitor. 
The fact that they are preferentially found in 
the most luminous regions of galaxies suggests that they are 
the explosions of the most massive stars \citep[e.g.][]{kelly08}.
Interestingly, also
long duration gamma-ray
bursts are associated with Type Ic supernovae \citep[e.g.][]{galama98,stanek03}.

The sub-division of the Type II class in Type IIP (plateau), 
Type IIL (linear light curve) and Type IIb
is based on a combination of two observational criteria: optical spectroscopy
and light-curve shape ( Fig.~\ref{fig:sntypes}).
Type IIP are the most common type of core collapse supernovae,
and optical studies of potential progenitor stars  confirm that their 
progenitors
have initial masses in the $\sim 8-17$~\msun\ range, and that they
explode in the red supergiant phase, while still having a substantial
hydrogen envelope \citep[see the discussions in][]{smartt09,chevalier05}.
Type IIL progenitors probably have a substantially less massive envelope, either
due to stellar wind mass loss, or due to binary interaction.
Type IIb supernovae are a class intermediate between Type Ib and Type II,
in that their spectra would initially identify them as
Type II explosions, but at late times their spectra evolve into
Type Ib spectra. Also this can be understood as the result
of substantial, but not complete,
removal of the hydrogen-rich envelope due to stellar wind mass loss,
or binary interaction.
The prototypal Type IIb supernova is SN 1993J 
\citep{podsiadlowski93,woosley94b}.
Interestingly,
recent identification and subsequent spectroscopy of the light echo
of the supernova that caused the extensively studied SNR Cassiopeia A (Cas A)
shows that it is the remnant
of a Type IIb supernova, as the spectrum shows both
hydrogen and weak helium line absorption \citep{krause08}.

Not listed in Fig.~\ref{fig:sntypes} is the Type IIn class.
Type IIn supernovae are characterized by narrow hydrogen emission
lines, which are thought to
come from a dense circumstellar environment, probably caused by
substantial mass lost by the progenitor.
Its place in the diagram is not quite clear, as
at least one Type IIn supernova, SN 2001ic, was observed to be
a Type Ia supernova whose spectrum 
subsequently evolved into a Type IIn supernova \citep{hamuy03}.

%% file: jvink_aarv_fig_sn_types.tex
\begin{figure}
\centerline{
\includegraphics[trim=20 100 50 20,clip=true,width=\medfig]{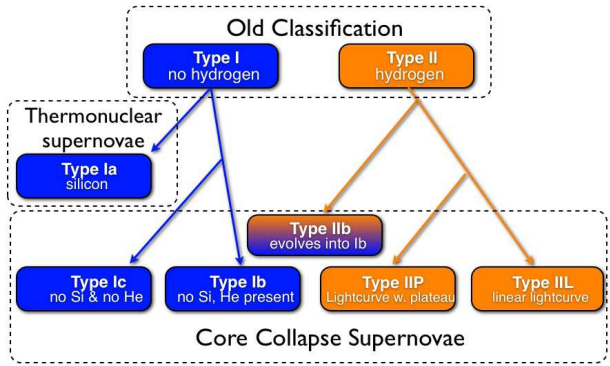}}
\caption{
The classification of supernovae, based on optical spectroscopy
and light-curve shape. 
\label{fig:sntypes}
}
\end{figure}

%% file: jvink_aarv_core_collapse.tex
\input{jvink_aarv_fig_yields}

\subsection{Core collapse supernovae}
Core collapse supernovae mark the end of the lives of
massive stars; that is, those stars with
main sequence masses $M \gtrsim 8$~\msun\ \citep[see][for a review]{woosley05}.
Just prior to collapse the star consists of different layers
with the products of the different consecutive burning stages. From the core
to the outside one expects:
iron-group elements in the core
(silicon-burning products),
then silicon-group elements (oxygen-burning products), 
oxygen (a neon burning product),
neon and magnesium (carbon-burning products), 
carbon (a helium-burning product), helium (a hydrogen-burning product),
and, finally, unprocessed hydrogen-rich material.

The creation of the iron-group core, which lasts about a day, is the beginning
of the end of the star, as no energy can be gained from nuclear fusion of
iron. The core collapses into a proto-neutron star, and for the most
massive stars into a black hole.
Most of the gravitational energy liberated 
($E\sim GM^2/R_{\rm ns} \sim 10^{53}$~erg, with $R_{ns}$ the neutron star radius)
is in the form of neutrinos.
This has been confirmed with the detection of neutrinos from SN1987A
by the
Kamiokande \citep{hirata87} and Irvine-Michigan-Brookhaven \citep{bionta87}
water Cherenkov neutrino detectors.

The supernova explosion mechanism itself, which requires that  $\gtrsim
10^{51}$~erg
of energy is deposited in the outer layers, is not well understood.
The formation of a proto-neutron star suddenly terminates
the collapse, and drives a shock wave through
the infalling material. However, numerical simulations show that the shock
wave stalls. It is thought that the shock wave may be energized by
the absorption of a fraction of the neutrinos escaping the
proto-neutron star, but
most numerical
models involving neutrino absorption are still unsuccessful in
reproducing a supernova explosion \citep{janka07}.
The most recent research, therefore, focusses on the role of accretion
instabilities for the supernova explosion, as these instabilities
help to enhance the neutrino
absorption in certain regions just outside the proto-neutron star.
A promising instability is the so called, 
{\em non-spherically symmetric
standing accretion shock instability} \citep[SASI,][]{blondin03}, which may
also help to explain pulsar kicks and rotation \citep{blondin07}.
Alternatively, simulations by \citet{burrows07} suggest that acoustic
power, generated in the proto-neutron star due to g-mode oscillations, 
ultimately
leads to a successful explosion \citep[but see][]{weinberg08}.
Finally, there have been suggestions that neutrino
deposition is not the most important ingredient for a successful explosion,
but that amplification of the stellar magnetic field, due to differential
rotation and compression, may lead to magneto-centrifugal
jet formation, which drives the explosion \citep{wheeler02}.

SASI, acoustic power, and magneto-centrifugal 
models all predict deviations from spherical symmetry.
In the magneto-rotational models one even
expects a bipolar symmetry.
Another reason why deviations from spherical symmetry has received more
attention is that long duration gamma-ray bursts are associated
with very energetic Type Ic supernovae (hypernovae).
Given the nature of gamma-ray burst these explosions are
likely jet driven. This raises the possibility that also
more normal core collapse supernovae have jet components \citep{wheeler02}.
There is indeed evidence, based on optical polarimetry, 
that core collapse supernovae, especially Type Ib/c, are aspherical
\citep[e.g.][]{wang01}. 
In \sect~\ref{sec:orich} I will discuss evidence that also the remnants
of core collapse supernovae show deviations from spherical
symmetry.

The ejecta of core collapse supernovae consist primarily of stellar material,
except for the innermost ejecta, which consist of explosive nucleosynthesis
products, mostly Fe and Si-group elements.
These elements are synthesized from protons and alpha-particles,
which are the remains of the heavy elements that have disintegrated 
in the intense
heat in the the innermost regions surrounding to the collapsing
core \citep{arnett96}.
Some of the explosive nucleosynthesis products
are radioactive, such as $^{56}$Ni, and $^{44}$Ti. 
In particular the energy generated by the decay of $^{56}$Ni into $^{56}$Co, 
and finally $^{56}$Fe, heats the ejecta, which leaves a major imprint
on the evolution of the supernova light curve.
The yields of these elements depend sensitively 
on the details of the explosion, such as the mass cut (the boundary between
material that accretes on the neutron star and material that is ejected),
explosion energy, and explosion asymmetry. 
Since 
the mass of the neutron star/black hole,
the location of energy deposition and the presence of asymmetries are not well
constrained, the expected yields of these elements 
are uncertain, and vary substantially from one set of models to
the other \citep[Fig.~\ref{fig:yields},][]{ww95,thielemann96,chieffi04}. 

Overall the yields of core collapse supernovae are dominated by 
carbon, oxygen,  neon and magnesium, which are products of the various stellar
burning phases
\citep[e.g.][]{ww95,thielemann96,chieffi04}. 
These yields are a function of the initial  mass
of the progenitor (Fig.~\ref{fig:yields}). 
It is for this reason that oxygen-rich SNRs
(\sect~\ref{sec:orich}) 
are considered to be the remnants of the most massive stars.

%% file: jvink_aarv_fig_yields.tex
\begin{figure*}
\centerline{  
\includegraphics[angle=-90, width=\twofig]{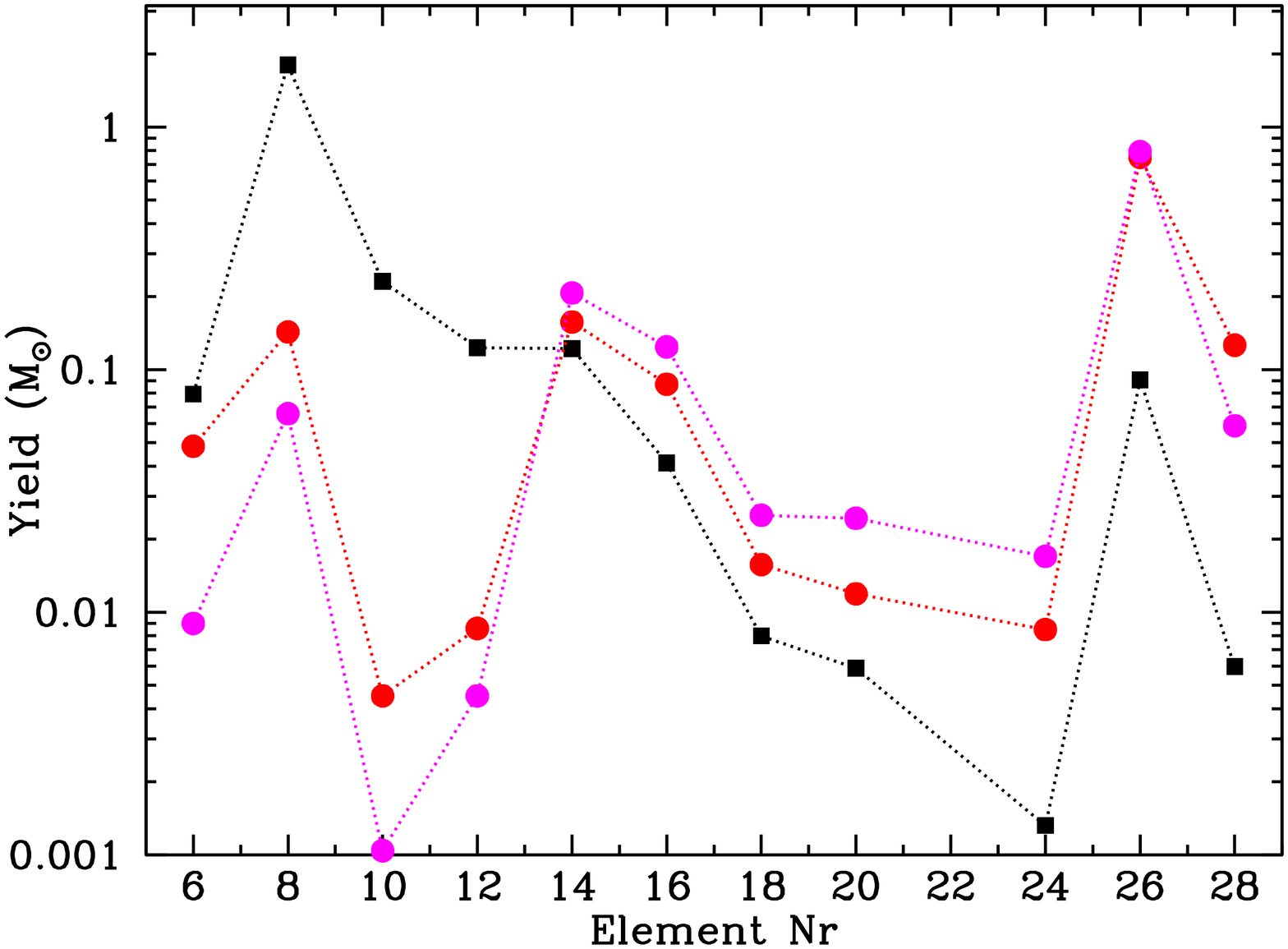}
\includegraphics[angle=-90, width=\twofig]{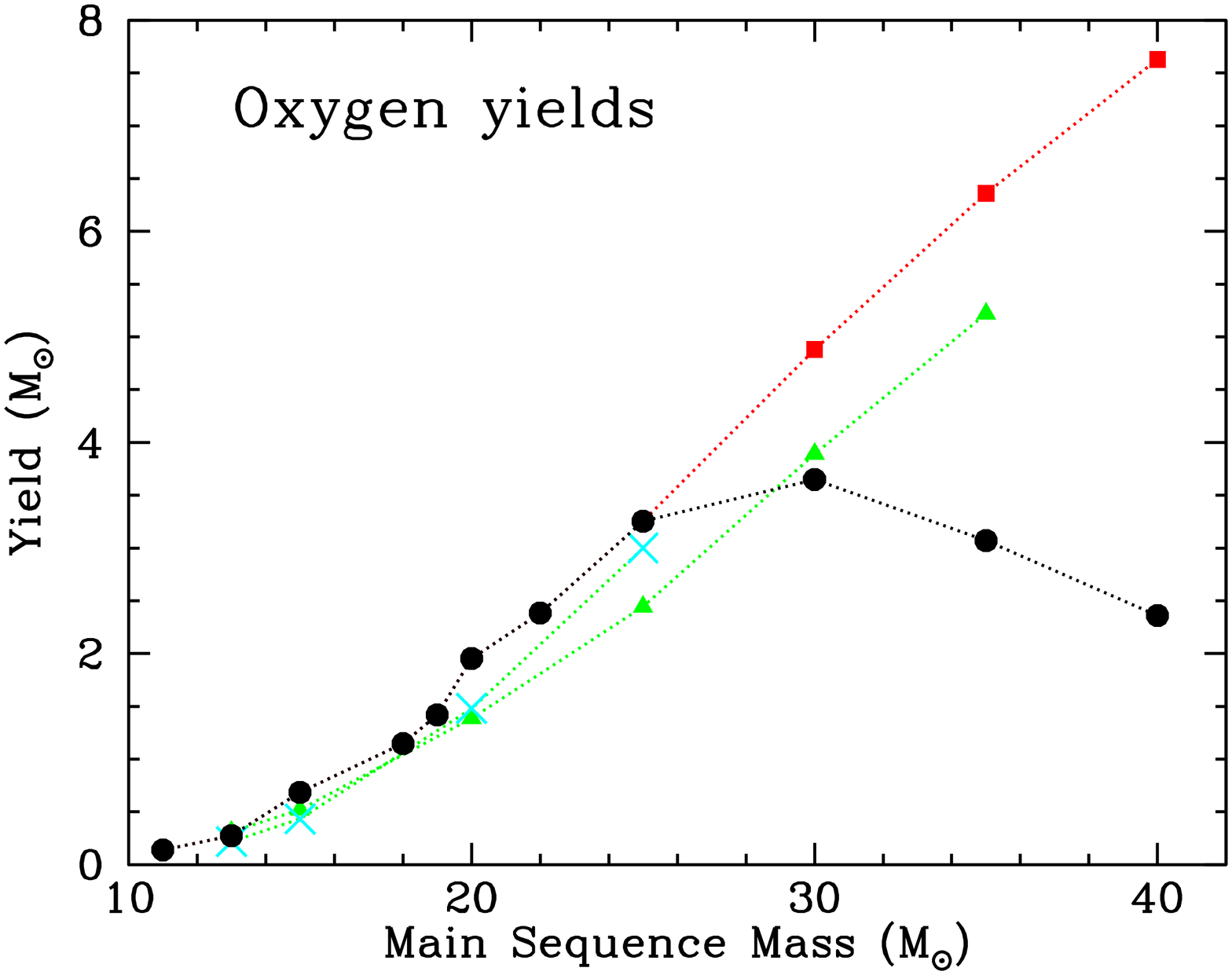}
}
\caption{
Left: Supernova yields for the most abundant X-ray emitting elements.
The squares/black line indicates the mean yield for core collapse supernovae,
whereas the circles indicate thermonuclear supernovae 
(the W7 deflagration model in red 
and the WDD2 delayed detonation model in magenta).
The model yields were taken from \citet{iwamoto99}.
Right:
Oxygen yield of core collapse supernovae 
as a function of main sequence mass.
The circles and squares are the predictions of \cite{ww95}, 
the triangles are predictions of \cite{chieffi04}, 
and the crosses of \cite{thielemann96}.
In general the oxygen yields obtained by the various groups
are very similar, but above
30~\msun\, one sees that certain models \citep[][using $10^{51}$~erg explosion
energies]{ww95} predict a diminishing
oxygen yield. The reason is that above 30~\msun\ stellar cores
may collapse into black holes, and part of the oxygen
falls onto the black hole. The amount of fall-back
is governed by the explosion energy and the amount of pre-supernova
mass loss, but it is also sensitive to
the numerical treatment of the explosion.
}
\label{fig:yields}       
\end{figure*}

%% file: jvink_aarv_thermonuclear.tex
\subsection{Thermonuclear supernovae}
\label{sec:sn1a}
Type Ia supernovae 
are generally thought to be thermonuclear
explosions of C/O white dwarfs, i.e. the explosion energy
originates from explosive nuclear burning, rather than from
gravitational energy liberated during the collapse of a stellar core.

Although there is some variation in peak brightness 
of Type Ia supernovae, the variation is much less than that of Type II
supernovae. This is in line with the idea that all Type Ia supernovae
are explosions of similar objects: C/O white dwarfs with masses
close to the Chandrasekhar limit ($1.38$~\msun).
Moreover, an empirical relation exists between their peak brightness 
and the post peak decline rate of the light curve \citep{phillips92},
which can be used to calibrate the absolute peak brightness of each event. 
This makes
Type Ia supernovae
excellent distance indicators, on which much of the evidence rests
that the expansion of the
Universe is accelerating \citep{perlmutter98,garnavich98,riess07}.

There is no direct observational
evidence that Type Ia progenitors are white dwarfs,
but the fact that only Type Ia supernovae can occur among old stellar 
populations
indicates that massive stars cannot be their progenitors.
Their relative uniformity can be best explained by assuming
a progenitor type with a narrow mass range \citep{mazzali07}.
Moreover, C/O white dwarfs close to the Chandrasekhar mass limit are
very likely
Type Ia progenitors,
since their high density makes for an ideal 
``nuclear fusion bomb'' \citep{arnett96}:
Once a nuclear reaction in the core is triggered, 
it will result in an explosion.
The trigger mechanism itself is, however, not well understood.

A more serious problem is that we do not know
what the most likely progenitor systems are. 
It is clear that C/O white dwarfs have to accrete matter in order 
to reach the Chandrasekhar limit, so thermonuclear supernovae 
must occur in a binary system. However,
this still leaves several possibilities for the progenitor systems: 
double degenerate systems (two white dwarfs),  
or white dwarfs with either 
a main sequence star, or an evolved companion (i.e. single degenerate). 
Only a limited range of mass transfer rates, $\sim 4\times 10^{-8}- 7\times 10^{-7}$~\msun\,yr$^{-1}$,
lead to stable growth of the white dwarf \citep{nomoto82,shen07}.
Novae, for instance, are relatively slow accretors, which are
thought to blow off more mass than they accrete. 
The only stable white dwarf accretors seem to be supersoft
sources\footnote{Soft X-ray point sources that are thought to
be white dwarfs that stably accrete material from a companion \citep{kahabka97}.},
but their population seems too small to account for the 
observed supernova rate
\citep[e.g.][]{ruiter09}.
Depending on the progenitor evolution and accretion scenarios, the progenitor
binary system may affect its circumstellar medium. For example,
in the case of wind accretion, not all the mass lost from the donor
will end up on the white dwarf. \citet{hachisu96,hachisu99} proposed
a scenario in which Roche-lobe overflow from the donor star is stabilized
through a fast, optically thick, wind from the white dwarf, which will also affect the immediate
surroundings of the SNR.

For some time it was thought that Type Ia supernovae were associated with
population II stars, in which case they must evolve
on time scales  $\gtrsim 10^9$~yr,
but it has recently been established that also a short ``channel'' exists, 
with an evolutionary time scale
of $\sim 10^8$~yr \citep[e.g.][]{mannucci06,auburg08}.

Models for thermonuclear explosions come in three classes: detonation,
deflagration, and delayed detonation models.
In detonation models the explosive nucleosynthesis occurs due the 
compression and heating of
the plasma by a shock wave moving through the star. 
In deflagration models the burning front
proceeds slower than the local sound speed. 
The nuclear fusion in the burning front is sustained
by convective motions that mixes unburnt material into the hot burning zone.
The classical deflagration model is the W7 of \citet{nomoto84}.

Pure detonation models predict that almost all of the white dwarf matter 
will be transformed into
iron-group elements, whereas optical spectroscopy of Type Ia supernovae 
show that the ejecta contain significant amount of
intermediate mass elements \citep{branch82}. 
On the other hand, the pure deflagration models
overpredict the production of $^{54}$Fe with respect to  
$^{56}$Fe and predict  too narrow a
velocity range for the intermediate mass elements compared to observations.
For these reasons, the currently most popular model for Type Ia supernovae are 
the delayed detonation
(DDT) models \citep{khokhlov91}, in which the explosion starts as a 
deflagration, but changes
to a detonation wave burning the remainder of the white dwarf into 
intermediate mass elements (IMEs), such as silicon. 
There is observational evidence
that the fraction of C/O that is burned is roughly constant $M=1.1$~\msun,
but that the variation in peak brightness is either caused by the
ratio of iron-group elements over IME products \citep{mazzali07}, or by
the ratio of stable iron over $^{56}$Ni \citep{woosley07}.
$^{56}$Ni decays into $^{56}$Fe, and the heat that this generates determines
the brightness of a Type Ia supernova (see \sect~\ref{sec:radioactivity}).

Figure~\ref{fig:yields} shows the predicted 
overall abundance pattern of deflagration and DDT models. 
Compared to core collapse supernovae,
thermonuclear supernovae produce much more iron-group elements 
($\sim 0.6$~\msun). The explosion energy, $E$, of thermonuclear explosions
is determined by the mass of their burning products \citep{woosley07}:
\begin{equation}
E_{51} = 1.56 M_{\rm Ni} + 1.74 M_{\rm Fe} + 1.24 M_{\rm IME} - 0.46,
\end{equation}
with the $E_{51}$ the final kinetic 
energy in units of $10^{51}$~erg, and the masses of
stable Fe, $^{56}$Ni and IME in solar units.

%% file: jvink_aarv_snr_classes.tex
Given the fact that supernovae can be broadly classified in core collapse and 
thermonuclear/Type Ia supernovae, one would hope that SNRs would be classified 
as core collapse or Type Ia SNRs.
Although it is possible now to determine the supernova origin of a given SNR 
using X-ray spectroscopy (see \sect~\ref{sec:spectra} and \ref{sec:typeia}), 
for many old SNRs, whose emission is mainly coming from swept-up gas,
one has to rely on secondary indicators for determining their
supernova origin.
The most reliable indicator is of course the presence of a neutron star inside 
the SNR, which makes
it clear that the SNR must have a core collapse origin. But even here one has 
to be aware of chance alignments \citep{kaspi98}, 
in particular for SNRs with large angular diameters.
A  secondary indicator for a SNR to originate from a core collapse event is 
whether  a SNR is located in a star forming region or inside
an OB association \citep[e.g][]{westerlund69}. But this does
not constitute proof for such an origin. In contrast, a position of a SNR high above
the Galactic plane can be taken as supporting evidence for a Type Ia origin. Such is the
case for, for example, SN\,1006 \citep{stephenson02}.

\input{jvink_aarv_fig_snr_types}

Because the supernova origin of SNRs is often difficult to establish,
SNRs have a classification of their own, which is mostly based on their 
morphology.  Traditionally, this classification recognized three
classes: 
{\em shell type} SNRs,
{\em plerions}, and  {\em composite SNRs}.
As the blast wave sweeps through the interstellar medium 
(\sect~\ref{sec:hydro} ) a shell
of shock heated plasma is created. 
Therefore, in many cases the morphology of a SNR is characterized
by a limb brightened shell, which classifies the SNR as a shell-type SNR.

However, in case of a core collapse one may expect the
presence of a rapidly rotating neutron star. It
loses energy with a rate of
$\dot{E} = I\Omega \dot{\Omega} = 4\pi^2 I \dot{P}/P^3$,
with $\Omega$, the angular frequency, $P$ the rotational period and $I\approx 10^{45}$ ~g\,cm$^{2}$ 
the moment of inertia.
This energy loss produces a wind of relativistic electrons and positrons, 
which terminates in a shock, where the electrons/positrons are accelerated to 
ultra-relativistic energies. These particles advect and diffuse away from the 
shock creating a nebula of relativistic electron/positrons which emit 
synchrotron radiation from the radio to the soft \gray\ bands, and 
inverse Compton scattering
from soft \gray\ to the TeV band \citep[][for a review]{gaensler06}. 
Such a nebula is aptly named a {\em pulsar wind nebula}. The most famous 
pulsar wind nebula is the Crab Nebula (also known as M1 or Taurus A), which 
is powered by the pulsar B0531+21. This nebula is associated with the 
historical supernova of 1054 \citep{stephenson02}. As a result the 
Crab Nebula, but also similar objects like 3C58, have the status of a SNR. 
Since the nebula has a morphology that is bright in the center, and does not 
show a shell, they are called {\em filled center} SNRs or
{\em plerions}, with the name plerion derived from the Greek word for "full", 
{\em pleres} \citep{weiler78}.

It is arbitrary whether one should call the Crab-nebula and related objects 
plerions or pulsar wind nebula. The radio and X-ray emission from plerions is 
powered by the pulsar wind, 
not by the supernova explosion, and as such the nebula should
not really be referred to as a SNR. To complicate matters, 
the Crab nebula does show optical line emission from supernova ejecta, 
which could be referred to as a SNR.
Pulsar wind nebulae can be found around both young and old pulsars,
even a few millisecond pulsars have small nebulae around them
\citep{kargaltsev08}.
Since these older pulsars do not have a
connection with recent supernova activity, one should not 
call them plerions. So the name pulsar wind nebula is more generic  
and informative than plerion, and is, therefore, preferable.

Energetic pulsars with ages less than $\sim 20,000$~yr 
are expected to have blown  a pulsar nebula
while they are still surrounded by the SNR shell. One expects then a radio and 
X-ray morphology that consists of a pulsar wind nebula surrounded by a shell.
Indeed a few SNRs have this characteristic (Fig.~\ref{fig:morphology}c)
and are classified as {\em composite SNRs}.
In fact, it is still puzzling why a young 
object like the Crab Nebula 
does not show a SNR shell \citep[see][for a discussion]{hester08}.

Since the 1980s the SNR classification has become broader.
Due to the imaging capabilities of X-ray satellites like 
\einstein\ and \rosat\ it became clear that
there are many SNRs that display a shell-type morphology
in radio emission, but whose X-ray emission mainly
comes from the center of the SNR
\citep{white91,rho98}, as illustrated in
Fig.~\ref{fig:morphology}d. In general, these SNRs are relatively old,
and are associated with dense interstellar medium.
The X-ray emission from the center of these SNRs was 
not powered by a pulsar, but consisted of thermal emission from a 
hot plasma.
These SNRs are referred to
as {\em mixed-morphology SNRs} \citep{rho98} or as {\em thermal-composite SNRs} \citep{shelton99}.
I will discuss their properties in \sect~\ref{sec:mixed}.

X-ray spectroscopy has greatly enhanced our ability to derive the origin of SNRs
 (\sect~\ref{sec:spectra} and \ref{sec:typeia}). So one now also encounters 
classifications such as {\em Type Ia SNR}. 
Other SNRs show optical
and X-ray evidence for enhanced oxygen abundances;
these so-called {\em oxygen-rich SNRs}  (\sect~\ref{sec:orich})
are likely the remnants of the most massive stars. 
Depending on the context these designations are often more helpful
than the morphological classifications. 

Note that these new classifications
are not mutually exclusive with the traditional morphological classifications.
For example  oxygen-rich SNRs such as G292.0+1.8 
and the Large Magellanic Cloud SNR B0540-69.3,
also harbor a pulsar wind nebulae. They can therefore be classified
both as composite SNRs and as oxygen-rich SNRs.

%% file: jvink_aarv_fig_snr_types.tex
\begin{figure*}
\centerline{
\includegraphics[width=0.47\textwidth]{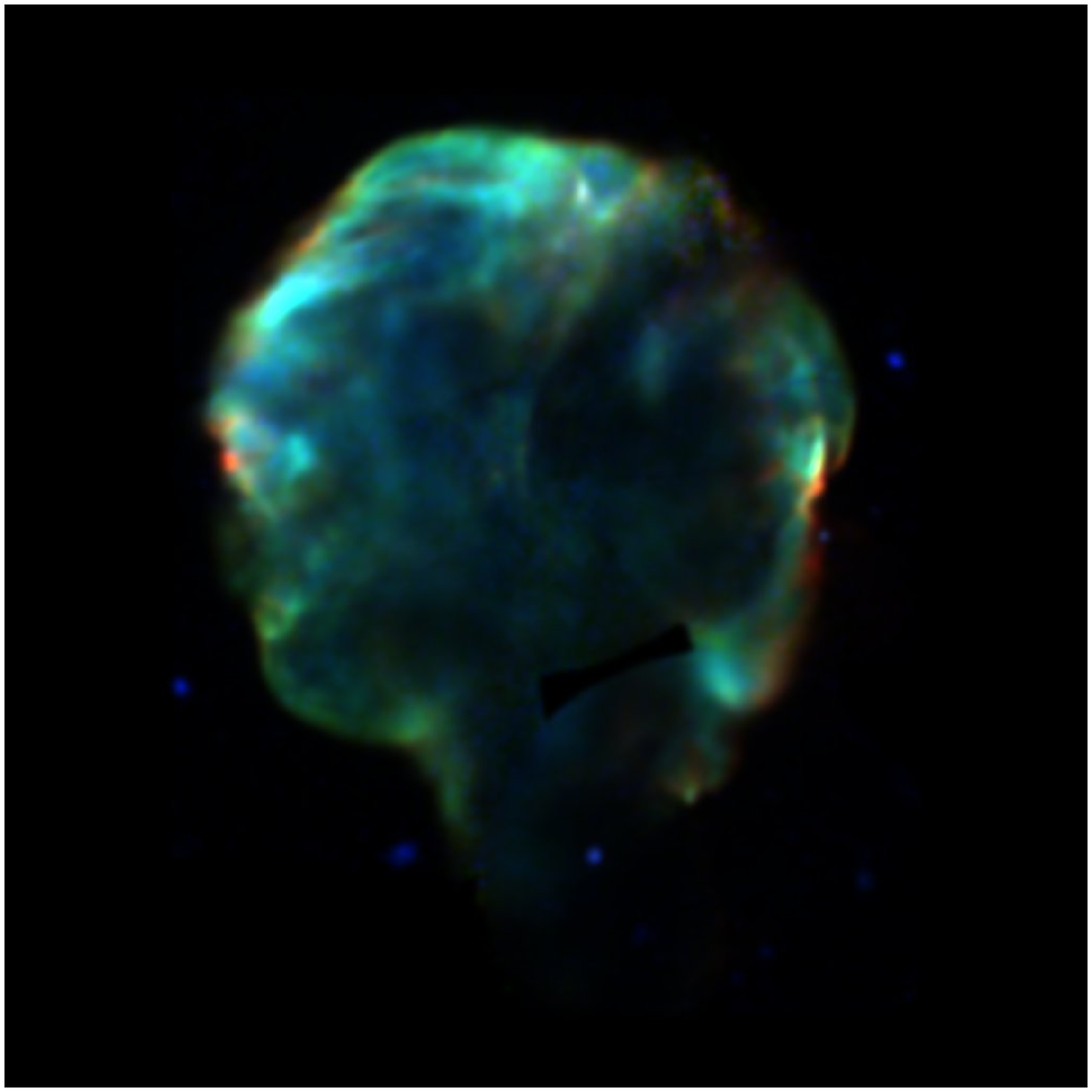}
\includegraphics[width=0.47\textwidth]{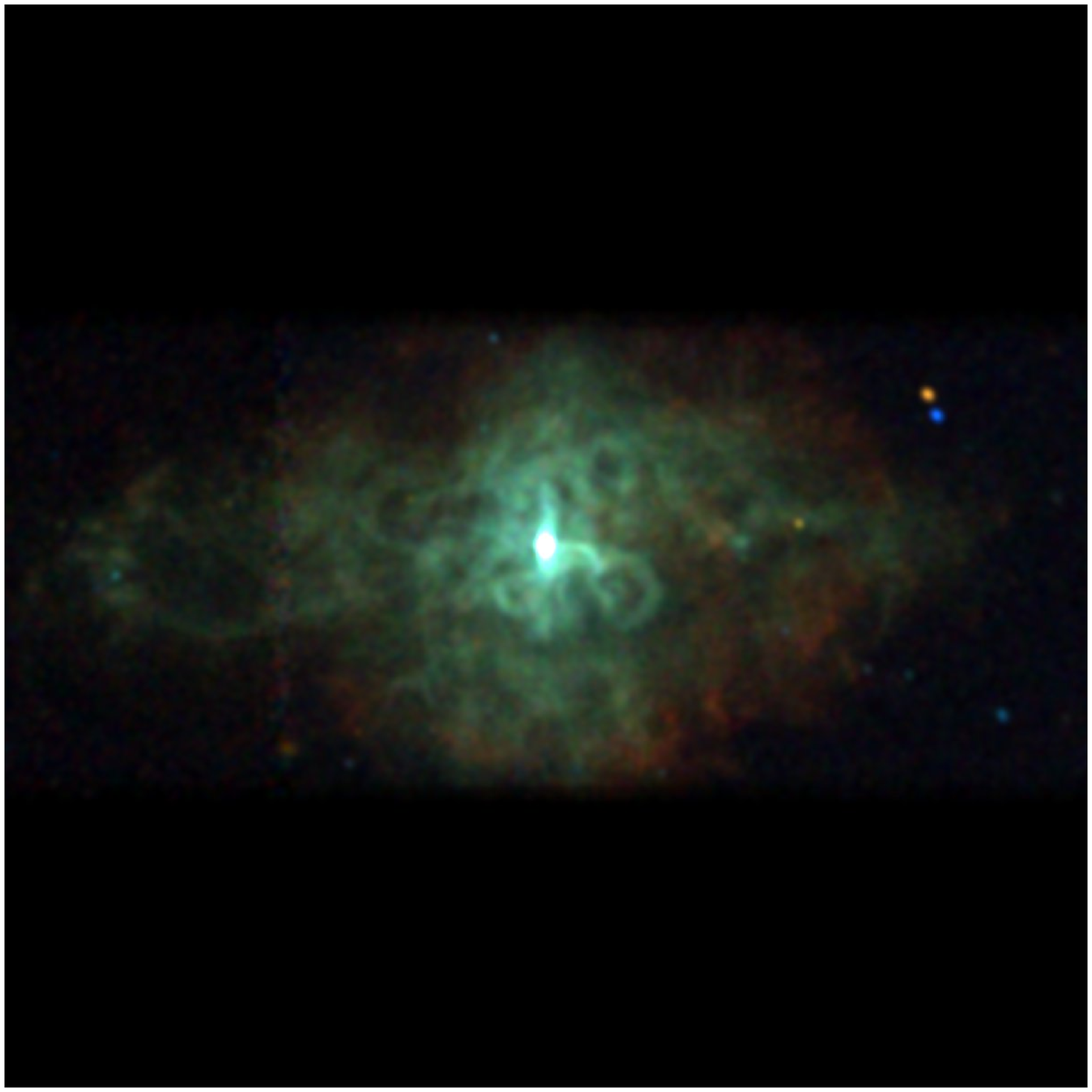}
}
\vskip0.7mm
\centerline{
\includegraphics[width=0.47\textwidth]{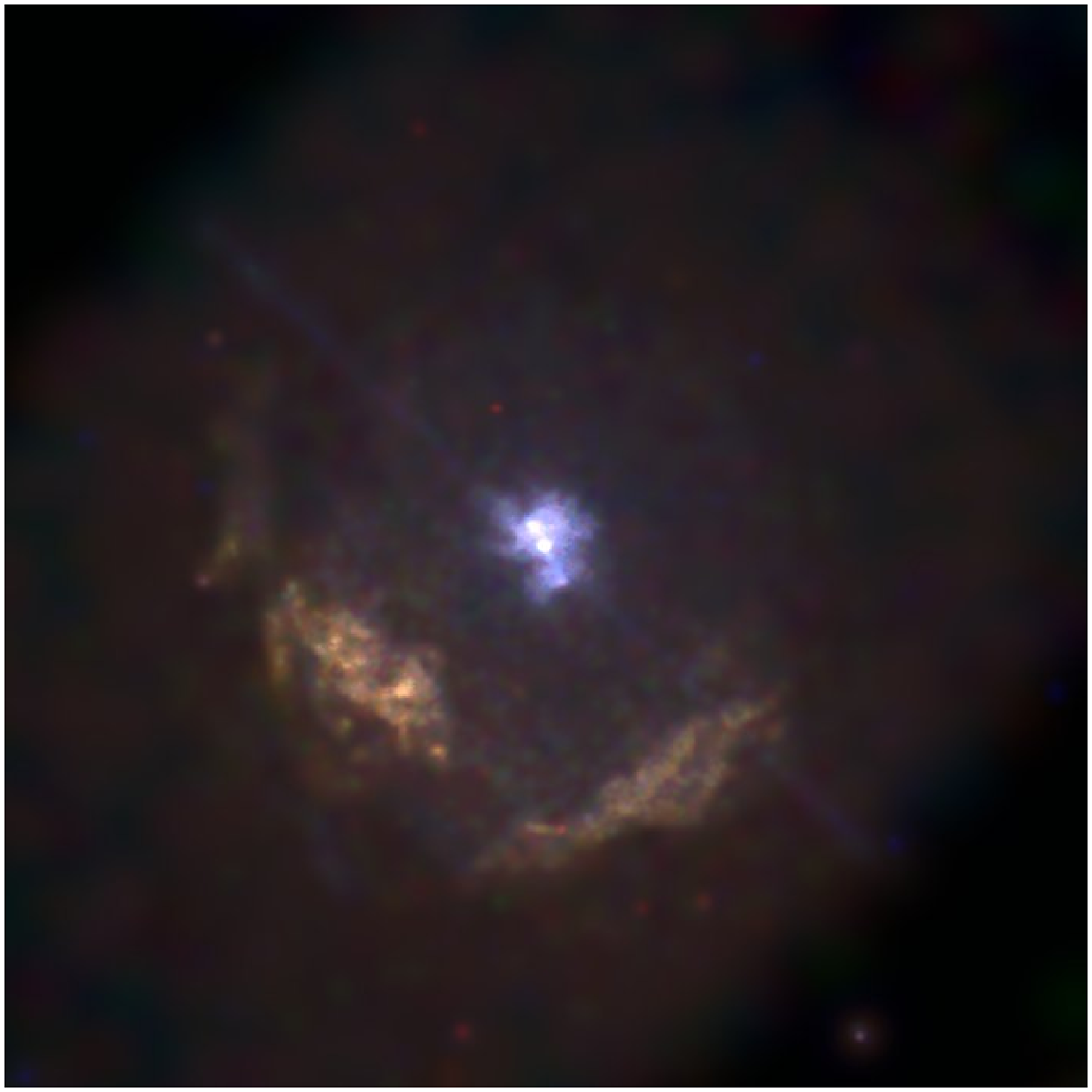}
\includegraphics[width=0.47\textwidth]{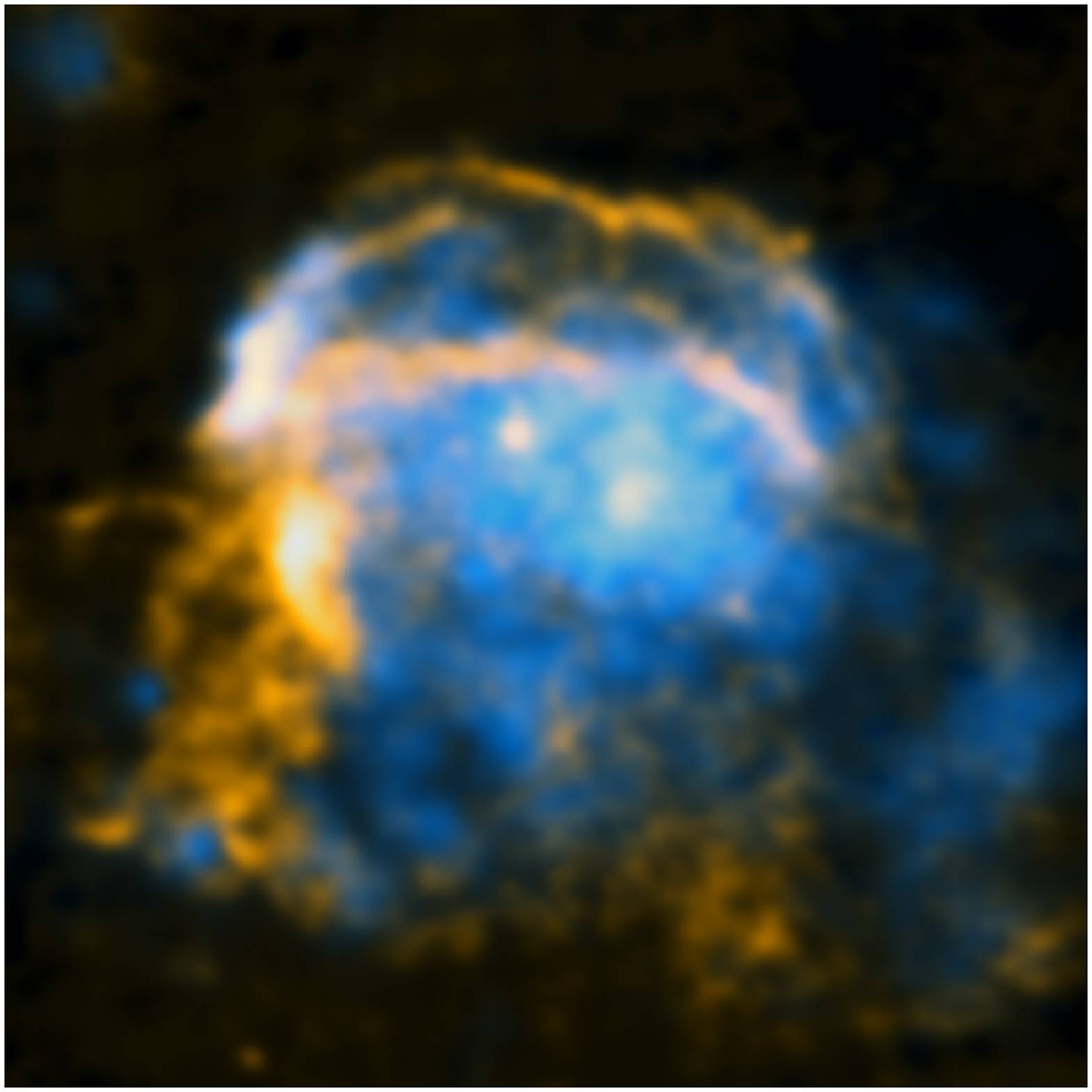}
}
\caption{
The SNR morphological classification illustrated with examples. 
From top left to bottom right:
a) The Cygnus Loop, a shell-type SNR with a diameter of 3\deg, 
as observed by the \rosat\ PSPC instrument
\citep{levenson98}, red is very soft emission from
$\sim$0.1-0.4 keV, green $\sim 0.5-1.2$~keV,
and blue $\sim 1.2-2.2$~keV.
b) 3C58, a plerion/pulsar wind nebula, as observed by
\chandra\ \citep{slane04}. The long axis of this object is $\sim$7\arcmin.
c) The composite SNR Kes 75 as observed by \chandra\ \citep{helfand03} 
with the inner pulsar wind nebula,
which has a hard X-ray spectrum, powered
by the pulsar J1846-0258. The partial shell has a radius of $\sim$1.4\arcmin.
The colors indicate 1-1.7 keV (red, Ne and Mg emission), 1.7-2.5 keV (green,
Si/S), and 2.5-5 keV (blue, mostly continuum emission from the pulsar wind nebula). 
d) 
The ``thermal-composite'' 
SNR W28 as observed in X-rays by the ROSAT PSPC (blue) and 
in radio by the VLA \citep{dubner00}. (Image credit: Chandra press office,
\url{http://chandra.harvard.edu/photo/2008/w28/more.html})
\label{fig:morphology}
}
\end{figure*}

%% file: jvink_aarv_hydro.tex
\subsection{Evolutionary phases}
\label{sec:phases}
The evolution of SNRs is usually divided in four phases \citep{woltjer72}: 
I) the ejecta dominated phase, in which the mass of the supernova ejecta, $M_{\rm ej}$, 
dominates over the swept-up mass, $M_{\rm sw}$; II) the Sedov-Taylor phase, for
which $M_{\rm sw} > M_{\rm ej}$, but for which radiative losses are not energetically 
important;
III) the pressure-driven, or ``snow-plough'' phase, 
in which radiative cooling has become energetically important;
the evolution of the shock radius is now best described by using momentum 
conservation;
 IV) the merging phase, in which the shock velocity
and temperature behind the shock become comparable to, respectively,
the turbulent velocity and temperature of the interstellar medium.

Although these discrete phases provide a useful framework to think of
SNRs, it should be kept in mind that it is an oversimplification,
and the phase of an individual SNR is not always that easily labeled.
Moreover, different parts of a SNR may be
in different phases. For example, a remnant like RCW 86 has
radiative shocks  in the southwest (phase III), 
whereas in the northeast it has very fast, non-radiative, 
shocks (phase I). This
is the result of the  complexity of the medium it is involving in, which
is probably shaped by the stellar wind of the progenitor that created
a cavity, surrounded by a dense shell, with which part of the shock wave is 
interacting \citep[e.g.][]{rosado94,vink97}.

Phase I is sometimes called the free-expansion phase, and 
phase II the adiabatic phase. Both names are somewhat misleading. 
Free expansion suggests that the shock velocity is described by $V_{\rm s} = R_{\rm s}/t$,
with $R_{\rm s}$, the radius of the outer shock, and $t$ the age of the SNR.
However, as described below, even in phase I one has $V_{\rm s} < R_{\rm s}/t$. 
Since energy losses are according to standard
models  not important in both phase I and II, it is also misleading to
call exclusively phase II the adiabatic phase. Moreover, in the very early evolution
of SNRs (almost the supernova phase) SNRs may go through a short radiative loss
phase \citep[e.g.][]{truelove99,sorokina04}. Also escaping cosmic rays
may add to the energy losses in phase I and II (see \sect~\ref{sec:kT}).
Despite these shortcomings, the labels have proven to be of some value, as they
provide some framework to characterize the evolutionary phase of a given SNR.

In the literature one also 
often finds designations for SNRs like ``young'', ``mature'' and ``old'' \citep{jones98}. 
These designations do not have a very precise meaning,
but a general guideline is that young SNRs are less than $\sim 1000-2000$~yr 
old, and are in phase I or early in phase II, mature SNRs are in late
phase II, or early phase III, whereas the label old SNRs is usually given to
the very extended structures associated with SNRs in phase IV. These
old SNRs hardly produce X-ray emissions, so in this review we will only
encounter young and mature SNRs.

\subsection{Analytical models}
\label{sec:hydro_analytical}
Several analytical models for the shock evolution of SNRs exist.
The most widely used is the Sedov-Taylor self-similar solution 
\citep{sedov59,taylor50}. 
It assumes that the explosion energy $E$ is instantaneously 
injected
into a uniform medium with uniform density $\rho_0$
(i.e. a point explosion),
and that no energy losses occur. In that case the shock radius $R_{\rm s}$
and velocity $V_{\rm s}$ will develop as
\begin{equation}
R_{\rm s} =  \Bigl( \xi \frac{E t^2}{\rho_0} \Bigr)^{1/5},\label{eq:sedovr}
\end{equation}
\begin{equation}
V_{\rm s} =  \frac{dR_{\rm s}}{dt} = 
\frac{2}{5}\Bigl( \xi \frac{E}{\rho_0} \Bigr)^{1/5}t^{-3/5} =  
\frac{2}{5}\frac{R_{\rm s}}{t}.\label{eq:sedovv}
\end{equation}
The dimensionless constant $\xi$ depends on the adiabatic index; $\xi = 2.026$
for a non-relativistic, monatomic gas ($\gamma = 5/3$).
An analytical solution exists for the density, pressure, 
and velocity profiles inside the shocked medium, which is shown in Fig.~\ref{fig:chevalier}.
The Sedov-Taylor solution can be generalized to a gas medium with a power-law 
density profile $\rho(r) \propto r^{-s}$:  $R_{\rm s} \propto t^\beta$,
$V_{\rm s} = \beta R_{\rm s}/t$, with $\beta = 2/(5-s)$, the expansion parameter.
An astrophysically relevant case is $s=2$, corresponding
to a SNR shock moving through the progenitor's stellar wind (see below); 
a situation that likely applies to
the young Galactic SNR Cas A \citep[e.g.][]{vanveelen09}.
This gives $\beta = 2/3$, which should be compared to
the experimental value found for Cas A based on the expansion
found in X-rays: $\beta =0.63 \pm 0.02$ 
\citep{vink98a,delaney03,patnaude09}.

The Sedov-Taylor solutions do not take into the account structure of the supernova
ejecta itself. This is a good approximation once the swept-up
mass exceeds the ejecta mass. 
However, in the early phase,
only the outer layers of the supernova transfer
their energy to the surrounding medium. As time progresses, more energy
is transferred from the freely expanding ejecta to the SNR shell. This 
takes place at a shock separating hot ejecta from freely
expanding (cold) ejecta, the so-called reverse shock \citep{mckee74}.

\input{jvink_aarv_fig_chevalier}

Two analytical models exist to describe the structure and evolution
of SNRs taking into account the initial velocity
structure of the ejecta. The first one, by \citet{chevalier82},
describes the early evolution of SNRs, in which the freely expanding
ejecta have approximately a  power-law density distribution $\rho_{ej} \propto v_{ej}^{-n}$.
This is a reasonable approximation for the outer ejecta structure 
as found in
numerical models of supernova explosions, with $n=7$ a situation that
describes reasonably well the ejecta structure of Type Ia supernovae,
whereas $n=9-12$ is a valid approximation for the density structure of 
core collapse supernovae. 

As shown by \citet{chevalier82} the SNR
evolution can be described by a self-similar solution of the form:
\begin{equation}
R_{\rm s} \propto t^\beta, 
\end{equation}
with $\beta$ the expansion parameter given by\footnote{The expansion parameter
is often, but not uniformly,
 indicated with the symbol $m$, but since the characters $i,...,n$ have an integer
 connotation, I opt here for $\beta$. Note that \citet{truelove99} use $\eta$ (but this
 conflicts with its use in acceleration theory, \sect~\ref{sec:cr}), and
 \citet{chevalier82} uses $1/\lambda$.
 }

\begin{equation}
\beta = \frac{n-3}{n-s}. \label{eq:chevalier}
\end{equation}
For $s=0, n=7$ this gives $\beta=0.57$, and for $n=9,s=2$ this gives
$\beta=0.86$.
The models describe a self-similar velocity, pressure, and
density structure. I will not reproduce them here, but show
as an example the density, pressure, and velocity structure of a $n=7, s=2$ model (Fig.~\ref{fig:chevalier}).
It important to realize that the Chevalier solutions describe the early evolution of SNRs, when
the reverse shock has not yet reached the inner most ejecta. Once the inner
ejecta are reached by the reverse shock the expansion parameter will
evolve toward the Sedov solution (thus the expansion parameter
evolves from $\beta=(n-3)/(n-s)$ to $\beta=2/(5-s)$.
It is clear from the Chevalier solutions that from a very early stage on $\beta < 1$.
There is also observational evidence for this. For example,
the initial expansion parameter of SN 1993J in the galaxy M81
has recently been determined to be $\beta=0.85$ \citep{marcaide09}; assuming an interaction of the shock with
a circumstellar wind ($s=2$) this implies $n\approx 8.5$.

\input{jvink_aarv_fig_truelove}
An analytical model that takes into account the smooth transition
from phase I to phase II was obtained by \citet{truelove99}.
Their models employ the following
characteristic length, time and mass scales:
\begin{align}
R_{ch} &\equiv M_{\rm ej}^{1/3}\rho_0^{-1/3}, \nonumber\\
t_{ch} &\equiv E^{-1/2}M_{\rm ej}^{5/6}\rho_0^{-1/3},\\ 
M_{ch} &\equiv M_{\rm ej}, \nonumber
\end{align}
with  $M_{\rm ej}$ the ejected mass, $E$ the explosion energy, and
$\rho_0$ the circumstellar medium density. These numbers can be used
to construct a set of solutions, which now only depend
on $n$ and $s$, and the dimensionless variables $R^* = R/R_{ch}$
and $t^* = t/t_{ch}$. The models are continuous, but consist of two parts;
one for the evolution in the ejecta dominated phase and one for
the Sedov-Taylor phase,
with $t_{ST}$ the dimensionless transition age.
For example, the blast wave trajectory
in the $n=7, s=0$ model is 
$R_{\rm s}^*1.06={t^*}^{4/7}$ for $t^*<t_{ST}$ and $R_{\rm s}^*=(1.42t^* - 0.312)^{2/5}$  for 
$t^*>t_{ST}$, with $t_{ST} = 0.732$. These solutions show that initially
the expansion parameter is identical to the one derived by 
\citet{chevalier82} (Eq.~\ref{eq:chevalier}), while it asymptotically 
approaches the Sedov-Taylor solution, $\beta=2/5$.

Fig.~\ref{fig:truelove} illustrates the evolution of SNR
shocks as given by  one of the Truelove
\& McKee solutions.
It shows that the reverse shock initially 
expands outward, despite its name, with $\beta=(n-3)/(n-s)$. But
as soon, as the shock heated shell has more pressure than the ram pressure of
the freely expanding ejecta, the reverse shock will move toward the center; i.e.
its velocity in the observers frame will become negative. For the
plasma temperature the shock velocity in the frame of the freely expanding
ejecta is important (\sect~\ref{sec:kT}). This is
is given by
\begin{equation}
V_{\rm rev,phys} =  \frac{R_{\rm rev}}{t} - V_{\rm rev,obs},
\end{equation}
with  $R_{rev}/t$ the velocity of the freely expanding ejecta just before
entering the reverse shock, and $V_{\rm rev,obs} = dR_{\rm rev}/dt$, the reverse
shock velocity in the frame of the observer.
Fig.~\ref{fig:truelove} shows that initially $|V_{\rm rev,phys}|< V_{\rm s}$,
in which case one expects the shocked ambient medium to be hotter
than the shocked ejecta. In later phases this situation reverses,
as $|V_{\rm rev,phys}|> V_{\rm s}$, resulting in a hot core enclosed by a cooler shell.
Once the reverse shock has reached the center of the SNR, it consists
of a hot shell surrounding an even hotter, but very tenuous interior.

It is worth mentioning that efficient cosmic-ray acceleration 
(\sect~\ref{sec:cr}) may
alter the hydrodynamics of SNRs, because it alters the equation of state, and
because escaping cosmic rays may result in energy losses. This can result
in a reverse shock radius at any given time that is closer to the outer shock
than indicated by models like that of Truelove \& McKee  
\citep{decourchelle00}. I will discuss in \sect~\ref{sec:efficient_crs} the observational evidence
for this.

In phase III, the  momentum conservation or ``snow-plough'' phase,
radiative energy losses have become dynamically important. Instead of
energy conservation, the radial expansion is governed by 
momentum conservation, $MV_{\rm s} = 4\pi/3\, R_{\rm s}^3\rho_0 \, dR_{\rm s}/dt = {
\rm constant}$. For the constant one can take the momentum
at the time $t_{\rm rad}$, for which
radiative losses have become imporant: 
$t_{\rm rad}=4\pi R_{\rm rad}^3/3\rho_0 V_{\rm rad}$.
Integration yields the following expression for the age, $t$, of the SNR
as a function of radius \citep[e.g.][]{toledo-roy09}:
\begin{equation}
t = t_{\rm rad} +
\frac{R_{\rm rad}}{4V_{\rm rad}}\Bigl[\Bigl(\frac{R}{R_{\rm rad}}\Bigr)^4 - 1\Bigl].
\label{eq:snowplow}
\end{equation}

As can be seen, the expansion parameter in this phase is $\beta \approx 1/4$.
Generally speaking, radiative losses become important when the post-shock
temperature falls below $\sim 5\times 10^5$~K, 
in which case oxygen line emission
becomes an important coolant \citep[e.g.][]{schure09}. 
This occurs when the shock velocity is
$V_{\rm s} = V_{\rm rad}=200$~\kms\ \citep[][see also \sect~\ref{sec:kT}]{woltjer72}. 
One can use the Sedov self-similar solution (Eq.~\ref{eq:sedovr} and  \ref{eq:sedovv} ) 
to calculate when $V_{\rm s}=V_{\rm rad}$,
thereby estimating $R_{\rm rad}$ and $t_{\rm rad}$.
Eq.~ \ref{eq:sedovv} leads to 
expression for $t_{\rm rad}$ in terms of $R_{\rm rad}$:
$t_{\rm rad} = R_{\rm rad}/V_{\rm rad} = 2.0\times 10^{-8} R_{\rm rad}$~s.
Eq.~\ref{eq:sedovr},
then provides the following estimate for the age and radius at 
which the SNR becomes radiative:
\begin{align}
\label{eq:trad}
t_{\rm rad} = & 1.5\times 10^{-13}\Bigl(\frac{\xi E}{\rho_0}\Bigr)^{1/3} =\\\nonumber
&1.4\times 10^{12} \Bigl(\frac{E_{51}}{n_{\rm H}}\Bigr)^{1/3}{\ \rm s} \approx 
44,600 \Bigl(\frac{E_{51}}{n_{\rm H}}\Bigr)^{1/3}{\ \rm yr} \\
R_{\rm rad} = & 7.0\times 10^{19} \Bigl(\frac{E_{51}}{n_{\rm H}}\Bigr)^{1/3} {\rm \ cm} \approx
23 \Bigl(\frac{E_{51}}{n_{\rm H}}\Bigr)^{1/3} {\rm pc}, \label{eq:rrad}
\end{align}
with $n_{\rm H}$ the pre-shock hydrogen density, $\xi=2.026$, and $E_{51}$ the explosion energy $E$
in units of $10^{51}$~erg.

\input{jvink_aarv_fig_wind_bubble}
\subsection{Supernova remnants evolving inside wind-blown bubbles}
\label{sec:csm}
 The structure of the circumstellar medium is usually not as simple
as used by the analytical solutions discussed above. The interstellar
medium itself is not homogeneous, but also stellar winds from the 
supernova progenitors impose a specific structure on the ambient media.
For example, a massive O-star will spend $\sim 90$\%
of its life on the main sequence, during which it blows a fast,
tenuous wind with velocity $v_w\approx 1500-3000$~\kms, with 
$\dot{M} \approx 5\times 10^{-7} - 10^{-5}$~\msun\,yr$^{-1}$\ 
\citep[e.g.][]{mokiem07}.
This wind carves  a low density region into the interstellar
medium, surrounded by a shock heated shell 
containing the swept-up interstellar medium and wind material.

In the subsequent red supergiant  (RSG) phase the mass loss rate 
increases to as much
as $\dot{M} \approx 10^{-4}$~\msun\,yr$^{-1}$ \citep[e.g.][]{vanloon05}, which, together with a much
slower wind velocity $v_w\sim 10$~\kms, results in a very dense wind, with
a density at radius $r$ given by
\begin{equation}
\rho_w (r)= \frac{\dot{M}}{4\pi r^2v_w},
\end{equation}
a relation that follows from mass conservation. 
Stars more massive than $60 $~\msun\ can even have extremely high mass
loss rates, often accompanied
by violent bursts of mass loss. These are so-called Luminous Blue Variables 
(LBVs).
An example is Eta Carinae \citep{davidson97}.

Most massive stars will explode in the RSG phase, but the highest mass stars
may enter the Wolf-Rayet star phase. In this phase the stars have lost all
of their hydrogen envelopes. Their mass loss rates are high,
$\dot{M} \approx 10^{-5}-10^{-4}$~\msun\,yr$^{-1}$, with a high wind velocity $1000-2000$~\kms\ 
\citep{nugis00}.
This results in a tenuous wind, which will sweep up the RSG wind into
a dense shell. It is not clear whether LBVs will become Wolf-Rayet stars, or
whether they are the last stage of the stellar evolution.

Progenitors that were part of a binary system may
even have had more complicated
mass loss histories. In some cases a common envelope
phase may have existed, which could also have
lead to a stellar merger. 
In these case the circumstellar medium structure may be rather complex.
See for example the binary evolution model that can explain some of the features of
the circumstellar medium of SN 1987A \citep[][\sect~\ref{sec:sn1987a}]{morris07}.

So core collapse supernovae are, even in the simplest cases, surrounded
by regions with different densities and temperatures, created and shock heated
during the various wind phases 
\citep[][Fig.~\ref{fig:windbubble}]{weaver77,dwarkadas05}.
As long as the SNR shock moves
through the unshocked stellar wind of the progenitor, the analytical
models are valid, but if this is not the case, then
the evolutionary path 
may differ quite substantially from the standard SNR evolutionary models
discussed above.

These more complicated situations have been studied with numerical 
simulations. See for example 
\citet{tenorio91}, \citet{dwarkadas05} and Fig.~\ref{fig:windbubble}.
In the case that the shell swept up by the wind has a mass that is higher than
the ejecta mass, the shock wave will first move through the
low density region without much deceleration ($\beta = 0.87$,
for an $n=9$ ejecta density profile, Eq.~\ref{eq:chevalier}).
In this phase the X-ray emission from the SNR is low, because of the low
density inside the stellar wind. Once
the shock reaches the dense shell, transmitted and 
reflected shocks develop. The transmitted shock heats  the shell and pushes it outward,
whereas the reflected shock further heats the interior of the SNR. In this phase,
the X-ray luminosity increases rapidly by a factor ten to hundred, due to
the high density in the shell \citep{dwarkadas05}.

%% file: jvink_aarv_fig_chevalier.tex
\begin{figure*}
\centerline{
\includegraphics[angle=-90,width=0.5\textwidth]{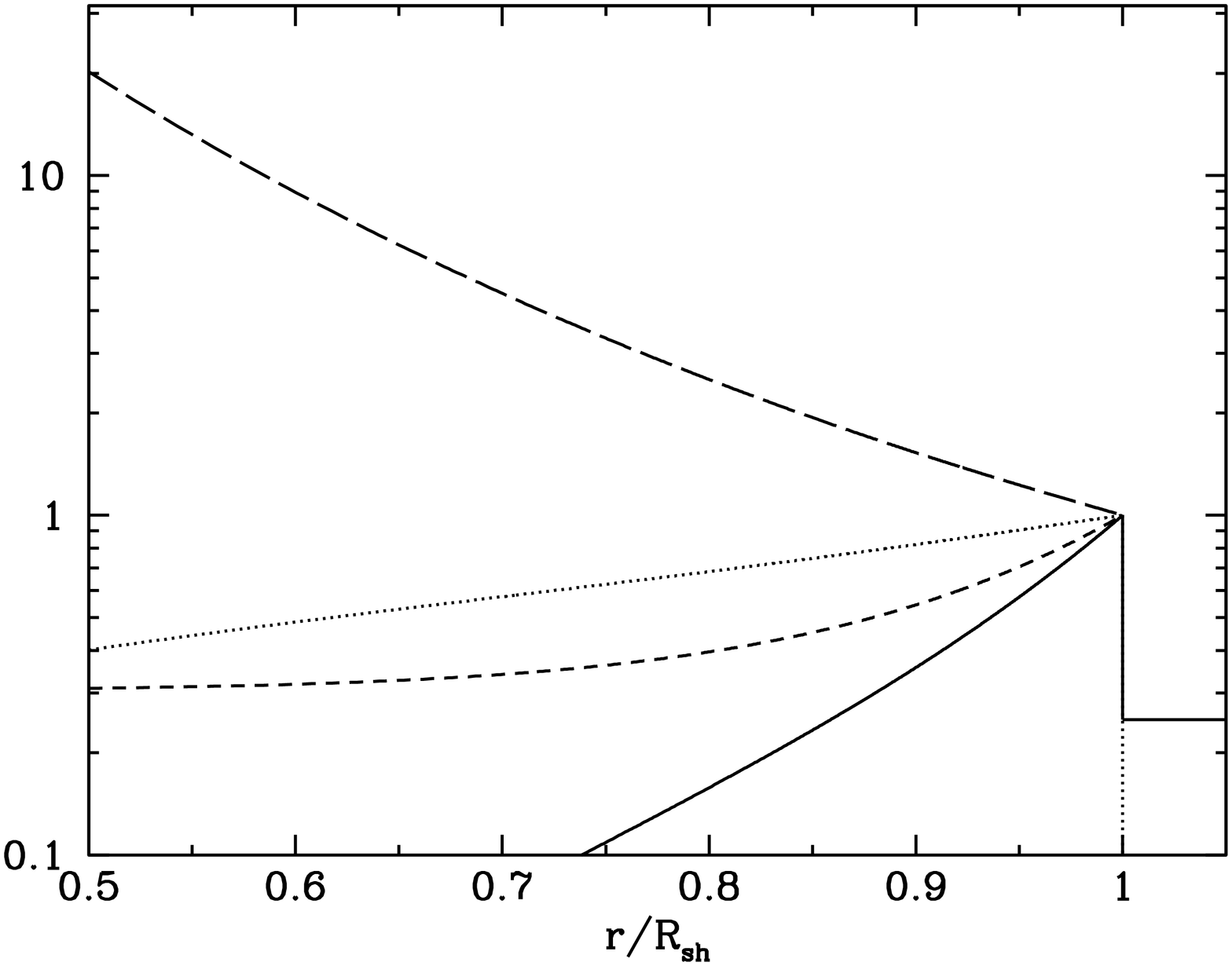}
\includegraphics[angle=-90,width=0.5\textwidth]{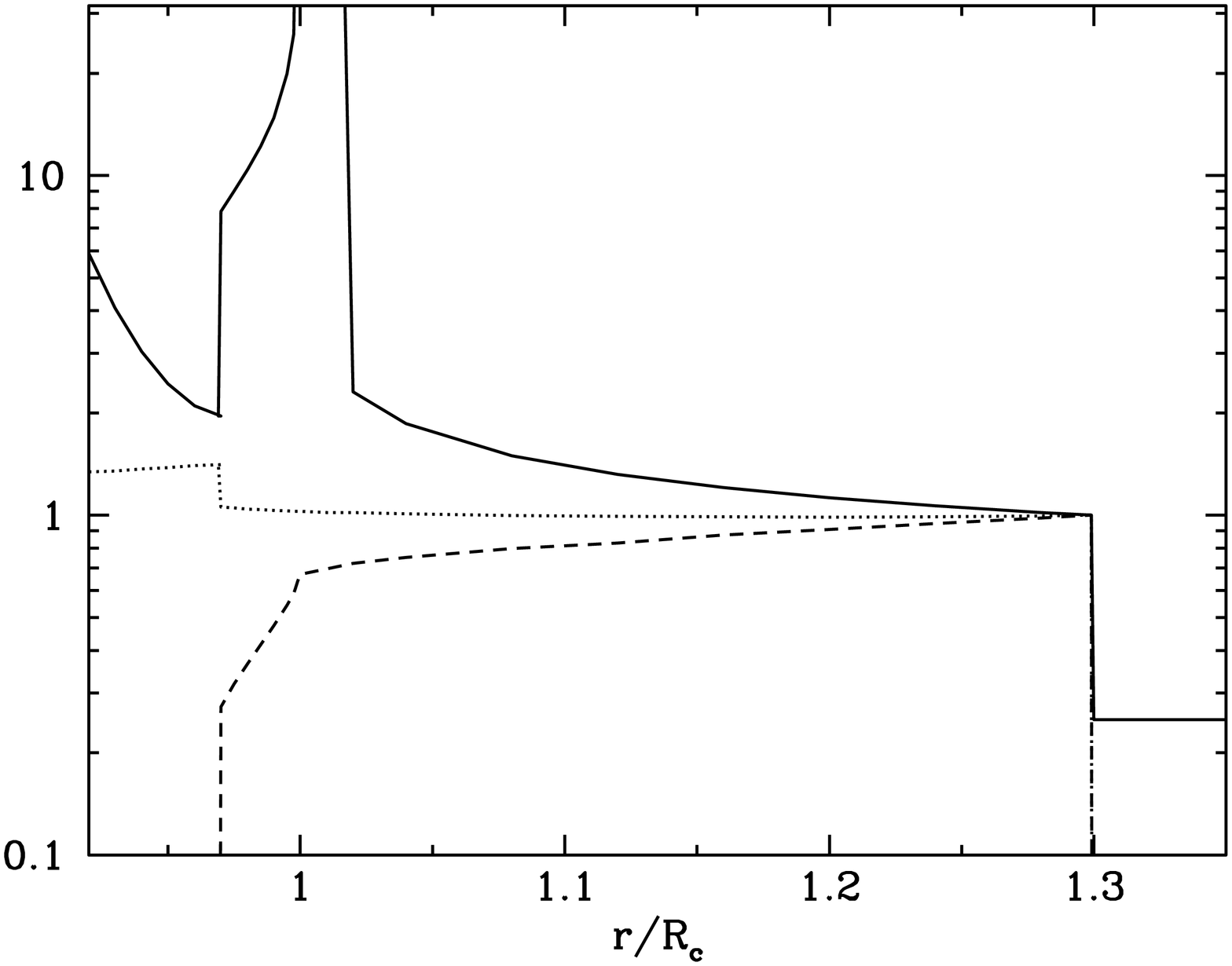}
}
\caption{
The structure of a SNR as given by the self-similar models of
\citet{sedov59} (left) and \citet{chevalier82} ($n=7, s=2$).
The values for the parameters have been
normalized to the values immediately behind the forward shock.
For the Sedov model the radius is expressed in units
of the shock radius, for the Chevalier model in units of
the contact discontinuity $R_c$, the border between swept-up and
ejecta material.
The solid lines show the density, the dotted lines the velocity,
and the short-dashed lines the pressure profiles.
For the Sedov model also the temperature is indicated (long-dashed line).
For this model the temperature goes to infinity
toward the center. For the Chevalier model the density goes to infinity
for $r=R_c$. Note that in the observer frame the velocity drops at
the reverse shock.
\label{fig:chevalier}
}
\end{figure*}

%% file: jvink_aarv_fig_truelove.tex
\begin{figure}
\centerline{
  \includegraphics[angle=-90,width=\medfig]{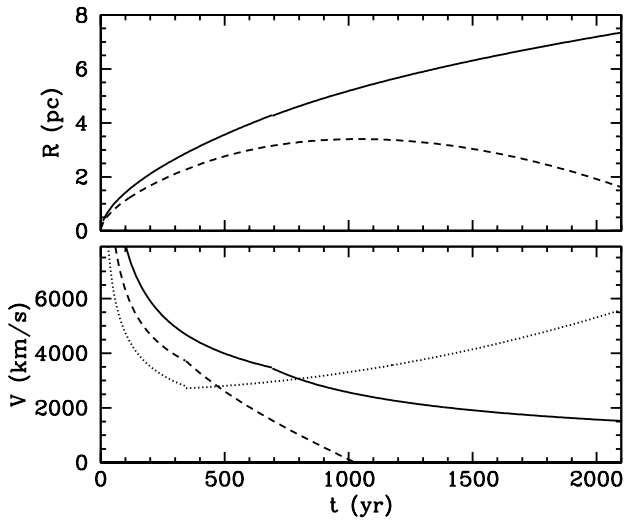}
}
\caption{Shock radii and shock velocities as a function of time
according to the Truelove-McKee $n=7, s=0$\ model \citep{truelove99}.
The solid line represents the forward shock, whereas the dashed line represents
the reverse shock. In the bottom panel the reverse shock velocity in
the frame of the ejecta is shown as a dotted line.
The dimensionless model was adjusted to fit the properties of Kepler's 
SNR \citep{vink08b}.
\label{fig:truelove}
}
\end{figure}

%% file: jvink_aarv_fig_wind_bubble.tex
\begin{figure*}
\centerline{
\includegraphics[width=0.44\textwidth]{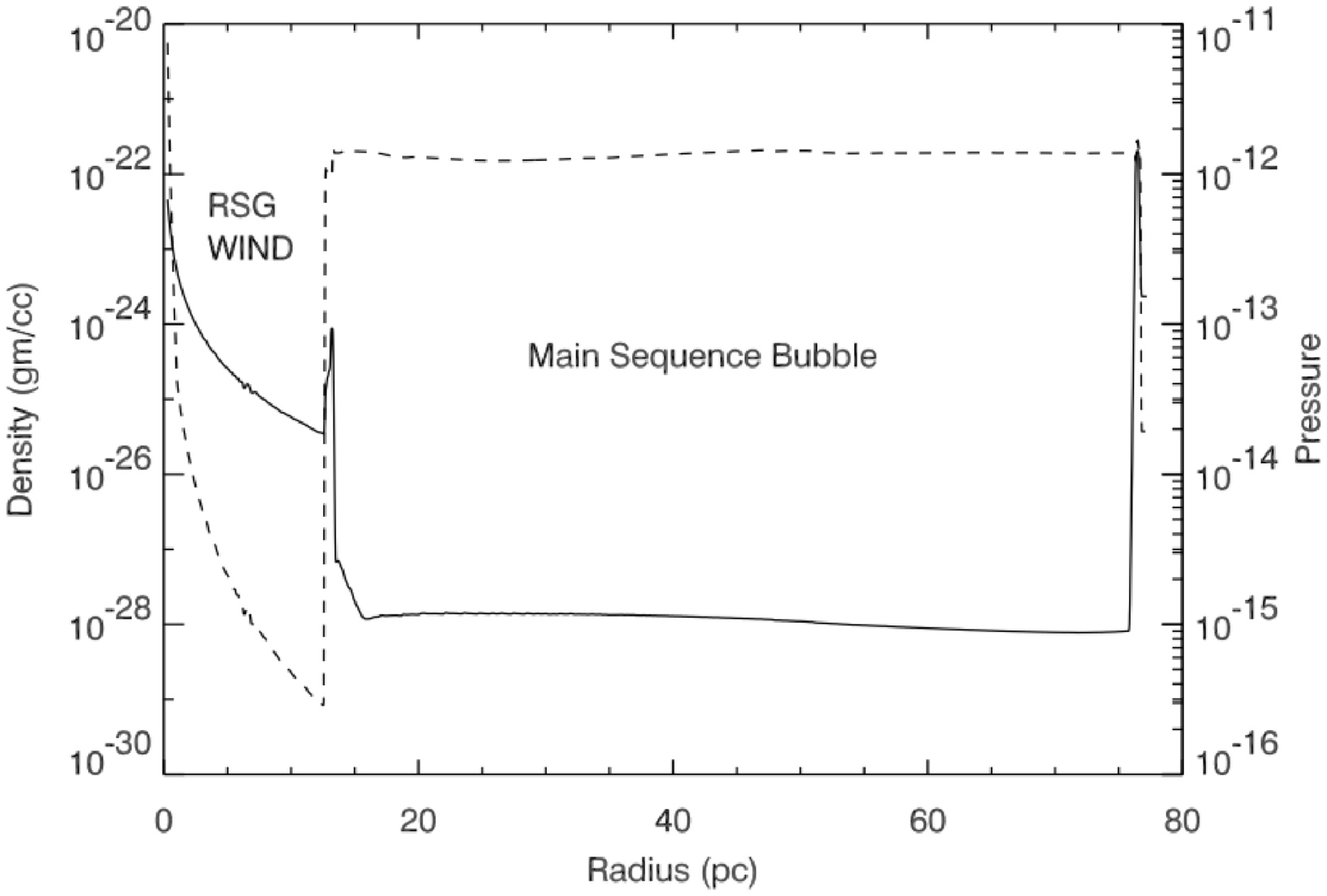}
\includegraphics[trim=10 0 310 0,clip=true,width=0.28\textwidth]{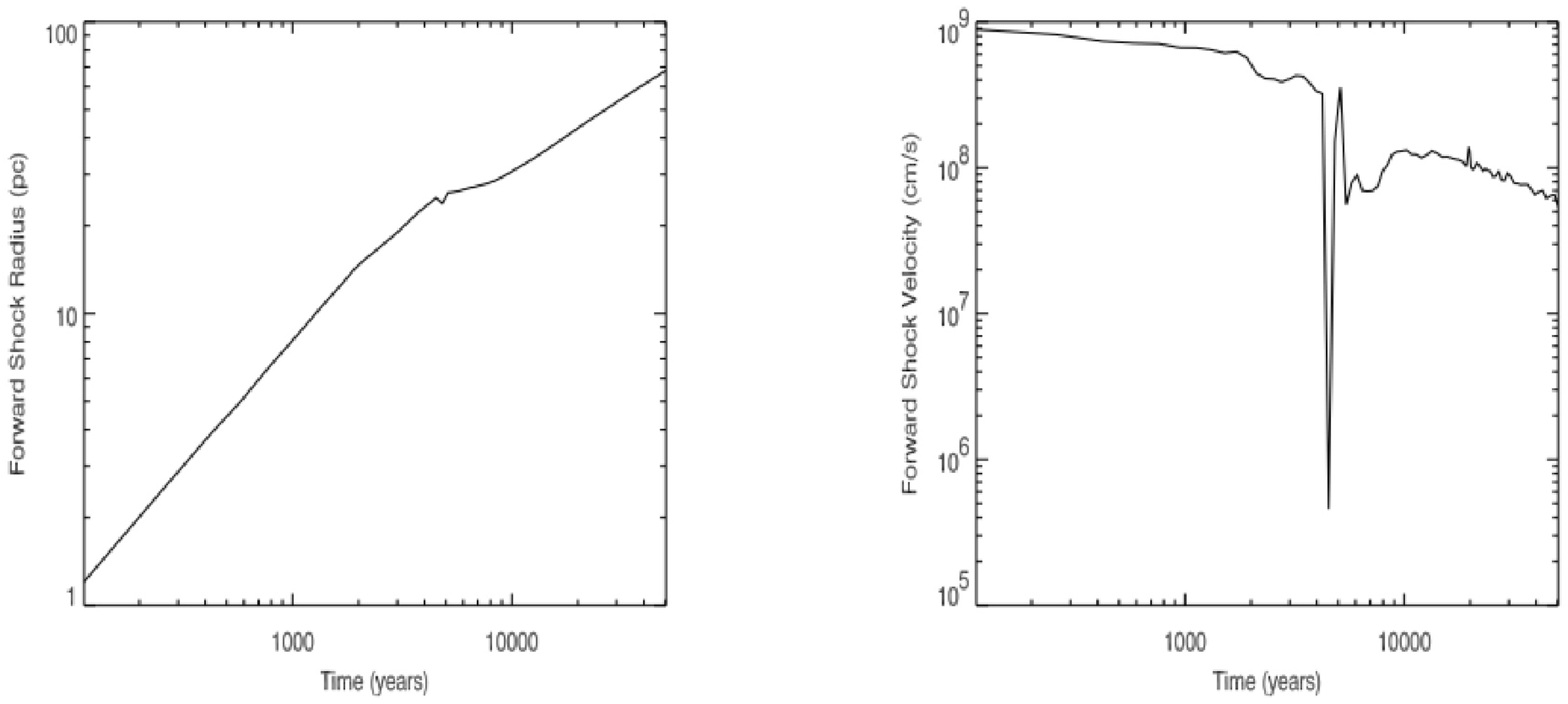}
\includegraphics[trim=310 0 5 0,clip=true,width=0.28\textwidth]{Fig7bc.ps}
}
\caption{
Left:
The density (solid line) and pressure (dashed line)
of the circumstellar medium around a red supergiant.
Middle and right: The evolution of the shock radius and velocity
for a SNR in a wind-blown cavity, surrounded by a dense, swept-up shell.
All figures taken from \citet{dwarkadas05}.
\label{fig:windbubble}
}
\end{figure*}

%% file: jvink_aarv_shock_heating.tex
Shock waves transform part of 
bulk kinetic energy into thermal energy. Astrophysical shocks
are characterized by low densities, which makes that
the actual heating process cannot be established by particle-particle
interaction (\sect~\ref{sec:collisionless}). For SNRs 
 part of the kinetic energy may not only be used for
shock heating the plasma, but also for cosmic-ray acceleration.

Irrespective of the microphysical details of particle heating and acceleration,
the collisionless shocks still have to obey
the conservation of mass, momentum, and energy flux across the shock
\citep[e.g.][]{zeldovich66,mckee80}.
However, one should realize that SNR shocks are not the idealized,
infinitely narrow temperature transition zones treated in standard 
physics textbooks.
Instead, collisionless heating, and cosmic-ray acceleration
imply that there is a broad 
transition  zone from the unshocked medium (or upstream) 
to the shocked (downstream) region. In the case of efficient cosmic-ray 
acceleration (\sect~\ref{sec:cr}), this transition zone, the so-called 
{\em cosmic-ray shock precursor}, has a width
of approximately the diffusion length scale of the cosmic rays 
with energies
near the peak in the energy spectrum \citep[e.g.][]{blasi05,vladimirov08}. 

Irrespective of whether cosmic-ray acceleration is important,
mass conservation across the shock remains valid, and for a plane
parallel case implies:
\begin{equation}
\rho_0 v_0 = \rho_2 v_2 \ \leftrightarrow  \
\rho_2 =  \frac{v_0}{v_2} \rho_0 \equiv \chi \rho_0, \label{eq:massflux}
\end{equation}
with $\chi$ the shock compression ratio, and $v_0$ and $v_2$ taken to
be in the frame of the shock.
The subscript $2$ refers to the downstream plasma, and
the subscript $0$ the (far) upstream medium,
thereby symbolically ignoring ``region $1$'', 
the transition zone containing the cosmic-ray shock precursor.

Cosmic-ray acceleration also hardly affects the conservation of momentum
flux:
\begin{equation}
P_2 + \rho_2 v_2^2  = P_0 + \rho_0 v_0^2 \ \leftrightarrow \ 
P_2 = P_0 + \Bigl(1 - \frac{1}{\chi}\Bigr) \rho_0 V_{\rm s}^2,\label{eq:momentum}
\end{equation}
with the  second part following from substituting Eq.~\ref{eq:massflux},
and relating $v_0$ to the shock velocity, $v_0 = -V_{\rm s}$.
For high Mach number shock waves the pressure far upstream can be neglected 
($P_0=0$).
And if we make the assumption that the downstream plasma pressure
consists of a thermal and a non-thermal (cosmic-ray) part, with
the parameter $w$ giving the non-thermal (NT) pressure contribution,
\begin{equation}
w \equiv  \frac{P_{2,NT}}{P_{2,T} + P_{2,NT}},\label{eq:w}
\end{equation}
we find from Eq.~\ref{eq:momentum} and \ref{eq:w}
that the post-shock plasma must have a temperature
\begin{equation}
kT_2 = (1 -w ) \frac{1}{\chi}\Bigl( 1 - \frac{1}{\chi}\Bigr) \mu m_{\rm p} V_{\rm s}^2.
\label{eq:kT1}
\end{equation}

One of the consequences of efficient cosmic ray acceleration is
that the highest energy cosmic rays may escape far upstream (i.e.
into the unshocked medium), whereas 
downstream (in the shock-heated plasma)
the dominant flux of cosmic-ray energy is through
plasma transport.
This means that conservation of energy flux is not necessarily valid.
A similar situation holds for so-called {\em radiative shocks},
for which radiative energy losses are important, a situation that
can occur for shocks with $V_{\rm s} \lesssim 200$~\kms.

We can take into account of the energy flux losses, $F_{esc}$,
by introducing \citep{berezhko99}:
$\epsilon_{esc} \equiv \frac{F_{esc}}{\rho_0 V_{\rm s}^3}$.
With this equation the conservation of energy flux can be written as
\begin{equation}
(u_2 + P_2 +\frac{1}{2} \rho_2 v_2^2) v_2 = 
(1 - \epsilon_{esc})\frac{1}{2} \rho_0 v_0^2 v_0, \label{eq:energyflux}
\end{equation}
with 
$u = \frac{1}{\gamma -1 } P$,
the internal energy, and $\gamma$ the adiabatic index.
Taking for the non-thermal (cosmic ray) contribution $\gamma=4/3$ and
for the thermal plasma $\gamma=5/3$ we obtain
\begin{equation}
u_2 + P_2 = \Bigr(\frac{3}{2}w + \frac{5}{2}\Bigl)P_2 \equiv \Gamma P_2.
\label{eq:energy_flux2}
\end{equation}
Using Eq.~\ref{eq:momentum}, we can rewrite Eq.~\ref{eq:energyflux} as
\begin{equation}
2\Gamma \Bigl( 1 - \frac{1}{\chi}\Bigr) + \frac{1}{\chi} =  
(1 - \epsilon_{esc})\chi,
\end{equation}
which has the non-trivial root \citep[e.g.][]{berezhko99,vink08d}:
\begin{equation}
\chi = \frac{\Gamma + \sqrt{\Gamma^2  - (1-\epsilon_{esc})(2\Gamma-1) }}
{1 - \epsilon_{esc}}. \label{eq:chi}
\end{equation}

In most textbooks only the case  of a single fluid gas
without energy losses is considered. This corresponds to $w=0$ and $\epsilon_{esc}=0$.
This gives a compression ratio of $\chi=4$. 
Substituting this in Eq~\ref{eq:kT1} gives
\begin{equation}
\bar{kT} = \frac{3}{16} \mu m_{\rm p} V_{\rm s}^2 \approx 
1.2 \Bigr( \frac{V_{\rm s}}{1000\ {\rm km\,s}^{-1}} \Bigl)^2 {\rm \ keV},
 \label{eq:kT_std}
\end{equation}
with $\mu$ the average particle mass ($\mu\approx 0.6$ for a plasma
with solar abundances).

I used here $\bar{kT}$ in order to indicate that this refers to the
average temperature, as different particle species can have different
temperatures. The reason is that for collisionless shocks it is not quite
clear whether the microphysics that governs the actual heating 
results in temperature equilibration among different, 
or whether different species
will have different temperatures proportional to their mass:
$kT_i \propto m_i V_{\rm s}^2$ (\sect~\ref{sec:collisionless}).
The latter can be expected if heating is predominantly governed by
scattering isotropization of the incoming particles by plasma waves
in the shock region. 

\input{jvink_aarv_fig_hugoniot}

Finally, if cosmic-rays contribute substantially
to the post-shock pressure ($w \neq 0$) and, if cosmic-ray
escape plays an important role, then the compression ratios 
can be much higher than $\chi=4$.
In this so-called {\em non-linear cosmic-ray acceleration} case,
there is no longer a unique value for $kT$ and $\chi$, but a range
of values is possible.
The possible values of the compression ratio and post-shock
plasma temperature as a function of post-shock cosmic-ray pressure fraction,
and escape flux are summarized in Fig.~\ref{fig:hugoniot}.
In this figure the mean plasma temperature $\bar{kT}$ is expressed
in terms of the temperature for the $w=0$/$\epsilon_{esc} = 0$ case, using
$\beta \equiv \bar{kT}/(3 \mu m_{\rm p} V_{\rm s}^2/16)$.

In all cases non-linear cosmic ray acceleration leads to lower
plasma temperatures, in the most extreme cases perhaps even quenching
thermal X-ray emission \citep{drury09}.
On a macroscopic level this can be understood
by considering that part of the incoming energy flux is used
for particle acceleration. On a more detailed level it can be
understood by considering that directly upstream of the 
shock the cosmic rays provide a non-negligible pressure, which
pre-compresses and pre-heats the plasma.
The presence of the cosmic-ray precursor results therefore
in a lower Mach number gas-shock \footnote{This is the actual shock that
heats the plasma,
but it is often called {\em sub-shock} in order to set it apart
from the overall shock structure that includes both the precursor
and the gas-/sub-shock.}
For a more thorough treatment of the shock equations, including a cosmic ray precursor, see \citet{vink10a} for
an analytical treatment, or \citet{vladimirov08,caprioli08} for
a kinetic approach. In the latter two references also 
the effects of non-adiabatic precursor heating is taken into account.

%% file: jvink_aarv_fig_hugoniot.tex
\begin{figure}
\centerline{
\includegraphics[width=\medfig]{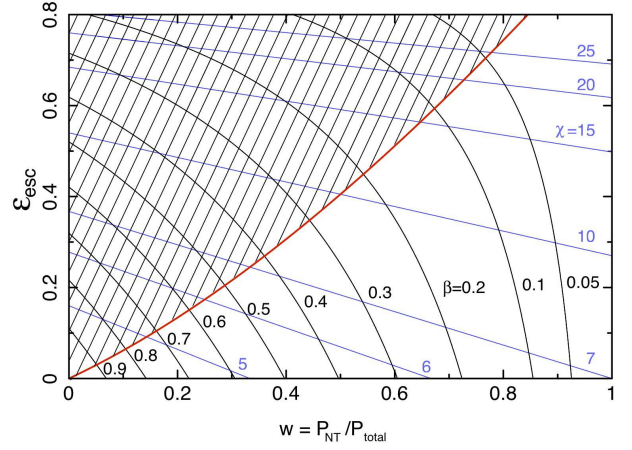}}
\caption{
The impact of the partial pressure by non-thermal components (cosmic rays)
and the fraction of energy flux escaping as cosmic rays 
\citep{vink08d,helder09}.
The black lines indicate constant values of temperature (Eq.~\ref{eq:kT1}), 
given as a correction factor with respect to post-shock temperature for the 
standard case (i.e. $w=0, \epsilon_{esc}=0$, Eq.~\ref{eq:kT_std}): 
$\beta \equiv \bar{kT}/(\frac{3}{16} \mu m_p V_{\rm s}^2)$.
Also indicated are lines of constant post-shock compression
ratio $\chi$. 
The red line depicts the solution cosmic-ray acceleration
with the highest possible efficiency \citep{helder09,vink10a}; 
the combination of $w$ and $\epsilon_{esc}$ to
the left of this line are considered unphysical.
\label{fig:hugoniot}
}
\end{figure}

%% file: jvink_aarv_collisionless_shocks.tex
Shock waves
in the interstellar medium (ISM) are characterized by very low densities, typically 
$n \sim 1$ cm$^{-3}$.
In high density shocks, such as in the earth's atmosphere,
the shock heating process is
accomplished by particle-particle interactions. However, 
the mean free paths for particle-particle interactions
in shocks in the ISM are long compared to the typical sizes
of SNRs.
This can be illustrated as follows.
For charged particles the energy exchange is governed 
by Coulomb interactions. A particle is deflected
by 90\deg\ in the center of mass frame,
if the following condition for the impact
parameter  $b$ is met:
\begin{equation}
\frac{1}{2} \frac{m_1 m_2}{m_1 + m_2} v^2 = \frac{ Z_1 Z_2 e^2}{b},
\end{equation}
with $m_1$, $m_2$ the masses of the two particles, 
$Z_1e$, $Z_2e$ their charges, and $v$ their relative velocity.
The cross section for such a deflection by a single scattering
is ($\sigma_{\rm Coulomb} = \pi b^2$):
\begin{equation}
\sigma_{\rm Coulomb} \approx 4 \pi \frac{Z_1^2Z_2^2 e^4}{v^4} 
\Bigl( \frac{m_1 + m_2}{m_1 m_2}\Bigr)^2.
\label{eq:coulomb}
\end{equation}

The corresponding collision time scale is
$\tau_{\rm Coulomb} = \\
1/(n \sigma v) \propto v^{-3} \propto E^{-3/2}$.
Inserting for the $m_1$ and $m_2$ the proton mass
and a typical velocity of $v = 1000$~\kms, one finds for 
proton-proton collisions
$\tau_{\rm pp} \approx 10^{12} n_{\rm p}^{-1}$~s (about 32,000 $n_{\rm p}^{-1}$\ yr), 
and a mean free path $\lambda_{\rm p} \approx 10^{20}n_{\rm p}^{-1}$~cm
(32 pc). These are much larger than the ages and radii of young SNRs.
As these young SNRs clearly do have shocks and hot, X-ray emitting, plasma,
as beautifully shown by Fig.~\ref{fig:tycho}, 
the formation of the shock and the plasma heating 
cannot be the result of Coulomb interactions. Instead 
``collective interactions'', occurring through
fluctuating electric and
magnetic fields,
must be responsible for the plasma heating.
Simulations \citep[e.g.][]{bennett95}
show that the heating in such shocks takes place over a distance
of typically 10-100 times
$c/\omega_{p\rm e}$, with $\omega_{p\rm e} = (4\pi e^2n_{\rm e}/m_{\rm e})^{1/2}$\ 
the plasma frequency.
This corresponds to a shock thickness of roughly 
$\Delta x= 10^7n_{\rm e}^{-1/2}$~cm, thirteen orders of magnitude
smaller than the range over which Coulomb collisions operate.
For this reason SNR shocks and shocks in other low density
media, such as the interplanetary medium, are called {\em collisionless shocks}.

The heating mechanism is somewhat analogous to  the process of ``violent relaxation''
\citep{lyndenbell67}
in the formation of bound gravitational systems, such as galaxies and clusters
of galaxies; i.e. large scale gravitational potential fluctuations
rather than two-body interactions are important for the relaxation of the system.

\subsection{Temperature equilibration}
\label{sec:equilibration}
It is not a priori clear, whether collisionless shocks will result in
temperature equilibration of all types of particles 
immediately behind the shock front
 \citep[see][for a review]{bykov08a}.
The two opposite cases are
a) complete equilibration, in which all different particle
populations (electrons, protons, other ions) have the same temperature, which
for a strong shock without cosmic-ray acceleration would
mean $T \propto \mu m_{\rm p}V_{\rm s}^2$\ (Eq.~\ref{eq:kT1}), 
or b) full non-equilibration, in which case 
each particle 
species thermalizes the bulk kinetic energy independently
and one expects $T_i \propto m_i V_{\rm s}^2$ (corresponding to
replacing $\mu m_{\rm p}$ with $m_i$ in Eq.~\ref{eq:kT1} and \ref{eq:kT_std}).\footnote{
The latter case corresponds to a velocity distribution that is 
independent of the particle mass.
This is again similar to the outcome of violent relaxation
for bound gravitational systems. For example,
in galaxy clusters the velocity distribution is similar
for massive and dwarf galaxies
(with the exception of the giant ellipticals that reside in a
large fraction of cluster centers).
Also in this case the kinetic energy (``temperature'')  distribution
is proportional to the mass of the galaxy. 
}
There are many plasma instabilities
that act to equilibrate the energies of different plasma constituents.
Both hybrid simulations \citep{cargill88} and particle in cell (PIC)
simulations \citep[e.g.][]{shimada00} show that  high Mach number shocks
may substantially heat the electrons, but not completely
up to the same temperature as the protons. 
So one may expect $m_{\rm e}/m_{\rm p} \ll T_{\rm e}/T_{\rm p} < 1$.
For example, \citet{shimada00} report $T_{\rm e}/T_{\rm p}\approx 17$\%.
Optical spectroscopy of SNR shocks show mixed results
with indications that full equilibration is important
fo $V_{\rm s} < 500$~\kms, and non-equilibration for higher velocities
\citep{ghavamian07,helder11}. 
In \sect~\ref{sec:thermaldoppler} I discuss the implications for
X-ray spectroscopy.

Note that these simulations have several limitations. For example,
PIC simulations often have $m_{\rm p}/m_{\rm e} \approx 0.01-1$, for 
computational reasons. In contrast, hybrid simulations treat the
electrons as a massless fluid.
Moreover, the amount of equilibration of electron/proton temperatures
may depend on the initial conditions, such 
as Mach number, the ratio of magnetic over kinetic pressure and 
the magnetic-field orientation. This means that some variation in shock
equilibration can be expected in SNRs.

An important ingredient that affects electron heating, and one
that is difficult to model,
is cosmic-ray acceleration (see \sect~\ref{sec:cr}), because the
process of acceleration to high energies takes place on a longer
time scale than the formation of strong shocks, and longer than the
typical time scales that can be modeled with PIC simulations.

Nevertheless, hybrid and PIC 
simulations \citep[e.g.][]{bennett95,dieckmann09}
do show evidence for the initial stages of particle acceleration: 
often the particle distribution has non-thermal components,
i.e. the particles have roughly a Maxwellian energy
distribution at low energies, but 
a non-thermal, quasi-power-law distribution at high energies.
This non-thermal tail, may be the first stage of particle acceleration,
but part of this non-thermal distribution, may disappear again
further behind the shock, as the particles are slowed
down through Coulomb interactions. This is especially relevant
for the low energy part of the non-thermal electron distribution
\citep[\sect~\ref{sec:nonthermal_brems},][]{vink08a}, and amounts to a source of late time electron heating.

At the shock front plasma wave coupling between electrons and accelerated ions
may also result in heating of the electrons \citep[][]{rakowski08}.
Finally, very efficient particle acceleration by the Fermi process may alter
the shock structure altogether, as it results in the formation of 
a cosmic-ray shock precursor
that may pre-heat both electrons and protons before they enter the main
shock (see \sect~\ref{sec:cr} and
\sect~\ref{sec:kT}).
In \sect~\ref{sec:thermaldoppler} I review the observational measurements
concerning electron/ion equilibration.

If the shock heating does not immediately lead to equal electron and ion temperatures, Coulomb
interactions will slowly equilibrate them. It turns out
that Eq.~\ref{eq:coulomb} is not a good approximation for the relevant
cross sections, as it ignores the contributions for long range
interactions. These result in less energy exchange per encounter, but they
occur much more frequently. The relevant expressions for energy
exchange between charged particles of various mass ratios 
can be found in \citet{nrlplasma}, which gives for the temperature
equilibration rate
\citep[see also][]{zeldovich66,itoh84}:
\begin{align}
\label{eq:equil}
\frac{dkT_1}{dt} = &  1.8\times 10^{-19} \times \\ \nonumber
&\sum_i \frac{ (m_1m_i)^{1/2}Z_1^2Z_i^2e^4 n_i \ln \Lambda_{1i}}{(m_ikT_1 + m_1kT_i)^{3/2}}
(kT_i - kT_1)\ {\rm eV\,s^{-1}},
\end{align}
with label $1$ for the particle species of interest, $i$ for all the other species,
and $kT$ measured in units of eV.
$\ln \Lambda$ is the Coulomb logarithm, which is roughly similar
for electron proton interactions and for electron-electron interactions:
\begin{equation}
  \ln \Lambda  = 30.9 - 
  \ln\Bigl[ n_{\rm e}^{1/2} \Bigl(\frac{kT_{\rm e}}{1\,{\rm keV}} \Bigr)^{-1} \Bigr].
\end{equation}
For electron-proton equilibration Eq.~\ref{eq:equil} implies an equilibration
time scale:
\begin{equation}
\tau_{\rm ep} \approx 3.1 \times 10^{11} n_{\rm p}^{-1} \Big( \frac{kT}{1\,{\rm keV}} \Big)^{3/2}
\Bigl(\frac{\ln\Lambda}{30.9}\Bigr)^{-1}\ {\rm s},\label{eq:tau_ep}
\end{equation}
with $T$ the mean temperature. 
Similarly, the electron self-equilibration time, which determines how
fast the electrons can establish a Maxwellian distribution, is
\begin{equation}
\tau_{\rm ee} \approx 4.9 \times 10^{8} n_{\rm e}^{-1} \Big( \frac{kT}{1\,{\rm keV}} \Big)^{3/2}
\Bigl(\frac{\ln\Lambda}{30.9}\Bigr)^{-1}\ {\rm s}.
\label{eq:tau_ee}
\end{equation}

Eq.~\ref{eq:tau_ep} shows that it can easily take 10,000
years for electrons and protons to equilibrate, much longer than
the ages of many of the young 
SNRs that will be discussed in the second part of this review.
Note that the equilibration time is 
roughly inversely  proportional to $n_{\rm p} (\approx n_{\rm e})$.
A relevant parameter is, therefore, not so much the age of the
SNR, but the average \net, a parameter that will resurface
in \sect~\ref{sec:nei}, where non-equilibrium ionization is discussed.
In fact, X-ray spectroscopy provides a direct 
way of measuring \net, thereby providing the means to estimate whether
non-equilibration of temperatures could be important.

\input{jvink_aarv_fig_equilibration}

The possible absence of full equilibration has important 
consequences for the interpretation
of  X-ray observations \citep[see][for an early discussion]{itoh78}, 
because in general the X-ray spectra reveal
the electron temperature (\sect~\ref{sec:continuum}), and not
the mean plasma temperature.
The mean plasma temperature is energetically the relevant parameter. It
is for example directly related to the shock velocity
(\sect~\ref{sec:kT}). Using the electron temperature,
as an estimate for the mean temperature
may lead to a serious underestimation of the shock velocity.
But, as shown in \sect~\ref{sec:kT}, efficient cosmic-ray acceleration
will lead to lower mean temperatures. So even having a good
estimate of both the electron and ion temperatures may not
be sufficient to estimate the shock velocity.

Fig.~\ref{fig:equilibration} illustrates the effects of non-equilibration
of temperatures at the shock front. The equilibration time
depends on the density, the mass ratio and the square of the charge
of the particles 
(Eq.~\ref{eq:equil}). 
The ion-ion equilibration, therefore,
proceeds faster than electron-ion equilibration. Interestingly,
iron-proton equilibration proceeds faster than proton-helium equilibration,
due to the charge dependence. This becomes important for
\net$> 10^9$~\netunit, when iron has become
stripped of all m-shell electrons. Broadly speaking one can 
recognize three phases: 
\begin{enumerate}
\item {\em Full non-equilibration}; all species have different
temperatures (\net $\lesssim 5\times 10^{10}$~\netunit), 
\item {\em Partial non-equilibration};
all ions are equilibrated, but without electron-ion equilibration \\
($5\times 10^{10} \lesssim $ \net $\lesssim 10^{12}$~\netunit),
\item {\em Full equilibration} (\net $\gtrsim 10^{12}$~\netunit).
\end{enumerate}

It is worth noting that non-equilibration of electron-ion temperatures may
also be caused by ionization \citep{itoh84}:
Due to the fact that the
cross sections for ionization peak near the ionization energy, 
the electrons originating from post-shock ionization
will start relatively cool. The electron-electron equilibration
is relatively fast, so this results in a relatively cool
electron population, even if the free electrons that entered the shock
were equilibrated with the ions.
This may be an important
effect in metal-rich plasmas, which have been shock heated
by the reverse shocks in young SNRs (\sect~\ref{sec:hydro}),
since in that case most of the electrons will originate from ionizations,
rather than from entering the shock as free electrons.

%% file: jvink_aarv_fig_equilibration.tex
\begin{figure}
\centerline{
\includegraphics[angle=-90,width=\medfig]{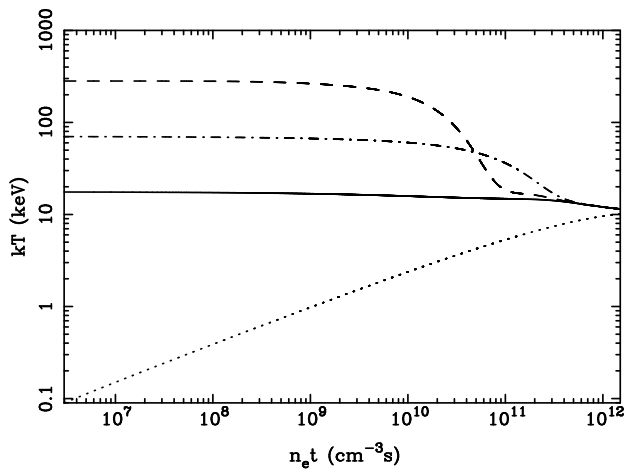}
}
\caption{An illustration of the effect of temperature
non-equilibration at the shock front for a shock velocity of
$V_{\rm s}=3000$~\kms. 
Shown is the temperature
of electrons (dotted), protons (solid), helium
(dashed-dotted) and oxygen ions (dashed) as a function of \net,
assuming that heating at the shock front is proportional to the mass
of the particle (\sect~\ref{sec:kT}).
The oxygen-proton equilibration is
faster than the helium-proton equilibration,
as the cross sections scale linearly with particle mass, 
but quadratically with charge (Eq.~\ref{eq:coulomb}).
(Figure previously published in \citet{vink06b}.)
\label{fig:equilibration}
}
\end{figure}

%% file: jvink_aarv_conduction.tex
Thermal conduction is a topic that every now and then enters
the discussion of SNRs, as significant thermal conduction may
alter the temperature and density structures of
SNRs \citep[see for example ][and \sect~\ref{sec:mixed}]{cui92}. 

The role of thermal conduction in SNR structure and evolution,
 like that of temperature non-equilibration, 
 has never been satisfactory resolved.
The reason is that thermal conduction in a plasma is likely to
be anisotropic due to the inhibiting effects of magnetic fields.
Note that the processes of thermal conduction and temperature equilibration are
related, because thermal conduction is a combination of transport and exchange of heat.
Without magnetic fields, or along magnetic fields, thermal conduction is usually
mediated through electrons, because they have higher thermal speeds and
are more strongly coupled to each other. Across magnetic-field lines the
particles with the largest gyroradius, i.e. the ions, 
will dominate thermal conduction.

The process of thermal conduction is described by Fourier's law:
\begin{equation}
{\bf F}_{\rm heat}= -\kappa {\bf \nabla} T.\label{eq:fourier}
\end{equation}
In the absence of magnetic fields, or along magnetic field lines, the coeffcient $\kappa$
depends on the typical mean free path for energy exchange, 
$\lambda \approx \tau_{\rm ee} v_{\rm th}$, with $v_{\rm th}=\sqrt{kT_{\rm e}/m_{\rm e}}$ and $\tau_{\rm ee}$
the self-equilibration time (Eq~\ref{eq:tau_ee}). Numerically, $\lambda\approx
0.2 (kT/1\ {\rm keV})^2$~pc.
The amount of heat being transported
relates to the energy density $u=3/2 n_{\rm e}kT_{\rm e}$ 
and the transport
velocity $v_{\rm th}$, so that approximately 
$\kappa/k \approx 3/2 n_{\rm e}kT \lambda v_{\rm th}= 3/2 n_{\rm e} kT  \tau_{\rm ee} v_{\rm th}^2$. This is correct up to a small
factor that is needed to ensure zero electric current \citep[e.g.][]{spitzer62}. Taking this into account
one obtains \citep{nrlplasma}\footnote{Correcting for  a slightly different definition of $\tau_{\rm ee}$.}:
\begin{equation}
\kappa_{\parallel}/k = 2.3 \frac{n_{\rm e} kT_{\rm e} \tau_{ee}}{m_{\rm e}}=
2.0\times 10^{27} 
\Bigl( 
\frac{kT_{\rm e}}{1\,{\rm keV}}
\Bigr)^{5/2} \ {\rm cm^{-1}s^{-1}}.
\label{eq:kappa_par}
\end{equation}
This shows that thermal conduction is strongly temperature dependent.

Eq.~\ref{eq:fourier} is only valid when the temperature scale height
is larger than the mean free path for energy exchange. If this is not the
case then the heat flux is described by the so-called saturated heat flux
\citep{cowie77} is
\begin{equation}
F_{\rm sat}= 0.4 \Bigl(\frac{2kT_{\rm e}}{\pi m_{\rm e}}\Bigr)^{1/2}
n_{\rm e} kT_{\rm e}.
\end{equation}
In this case heat is transported by electrons streaming away from hot regions, but
heat exchange between electrons is not taken into account, i.e. the electron
distribution will not be Maxwellian. The factor 0.4 takes into account that the
heat flux should be electrically neutral, reducing the heat flow. Unlike the
classical heat flux, the saturated heat flux is density dependent.
The ratio of the classical over saturated heat flow, involving a temperature scale height
parameter $R_{\rm s}$ is \citep[c.f][]{cowie77}:
\begin{equation}
\sigma \equiv \frac{ 7.2 \tau_{\rm ee}\sqrt{kT_{\rm e}/m_{\rm e}}}
{R_{\rm s}} = 1.5 n_{\rm e}^{-1}\Bigl(\frac{R_{\rm s}}{1\ {\rm pc}}\Bigr)^{-1} \Big( \frac{kT_{\rm e}}{1\,{\rm keV}} \Big)^{2}
\Bigl(\frac{\ln\Lambda}{30.9}\Bigr)^{-1}.
\end{equation}
The saturated heat flux is, therefore, dominant for small scale sizes ($R_{\rm s}<< 1\ {\rm pc}$).

The thermal conduction across field lines, which is predominantly
mediated by ions (protons), is \citep{spitzer62,nrlplasma}:
\begin{equation}
\kappa_{\perp}/k= \frac{2.8 n_{\rm p}kT_{\rm p}}{m_{\rm p }\omega_{\rm cp}^2\tau_{\rm pp}}=
1.4\times 10^7 \Bigl(\frac{B}{10\,{\rm \mu G}}\Bigr)\ \rm cm^{-1}s^{-1},
\label{eq:kappa_perp}
\end{equation}
with $\omega_{\rm cp}=eB/(m_{\rm p}c)$ the proton gyrofrequency, and $\tau_{\rm pp}$ the
proton self-equilibration time scale.
This shows that there is a very large difference in conduction parallel and across field lines.
Only for $B\sim 10^{-15}$~G does the perpendicular and parallel conduction become comparable (for $n=1$~cm$^{-3}$).

To get an order of magnitude estimate 
for the importance of thermal conduction one should
compare the advective enthalpy flux $F_{\rm E}=v \frac{5}{2}n kT$ (c.f. Eq~\ref{eq:energyflux}) with the thermal conduction flux, approximating the temperature gradient with $kT/R_{\rm s}$ and assuming a typical flow speed $v$:
\begin{equation}
\frac{F_{\rm  E}}{F_{\rm heat}} \approx
 \frac{5 n R v} {2 \kappa}=
0.4 n 
\Bigl(\frac{R_{\rm s}}{1\ {\rm pc}}\Bigr)
\Bigl(\frac{v}{1000\ {\rm km\,s^{-1}}}\Bigr)
\Bigl( \frac{kT_{\rm e}}{1\,{\rm keV}}\Bigr)^{-5/2},\label{eq:peclet}
\end{equation}
where for $\kappa$ the value along the magnetic field has been used. The fact that
the numerical value is around one suggests that in SNRs the issue of thermal conduction is critical.
This in turn makes the regularity of the magnetic field a crucial parameter. 
Note that the advected over conductive heat flow ratio (Eq~\ref{eq:peclet}) 
is proportional to $(kT)^{-5/2}$, if in addition one assumes 
$v\propto \sqrt{kT}$, this ratio will 
be proportional to $(kT)^2$. 
This means that thermal conduction is more important for hotter plasmas.
However, there are indications that the magnetic fields in young SNRs, 
which have hotter plasmas, are more turbulent (\sect~\ref{sec:efficient_crs}),
reducing the thermal conductivity.
On the other hand little is known about the magnetic fields
in the ejecta components of SNRs. 
This may, as we see be of interest for the interior components
of some mature SNRs (\sect~\ref{sec:mixed}).

%% file: jvink_aarv_cr_acceleration.tex
SNRs are considered to be the prime candidate
sources for cosmic rays \citep[e.g.][]{ginzburg69b}, 
at least for energies up to $3\times10^{15}$~eV,
at which energy the cosmic-ray spectrum steepens from a power law
with slope -2.7 to a slope of -3.1. This spectral feature is often
referred to as the ``knee''. 
There is evidence that around the ``knee'' the composition changes
\citep[][for a review]{hoerandel08},
suggesting a rigidity\footnote{The rigidity of a particle
is $R\equiv p/q$, with $p$ the momentum and $q=Ze$ the charge.
Massive particles, having a higher charge, will have
a smaller gyroradius, and will therefore less easily diffuse away from
the shock.} dependent maximum cosmic-ray energy at the source, 
with more massive particles having their spectral breaks at higher energies.
It is thought that only cosmic rays above $\sim 10^{18}$~eV may have
an extra-galactic origin.

The main reason to consider SNRs as the dominant sources of Galactic cosmic
rays 
is that the inferred
Galactic cosmic-ray energy density 
\citep[$\sim 1-2$ eV cm$^{-3}$,][]{webber98}, 
combined with the inferred time an average cosmic ray spends in the Galaxy,
requires a Galactic cosmic-ray production
with a total power of about $10^{41}$~erg\,s$^{-1}$ \citep[e.g.][]{ginzburg67}.
Supernovae are the only sources known to provide such a power:
with their average explosion energy of $\sim 10^{51}$~erg and an inferred
Galactic supernova rate of 2-3 per century, they provide a total power of
$\sim 10^{42}$erg\,s$^{-1}$.
This requires that on average about 10\% of the initial explosion energy
is used for accelerating cosmic rays. Note, however, that the cosmic-ray energy
density is very much dependent on the lowest energy part of the cosmic-ray spectrum,
of which is little known, as low energy cosmic rays are shielded from
the solar system by the solar wind.

Although supernovae are the most likely energy source for cosmic-ray acceleration,
there is a debate on how and when their kinetic energy is used to accelerate
cosmic rays.
The general view is that
cosmic rays are mostly accelerated in the (early) SNR stage.
An alternative view is that the collective effects of multiple SNR and stellar wind shocks 
inside star-forming regions are responsible for most of the
Galactic cosmic-ray acceleration, and that this provides a mechanism
to accelerate cosmic rays up to energies of $10^{18}$~eV
 \citep[e.g.][]{bykov92,parizot04}.

A long standing argument to consider the SNRs as the main source of
Galactic cosmic rays is that non-thermal radio emission from
SNRs provides clear evidence for particle acceleration, although
strictly speaking this only provides evidence for electron acceleration, whereas
the cosmic-ray spectrum observed on Earth consists for 99\% of protons and other ions.
Over the last 10-15 years the evidence for cosmic-ray acceleration in SNRs has
been reinforced by the
detection of X-ray synchrotron radiation
(discussed in \sect~\ref{sec:synchrotron}),
and TeV \gray s from several 
SNRs.  See \sect~\ref{sec:xray_gammarays} and
the reviews by \citet{hinton09} and \citet{reynolds08}.

The acceleration of particles in SNRs is likely the result of the so-called
first order Fermi process, also often
referred to as {\em diffusive shock acceleration}
\citep{axford77,bell78a,blandford78,krymskii77b}.
In this process some charged particles repeatedly scatter back and forth 
across the shock front.
On either side of the shock the particles scatter elastically, 
due to the presence
of a turbulent magnetic field. After each crossing the particles will
be pushed along with the plasma, resulting
in a net energy gain given by
\begin{equation}
\frac{\Delta E}{E} \approx \frac{4}{3}\frac{v_0 -v_2}{c} = \frac{4}{3} 
\Bigl(\frac{\chi - 1}{\chi}\Bigr)\frac{V_{\rm s}}{c},\label{eq:crgain}
\end{equation}
with $v_0$ and $v_2$ the plasma velocities upstream 
and downstream\footnote{To be more precise: if there is a cosmic-ray precursor 
one
should write $v(x)-v_2$ with $v(x)$ the velocity at coordinate $x$ within the precursor, and
$v_0<v(x)<v_1$, with subscripts as defined in
 \sect~\ref{sec:kT}. So particles of a given energy may not sample the total
change in shock velocity over the shock structure, but only a fraction of it.}, seen in the frame of the shock, and
$v_1= V_{\rm s} = \chi v_2$ (Eq.~\ref{eq:massflux}).
The particles downstream will be advected away from the shock by the plasma
with a rate $n_{CR}v_2 =n_{CR}V_{\rm s}/\chi$, but the rate at which particles
scatter back into the upstream medium  is $\frac{1}{4}n_{CR}c$ for an
isotropic particle distribution,\footnote{
$1/4 c$ is the projected velocity
to the shock normal for particles diffusing upstream.}
with $c$ the velocity of the particle. 
The ratio of these two rates gives the chance that the particles
escape downstream:
\begin{equation}
P_{\rm esc} = \frac{n_{\rm CR}v_2}{\frac{1}{4}n_{\rm CR}c}=4\frac{v_2}{c} = 4\frac{V_{\rm s}}{c\chi}.
\end{equation}
The chance that a particle will be scattered up and down at least $k$-times
is $P(n \geq k) =  (1-P_{\rm esc})^k$. A particle that has crossed the shock
$k$ time  will on average have an energy $E \approx E_0 (1 + \Delta E/E)^k$.
Eliminating $k$ gives
\begin{equation}
\ln P (n \geq k) = \frac{ \ln(E/E_0)}{\ln(1 + \Delta E/E)}\ln(1 - P_{\rm esc}) 
\end{equation}
The {\em integrated} 
probability function is therefore a power law with slope $-q+1$:
\begin{align}
-q+1 = &
\frac{d \ln P}{d\ln E} = \frac{\ln(1 - P_{\rm esc})}{\ln(1 +  \Delta E/E)} \\ \nonumber
&\approx -\frac{4V_{\rm s}/(c\chi)}{4/3 (\chi-1)/\chi V_{\rm s}/c} = -\frac{3}{\chi-1}.
\end{align}
The {\em differential} distribution function is then
\begin{equation}
n(E)dE \propto E^{-q}dE,\ q = \frac{\chi + 2}{\chi -1}.\label{eq:fermi}
\end{equation}
In order to give an idea about how many shock crossings are needed to reach
$10^{15}$~eV, assume
that a proton is accelerated by a shock with $V_{\rm s}=5000$~\kms, starting with an initial energy of 100~keV, and assuming a shock compression ratio of $\chi=4$. According to Eq.~\ref{eq:crgain}
the gain per shock crossing is 1.7\%, therefore the number of
scatterings is $\log(10^{15}\ {\rm eV}/10^5\ {\rm eV})/\log(1.017)\approx 1400$.

One of the outstanding uncertainties is how charged particles are injected
into the Fermi acceleration process, i.e. how do they reach sufficient
velocity to make it back to the unshocked medium.
It is
often assumed that these are the particles in the high velocity tail of the
Maxwellian distribution, but computer simulations show that 
immediately behind the shock a non-thermal distribution
of particles is present that may act as seed particles for further acceleration
by the Fermi process
\citep[e.g.][for the case of electrons]{bykov99,riquelme11}.

The typical time accelerate a particle 
from an initial momentum $p_i$ to a final momentum $p_f$ is
\citep[e.g.][]{malkov01}
\begin{equation}
t_{\rm acc}= \frac{3}{v_2-v_0}\int_{p_i}^{p_f} \Bigl( \frac{D_0}{v_0} + \frac{D_2}{v_2}\Bigr),
\end{equation}
with $p$ the momentum ($E=pc$ for ultra-relativistic particles), 
$D$ the diffusion coefficient, and $v$ the plasma
velocity, with subscripts $0$ and $2$ referring to the upstream and 
downstream regions.
The diffusion coefficients are usually parameterized in terms
of the so-called Bohm-diffusion coefficient, for which the particle mean free path 
corresponds to the gyroradius:
\begin{equation}
D=\eta  \frac{Ec}{3eB},\label{eq:diff_const}
\end{equation}
with $\eta$ expressing the deviation from Bohm diffusion.

This suggests a typical acceleration time scale of \citep[][]{parizot06}
\begin{align}
\tau_{\rm acc} &\approx  1.83 \frac{D_2}{V_{\rm s}^2}\frac{3\chi^2}{\chi-1}  \\ \nonumber
&=124
\eta  B_{-4}^{-1} \Bigl(\frac{V_{\rm s}}{5000\ {\rm km\,s}^{-1}}\Bigl)^{-2}\Bigl(\frac{E}{100\, {\rm TeV}}\Bigr) 
\frac{\chi_4^2}{\chi_4-\frac{1}{4}}~{\rm yr}, \label{eq:tau_cr}
\end{align}
with $B_{-4}$ the downstream magnetic field in units of 100~$\mu$G and $\chi_4$ the overall compression ratio
in units of 4. The factor 1.83 comes from taking into account the difference
in diffusion coefficient between the upstream  (ahead) and downstream (post-shock) region, under the assumption that upstream
the magnetic field is highly turbulent and isotropic,
whereas the downstream magnetic field is determined solely by the
compression of the magnetic-field component perpendicular to
the shock normal.
Note that a higher compression ratio corresponds to a longer acceleration time,
but that a high compression ratio itself is a result of efficient acceleration (\sect~\ref{sec:kT}).

The typical length scale over which diffusion dominates over advection\footnote{
To see this consider that the length scale associated with diffusion over a time $t$ is
$l_{\rm diff}=\sqrt{2D_2t}$, whereas for advection $l_{\rm adv}=v_2t$. These two are
equal for $l=2D_2/v_2=2D_2\chi/V_s$.}
 is given by
$l_{\rm diff}\approx 2D/v$,
which for the downstream (post-shock) region becomes
\begin{align}
l_{\rm diff}&\approx  2\frac{D_2 \chi}{V_{\rm s}} \\ \nonumber
&=5.3\times 10^{17} B_{-4}^{-1}
\eta \chi_4 \Bigl(\frac{V_{\rm s}}{5000\ {\rm km\,s}^{-1}}\Bigl)^{-1} \Bigl(\frac{E}{100\, {\rm TeV}}\Bigr)
\  {\rm cm}.\label{eq:ldiff}
\end{align}
The diffusion length scale gives approximately the region from which particles situated in the
downstream (shock-heated) region are still able
to cross the shock front.

Eq.\ref{eq:fermi} shows that a compression ratio of 4 
corresponds to a particle index of $q=2$.
Note that this result is independent of the diffusion properties (slow or
fast) as long as the diffusion is isotropic. This slope is close
to what is needed to explain the cosmic-ray spectrum observed on Earth
($q=2.7$), provided one 
takes into account that cosmic rays with higher energies escape faster out
of the Galaxy than low energy cosmic rays.
Note that for $q=2$ each decade in particle energy contributes the same amount
of energy to the overall cosmic-ray energy budget. For $q<2$ the highest energy
particles contain most energy, whereas for $q>2$ the lowest energy particles
contain most of the energy. For the lower energy limit one can take the 
particle's rest mass energy, as around that energy a break in the cosmic-ray 
spectrum is expected \citep{bell78b}.

Diffusive shock acceleration becomes more complicated when particle
acceleration becomes very efficient and the accelerated particles
contribute a significant fraction to the overall pressure. 
This situation is usually referred to as {\em non-linear cosmic-ray
acceleration}.
As explained in \sect~\ref{sec:kT} this results 
in a lower downstream plasma temperature 
and an overall compression ratio $\chi_{\rm tot}>4$. The
main shock (subshock) will have a compression ratio $\chi_{\rm gas}\leq 4$.
The compression ratio sampled by a diffusing particle may therefore depend
on its energy/diffusion length scale. The highest energy particles experience a compression
ratio of  $\chi_{\rm tot}$ and the low energy particles $\chi_{\rm gas}$.
As a result the spectrum at low energies is steep, with
$q>2$ and flattens at higher energies. For very efficient
acceleration Eq.~\ref{eq:fermi} 
is not valid, but instead the spectrum approaches
the limit $q=1.5$ \citep{malkov97}.

There is observational evidence that the particle spectrum is indeed steeper at low energies
than at high energies (\sect~\ref{sec:synchrotron}). However, there is no
proof yet that the spectrum at high energies becomes as flat as $q=1.5$.
Some physical processes may prevent that, such as a possible tendency for
Alfv\'en waves to move away from the shock, reducing the velocity gradient
that the particles experience \citep{zirakashvili08b}. Alternatively,
an additional pressure term in the precursor, for example caused by
amplified magnetic fields and cosmic ray induced heating, may prevent very high compression ratios
in the precursor, thereby limiting the overall compression ratio to
$\chi \approx 4-7$ \citep{morlino07}.

%% file: jvink_aarv_nei.tex
\subsection{Thermal X-ray emission}
\label{sec:continuum}
The hot, X-ray emitting, plasmas created by SNR shocks have two characteristic
properties:
they are to a very good approximation
optically thin, and the ionization distribution of atoms
is often out of equilibrium.

Optically thin, X-ray emitting, plasmas are also found in a clusters of galaxies 
\citep{boehringer10},
colliding stellar winds, and the coronae of cool stars like our sun
\citep[hence the name coronal plasmas, see][for details on emission processes]{mewe99}.
In some cases, when the lines are not too much broadened by thermal
or turbulent motion, line emission may be affected by optical depth effects,
caused by resonant line scattering \citep{kaastra95}.
The overall spectrum will not be affected by this process, as resonant line
scattering does not destroy photons, but merely changes 
the direction of photons.
In coronal plasma's excitations and ionizations are caused predominantly
by the electrons colliding with the ions.
Most of the spectral characteristics of the thermal
emission (continuum shape, emission line ratios)
are determined by the {\em electron temperature}, which, as explained in
\sect~\ref{sec:collisionless} is not necessarily the same
as the proton/ion temperature.
The most important spectroscopic effect of the {\em ion} temperature
is thermal line broadening, which is difficult to measure with
the current generation of X-ray spectroscopic instruments (but see \sect~\ref{sec:thermaldoppler}).

Because SNR plasmas are optically thin for X-rays, 
X-ray spectroscopy is a powerful tool for measuring abundances
in SNRs, as no detailed radiative transfer models are needed. 
For old SNRs this can be used to reliably measure the abundances
of the interstellar medium \citep[e.g.][]{hughes98},
whereas for young SNRs X-ray spectroscopy is an important tool for measuring
abundances, and connect them to supernova ejecta yields for the various
types of supernova.
Here I discuss the basic radiation mechanisms. In depth reviews
of thermal X-ray emission are provided by \citet{mewe99,kaastra08}.

\subsubsection{Thermal continuum emission}
Thermal X-ray spectra consist of continuum emission caused by
bremsstrahlung  (free-free emission), recombination continuum 
(free-bound emission), and two-photon emission, the latter caused by the radiative
electron transition from a meta-stable quantum level.
For a Maxwellian energy distribution of the electrons, the
emissivity is given by
\begin{align}
\epsilon_{ff}  = & \frac{2^5\pi e^6}{3m_{\rm e}c^3}
\Bigl(\frac{2\pi}{3 k m_{\rm e}}\Bigr)^{1/2}
g_{\rm ff}(T_{\rm e}) T_{\rm e}^{-1/2}
\exp\Bigl( - \frac{h\nu}{kT_{\rm e}}\Bigr)\ 
 \\ \nonumber
&\times n_{\rm e}\sum_i  n_i Z_i^2\  
{\rm erg\, s^{-1} cm^{-3} Hz^{-1}},
\label{eq:bremss}
\end{align} 
with $g_{\rm ff} \approx 1$, the gaunt-factor, which has a frequency
dependence \citep[e.g.][]{rybicki}. The subscript $i$ denotes the
various ion species, each with charge $e Z_i$. 
The emissivity at a given temperature is therefore determined by
the factor $n_{\rm e}\sum_i  n_i Z_i^2$.
For solar or sub-solar abundances,
bremsstrahlung is dominated by electrons colliding with protons and helium ions. 
In spectral codes one therefore usually parametrizes the 
normalization factor in Eq.~\ref{eq:bremss} ($n_{\rm e}\sum_i  n_i Z_i^2$) 
with  $n_{\rm e} n_{\rm H}$
or $n^2_{\rm e}$. Since one observes the emission from plasma
of a given volume, spectral fitting codes usually use as a normalization factor for
the plasma  $\int n_{\rm e} n_{\rm H}dV$
or $\int n^2_{\rm e} dV$. This quantity is called the {\em emission measure} or $EM$. 
Of course one should also
take into account the distance, in order to convert to flux units. A popular X-ray spectral
fitting code like {\em xspec} uses the normalization factor
$\int  n_{\rm e} n_{\rm H}dV/(4\pi d^2)$.  

Using this type of normalization is valid as long as 
the helium to proton ratio is close to the cosmic ratio, and the continuum contribution for
elements beyond $Z=2$ can be neglected.
However,
for shocked supernova {\em ejecta}  electrons colliding with heavy ions can also be an important or even dominant
source of continuum radiation. 
If one does not take this into account one
can derive erroneous density and mass estimates from the bremsstrahlung
emissivities \citep[e.g.][]{vink96}. Note that the higher emissivity of a 
metal-rich 
plasma is not only caused by the factor $Z_i^2$, but also by the effect
that ionized metals produce more free electrons. For example,
each hydrogen atom contributes only 1 electron, whereas fully ionized
oxygen produces 8 electrons. 

\input{jvink_aarv_fig_continua}

Free-bound and two-photon emission are not always mentioned
as sources of
continuum radiation, but they can be a dominant source of continuum emission
\citep{kaastra08}, in particular for the metal-rich plasmas in young SNRs.
Free-bound emission arises as an electron is captured into one of the atomic
shells. The energy of the emitted photon is then 
$h\nu_n = E_{\rm e} + \chi_n$, with $E_{\rm e}$ the energy of the free electron,
and $\chi_n$ the ionization potential for an electron in level $n$.
An electron is more likely to be captured in a shell with high 
principal quantum number $n$, due to the high statistical weights of these shells.
As the energy difference for levels with high $n$ are small, 
free-bound emission shows a sharp edge near the series limits.
For a given energy level  the emissivity is given by 
\begin{align}
\epsilon_{\rm fb} = &
\Bigl(\frac{2}{\pi}\Bigr)^{1/2} n_{\rm e} n_{z+1} 
\frac{g_i}{g_{i+1}} \times \nonumber \\
&c \sigma(h\nu) \Bigl(\frac{h\nu}{{\chi_n}}\Bigr)
\Bigl(\frac{\chi_n^2}{m_{\rm e}c^2 kT}\Bigr)^{3/2}   \\ \nonumber
&\times \exp\Big( -\frac{(h\nu - \chi_n)}{kT_{\rm e}}\Bigr)\ 
{\rm erg\,s^{-1} cm^{-3} Hz^{-1}},
\end{align}
with $n_{z+1}$ the density of a ion with charge $z+1$, $g_{z+1}, g_z$ the statistical weights of the ion before and after
recombination, and $\sigma(h\nu)$ the photo-ionization cross section of the
ion in its final state. Since $\sigma(h\nu) \propto \nu^{-3}$, the 
spectral shape for $h\nu \gg \chi_n$ resembles that of thermal bremsstrahlung.
However, if 
$kT \ll h\nu$, a situation that will occur in photo-ionized  
or overionized plasmas,
free-bound emission results in narrow emission peaks near the
series limits of lines. These line-resembling features are called
radiative-recombination continua, which is usually shortened to RRCs.
RRCs have recently been identified in the X-ray spectra of a few mature SNRs,
suggesting the presence of overionized plasmas (\sect~\ref{sec:mixed}).

Two-photon emission results from electrons in meta\-stable states,
such as the $2s$ state of a hydrogen-like atom. Since decay to
the $1s$ level is forbidden (because $\Delta s=0$), it can either be collisionally
de-excited (unlikely in the rarified plasmas of SNRs),
or it can de-excite by emitting two photons, with the associated energy
distributed over two photons. 

The different types of thermal continuum processes are illustrated
in Fig.~\ref{fig:continua} for the case of a silicon-rich plasma out of
equilibrium with  $kT_{\rm e} = 1 $~keV.

\input{jvink_aarv_fig_nei}

\subsubsection{Non-equilibrium ionization}
\label{sec:nei}
The discussion of thermal emission so far pertains to all coronal
plasmas, whether they are found in clusters of galaxies, SNRs, or cool
stars, with the additional complication that 
young SNRs can have very metal-rich plasmas.
But there is another
important difference between the optical emission from SNRs and other
hot astrophysical plasmas:  SNR plasmas are often out of ionization
equilibrium. This is usually indicated with the term {\em non-equilibrium
ionization}, or {\em NEI}. The plasmas of  cool stars and clusters of galaxies are referred
to as  {\em collisional ionization equilibrium}, or {\em CIE}.

The reason that SNR
plasmas are in NEI is simply that, for the low densities involved, not enough
time has passed since the plasma was shocked, and per ion only a few
ionizing collisions have occurred for any given atom
\citep{itoh77}. 

The number fraction of atoms in a given ionization state $F_i$
is governed by the following differential equation:
\begin{equation}
\frac{1}{n_{\rm e}}\frac{dF_i}{dt} = 
\alpha_{i-1}(T) F_{i-1} - 
\big[\alpha_i(T)  + R_{i-1}(T)\big] F_i + R_i(T) F_{i+1},\label{eq_nei}
\end{equation}

with $\alpha_i(T)$ being the ionization rate for a given temperature, and
ion $i$, and $R_i$ the recombination rate.\footnote{For this equation it is assumed that
the fractions only increase by ionization from a lower ionization state
or recombination from a higher ionization state. This may not be strictly
true, if one takes into account inner shell ionizations, which may result
in multiple electrons being liberated.}

For CIE the ion fraction of a given state remains constant, thus
$dF_i/dt=0$. For NEI $dF_i/dt \neq 0$, and the ionization fractions have to be
solved using  the coupled
differential equations of all ion species $i$
as a function of time, or as indicated by Eq.~\ref{eq_nei} 
as a function of \net. To complicate matters, 
in reality neither $kT_{\rm e}$, nor
$n_{\rm e}$ are expected to be constant in time. 
The parameter \net\ is often referred to as the {\em ionization age}
of the plasma.

Equation~\ref{eq_nei} can be solved by direct integration, but
this is CPU intensive and not very practical when it comes to fitting
X-ray spectral data. A faster approach \citep{hughes85,jansen93,smith10} is
to rewrite Eq.~\ref{eq_nei} in matrix notation,
\begin{equation}
\frac{1}{n_{\rm e}}\frac{d{\bf F}}{dt} = {\bf A}\cdot {\bf F},
\end{equation} 
determine the
eigenvalues and eigenvectors of ${\bf A}$ and then solve for the uncoupled
equations 
\begin{equation}
\frac{1}{n_{\rm e}}\frac{d{\bf F'}}{dt} = {\bf \lambda}\cdot {\bf F'},
\end{equation}
with ${\bf \lambda}$ the diagonal matrix containing the eigenvalue, and
${\bf F'} = {\bf V}^{-1} {\bf F}$, with ${\bf V}$ containing the eigenvectors.

The main effect of NEI in young SNRs is that the ionization states at a given temperature 
are lower than in the CIE situation.
In Fig. \ref{fig:nei} the effect of NEI in young SNRs is illustrated. 
As an example consider the presence of
O VIII in Fig.~\ref{fig:nei}a, 
which peaks in CIE at $kT_{\rm e} = 0.2$~keV. Seeing a plasma
in a SNR with most of the oxygen in O VIII may lead to the conclusion
that the plasma is relatively cool. However, as 
Fig.~\ref{fig:nei}b shows, the temperature may very well be 
$kT_{\rm e} = 1$~keV, but with a low ionization parameter  of 
$n_{\rm e}t \approx 8\times 10^{9}$~\netunit.
For temperatures relevant for young SNRs ($kT_{\rm e} =0.5-4$~keV)
the time needed to reach CIE is around $n_{\rm e} t \approx 10^{12}$~eV.
Typical values for \net\ in SNRs are between $10^9- 5\times 10^{12}$~\netunit,
i.e. the plasmas of most SNRs are not in CIE. Note that the relevant
time scales for reaching CIE is similar to that for
reaching electron-proton temperature equilibration (\sect~\ref{sec:equilibration}).

Once a plasma has reached CIE one can have a reversal of the NEI
situation; i.e. the plasma can become {\em  overionized} instead of 
underionized,
if the cooling rate is faster than the recombination rate. 
This will be discussed
in connection to mixed-morphology SNRs in \sect~\ref{sec:mixed}.

NEI models as found in X-ray spectral codes can have various levels
of refinement, both concerning the atomic data \citep[a strong point
of the SPEX NEI code][]{kaastra03}, or in the additional physics taken
into account, such as non-equilibration of electron and ion temperatures
(\sect~\ref{sec:collisionless}),
and the ensuing gradients in the electron temperature behind the
shock front \citep[e.g. the {\em vpshock} model in XSPEC,][]{borkowski01b}.
Some models \citep[e.g.][]{hamilton83,jansen93,borkowski01b} 
also take into account the temperature structure
within a SNR in the Sedov-Taylor phase of its evolution 
(\sect~\ref{sec:hydro}).

%% file: jvink_aarv_fig_continua.tex
\begin{figure}
\centerline{
\includegraphics[angle=-90,width=\medfig]{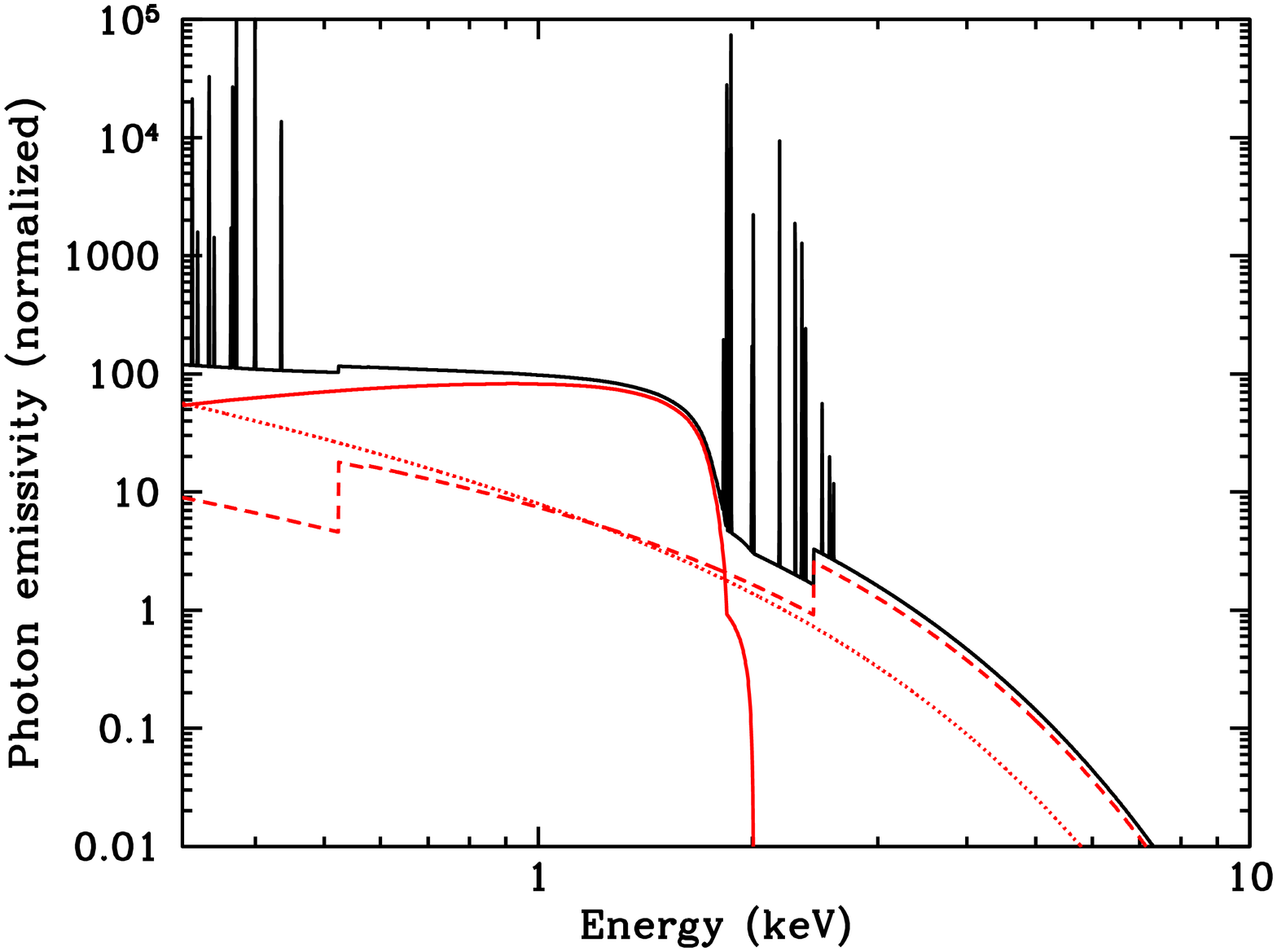}
}
\caption{
The emissivity of a pure silicon plasma out of ionization
equilibrium ($kT_{\rm e}=1$~keV, \net $=5\times 10^{10}$~\netunit).
Shown are the contributions of two-photon emission (red solid line), 
free-bound continuum (red dashed line)
and bremsstrahlung (free-free emission, red dotted line). The total emissivity
is also shown, including Si-L and Si-K shell line emission
\citep[based on calculations made with the spectral code \spex,][]{kaastra03}.
\label{fig:continua}
}
\end{figure}

%% file: jvink_aarv_fig_nei.tex
\begin{figure*}
\centerline{
  \includegraphics[width=\textwidth]{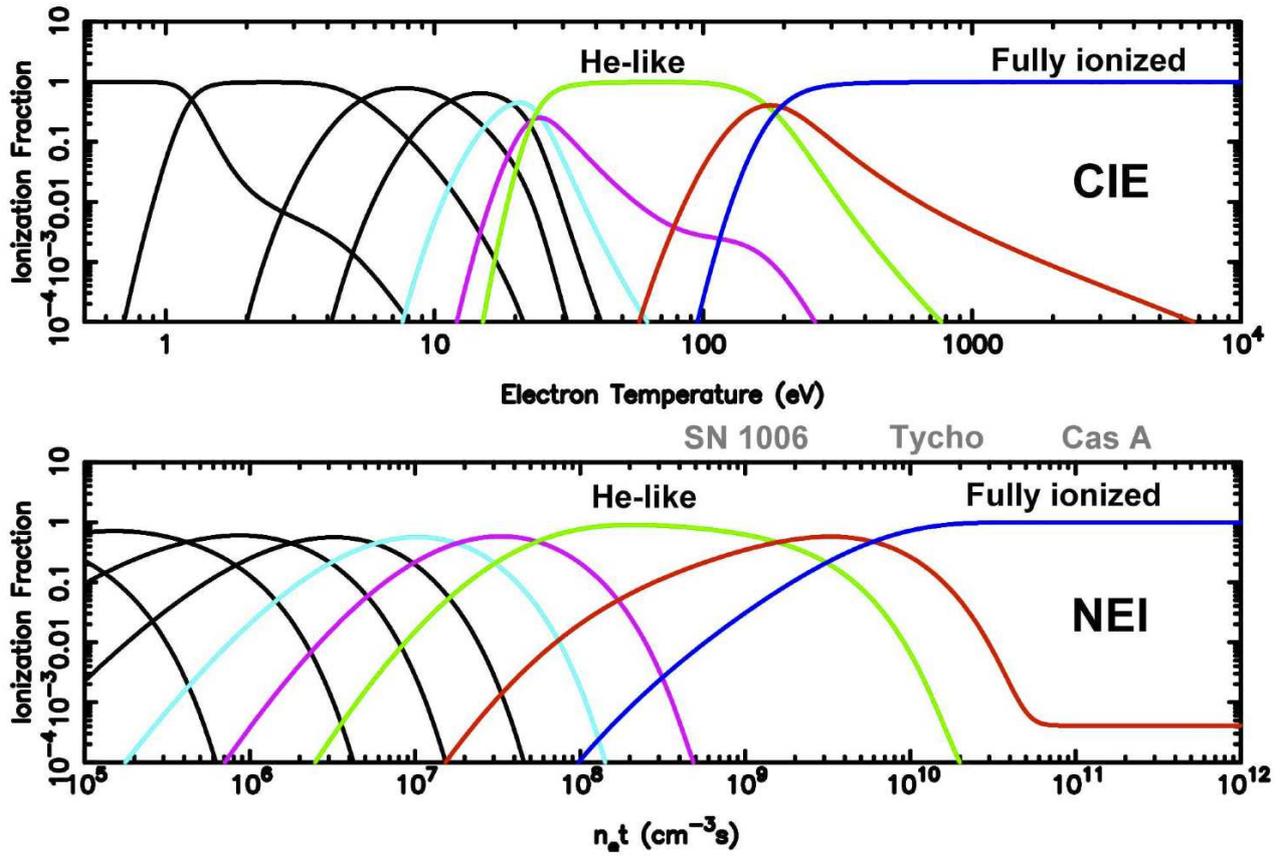}
}
\caption{The effects of non-equilibration ionization (NEI)
illustrated for oxygen.
Both panels look very similar, but the top panel
shows the oxygen ionization fraction as a function of electron
{\em temperature} for collisional ionization equilibrium (CIE),
whereas the bottom panel shows the ionization fraction as function
of \net, and for a fixed  temperature of \kte$=1.5$~keV
\citep[ionization/recombination rates based on][]{shull82}.
Approximate, mean \net\ values for the plasma in the young SNRs
Cas A, Tycho and SN\,1006 are indicated. \citep[Figure earlier published
in][]{vink06b}
\label{fig:nei}}
\end{figure*}

%% file: jvink_aarv_lines.tex
\subsubsection{Line emission diagnostics}
\label{sec:linediag}
In SNRs line emission results from collisional excitation of ions, a process dominated
by electron-ion collisions. This is either direct
excitation or through recombination, which usually results in an ion
in an excited energy level. Since the density is very low, most
ions can be assumed to be in the ground state; and, once an excitation occurs,
collisional de-excitation or further excitation or ionization can be neglected.
This also means that the ionization balance can be treated independently
of the line emission properties \citep[see][for a full treatment]{mewe99}.

An important aspect of line emission that is characteristic
for NEI plasma 
\citep[and under certain circumstances photo-ionized plasmas:][]{liedahl99,kallman04}  is inner shell ionization. 
In this process an electron from an inner shell, for
example the K-shell ($n=1$) is
removed, whereas higher level atomic shells (L- or M-shells, $n=2, n=3$)
are still filled.
This results in an ion with a hole in the K-shell, but plenty of electrons
around from L, or M shells.
The ion can then either de-excite without radiation,
by filling the hole, and using the energy from that transition for
further ionization (called Aug\'er transitions),
or it can adjust radiatively, which is called fluorescence.
The likelihood for a radiative transition is
called the {\em fluorescence yield}, and is
higher for atoms with larger nuclear charge. For example the fluorescence
yield for a K-shell ionization of 
neutral iron (Fe I) for a K-shell transition is 34\%, for
neutral silicon it is 5\% and for neutral oxygen it is only 0.8\%
\citep{krause79,kallman04}. 
Because the radiative transition occurs in the presence
of L-shell and M-shell electrons, the effective
charge of the nucleus is reduced as compared to a K-shell
transition in the helium-like state. As a result the K$\alpha$
line energy is smaller, and rises slowly
as a function of ionization state (Fig.~\ref{fig:fek}).

\input{jvink_aarv_fig_fek}

Iron line emission is an important diagnostic tool for the state
of a SNR plasma, i.e. the average electron temperature and
ionization age, \net. This is even true for medium energy resolution
spectroscopy as provided with the CCD instruments on board \chandra, \xmm, and
\suz.
Because of its high fluorescence yield and high abundance,
Fe K-shell emission can be
observed for all ionization states of iron, provided that the
electron temperature is high enough ($kT_{\rm e} \gtrsim 2$~keV).
The average line energy of the Fe-K shell emission provides
information about the dominant ionization state (Fig.~\ref{fig:fek}).
For ionization states from Fe I to  Fe XVII the average Fe-K shell line
is close to 6.4 keV.

Fe has also prominent Fe-L-shell transitions in the 0.7-1.12 keV range.
These transitions occur for ionization states with electrons present
in the L-shell ($n=2$), i.e. Fe XVII to Fe XXIV,
with each ionization state having its own specific line transitions, which
increase on average in line energy for higher ionization states.
Fe-L-line emission occurs  for lower temperatures/ionization ages 
than Fe-K, i.e. $kT_{\rm e} \gtrsim 0.15$~keV.
Taken together, Fe-L- and Fe-K-shell emission 
can be used to accurately determine the ionization state of the plasma.
For example, Fe-K emission around 6.4~keV could be
caused by Fe XVII-XIX, or by lower ionization states, but in
the latter case no Fe-L emission should be present. 
The presence of Fe-K-line emission around
6.7~keV indicates the presence of Fe XXV (He-like Fe), and
around 6.96 keV Fe XXVI (H-like). High resolution X-ray spectroscopy
greatly improves the value of Fe-line diagnostics, as it allows one to resolve 
the Fe-L shell emission in individual lines. And for Fe-K shell emission 
it is possible to detect different ionization stage individually instead
of relying line emission centroids.

It is worth mentioning that K$\alpha$ line emission from low
ionization states of iron can be the result of
dust grains embedded in hot SNR plasmas \citep{borkowski97}.
Hot electrons can penetrate dust grains, giving rise to inner shell ionizations
inside the grains, whereas the emitted photon can escape
from small grains. In addition, dust grains are slowly destroyed in
hot plasmas due to dust sputtering, 
on a time scale of $\sim 10^{13}/n_{\rm e}$~s.
This results in a slow release of near neutral iron
into the hot plasma. This should give rise to
the presence of a broad range of ionization stages of
Fe inside the plasma.

\input{jvink_aarv_fig_triplet}
\input{jvink_aarv_fig_gratio}

Another line diagnostic, available for high resolution spectroscopy
($E/\Delta E \gtrsim 200$), 
is provided by the ``triplet'' line emission from He-like ions,
consisting of three prominent (blended) lines designated $w$ (resonance), $x+y$ (intercombination),
and $z$ (forbidden)
\citep[Fig.~\ref{fig:triplet}, see ][]{gabriel69,mewe99,liedahl99,porquet10}.\footnote{
The word ``triplet'' denotes here the presence of three closely spaced lines,
 and is not used in the usual sense indicating a mixed quantum state.}
The ratio $G \equiv (z+\, x+y)/w = (f+i)/r$ is of interest 
for SNRs, as it is sensitive to the temperature and ionization state of
the plasma \citep{mewe78,liedahl99}. 
For example, inner shell ionization of the Li-like state,
which has ground state configuration $1s^22s$, will result in an excited
He-like ion in the $1s2s ^3S_1$ state. Since de-excitation to
the ground state is forbidden for electric dipole radiation ($\Delta l = 0$),
this gives rise to forbidden, magnetic dipole, radiation. Thus, inner
shell excitation enhances the forbidden line transition, and acts as
a measure for the fraction of Li-like ions. 
This can be seen in Fig.~\ref{fig:triplet}, showing that
the G-ratio is high
for $n_{\rm e}t < 3\times 10^9$~\netunit\ for oxygen and
$n_{\rm e}t  < 2\times 10^{10}$~\netunit\ for silicon.
This figure also shows that 
for higher \net\ values $G$-ratio may fall below the CIE values.
The reason is that Li-like ions are no longer present, but at the same
time the recombination rate to the He-like state is low, because the atoms 
are still underionized, and collisions are more likely to result in ionizations
than recombinations.
Only when \net$\gtrsim 10^{11}$\netunit\ are the G-ratios  similar to those
for CIE at the same temperature.

Another potential influence on the $G$-ratio is resonant line scattering.
Resonant line scattering only influences the intensity of the resonant line,
as it scatters the resonant photon out of the line of sight. A very high
$G$-ratio (i.e. relatively strong forbidden line emission)  may therefore help to diagnoze
resonant line scattering, and its importance can be estimated,
if  \net\ and $kT_{\rm e}$ can be estimated independently from the $G$-ratio.
Note that, since resonant line scattering does not destroy the photon, from the other parts
of the SNR, in particular from the edges, there may be enhanced
resonant line emission.

The ratio $R\equiv z/(x+y)= f/i$ is for CIE plasmas a
sensitive diagnostic tool to measure electron densities, if the densities
are in the range $n_{\rm e} = 10^8-10^{13}$~\cc, which is much higher than
encountered in SNRs.
For SNRs
it is of more interest that the R-ratio is 
also sensitive to the ionization age, and to a lesser
extent on temperature (Fig.~\ref{fig:gratio}, right). The reason is that the R-ratio
in SNRs is mainly determined by inner shell ionization, which enhances
the forbidden line emission for low \net. Once the fraction of Li-like ions is negligible,
and inner shell ionizations are therefore no longer important,
the R-ratio becomes relatively flat. Note that the R-ratio may help to disentangle
resonant line scattering effects from pure NEI effects; resonant line scattering
does not influence the R-ratio.

%% file: jvink_aarv_fig_fek.tex
\begin{figure}
\centerline{
\includegraphics[angle=-90,width=\medfig]{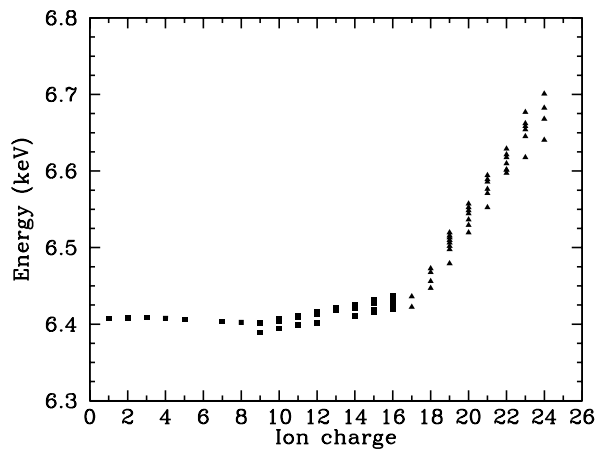}
}
\caption{
Fe-K shell line emission energies, as determined theoretically
\citep[squares, Fe II-Fe XVII,][]{mendoza03,mendoza04} 
and observationally \citep[triangles, Fe XVIII-Fe XXV,][]{beiersdorfer93}.
\label{fig:fek}
}
\end{figure}

%% file: jvink_aarv_fig_triplet.tex
\begin{figure}
\centerline{
\includegraphics[width=\medfig]{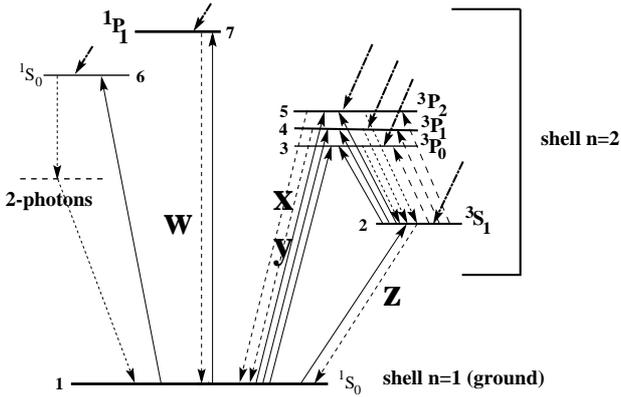}
}
\caption{
Simplified Grotrian diagram of He-like ions, with the transitions from
an excited $n=2$ state to the ground state \citep[taken from][]{porquet01}.
The triplet consists of the resonant line $r$ or $w$, the intercombination
line, $i=x+y$, and the forbidden line $f$ or $z$.
\label{fig:triplet}
}
\end{figure}

%% file: jvink_aarv_fig_gratio.tex
\begin{figure*}
\centerline{
\includegraphics[angle=-90,width=\twofig]{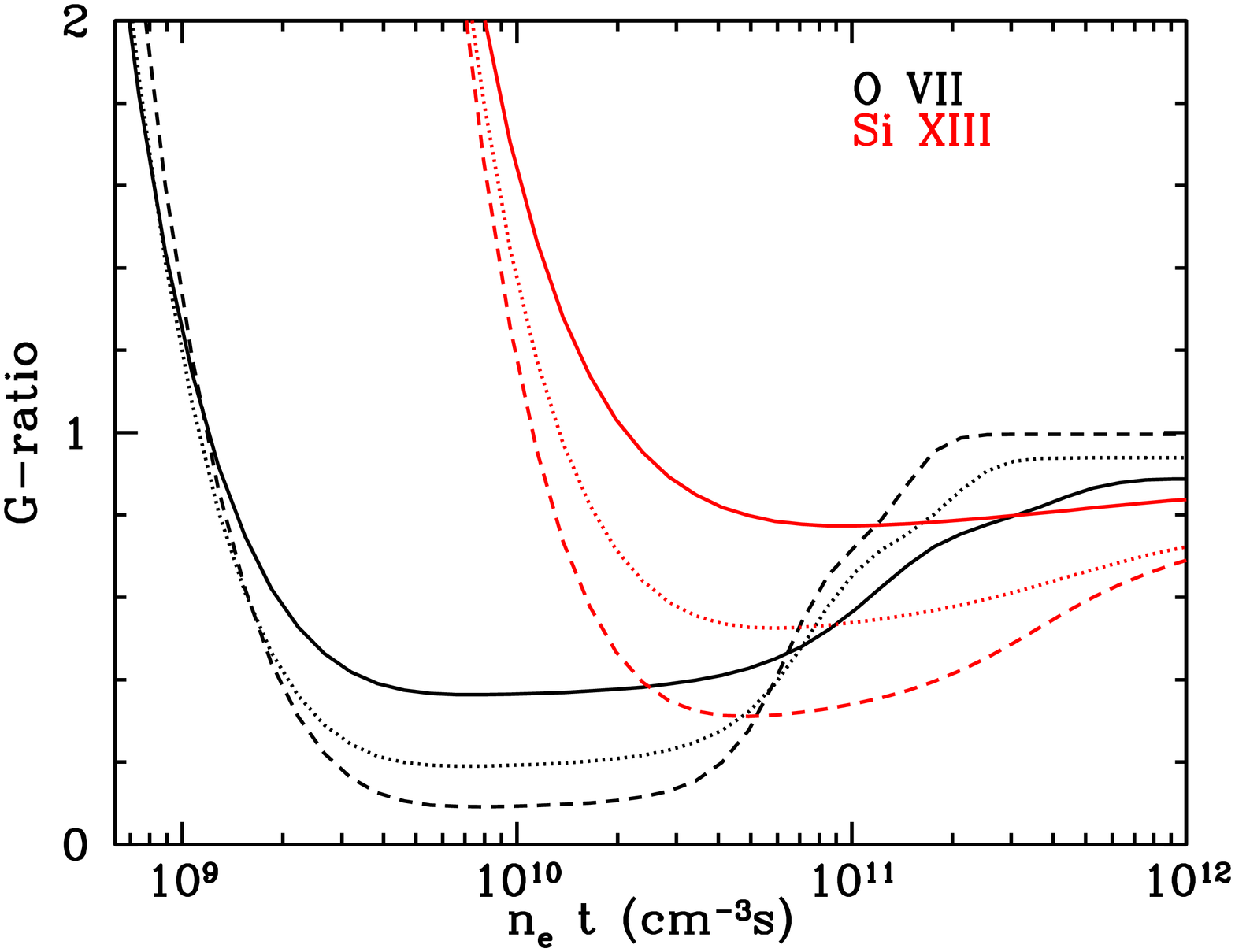}
\includegraphics[angle=-90,width=\twofig]{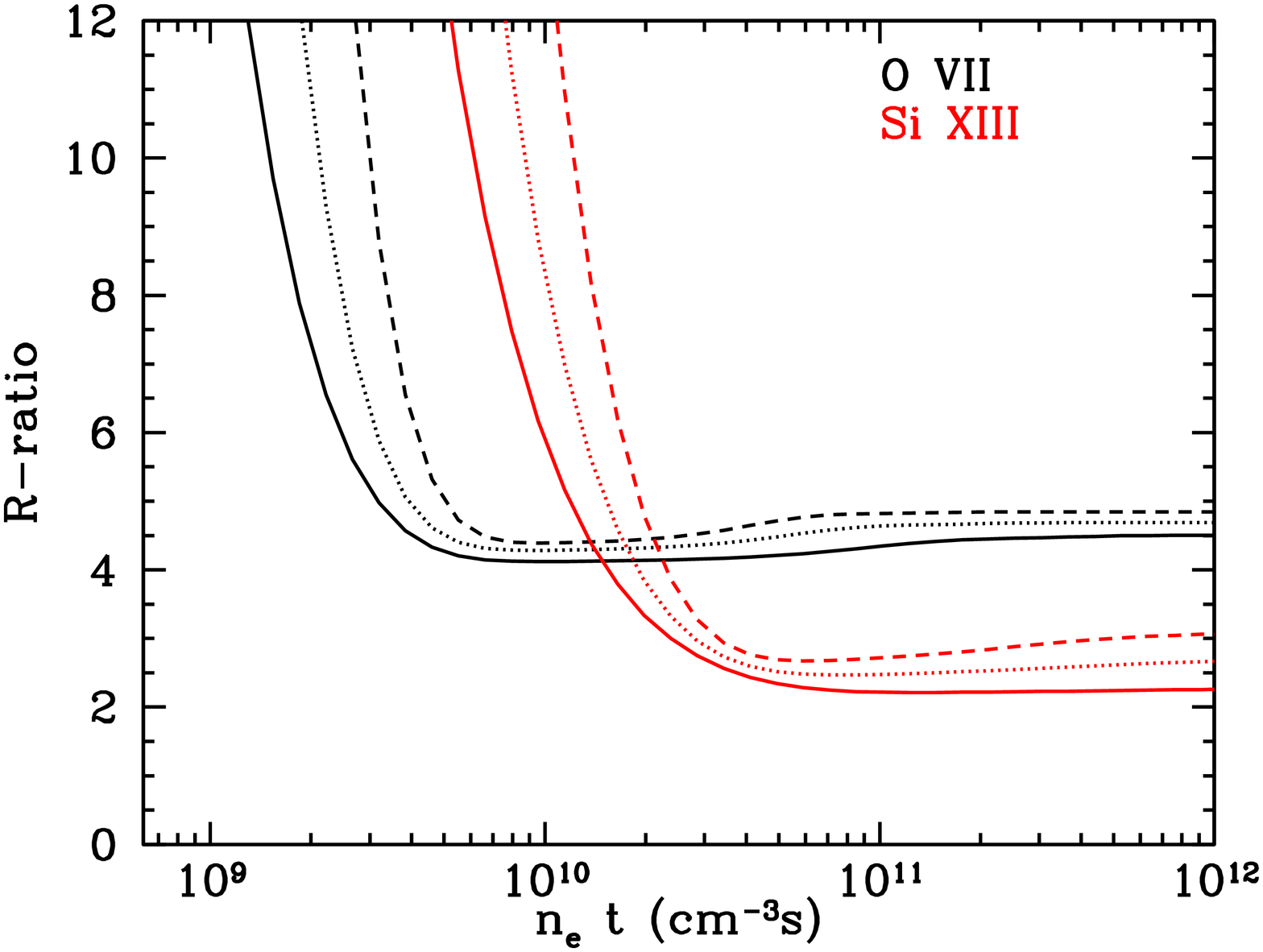}
}
\caption{
Left: The G-ratios $G=(f+i)/r$ for He-like oxygen and  silicon.
Three different temperatures are show 0.5~keV (solid line),
1~keV (dotted) and 2~keV (dashed). Right: Idem, but now
for the R-ratios $R=f/i$.
Note that these ratios are based on pure He-like ion
lines only,
in reality some blends may occur with L-like satellite lines
\citep[e.g.][]{vink03b}, which tend to
increase the G-ratio at low \net\ values.
In these figures the effects of blends with  line emission
from Li-like ions are not taken into account. These enhance the
G-ratio and R-ratio at low \net.
\citep[This figure is based on calculations by the author using the \spex\ code,][]{kaastra03}.
\label{fig:gratio}
}
\end{figure*}

%% file: jvink_aarv_radioactivity.tex
\subsection{Line emission associated with radioactivity}
\label{sec:radioactivity}

Apart from collisional processes one may also have line emission caused by radioactivity. 
Radioactive elements are produced during the lifetime of stars, and, of particular interest here, during the supernova explosion. 
The most important radioactive element is
$^{56}$Ni, which decays in 8.8 days\footnote{The decay time $\tau$ is the e-folding time, as opposed to
the half-life $\tau_{1/2}=\tau \ln 2$.} into $^{56}$Co, which subsequently decays into $^{56}$Fe
\citep[see Table~\ref{tab:radioactivity} for the details, and][for a review]{diehl98}.
This decay chain forms the dominant source of energy for the expanding supernova ejecta during the
first year after the explosion. It has a significant imprint on the light curves of supernovae. Moreover,
its final product, $^{56}$Fe,
is the most abundant Fe isotope in the Universe. A large part of the production comes
from Type Ia supernovae (\sect~\ref{sec:sn1a}), which produce typically 0.6~\msun\ per explosion. 

An element that is less abundantly produced is $^{44}$Ti, with typical expected yields in core collapse supernovae
of $10^{-5}-10^{-4}$~\msun\ \citep[e.g.][]{prantzos11b}. Type Ia supernovae probably produce less, unless the explosion
is triggered by an explosion at the surface of the white dwarf \citep[the so-called double detonation sub-Chandrasekhar model, e.g.][]{fink10}.
$^{44}$Ti has a much longer decay time than $^{56}$Ni, namely  85~yr \citep{ahmad06}. This
makes it of interest also for the SNR phase. $^{44}$Ti is an alpha-rich freeze-out product,
which means that its yield is determined by the amount of $\alpha$-particles left over after the initial
stages of explosive nuclear burning \citep{arnett96}. It is therefore sensitive
to the speed of expansion of  the inner layers of the ejecta, as this determines how rapidly
the density drops, which prevents further build up to more massive elements like $^{56}$Ni.
In addition,
the $^{44}$Ti yield is sensitive to the mass cut (the boundary between what is accreted onto the proto-neutron star and what will be ejected) and explosion asymmetries \citep{nagataki98}. This sensitivity to explosion conditions explains the astrophysical interest in $^{44}$Ti.

\input{jvink_aarv_tab_radioactivity}

\input{jvink_aarv_fig_ti44}

The most unambiguous, direct signature of radioactivity are the \gray\ lines
associated with their decay \citep[as predicted by][]{clayton69}.  The daughter products of radioactive elements are usually not in their nuclear ground states, but rapid de-excitation occurs through the emission of \gray\ lines (Table~\ref{tab:radioactivity}).
The nuclear decay lines of $^{56}$Co at  847~keV and 1238~keV  have been detected from SN\,1987A by several balloon and satellite experiments \citep[][see also \sect~\ref{sec:sn1987a}]{cook88,mahoney88,sandie88,teegarden89}. 
\citet{kurfess92} later report the detection of 122~keV line emission from $^{57}$Co by
the Oriented
Scintillation Spectrometer Experiment (\osse) on board the Compton
Gamma Ray Observatory (\cgro).
No supernova Type Ia has exploded near enough in modern times to firmly detect \gray\ lines,
but there was a hint of $^{56}$Co emission in \comptel\ observations of the bright
Type Ia supernova SN1991T \citep{morris97}, which had a distance of about 13.5~Mpc.

Apart from \gray\ line emission, the decay chain of $^{44}$Ti also results in hard X-ray line emission at 67.9~keV and 
78.4~keV, which are caused by the nuclear de-excitation of $^{44}$Sc (Fig.~\ref{fig:ti44}). 
This line emission, and the
the 1157 keV \gray\ line of $^{44}$Ca, have been detected for the young SNR Cas A (Sect.~\ref{sec:orich}) with
the \comptel, \sax-PDS, and \integral-IBIS instruments
\citep[respectively,][]{iyudin94,vink01a,renaud06}.
The flux per line of $(2.5\pm0.3)\ 10^{-5}$~cm\ s$^{-1}$cm$^{-3}$ 
 implies a rather large $^{44}$Ti yield of  $(1.6\pm0.3)\ 10^{-4}$~\msun \citep{renaud06}. This yield
 is comparable to what has been inferred for SN\,1987A, based on the late time light curve
 of SN\,1987A \citep{jerkstrand11}.  
 A $^{44}$Ti detection was also reported for the SNR RX J0852-4622 ("Vela jr"), based
 on measurements of the 1157~keV line with \comptel\  \citep{iyudin98x}. But \integral-IBIS
 measurements of the 68~keV and 78~keV lines are inconsistent with this detection \citep{renaud06b}.
 
 The radioactive processes in the decay chains of $^{56}$Ni, $^{57}$Ni and $^{44}$Ti are
 electron capture (EC) or beta-decay (e$^+$ in Table~\ref{tab:radioactivity}). The beta-decay
 process is a source of positrons and may be partially responsible for the 511~keV electron-positron annihilation
 line emission from the Galaxy \citep{prantzos11}. Most of the positrons released by the decay 
 of $^{56}$Co probably do not survive the supernova phase, but in the subsequent SNR phase 
 not many positrons may annihilate due to the low densities and high plasma temperature. No 511~keV
 line emission has yet been detected from a SNR \citep{kalemci06a,martin10}.
 
Apart from {\em nuclear} decay line emission, radioactivity also gives rise to X-ray line emission. The reason is that electron-captures results in a daughter product with a vacancy in the K-shell.
This vacancy may result in a K-shell transition that could be detected using more sensitive future X-ray telescopes such as \ixo/\athena\ 
\citep[][]{bookbinder10} .
\citet{leising01} lists in total 17 radioactive elements, but the most promising of them are 
$^{55}$Fe ($\rightarrow ^{55}$Mn, $\tau=3.9$~yr, K$\alpha$ energy 5.888 \& 5.899 keV), 
$^{44}$Ti ($\rightarrow ^{44}$Sc, $\tau=85$~yr,  4.086 \& 4.091 keV), 
$^{59}$Ni ($\rightarrow ^{59}$Co, $\tau=108$~kyr, 6.915 \& 6.930 keV) , 
and $^{53}$Mn ($\rightarrow ^{53}$Cr, $\tau=5.4$~Myr,  5.405 \& 5.415 keV).

For SNRs, K$\alpha$ line emission of $^{44}$Ti  and $^{59}$Ni is of particular interest, since the decay times 
are compatible with the lifetimes of SNRs. For $^{44}$Ti  the K-shell transitions form a complement
to the nuclear decay line emission. The advantage of the K-shell transition measurements is that
the spatial resolution of soft X-ray telescopes is in general  superior to  hard X-ray or \gray\ telescopes.
But it should be noted that the line transitions are not unique. 
In principle, other Sc isotopes may also produce the K-shell lines. But stable Sc is not a very abundant element. 
A more severe problem is that the K-shell transitions are also
a function of the ionization state of the atom (\sect~\ref{sec:linediag}), which may not
be a priori known, complicating the search for line emission. 
Nevertheless, future more sensitive telescopes  are expected to detect these lines \citep{leising01,hughes09}.
And one may then be able to determine the ionization state of the radioactive elements, which
allows one to position the $^{44}$Ti in the shocked or unshocked
parts of the ejecta. 

\citet{borkowski10} reported the detection of velocity broadened K$\alpha$ line emission from $^{44}$Ti
(or technically speaking from its daughter product Sc) for the youngest known Galactic SNR
G1.9+0.3 \citep[$\sim 100$~yr][]{carlton11}. The measured flux of $(0.35-2.4) \times 10^{-6}$\ ph cm$^{-2}$ s$^{-1}$ implies a $^{44}$Ti production of $(1-7)\times 10^{-5}$\msun. G1.9+0.3 is possibly a
Type Ia SNR (\sect~\ref{sec:typeia}), but the $^{44}$Ti yield is consistent with either a core collapse
or Type Ia origin.
A search for $^{44}$Ti associated K$\alpha$ line emission from Cas A \citep{theiling06} and SN\,1987A 
 \citep{leising06} did not yet result in detections.

%% file: jvink_aarv_tab_radioactivity.tex
\begin{table*}
\label{tab:radioactivity}
\begin{center}
\parskip 0mm
\caption{
Decay chains of the most important
radioactive explosive nucleosynthesis products and some of their
hard X-ray and
\gray\ signatures.\label{decay}}
{
\begin{tabular}{lllrlr}
\noalign{\smallskip}\hline\noalign{\smallskip}
&         &  &  Decay time & Process &Lines (keV)\\
\noalign{\smallskip}
\hline
\noalign{\smallskip}
$^{56}$Ni $\rightarrow$ &                         &         &8.8 d &
EC& 158, 812 \\
                        &$^{56}$Co $\rightarrow$  &$^{56}$Fe& 111.3 d& 
EC, $e^+$ (19\%) & 847, 1238\\ 
\noalign{\smallskip}
$^{57}$Ni $\rightarrow$ &                         &         & 52 hr & 
EC &1370\\
                        &$^{57}$Co $\rightarrow$ & $^{57}$Fe & 390 d & 
EC & 122\\
\noalign{\smallskip}
$^{44}$Ti $\rightarrow$ & & &86.0 yr & EC &67.9, 78.4\\
  &$^{44}$Sc $\rightarrow$ &  $^{44}$Ca& 5.7 hr & 
 $e^+$, EC (1\%) &1157\\
\noalign{\smallskip}
\hline
\end{tabular}
}
\end{center}
\end{table*}

%% file: jvink_aarv_fig_ti44.tex
\begin{figure}
\centerline{
\includegraphics[width=\medfig]{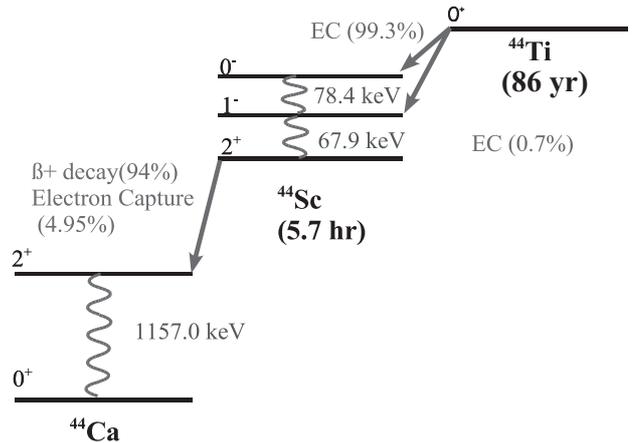}}
\caption{The decay chain of $^{44}$Ti with the indicated channels, electron capture (EC) and
beta-decay.\label{fig:ti44}}
\end{figure}

%% file: jvink_aarv_nonthermal.tex
\subsubsection{X-ray synchrotron radiation}
Relativistic electrons gyrating in a magnetic field emit
synchrotron radiation \citep{ginzburg67}.\footnote{For more massive charged particles synchrotron
radiation is still emitted, but at a  much lower efficiency.}
In most cases encountered in astrophysics, synchrotron radiation is caused by a
{\em non-thermal} population of electrons, which often well approximated over
a large energy range by a power-law distribution, certainly if first order Fermi acceleration
is responsible for the production of the relativistic electron populations. 

Sources of synchrotron radiation usually produce this radiation over a very
wide range in frequency, from the low frequency radio bands 
($\sim 10$~MHz) up to X-rays ($\sim 10^{18}$~Hz).
Most SNRs are radio synchrotron sources, with synchrotron radiation
coming from the SNR shell, but for composite SNRs also from the embedded
pulsar wind nebulae \footnote{See for example the catalogue by Green D. A., 2009, `A Catalogue of Galactic Supernova Remnants', Astrophysics Group, Cavendish Laboratory, Cambridge, United Kingdom (available at "href{http://www.mrao.cam.ac.uk/surveys/snrs/}"}.
X-ray synchrotron emission from SNRs was traditionally associated with the
presence of a pulsar wind nebula, but only relatively recently has X-ray synchrotron emission from
young SNR shells been established 
 \citep[][see \sect~\ref{sec:synchrotron}]{koyama95}.

A relativistic electron with  energy $E$ moving in a magnetic field with strength $B$
will
emit synchrotron radiation with a maximum emission at the characteristic frequency
$\nu_{\rm ch}$ given by  \citep{ginzburg65}
\begin{align}
\nu_{\rm ch}  &= 1.8 \times 10^{18} B_\perp \Bigl(\frac{E}{1\, {\rm erg}}\Bigr)^2\ {\rm Hz}, \nonumber\\
h\nu_{\rm ch} &= 13.9 \Bigl(\frac{B_\perp}{100\, {\rm \mu G}}\Bigr) \Bigl(\frac{E}{100\, {\rm TeV}}\Bigr)^2\ {\rm keV},\label{eq:nu_char}
\end{align}
with $B_\perp\approx \sqrt{2/3}B$ the magnetic-field component perpendicular to the 
motion of the electron.
For the typical magnetic fields inside SNRs, $B=10-500~\mu$G, 
the emitting electrons have energies of 10-100 TeV. These electrons
may therefore also be responsible for many of the detected TeV
gamma-ray emission \citep[e.g.][]{aharonian01,aharonian04}. 
However,
pion production by high energy ions also gives rise to TeV gamma-ray emission,
and there is an on going debate on which radiation mechanism
dominates the  TeV emission from SNRs \citep[see \sect~\ref{sec:xray_gammarays}, and][for a review]{hinton09}.

An electron emitting synchrotron radiation suffers radiative losses
with a rate:
\begin{align}
\label{eq:syn_em}
\frac{dE}{dt} &= -\frac{4}{3}\sigma_{\rm T} c \Bigl(\frac{E}{m_{\rm e}c^2}\Bigr)^2\frac{B_\perp^2}{8\pi} \\ \nonumber
&=
-4.05\times 10^{-7}\Bigl( \frac{E}{100~{\rm TeV}}\Bigr)^2 \Bigl(
\frac{B_\perp}{100\, {\rm \mu G}}
\Bigr)^2~{\rm erg\ s^{-1}},
\end{align}
with $\sigma_{\rm T}$ the Thomson cross section. 
Usually one includes into $B_\perp$  a term
to include the effect of losses due to inverse Compton scattering, replacing $B_\perp$ by
$B_{\rm eff}^2= B_{\rm rad}^2+B_\perp^2$,
with $B_{\rm rad}=3.3~\mu$G, the value for which the magnetic-field 
energy density
equals the radiation energy density of the cosmic microwave background.\footnote{
$B_{\rm rad}^2/(8\pi)=u_{\rm CMB}=4.2\times 10^{-13}$~erg\, cm$^{-3}$.}
This corresponds to a synchrotron loss time scale of
\begin{equation}
\tau_{\rm syn}= \frac{E}{dE/dt}= 12.5\  \Bigl( \frac{E}{100~{\rm TeV}}\Bigr)^{-1} \Bigr(\frac{B_{\rm eff}}{{100 \rm \mu G}}\Bigr)^{-2} {\ \rm yr}.\label{eq:tau_syn}
\end{equation}

For a population of relativistic electrons with 
a power-law energy distribution,
the spectral index, $\alpha$, of the synchrotron flux density spectrum is related to the 
spectral index of the electron energy distribution, $q$, as $\alpha = (q - 1)/2$.
The typical spectral radio index of young SNRs is
$\alpha \approx 0.6$,
indicating that electrons have an energy distribution with $q \approx 2.2$,
close to the predictions from first order Fermi acceleration 
(\sect~\ref{sec:cr}).
For X-ray emission it is more common to use the photon index, $\Gamma = \alpha +1$.

Typically, X-ray synchrotron spectra of young SNRs have rather
steep indices,
$\Gamma = 2-3.5$,
indicating a rather steep underlying electron energy distribution.
This steepness indicates that the synchrotron X-ray emission is caused
by electrons close to the 
maximum energy of the relativistic electron distribution. 
This maximum
is usually for energies at which 
acceleration gains are comparable to the radiative losses, the so-called
{\em  loss-limited case}. Alternatively, the electron energy spectrum breaks off,
because the shock acceleration process has not acted for long enough
time for the electrons to reach higher energies. This is the so-called
{\em age-limited case} \citep{reynolds98}.
In the age-limited case the underlying electron spectrum has an exponential
cut-off in energy $\propto \exp( -E/E_{\rm max})$, whereas in the 
loss-limited case
the cut-off is expected to be steeper $\propto \exp( -\{E/E_{\rm max}\}^2)$ \citep{zirakashvili07}.
Because the synchrotron emissivity function for a given electron energy
has a rather broad spectral
profile \citep{ginzburg65}, the rather steep cut-off in electron energies results in a much more gradual roll over of
the synchrotron spectrum:
\begin{equation}
n(h\nu) \propto (h\nu)^{-\Gamma} \exp\Bigl\{-\Bigl(\frac{h\nu}{h\nu_{\rm cut-off}} \Bigr)^{1/2}\Bigr\}.
\end{equation}

The cut-off photon energy for the loss-limited case can be calculated by combining 
Eq.~\ref{eq:tau_cr} and  Eq.~\ref{eq:tau_syn}, which gives
\begin{align}
\label{eq:emax}
\frac{E_{\rm max}}{100{\ \rm TeV}} \approx &
0.32 \eta^{-1/2} \Bigl(\frac{B_{\rm eff}}{100\ {\rm \mu G}}\Bigr)^{-1/2} \\ \nonumber
&\times \Bigl(\frac{V_s}{5000\ {\rm km\,s^{-1}}}\Bigr) \Bigl(\frac{\chi_4-\frac{1}{4}}{\chi_4^2}\Bigr)^{1/2}.
\end{align}
Inserting this in Eq.~\ref{eq:nu_char} shows that the cut-off photon energy, $h\nu_{\rm cut-off}$, is independent
of the magnetic field \citep{aharonian99}:
\begin{equation}
h\nu_{\rm cut-off} = 1.4 \eta^{-1} \Bigl(\frac{\chi_4-\frac{1}{4}}{\chi_4^2}\Bigr) \Bigl(\frac{V_s}{5000\ {\rm km\,s^{-1}}}\Bigr)^2\ {\rm keV}. 
\label{eq:syn_max}
\end{equation}
See \citet{zirakashvili07} for a more detailed derivation. This equation indicates that 
X-ray synchrotron emission requires the relatively high shock velocities ($\gtrsim 2000$~\kms) only
encountered in young SNRs.
Moreover, the detection of X-ray synchrotron emission from SNR shells
suggests that the diffusion coefficients cannot be too large compared
to the Bohm-diffusion case (\sect~\ref{sec:cr}), i.e. $\eta\lesssim 10$.

\subsubsection{Non-thermal bremsstrahlung}
\label{sec:nonthermal_brems}
Apart from synchrotron radiation, non-thermal X-ray emission can also
result from lower energy electrons through 
bremsstrahlung and inverse Compton scattering.
Inverse Compton scattering is for SNRs important in  GeV-TeV \gray\
band \citep{hinton09}, but for the magnetic fields inside
SNRs, $B \approx 5- 500~\mu$G, it is generally not expected to be important
in the soft X-ray band. 

Bremsstrahlung 
caused by the presence of
a non-thermal electron distribution has been considered as 
a possible source of X-ray continuum emission for some time
\citep[e.g.][]{asvarov90,vink97,bleeker01,laming01b}. Since the electrons that
produce X-ray bremsstrahlung have non-relativistic, to mildly
relativistic energies, identifying non-thermal bremsstrahlung would
help to obtain information about the low energy end of the electron
cosmic-ray distribution. This relates to the poorly known 
electron cosmic-ray injection process (\sect~\ref{sec:cr}).
However, it is unlikely that non-thermal bremsstrahlung contributes
enough radiation to identify it with the current generation of hard X-ray
telescopes. The reason is that the Coulomb interactions between
thermal and non-thermal electrons thermalizes the low energy tail
of the electrons on a relatively short timescale, \net$~\sim 10^{10}$~\netunit,
so that non-thermal bremsstrahlung can only be expected in a narrow
region close the shock front \citep[Fig.~\ref{fig:suprathermal}][]{vink08a},
or for those SNRs that have an overall low ionization age. 
Such a low ionization age is
for example encountered in SN\,1006, but here the X-ray continuum is overwhelmed by
synchrotron  radiation (\sect~\ref{sec:synchrotron}). The low ionization
age is caused by a low density of the plasma, which also makes that the
bremsstrahlung emissivity is not very large (Eq.~\ref{eq:bremss}).

\input{jvink_aarv_fig_suprathermal}

%% file: jvink_aarv_fig_suprathermal.tex
\begin{figure}
       \centerline{
\includegraphics[angle=-90,width=\medfig]{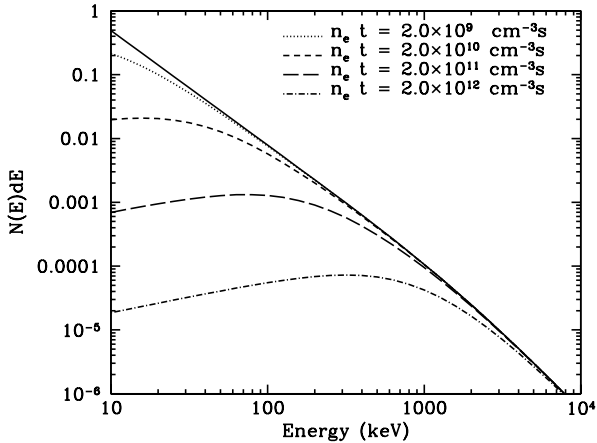}}
\caption{
The fate of a suprathermal electron distribution exchanging energy
with thermal electrons
through Coulomb collisions, as a function
of \net\ \citep{vink08a}.\label{fig:suprathermal}
}
\end{figure}

%% file: jvink_aarv_xray_instruments.tex
\input{jvink_aarv_fig_ccd_vs_rgs}

Before reviewing the most recent results obtained by the current generation
of X-ray missions capable of either high resolution spectroscopy and/or
imaging spectroscopy, \chandra, \xmm, and \suz, I describe some
of the characteristics of these telescopes and their instruments
in the context of studying SNRs.

All the aforementioned X-ray telescopes have multiple instruments on board,
with the CCD-detectors being the work horses 
for all three missions. X-ray detecting CCDs were first used
by the Broad Band X-ray Telecsope \citep[BBXRT,][]{petre93},
flown for ten days on the space shuttle Columbia, 
but only with the launch of the \asca\ satellite \citep{tanaka94}
did the results of
CCDs imaging spectroscopy become available to the broad community
of X-ray astronomers.
X-ray CCD-detectors 
are suitable for imaging spectroscopy with medium
spectral resolution ($R=E/\Delta E \approx 10-43$ FWHM for 0.5-6 keV)
and imaging resolution close to the telescope resolution.\footnote{
A predecessor of the CCD-detectors was the Solid State Spectrometer (SSS)
on board the Einstein satellite, which did not provide spatial information, but
an energy resolution similar to present day CCD-detectors \citep{joyce78}.}
The \chandra\ telescope provides 
the highest spatial resolution, which is slightly undersampled by the
CCD pixel size of 0.49\arcsec. The CCDs themselves
come in various types, even within one instrument. The CCD-detectors used
in current observatories go by the names
ACIS (AXAF CCD Imaging Spectrometer) on board \chandra, EPIC (European
Photon Imaging Camera) behind the \xmm\ telescopes, with EPIC-MOS 
\citep{turner01} behind two of the telescopes, and EPIC-pn behind the
third \citep{strueder01}, and XIS (X-ray Imaging Spectrometer) for
\suz. XIS consists also of four separate detectors behind
four independent telescopes.

CCD-detectors are excellent for studying the spatial distribution
of elemental abundances inside SNRs, 
or identifying regions with likely synchrotron
radiation contributions. They cannot be used to resolve Fe-L complexes,
or the He-like triplets (\sect~\ref{sec:linediag}), which offer more precise
diagnostic power to measure temperatures and ionization parameters.
They are also not good enough to measure Doppler shifts for velocities
below a few thousand km\,s$^{-1}$.

In addition to CCD camera's, \chandra\ and \xmm\ carry
grating spectrometers\footnote{\suz\ has another high
resolution X-ray spectrometer on board, XRS, which is based on the calorimetric
detection method. Unfortunately this instrument failed shortly after the launch
of \suz.},
which give high spectral resolution for point sources, but, compared
to the CCD instruments, have a smaller effective area.

\input{jvink_aarv_fig_g292_triplet}

\xmm\  has a Reflection Grating Spectrometer \citep[RGS,][]{denherder01},
consisting of two gratings behind two of the X-ray three telescopes. 
It covers the soft part of the X-ray spectrum from 6-38~\AA\
(0.32-2 keV).
About 50\%
of the X-rays go through the gratings, and the remaining fall on the EPIC-MOS
instruments. As a consequence, the RGS gathers data simultaneous with
EPIC. The spectral reflection angle for 
the RGS is large ($\sim 4.8$\deg)
compared to those of the transmission spectrometers of
\chandra. As a result an extended source in a monochromatic line
gets distorted (squeezed) in one direction. The design of the RGS is 
such that the spectrum is stretched over a large detecting area, with 
the spectrum being detected by nine CCDs for both RGS1 and RGS2.\footnote{
Both for RGS1 and RGS2 there is one CCD not working, luckily  intended to cover
different wavelengths. So the current
number of CCDs is 8 per grating.} 
The CCDs' intrinsic energy resolution are used to separate spectra of
different spectral orders.
The large physical size of the spectral extraction region has as a 
disadvantage that the background induced
by energetic particles is relatively large. But it has the 
advantage that the RGS is less sensitive to the
source extent, with the additional line width caused by source
extent, $\Delta \phi$,  
given by 
\begin{equation}
\Delta \lambda = 0.124 (\Delta \phi/1^\prime)~{\rm \AA}.\label{eq:rgs}
\end{equation}
So RGS spectra from typical young SNRs in the Magellanic Clouds, which have
$\Delta \phi \approx 0.5$\arcmin, are only mildly deteriorated compared
to those of  point sources with an additional broadening of
$\Delta \lambda= 0.06$~\AA. This is
comparable to the intrinsic resolution of the RGS, $\Delta \lambda= 0.06$~\AA.
Adding these numbers quadratically shows that the
spectral resolution for those SNRs is
$\lambda/\Delta \lambda =E/\Delta E \approx 265$ at $\lambda = 22.5$~\AA\ 
(i.e.around the O VII triplet, \sect~\ref{sec:linediag}).
Even for a source extent of 4\arcmin, one still obtains
a spectral resolution of 40, outperforming CCDs at long wavelengths
and comparable to the spectral resolution of
CCD-detectors at 6~keV.
If the extended object is dominated by some particularly bright regions, the spectral
resolution may be very good, as illustrated with the O VII triplet
line emission of SNR G292.0+1.8 in Fig.~\ref{fig:g292_triplet}.
The gain in spectral resolution obtained by using the RGS compared to
the CCD data is illustrated in Fig.~\ref{fig:mosrgs}.

\input{jvink_aarv_fig_hetg}

The \chandra\ transmission gratings come in two varieties, the High Energy
Transmission Grating Spectrometer \citep[HETGS,][]{canizares05} and the Low Energy Transmission Grating Spectrometer (LETGS), which provide spectral coverage in the range 
1.5-30~\AA\ (HETGS) and 1.2-175~\AA\ (LETGS).
The gratings for both instruments can be put into the 
optical path of the mirror assembly. The HETGS consists of
two grating arrays, a Medium Energy Grating (MEG) and High Energy Grating (HEG),placed at a small angle with respect to another. This
results in an X-shaped dispersion pattern on the detectors,
with the zero order image in the center of the X.
Since they are transmission gratings, the HETGS and LETGS have plus and minus
order spectra. Only one order is shown in Fig.~\ref{fig:hetg}.
The detectors onto which the spectra are dispersed are the normal
imaging detectors of \chandra, i.e. either the ACIS (CCD) instrument or
the High Resolution Camera (HRC),
a microchannel plate detector. For the HETGS the ACIS-S  is the instrument
of choice as it allows for the separation of different spectral orders.
The LETGS is either used with the ACIS-S array, if the long wavelength
range is not of interest, or with the 
HRC, which offers a longer wavelength
range and smaller pixel size, but does not offer an intrinsic energy
resolution and has lower quantum efficiency.

The dispersion angles of the HETGS and LETGS are much smaller than those
of the \xmm-RGS. As a result lines are much more narrowly spaced
on the detector, with smaller detector pixels to match.
However, for extended sources this means that the monochromatic image
of a source in a given line will easily start overlapping with the
images from nearby lines. This is even true for moderately
extended sources, such as the already mentioned Magellanic Cloud SNRs.
In numbers: the degradation in spectral resolution is (Eq.~\ref{eq:rgs})
0.67\AA/\arcmin, 1.33\AA/\arcmin, and
3.33\AA/\arcmin\ for the HEG, MEG and LETGS, respectively \citep{dewey02}.
The effect of source extent on the spectra 
is illustrated for the HETGS in Fig.~\ref{fig:hetg}, which gives an impression
of the challenges faced when analyzing HETGS SNR data.
Note that the complications are more severe for Type Ia SNRs (\sect~\ref{sec:typeia}), 
because they have bright Fe-L lines,
which, due to the richness of the line emission, easily 
blend together for extended sources. This is illustrated for
N103B in Fig.~\ref{fig:hetg} \citep[see also][]{dewey02}.
The situation is much better for the spectrum of the core collapse
SNR 1E0102.2-7219, whose intrinsic spectrum is less complex
(Fig.~\ref{fig:mosrgs}). 

Given that small SNRs are already difficult
to analyze, one would assume that
the situation is almost hopeless for large SNRs. However,
HETGS observations of a large SNR like Cas A (5\arcmin) still provide 
important results. But instead of obtaining high resolution spectra of the
SNR as a whole, one can obtain spectral properties of small, 
X-ray bright spots, embedded
in a background of diffuse emission from other parts of the SNR \citep{dewey07}.
In general, for extended sources, the RGS performs better for obtaining
high resolution spectra of the SNR as a whole, whereas the HETGS will provide
spectral information that relates to spectral properties
in localized regions of the SNR.

The problems of obtaining high resolution X-ray spectra
of extended sources will be over once the next generation of
spectrometers will be put into space. These will in all likelihood be
based on the calorimetric principle; an X-ray photon heats a small metal
element,
and the increase and subsequent decrease in temperature is measured with
high sensitivity, using, for example, Transition Edge Sensors (TESs). TESs
are
superconducting devices operated near the critical temperature, where
a small change in temperature leads to a strong response in conductivity.
The spectral resolution can be as good as $\Delta E= 2$~eV at 6 keV.
The small elements can be built into arrays, thereby allowing for
imaging, high resolution spectroscopy. Currently, microcalorimeters
are planned for the Japanese/US Astro-H mission \citep{takahashi08} 
and the proposed ESA mission Athena \footnote{Athena is a simplified
version of the International X-ray Observatory
\citep[IXO,][]{bookbinder10}}.

%% file: jvink_aarv_fig_ccd_vs_rgs.tex
\begin{figure*}
\centerline{
\includegraphics[angle=-90,width=\bigfig]{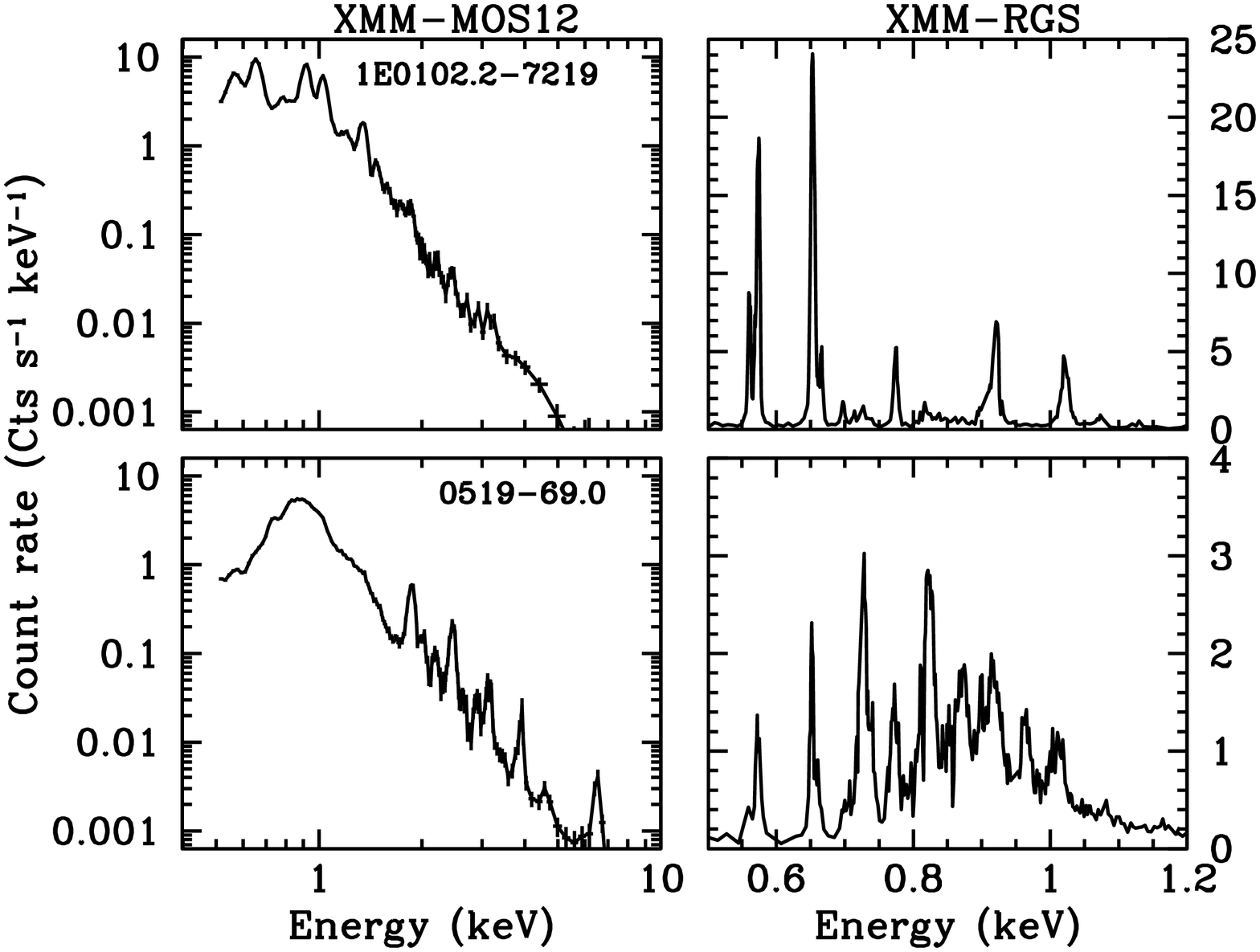}
}
\caption{
CCD (\xmm-MOS) and Reflective Grating Spectrometer (\xmm-RGS)
spectra of an oxygen-rich (1E0102.2-7219) and a Type Ia SNR (0519-69.0).
The figure illustrates the gain in spectral resolution of the RGS
instrument over CCD-detectors, even for mildly extended objects;
both SNRs have an extent of about 30\arcsec.
The figure also illustrates the  differences between
Type Ia spectra and core collapse SNR spectra, with the former
Type Ia SNR being dominated
by Fe-L emission (see \sect~\ref{sec:typeia} and \ref{sec:orich}).
\label{fig:mosrgs}
}
\end{figure*}

%% file: jvink_aarv_fig_g292_triplet.tex
\begin{figure}
\centerline{
\includegraphics[angle=-90,width=\medfig]{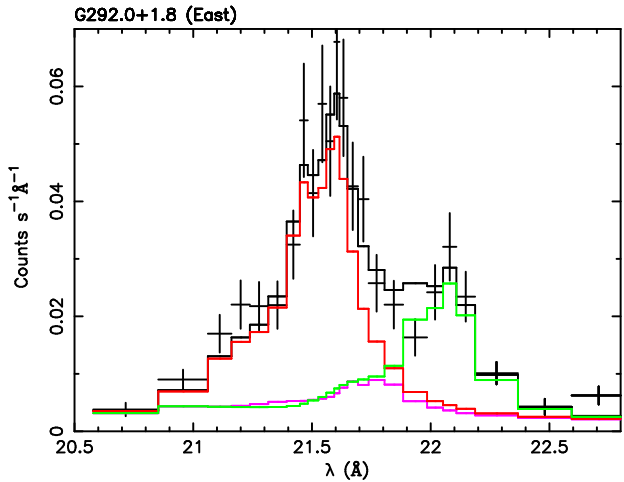}
}
\caption{
The O VII triplet (\sect~\ref{sec:linediag}) as observed with
the \xmm\ RGS in the oxygen-rich SNR G292.0+1.8 (\sect~\ref{sec:orich}). 
The RGS still offers
a good resolution despite the fact that G292.0+1.8 has an angular
size of nearly 8\arcmin. In this case the presence of a bright, narrow, bar-like
feature helped to obtain a good spectral resolution.
The red line indicates the emission from the resonance line, magenta
the intercombination line, and green the forbidden line.
(Figure first produced in \cite*{vink04f}.)
\label{fig:g292_triplet}
}
\end{figure}

%% file: jvink_aarv_fig_hetg.tex
\begin{figure}
\centerline{\includegraphics[width=\medfig]{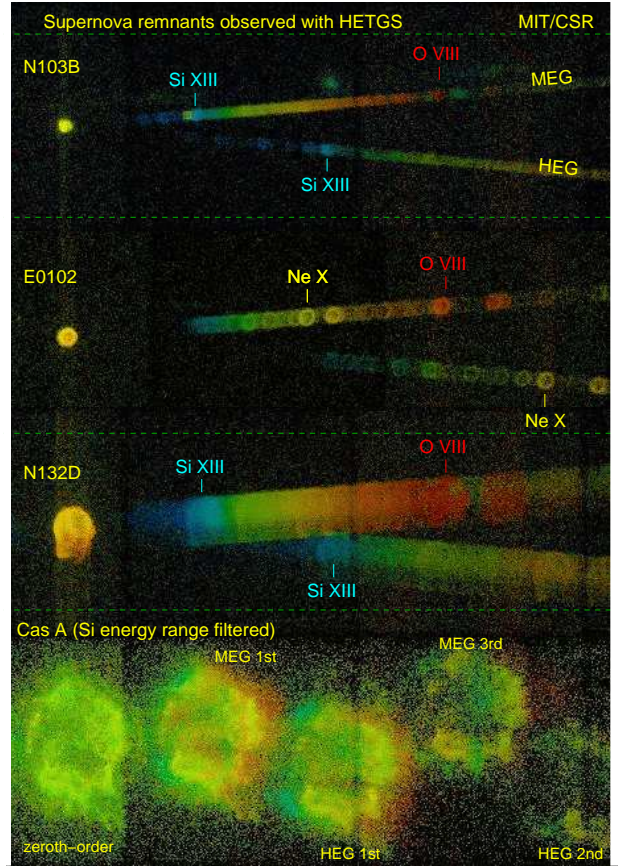}}
\caption{ 
An overview of \chandra\ high energy gratings 
observations of SNRs \citep{dewey02}.
(Figure kindly provided by D. Dewey.)\label{fig:hetg}
}
\end{figure}

%% file: jvink_aarv_typeia_snrs.tex
\input{jvink_aarv_tab_ia_snrs}

\input{jvink_aarv_fig_typeia_lmc}

As discussed in \sect~\ref{sec:sn1a}, 
thermonuclear or Type Ia
SNRs produce about a factor ten more mass in 
Fe-group elements than core collapse supernovae.
This is the reason  that Type Ia SNRs are often
characterized by strong Fe-L-shell emission, which cannot be resolved
with CCD instruments, but it
can be partially resolved by the current grating instruments 
\citep[\sect~\ref{sec:spectra}, Fig.~\ref{fig:mosrgs},  and ][]{hughes95}.
Many SNRs are now identified as Type Ia remnants based on the prominence
of the Fe-L emission (see Table~\ref{tab:Ia} for a list of likely
Type Ia SNRs and relevant references). 
In the case of two SNRs,
Tycho/SN1572 (Fig.~\ref{fig:tycho}) and SNR B0509-67.5 the Type Ia
identification has been confirmed by optical spectra of
the light echoes \citep[][truly spectacular results]{krause08b,rest08}.

Not all Type Ia SNRs display prominent Fe emission.
For example the X-ray spectra of SN\,1006 show hardly any evidence
for Fe-L emission \citep{vink03b}, although recently Fe-K emission
has been reported based on \suz\ observations 
\citep{yamaguchi06,yamaguchi08}.
For SN\,1006 the lack of Fe-L emission is partially due
to the low ionization age 
\citep[\net $\approx 2\times 10^9$~\netunit, e.g.][]{vink03b}, 
which makes that the dominant Fe ionization stage is below that of Fe~XVII.
In that case hardly any Fe-L line emission is produced,
but some Fe-K fluoresence line emission is still to be expected
(\sect~\ref{sec:linediag}).
However, optical/UV spectroscopy of a bright UV star behind SN\,1006
also shows a lack of absorption features from cool, unshocked,
Fe \citep{hamilton07}.
This is somewhat discomforting, but it may indicate that SN\,1006 produced
a relatively large amount of intermediate mass elements and relatively
small amounts of iron-group elements. 
This would still fit in with the 
optical results for Type Ia supernovae of \citet{mazzali07}, which allows
for the occurrence of fainter, Fe/Ni-poor  supernovae.
It is, however, more difficult to explain the lack of Fe with the models of
\citet{woosley07} that suggest that the total mass of
all Fe-group elements does not vary much among Type Ia SNe, but that there
is only a variation in the the relative amounts of radioactive $^{56}$Ni and stable Fe.
This is clearly of interest, and both models and the observational
data should be scrutinized further.

Another reason why a Type Ia SNR may not be very bright in Fe-L or Fe-K line emission is
that the Fe-rich layers has not yet been completely shock heated by the reverse shock.
For example, Tycho/SN\,1572, 
although far from devoid of Fe-L and Fe-K emission,
shows less Fe-L emission with respect to Si XIII emission than most other
Type Ia SNRs \citep[][]{hwang97,decourchelle01,hughes95}. 

The effects of the progression of the reverse shock into the ejecta
is nicely illustrated  using the Large Magellanic Clould (LMC) 
Type Ia SNRs in Fig.~\ref{fig:lmcia}. For SNR
B0509-67.5 \citep{warren04,kosenko08,badenes08a}
the Fe-L emission is confined to a small, partial shell that is
most pronounced in the west.
This is the youngest LMC remnant \citep{rest05}, and the one
with the largest X-ray line Doppler broadening, $\sigma_V\approx 5000$~\kms,
as measured using \xmm-RGS high resolution spectra
\citep{kosenko08}. These spectra also indicate that the
Fe-L emission is mainly coming from Fe XVII.
In SNR B0519-69.0 \citep{kosenko10} and N103B \citep{lewis03,vanderheyden02} the
Fe-L line emitting shell is more pronounced. The \xmm-RGS spectra of
B0519-69.0 show less line broadening,  $\sigma_V\approx 1900$~\kms,
and Fe-L emission is now also coming from the higher Fe ionization stages
Fe XVII-Fe XXI. The \cxo\ data 
show that the lower Fe ionization stage, Fe XVII, 
is located closer to the center.
This  is expected, because this is the location of the
most recently shock-heated Fe \citep[][see Fig.~\ref{fig:stratification}]{kosenko10}.

\input{jvink_aarv_fig_ia_stratification}

Concerning the more mature SNRs
Dem L71 \citep{hughes03,vanderheyden03}
and B0534-69.9 \citep{hendrick03}, one
observes that Fe-L emission is coming from the whole center of the SNR,
indicating that (almost) all ejecta have been shock-heated.
There is a clear indication from recent X-ray data from both Galactic and LMC
SNRs that the shocked ejecta  of Type Ia remnants are well stratified;
from the outside in one first encounters oxygen, then a silicon-rich layer,
and finally the iron-group elements (Fig.~\ref{fig:stratification}),
testifying of a more or less spherically symmetric explosion.
This is clearly the case in SNR B0519-69.0 \citep{kosenko10} and in 
the southwestern region of Kepler's SNR \citep{cassam04},
whereas the more mature SNR  Dem L71 has an Fe-rich core
\citep{vanderheyden02,hughes03}. Only Tycho's SNR seems less
stratified \citep{decourchelle01}. Note that the radial
X-ray line emission is also influenced by ionization age differences
in different layers, as indicated by \asca\ and \xmm\ observations
of Tycho \citep{hwang97,decourchelle01}, Kepler \citep{cassam04},
and the already discussed SNR B0519-69.0.

The more or less layered structure is unlike that of core 
collapse SNRs (Sect.~\ref{sec:orich}), where in some cases plasma with
more massive elements seems to have overtaken plasma with less massive
elements.
In addition, the morphologies of Type Ia SNRs are more symmetric than
those of core collapse remnants (more on this in \sect{sec:orich}). The reason for this may
be a more spherically symmetric explosion and that
the circumstellar medium (CSM) has been less disturbed, possibly
because Type Ia SNe occur more frequently outside turbulent starforming regions.
Nevertheless some imprints of CSM interaction can be found in Type Ia SNR
(see next section).

Both for core collapse and for Type Ia SNRs the variation of elemental
abundances inside the remnant poses a problem in modeling the overall X-ray
spectra
of SNRs. The reason is that one cannot realistically expect to fit the spectra
with one or two NEI (\sect~\ref{sec:nei})
plasma models. On the other hand, adding many more
NEI components leads to unconstrained results. Imaging spectroscopy
offers some help, as one can extract spectra from distinct regions and
apply single or double NEI components to these individual spectra. But
projections effects play a role as well. Recently \citet{kosenko10} 
stretched the capabilities of
multi-component fitting of SNR B0519-69.0 to the maximum, by using
seven NEI components, each corresponding to a layer that could also be
identified in the stratification inferred from narrow band \cxo\ imaging 
(Fig.~\ref{fig:stratification}): an O-rich layer, a Si/S-rich layer,
a Ar/Ca layer, a low ionized Fe layer, and a highly ionized Fe layer.
The continuum component was constrained by setting the abundance pattern to
that of the LMC. Such a procedure works better for the radially stratified
Type Ia SNRs than for the more chaotic spatial abundance distribution
of core collapse SNRs.

\input{jvink_aarv_fig_badenes}

An alternative method that has already given interesting results is
to model the whole hydrodynamical evolution of the SNR, from the explosion
to the SNR phase, including the temperature/ionization evolution and
X-ray emission, using the 
theoretically predicted initial stratification and velocity distribution of
the ejecta 
\citep{badenes03,badenes05,badenes08a,sorokina04,kosenko06,kosenko08,kosenko11}.
This procedure works relatively well for Type Ia SNRs, because the explosion
properties of thermonuclear supernovae are much better constrained and allow
for less variation than for core collapse supernovae. 
The grid of explosion models for Type Ia SNe is, therefore,
much more limited. The method has been applied to
Tycho/SN\,1572 
\citep[Fig.~\ref{fig:badenes}][]{badenes06,kosenko06} and SNR B0519-69.0
\citep{badenes06,kosenko06}. For the latter, \citet{badenes08a}
found that the most energetic Type Ia explosion models, which synthesize
more Fe, fit the data best. \citet{kosenko08} reported for this
SNR a very high plasma velocity based on the Doppler broadening measured
with the \xmm-RGS instrument. In addition, the RGS detected N VII line emission,
which helps to constrain the amount of shocked CSM, because
nitrogen is not a Type Ia nucleosynthesis product. In this case the inferred
circumstellar density is $n_{\rm H} = 0.4-0.8$~cm$^{-3}$. 

An uncertainty in modeling the combined hydrodynamics/X-ray emission from
Type Ia SNRs is the influence of efficient cosmic-ray acceleration,
as discussed in Sect.~\ref{sec:hydro} and \ref{sec:cr}. \citet{kosenko08}
incorporated the cosmic-ray acceleration effects by altering the equation
of state of the shocked plasma, 
which gave a good fit to the \xmm\ (EPIC-MOS) spectra for a 
cosmic-ray energy density that is about 40\% of the gas density \citep[see also][]{decourchelle00,kosenko11}.

\input{jvink_aarv_fig_tycho_cr_mn}

Abundance studies of SNRs in X-ray are usually confined to the 
alpha-elements (C,O, Ne, Mg, Si, S, Ar, Ca) 
and Fe-group elements (Fe/Ni). These elements are the most abundant elements in the Galaxy,
and are the dominant nucleosynthesis products of 
supernovae (Sect.~\ref{sec:supernovae}), and all have prominent line transitions in the
X-ray band.
Other elements, for example
those with uneven atom number, are usually too
rare to detect in X-rays. However, recently
\citet{tamagawa09} reported
the detection of chromium (Cr, $Z=24$) and  manganese (Mn, $Z=25$)
K-$\alpha$ line emission
from Tycho's SNR using \suz\ data 
\citep[Fig.~\ref{fig:tychocrmn}, see also][]{yang09}.
The nucleosynthesis of Mn is influenced by the presence of the neutron
rich element $^{22}$Ne, whose abundance is determined by the metallicity
of the white dwarf progenitor. An explanation for the
relatively high flux ratio of Mn/Cr is, therefore, that
SN1572 had a white dwarf progenitor that started as a main sequence
star with solar or supersolar metallicity \citep{badenes08b}. In other
words, SN 1572 was not the explosion of a Population II star.

%% file: jvink_aarv_tab_ia_snrs.tex
\begin{table*}
\caption{Likely Type Ia supernova remnants}
\label{tab:Ia}
\begin{center}
\begin{tabular}{llllll}\hline\hline\noalign{\smallskip}
Name            & Location & Distance    & Radius  & High Res. Spec. & Other studies\\
                &          & (kpc)  & (pc) \\
\noalign{\smallskip}\hline\noalign{\smallskip}
G1.9+1.3 $^{\rm a}$         & MW  &  8 (?)  & 2 & (1,2,3)
\\
Kepler (SN\,1604/G4.5+6.8) & MW  
&  6 (4,5,6) 
&  3 & & (7,8,9,10,5) 
\\
Tycho (SN\,1572/G120.1+1.4)& MW  & 3 & 3.8 &
(11,12)
& (13,14,15,16,12) 
\\
SN\,1006 (G327.6+14.6)     & MW  & 2.0 (17) 
& 9.3 & (18,19) )
& (20,21,22,23,24,25) 
\\
B0509-67.5                 & LMC & 50 (25) 
& 3.6 & (27) 
& (28,29) 
\\
N103B                      & LMC & 50 & 3.6 & (30) 
& (31) 
\\
B0519-69.0                 & LMC & 50 &  3.9 &
(32) 
\\
DEM L 71                   & LMC & 50 & 8.6 & (33) 
& (34) 
\\
B0548-70.4 & LMC & 50 & 12.5     & &(35) 
\\
DEM L316A$^{\rm c}$ & LMC & 50 & 15 & & (36,37) 
\\
B0534-69.9 & LMC & 50 & 16 & & (35) 
\\
DEM L238   & LMC & 50 & 21 & & (38)
\\
DEM L249   & LMC & 50 & 23 & & (38) 
\\
B0454-67.2 & LMC & 50 & 27 & & (39)
\\
IKT 5  (B0047-73.5)         & SMC & 60 (26) 
&  15$^{\rm b}$& & (40)
\\
IKT 25 (B0104-72.3)        & SMC & 60 & 18 &  & (40,41)
\\
DEM S 128 (B0103-72.4)     & SMC & 60 & 26 &  & (38) 
\\
\noalign{\smallskip}\hline\hline\noalign{\smallskip}
\end{tabular}\\
\end{center}
\scriptsize{
MW=Milky Way, LMC=Large Magellanic Cloud, SMC=Small Small Magellanic
Cloud. \\
References: 
1) \citet{reynolds08b}; 2) \citet{borkowski10} 3) \citet{carlton11};
4) \citet{sankrit05}; 5) \citet{vink08b}; 6) \citet{chiotellis11};
7) \citet{kinugasa99}; 8)\citet{cassam04}; 9) \citet{badenes07};
10) \citet{reynolds07}; 
11) \citet{ghavamian01}; 12) \citet{hayato10}; 13)\citet{hwang97};
14) \citet{decourchelle01}; 15) \citet{hwang02}; 16) \citet{tamagawa09}; 
17) \citet{ghavamian02}; 18) \citet{vink03b}; 19) \citet{vink04e};
20) \citet{long03}; 21) \citet{yamaguchi06};
22) \citet{acero07}; 23) \citet{yamaguchi08}; 24) \citet{katsuda09};
25) \citet{miceli09}; 
26) \citet{schaefer08}; 27) \citet{kosenko08}; 28) \citet{warren04};
29) \citet{badenes08a};
30) \citet{vanderheyden02x}; 31) \citet{lewis03}; 
32) \citet{kosenko10}; 
33) \citet{vanderheyden03}; 34) \citet{hughes03};
35) \citet{hendrick03}; 36) \citet{nishiuchi01}; 37) \citet{williams05};
38) \citet{borkowski06};
39) \citet{seward06}; 40) \citet{vanderheyden04};
41) \citet{lee11}\\
Notes:\\
$^{\rm a}$ Type Ia origin debatable, and largely based on morphology \citep{borkowski10}.\\
$^{\rm b}$ Only the Fe-L emitting interior is visible in X-rays, so the radius
is not necessarily the shock radius for IKT 5.\\
$^{\rm c}$ DEM L316A is part of two associated SNRs. 
Shell A is given its Fe abundance most likely a Type Ia SNRs,
whereas B is most likely a core collapse SNR \citep{williams05}.
The two shells are likely not physically connected \citep{toledo-roy09}.
}
\end{table*}

%% file: jvink_aarv_fig_typeia_lmc.tex
\begin{figure*}
\centerline{
\includegraphics[width=0.3\textwidth]{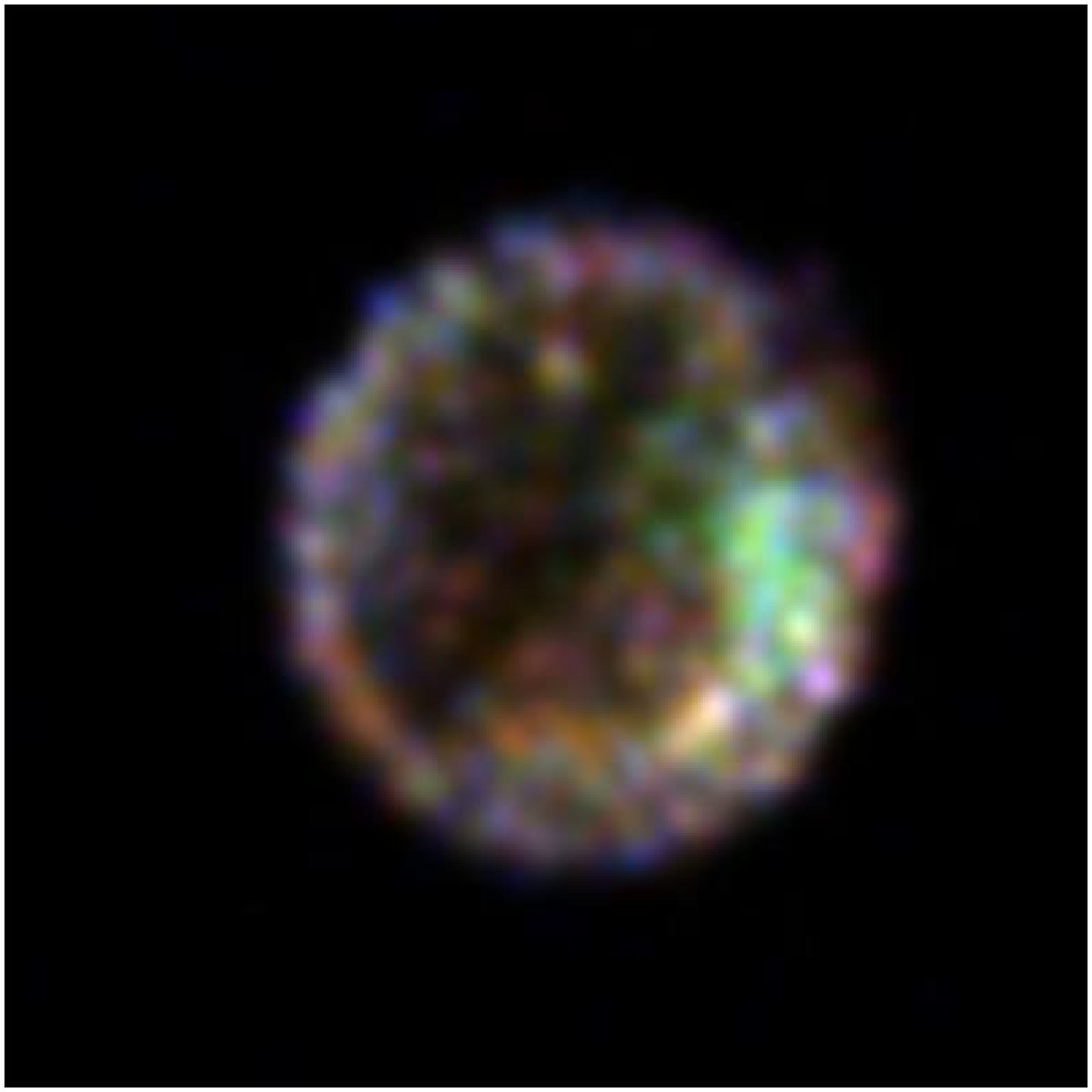}
\includegraphics[width=0.3\textwidth]{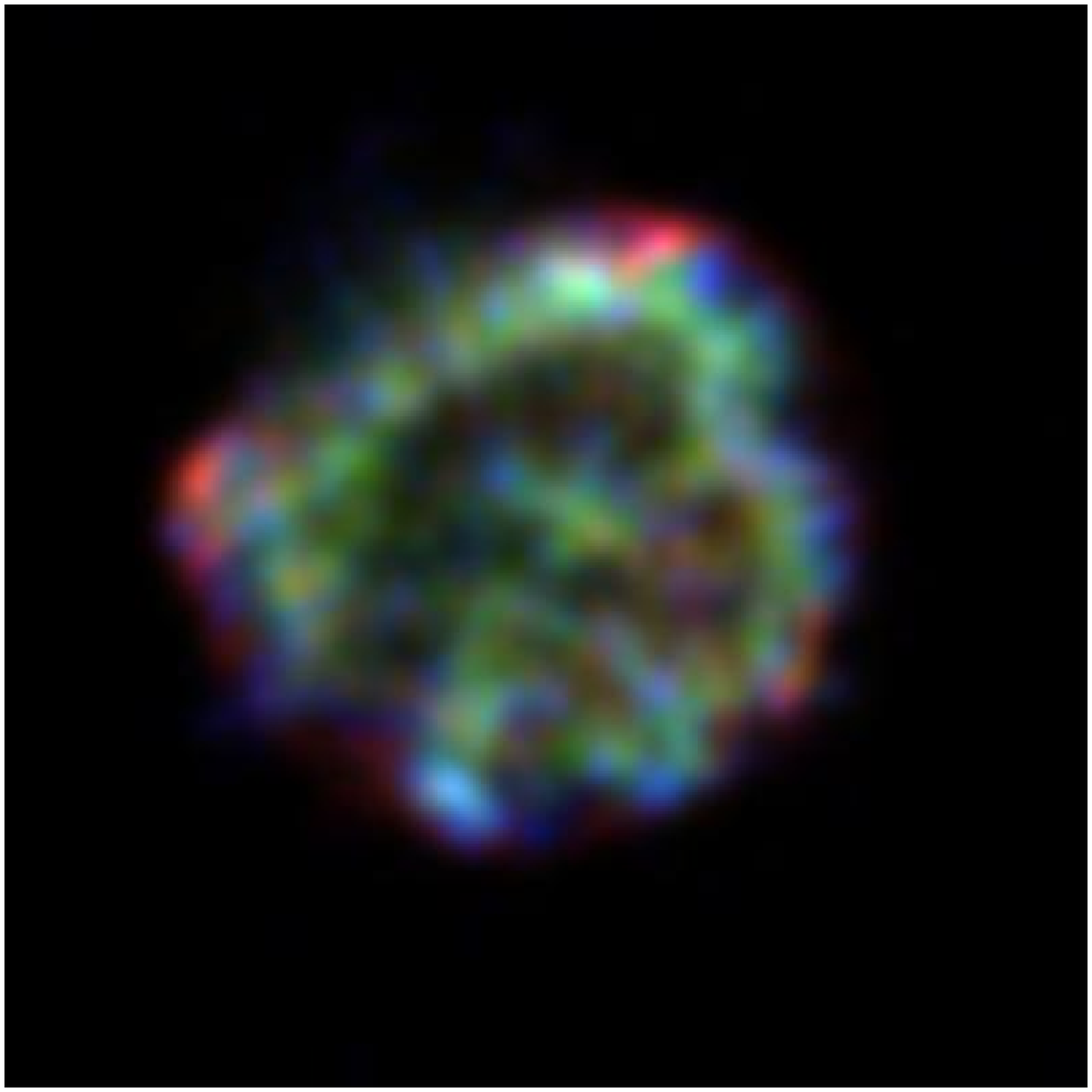}
\includegraphics[width=0.3\textwidth]{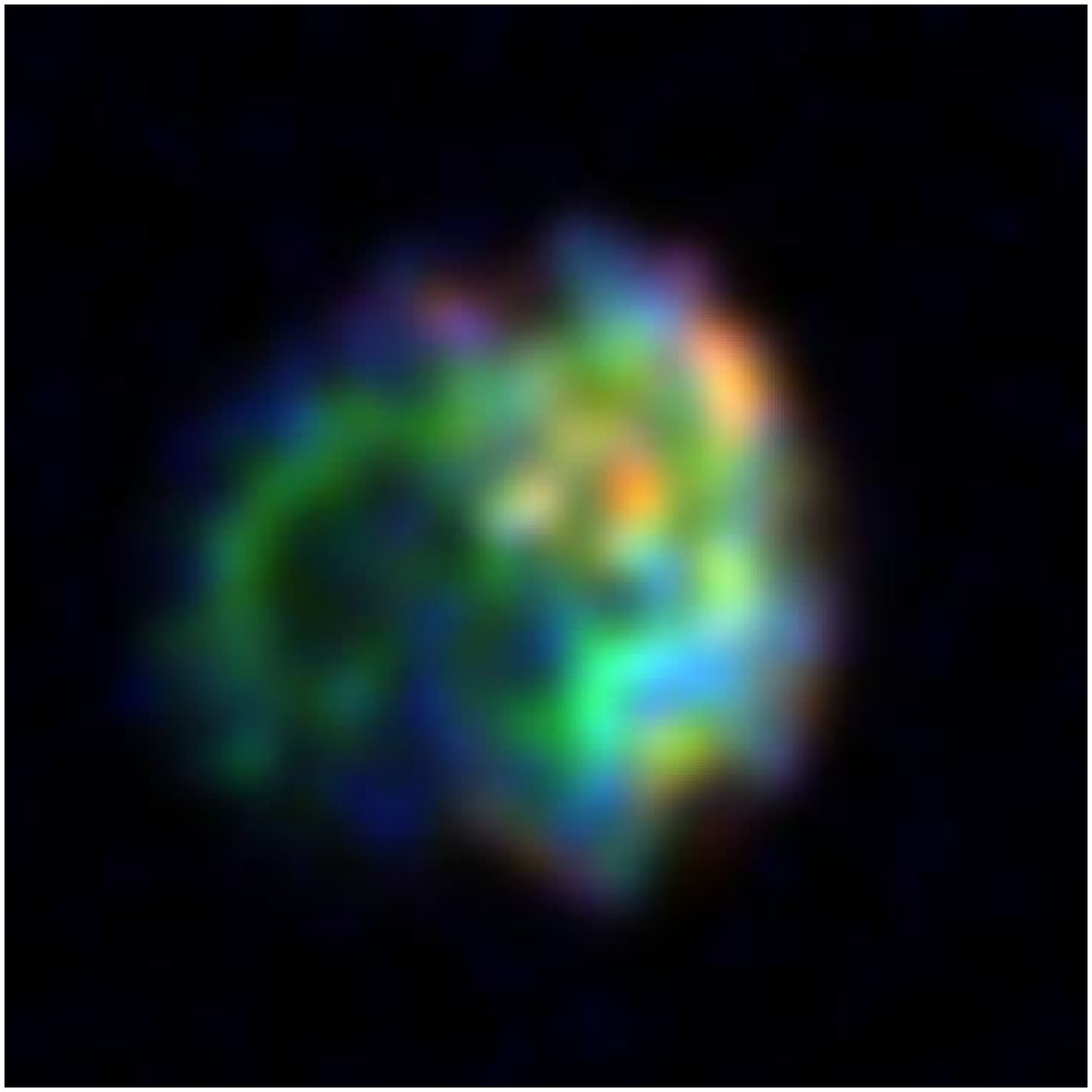}
}
\vskip 1mm
\centerline{
\includegraphics[width=0.45\textwidth]{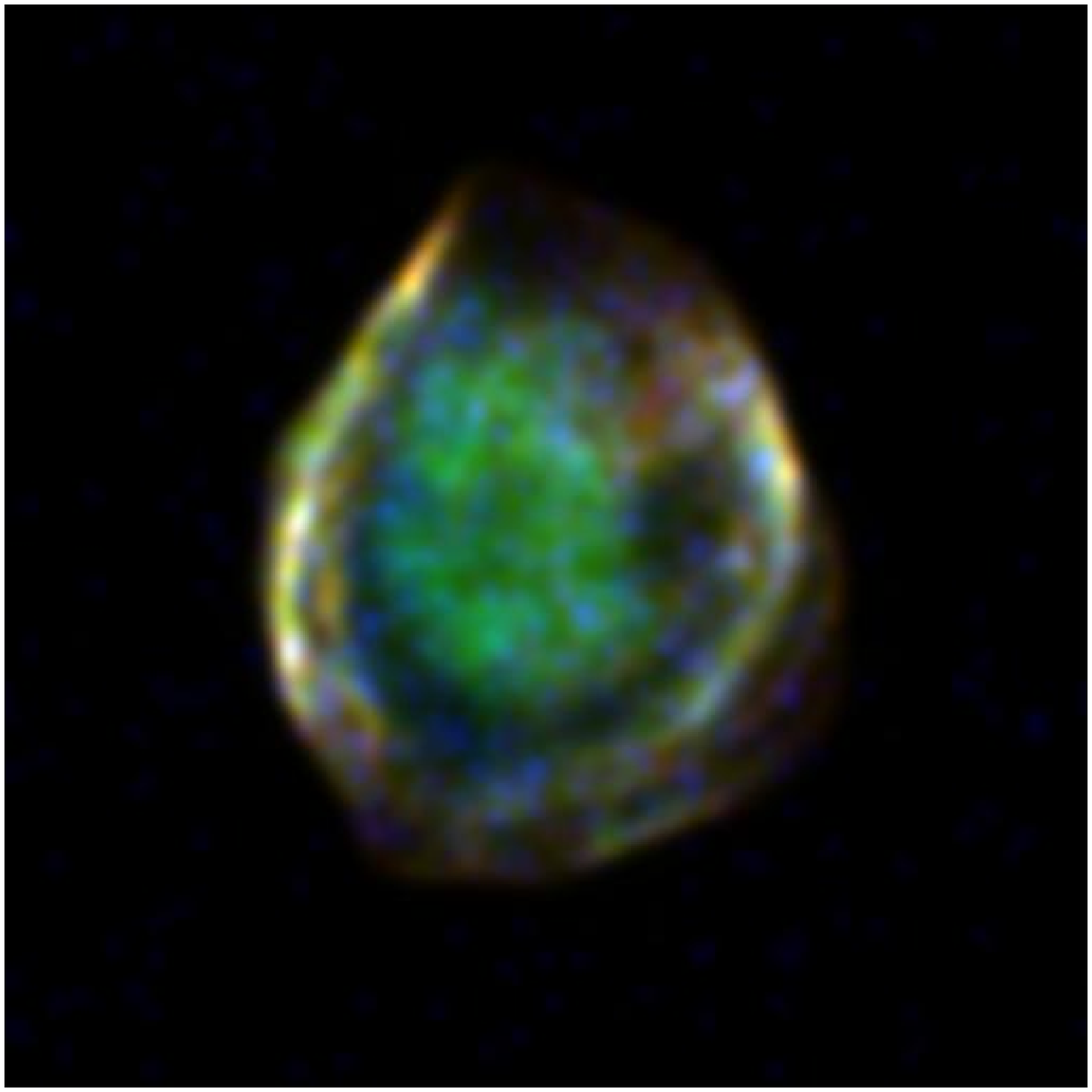}
\includegraphics[width=0.45\textwidth]{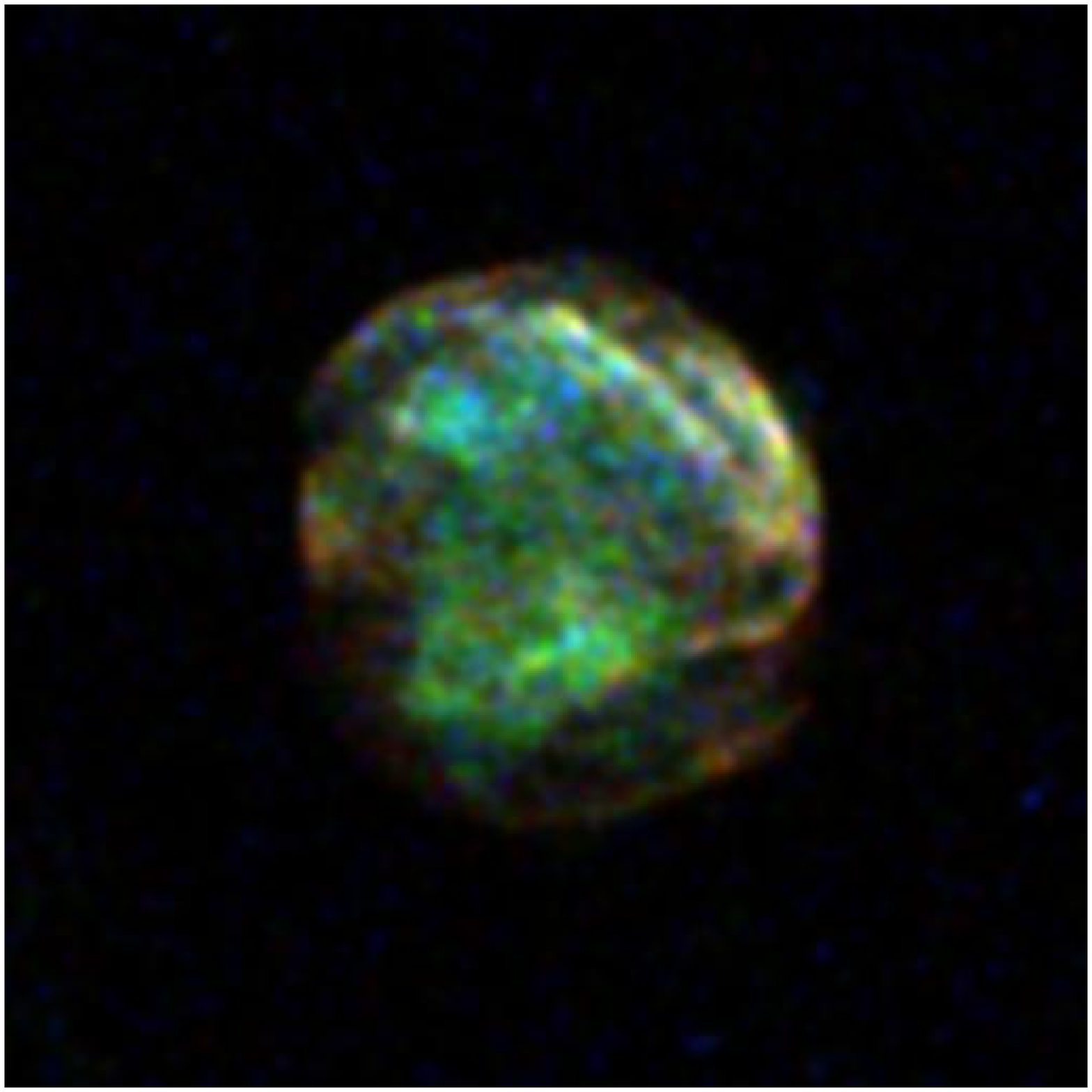}
}
\caption{
\cxo\ X-ray images of LMC Type Ia SNRs.
These are three-color images, and in all cases red represents
the O VII/O VIII band (0.5-0.7 keV), the color green
represents Fe-L emission line ($\sim 1$~keV).
From left to right, top to bottom: B0509-67.5 \citep{warren04}, 
B0519-69.0 \citep{kosenko10}, N103B \citep{lewis03}, 
Dem L 71 \citep{hughes03}, 
and B0534-699 \citep{hendrick03}.
The order of the figures is approximately indicative
of the relative dynamical age. (Images generated by the author
using the \cxo\ archive.)
\label{fig:lmcia}
}
\end{figure*}

%% file: jvink_aarv_fig_ia_stratification.tex
\begin{figure*}
\centerline{
\includegraphics[width=\bigfig]{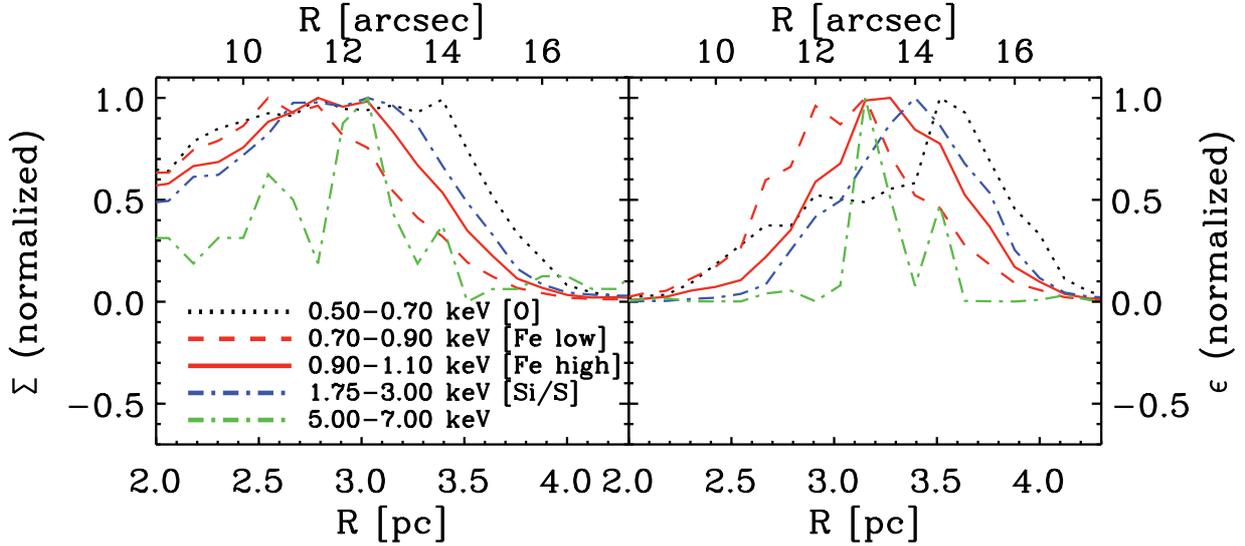}}
\caption{
The stratification of O VII-VIII, Si XII, and Fe XVII-XVIII (Fe low),
Fe XIX-XXI (Fe high) in SNR B0519-69.0, based on \cxo\ data.
The left panel shows the radial surface 
emission profile, whereas the right panel
shows the deprojected (emissivity) profile. 
\citep[Figure taken from][]{kosenko10}.
\label{fig:stratification}
}
\end{figure*}

%% file: jvink_aarv_fig_badenes.tex
\begin{figure*}
\centerline{
\includegraphics[width=\twofig]{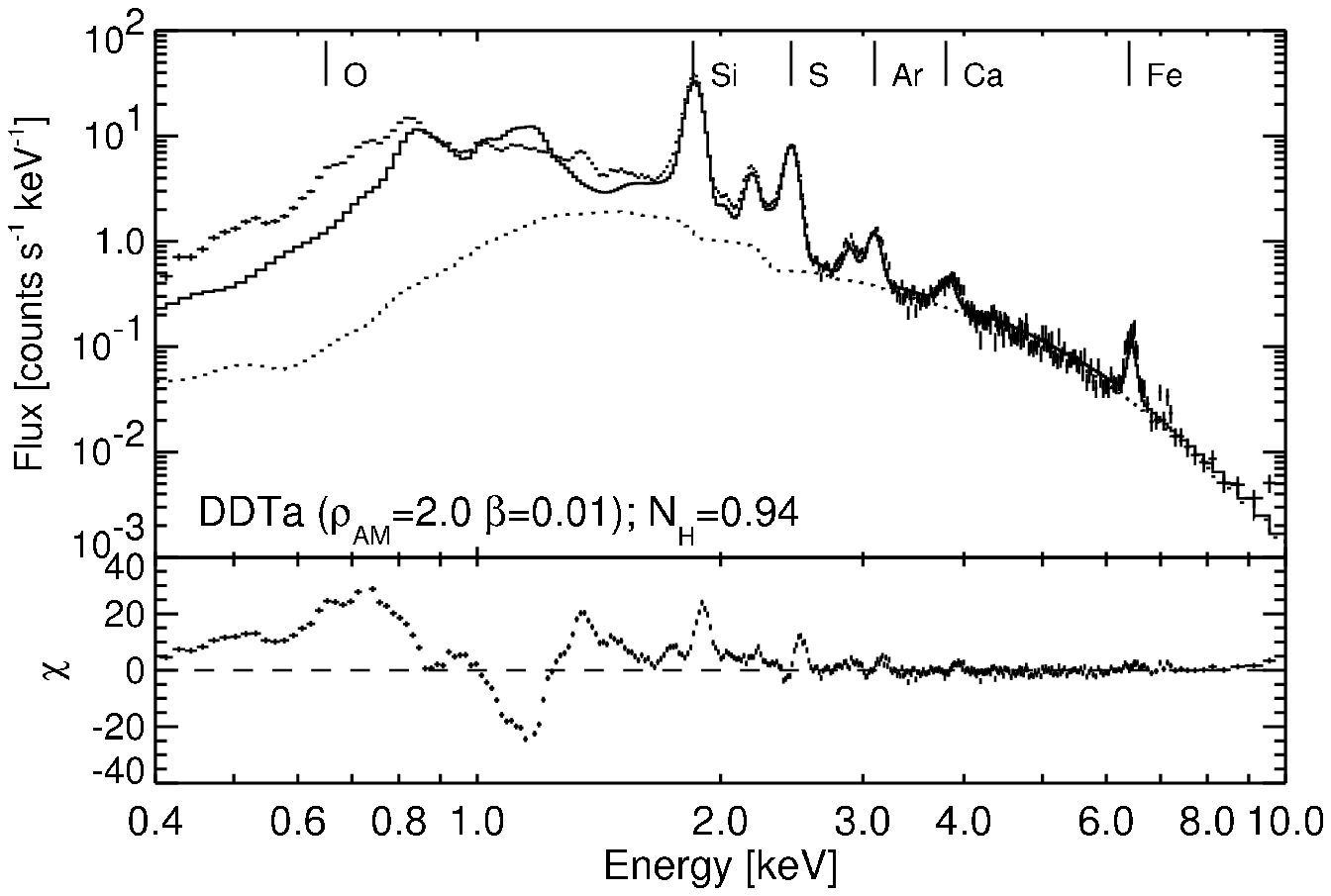}
\includegraphics[width=\twofig]{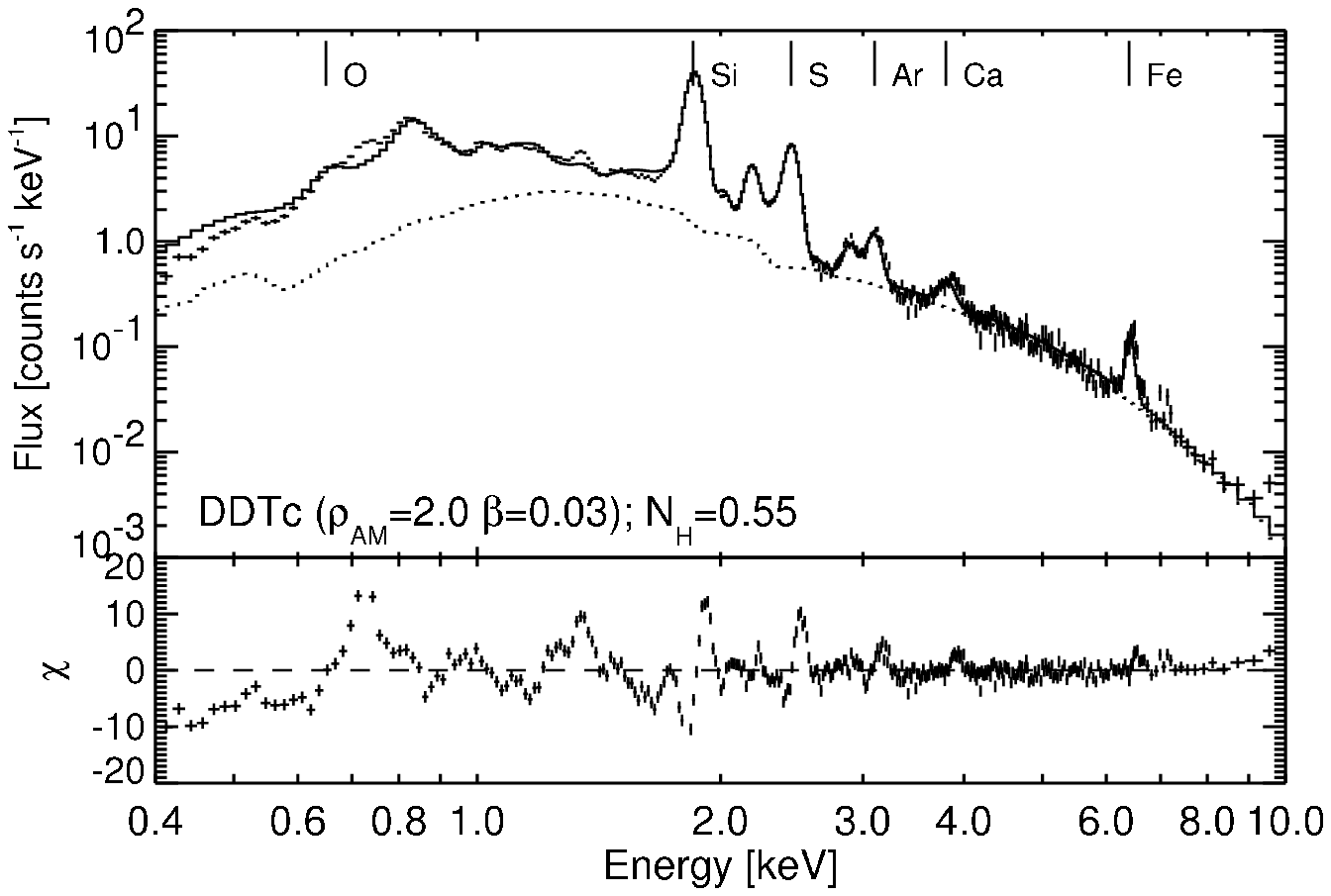}
}
\caption{
\xmm MOS1 spectrum of western part of Tycho/SN\,1572. The two panels
show two models (solid lines) from an extensive grid of Type Ia models 
\citep{badenes06}. The model characteristics are indicated:
the explosion model  (here DDTc and DDTc), the circumstellar density,
$\rho_{AM}$ in units of $10^{-24}$~g\,cm$^{-3}$ and absorption
column density, $N_{\rm H}$ in units of $10^{21}$~cm$^{-2}$.
(Courtesy C. Badenes)
\label{fig:badenes}
}
\end{figure*}

%% file: jvink_aarv_fig_tycho_cr_mn.tex
\begin{figure}
\centerline{
\includegraphics[width=\medfig]{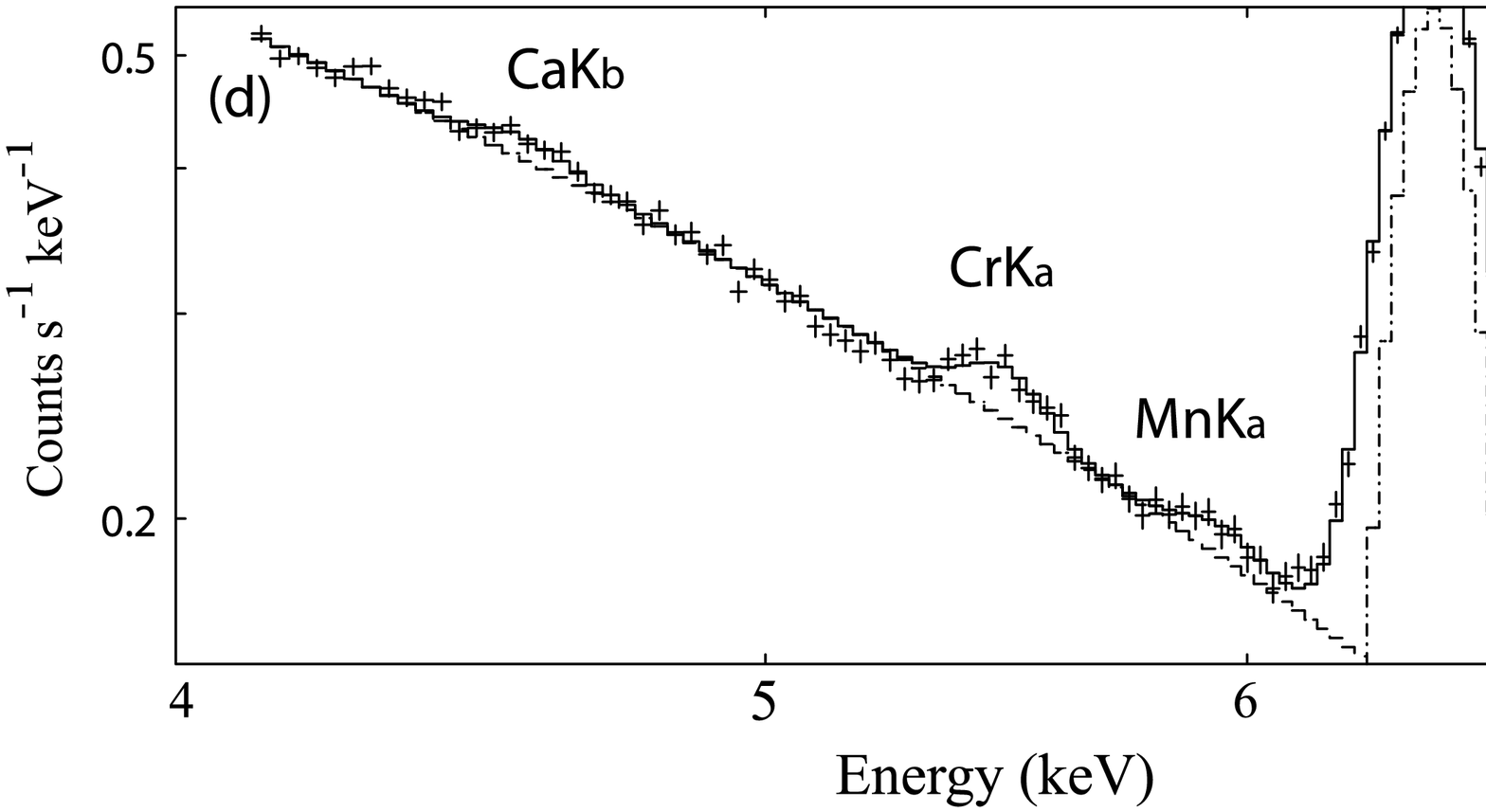}
}
\caption{
Detail of the \suz\ spectrum of Tycho/SN1572, showing K$\alpha$ emission
from Cr, Mn and Fe \citep[reprinted from][]{tamagawa09}.
\label{fig:tychocrmn}
}
\end{figure}

%% file: jvink_aarv_ia_csm.tex
As explained in \sect~\ref{sec:sn1a}, there is little doubt that 
Type Ia supernovae are thermonuclear explosions of white dwarfs, but there
is much uncertainty about the binary system that leads to the growth
of a C/O white dwarf close to the Chandrasekhar limit.
The mass transfer from the companion of the white dwarf
can occur during various phases in the evolution of the companion;
during the main sequence, red giant (RG), 
or asymptotic giant branch (AGB) phase, or when the
companion itself is a white dwarf. The mode by which the mass transfer
occurs may shape the CSM of the system.
For example, Roche-lobe overflow is considered to lead to
unstable mass transfer, but it may be stabilized by a fast, $\sim 1000$~\kms, 
wind
emanating from the white dwarf \citep{hachisu96,hachisu99}. This fast wind
will created a large ($>10$~pc), low density bubble surrounded by a dense
shell. Evolved stars, like RG and AGB stars, have slow winds ($\sim 10$~\kms) 
which result in a rather dense CSM with a density
profile scaling as $r^{-2}$ (\sect~\ref{sec:hydro}). 

The specific characteristics of the binary are, therefore, expected to
have an important imprint on the CSM, and, as a result,
on the evolution, morphology and X-ray spectra of the SNR. 
Models for Type Ia usually
ignore the possible effects of pre-supernova evolution, 
and instead assume a uniform CSM (i.e.
an $s=0$ model, in the notation of \citet{chevalier82}, 
see \sect~\ref{sec:hydro}).

One important constraint on the progenitor system provided by SNRs comes
from optical observations.
These show that
many of the SNRs mentioned in Table~\ref{tab:Ia} have so-called
Balmer-dominated shocks \citep[e.g.][]{smith91,sankrit05}.
Optical spectra from their shock regions show H$\alpha$
line emission, without [NII] line. This is caused by collisional
excitation of neutral hydrogen entering the shock region. In addition,
broad H$\alpha$ emission arises from charge transfer of neutral hydrogen with
shock heated ions \citep{chevalier80,vanadelsberg08}. The presence
of Balmer-dominated shocks is only possible when the neutral fraction in the CSM
is relatively high. It is therefore of interest that
most of the SNRs with Balmer-dominated shocks are Type Ia SNRs.
\footnote{The only exceptions are the
large SNRs the ``Cygnus Loop'' and RCW 86, which are both likely core collapse
SNRs \citep{ghavamian01}, but see \citet{williams11} for an alternative view on RCW 86.}
The reason for this is that progenitors of core collapse supernovae create
large Str\"omgenspheres \citep[$>10$~pc][]{chevalier90}.
And even if they explode as red supergiants the bright flash of ionizing
photons when the shock reaches the surface, creates a large ionized
region \citep{chevalier05}. Type Ia supernovae, on the other, do not produce
a UV flash at shock break-out bright enough to
ionize a large region.

The presence of neutral hydrogen
within a couple of parsecs from the explosion, therefore, is easy
to explain for Type Ia SNR, {\em provided that the progenitor itself was
not a powerful source of ionizing photons}. This excludes, therefore,
an important class of Type Ia candidate progenitors, namely the so-called
{\em supersoft sources}. These are soft X-ray sources that probably consist of
a white dwarfs accreting material from a companion at a rate
close to the Eddington limit, and burn the accreted material in stable
way at the surface \citep{li97}. But because supersoft sources
are luminous UV/X-ray sources, $10^{37} -10^{38}$ erg\,s$^{-1}$, they will easily
ionize a region of 30~pc in radius. Many of the Balmer-dominated
Type Ia SNRs, including Tycho, Kepler, B0509-67.5, B0519-69, DEML 71, 
can, therefore, not have had supersoft sources as progenitors
\citep{ghavamian03}.

There are several other recent studies that cast a
light on the structure
of Type Ia SNRs. In one of them \citet{badenes07} investigates whether
Galactic and LMC Type Ia SNRs show any evidence
for having evolved inside large, tenuous cavities, as implied by the fast-wind
model of \citet{hachisu96}. This study shows that this is not the
case, as the radii of the SNRs are too small and their
ionization age, \net, too large compared to model 
predictions that incorporate the effects of fast progenitor winds.
Possible exceptions are SN\,1006 and RCW 86. However,
SN\,1006 is located high above the Galactic plane, where the density
is expected to be low irrespective of the progenitor model, and
RCW 86 is not a bona fide Type Ia SNR, as it is located in an OB association
\citep{westerlund69}, in a region with a 
high star forming rate, as indicated by outbreaks of gas from the Galactic
plane region \citep{matsunaga01}.\footnote{However, recently \citet{williams11} argued,
nevertheless, for a Type Ia origin, because there is no evidence for a  neutron star
inside the SNR, and there is evidence for large amounts of Fe.}

Another study that may be revealing something about Type Ia progenitor
systems concerns an analysis of \chandra/\xmm\ of two old, $\sim 10^4$~yr,
LMC SNRs, DEM L238 and DEM L249 by \citet{borkowski06}. These two SNRs
are rather faint in X-rays, but are both characterized by a
relatively bright interior dominated by Fe-L emission, somewhat similar
to DEM L71 and SNR B0534-699 displayed in Fig.~\ref{fig:lmcia}. 
The most surprising result of the analysis is that 
of the spectra from the interior indicate high values for the 
ionisation parameter, \net, $\sim 10^{12}$\netunit\ for DEM L238 and
$\sim 4\times 10^{11}$\netunit\ for DEM L249. Such high values require
relatively high densities in the interior. This is in contrast to the
low densities implied by the X-ray spectra of the shell, and, indeed, the
overall faintness of these SNRs in X-rays.
\citet{borkowski06} argue that in order to account for the high central
density the blast wave must have encountered a high density CSM
 early during the SNR evolution. Such a dense medium is usually a signature
of a slow wind outflow, instead of the fast winds implied by
the models of \citet{hachisu99}, or a simple, uniform density medium.

\input{jvink_aarv_fig_kepler}

Finally, evidence that winds from Type Ia progenitors affect the CSM
is provided by the historical SNR of SN 1604 (also known as Kepler's SNR, or Kepler for short).
Kepler is located high above the Galactic plane, with latitude $b=6.8$\deg,
which translates to 446~pc for a distance of 4~kpc \citep{sankrit05}.\footnote{Or 670~pc for
a distance of 6~kpc, the distance preferred based on modeling the explosion, if the kinetic
energy is $>10^{51}$~erg \citep{aharonian08,vink08b,chiotellis11}.}
This high latitude would make Kepler an obvious Type Ia SNR, if it were not for the presence
of dense, nitrogen-rich material, mostly in the northern part of the remnant,
which indicates
the presence of a circumstellar shell \citep{vandenbergh77,bandiera87}.
During the 1990s a core collapse origin for SN 1604, was therefore 
preferred. X-ray spectroscopy, first with ASCA \citep{kinugasa99} 
and later with \xmm\ \citep{cassam04} and \chandra\ \citep{reynolds07},
revealed that the spectrum is dominated by Fe-L emission around 1~keV,
and bright Fe-K shell emission around 6.7~keV, characteristic for
Type Ia SNRs. Additional evidence for a Type Ia origin is the low oxygen
content, and the lack of an obvious X-ray point source that may be the neutron 
star that should 
in most cases result from a core collapse supernova \citep{reynolds07}.

The idea that the SNR encounters a dense shell in the north is
reinforced by two recent X-ray studies \citep{vink08b,katsuda08},
which confirm an earlier radio measurement \citep{dickel88}
that the expansion of this SNR in the north is relatively slow
with a small value for the expansion parameter  (Eq.~\ref{eq:chevalier})
of $\beta\approx 0.35$, whereas the average value is  $\approx 0.5$.
The value for the north is unusual in that it falls below the
value 0.4 for a SNR in the Sedov phase of is evolution.
On the basis of expansion measurements,
\cite{vink08b} estimated that the mass in the swept-up, nitrogen-rich
shell must have been $\sim 1$~\msun.

This result, and the 
morphology of Kepler, resembling a bow shock in the north 
(Fig.~\ref{fig:kepler} and Fig.~\ref{fig:chiotellis} for a model), 
is used by \citet{chiotellis11} to argue
that the  progenitor system was a binary consisting of
a C/O white dwarf with an asymptotic giant branch (AGB) star as a companion.
Such a system would be labeled as a symbiotic binary.
Part of the slow wind ($\sim 15$~\kms)
of the AGB star was accreted onto the white dwarf,
until it exploded. 
In order to account
for the nitrogen enrichment of the shell one has to assume a 
main sequence mass for the secondary star of 4-5~\msun.
The bow-shock shape of Kepler, and its height above
the Galactic plane ($\sim 400-700$ pc), can be explained by
a proper motion of the system with about 250~\kms \citep[c.f.][]{bandiera91,borkowski94}. 
Nevertheless, within a Type Ia scenario, it is not quite clear, why a binary
system was ejected with such a high velocity, as usually interaction of two binaries
leads to the ejection of fast single stars \citep{leonard90}. 
But it does occur, see for example the
relatively high velocity ($\sim 130$~\kms)
of the AGB star Mira A and its companion \citep{martin07}.

%% file: jvink_aarv_fig_kepler.tex
\begin{figure*}
\includegraphics[trim=10 75 10 60,clip=true,width=\twofig]{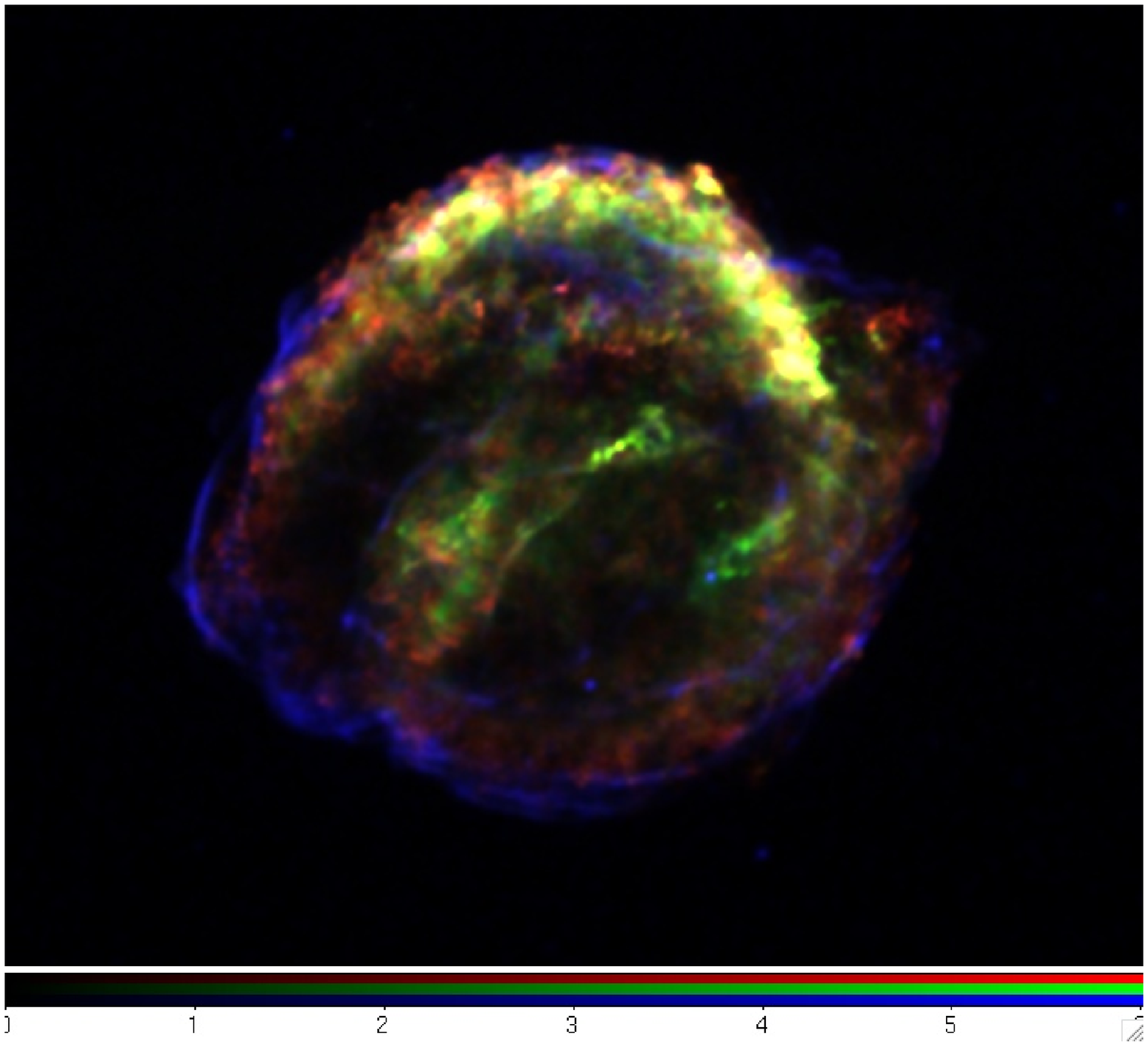}
\includegraphics[trim=0 0 0 0,clip=true,width=\twofig]{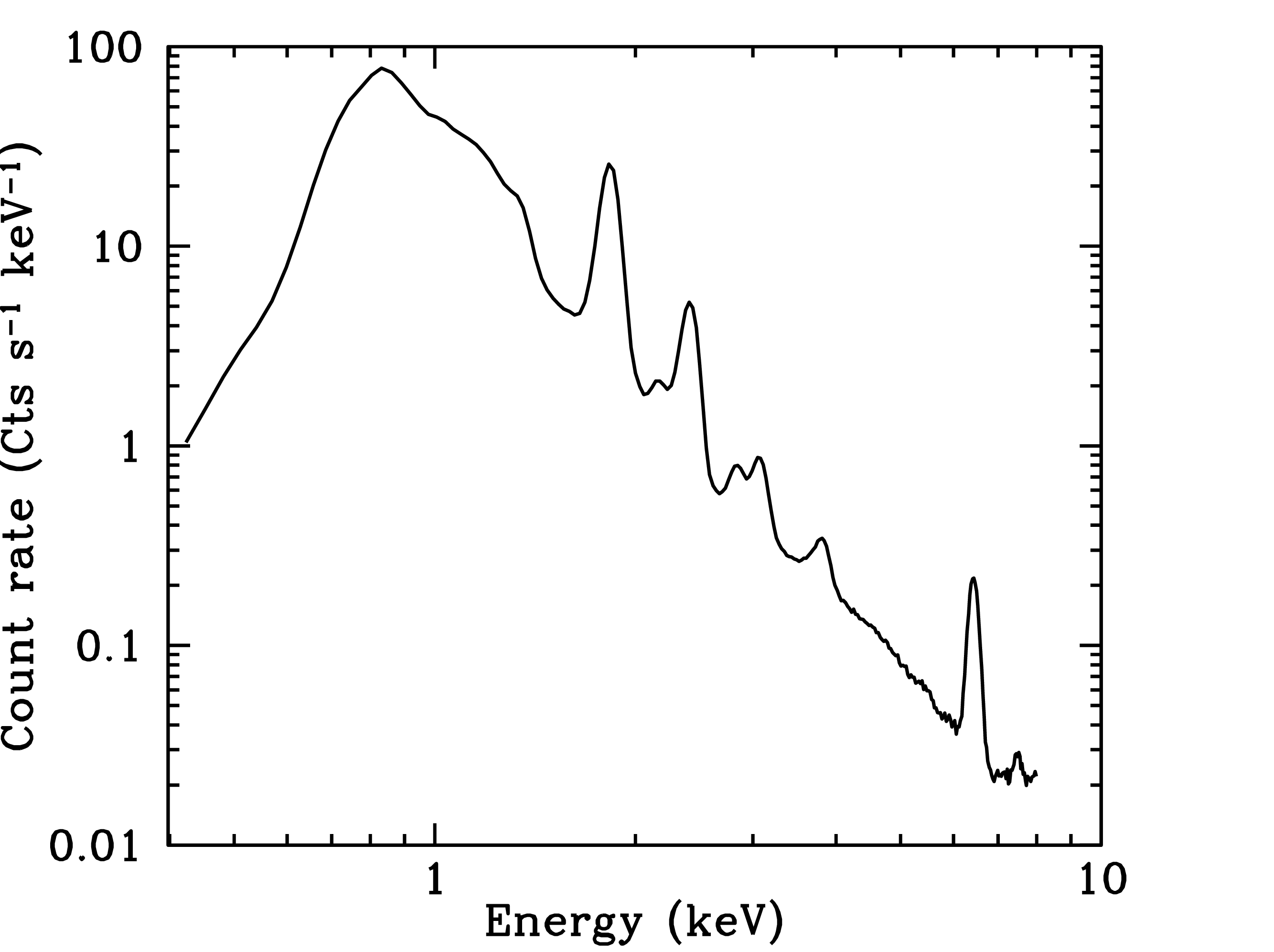}
\caption{
Left: \chandra\ image of Kepler's SNR, with red indicating Si-K $\alpha$
emission (1.75-1.95 keV), green Fe-L emission (0.8-1.6 keV),
and blue continuum emission (4-6 keV). The image is based on a deep, 750 ks,
\chandra\ observation \citep{reynolds07}.
Right: raw count rate spectrum from Kepler's SNR, based on the same
observation. Note the dominant Fe-L emission between 0.7-1.5 keV 
(c.f. Fig.~\ref{fig:mosrgs}c).
\label{fig:kepler}
}
\end{figure*}

\begin{figure}
\centerline{
\includegraphics[trim=20 20 200 100,clip=true,width=\medfig]{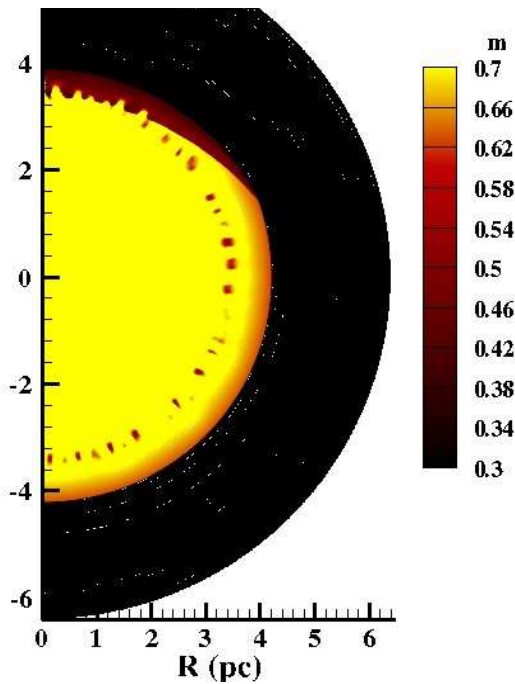}}
\caption{Results of the hydrodynamical study of Kepler's SNR 
\citet{chiotellis11},
showing the expansion parameter in the SNR for
a model in which the progenitor systems moved with respect to
the interstellar medium, and created a one side shell due
to interaction of the progenitor wind and the ISM. 
The expansion parameter in the 
densest regions of the shell are as low as $\beta=0.35$,
comparable to observations \citep{dickel88,vink08b,katsuda08}.
(Figure provided by A. Chiotellis.)
\label{fig:chiotellis}
}
\end{figure}

%% file: jvink_aarv_oxygen_rich.tex
\input{jvink_aarv_tab_orich}

As explained in \sect~\ref{sec:supernovae}, the nucleosynthesis yields
of core collapse supernovae peaks for the element oxygen. The oxygen yield
itself is strongly correlated with the main sequence mass of the progenitor,
although for masses $\gtrsim 30$~\msun\
the poorly known fraction of material that falls back onto the central
object (neutron star or black hole) may reverse this trend.
An important class of SNRs are the so-called oxygen-rich SNRs 
(Table~\ref{tab:orich}). 
As the name suggests, these
SNRs show large overabundances of oxygen, indicating that these
are remnants of the most massive stars ($M_{MS} \gtrsim 18$~\msun). 
Initially these sources were mainly
identified based on their optical oxygen-line emission properties
\citep[{[OIII]}, e.g.][]{goss79,mathewson80,lasker79,
dopita81},
but in many cases the X-ray spectral properties
confirm the elevated oxygen abundances, in particular for
young remnants like Cas A \citep{vink96,willingale02}, 
1E\,0102.2-7219 \citep{hughes94,gaetz00,rasmussen01,flanagan04},
and G292.0+1.8 \citep[MSH\,11-5{\it 4}, e.g.][] {park07}.
Even for some of the more mature oxygen-rich SNRs, like N132D and Puppis A,
it is still possible to identify regions with
oxygen-rich ejecta \citep{behar01,borkowski07,katsuda08b}. 

The neon and magnesium yields of core collapse supernovae 
trace the oxygen yield
(Fig.~\ref{fig:yields}), one therefore expects
also that the Ne and Mg abundances of oxygen-rich SNRs are high.
This appears to be indeed
the case for young oxygen-rich SNRs like 1E\,0102.2-7219 
\citep[][Fig.~\ref{fig:mosrgs}, \ref{fig:orich}, \ref{fig:flanagan}]{rasmussen01,gaetz00}
and G292.0+1.8 
\citep{hughes01b,gonzalez03,park07,vink04f}.
But an important exception to this rule
is Cas~A, which has Ne/Mg abundances
a factor 5-10 below the expected values \citep{vink96}. 
In general, the most 
striking feature of the X-ray spectrum of Cas A is the very bright silicon and
sulphur emission. It is usually assumed that the Cas~A ejecta 
are dominated by oxygen, which is almost fully ionized \citep{vink96,hwang03},
but \citet{dewey07} has argued that the most abundant element
in the ejecta is silicon. 

If the ejecta indeed are almost devoid of hydrogen, this
greatly affects the ejecta mass estimates
(\sect~\ref{sec:continuum}, Eq.~\ref{eq:bremss}).
For example, for Cas A it implies a rather low ejecta mass of
2-4~\msun, with about 1-2~\msun\ of oxygen \citep{vink96,willingale03}.
This oxygen mass is consistent with a progenitor main sequence mass
of $18\pm 2$\msun. This led \citet{chevalier03} to the conclusion
that Cas A must have been a Type IIb SNR, given the similarity between
the apparent mass loss and main sequence mass of the Type IIb supernova
SN1993J and Cas A. This conclusion has been spectacularly confirmed
by the optical light echo spectrum of the Cas A supernova \citep{krause08}.

\input{jvink_aarv_fig_orich}
\input{jvink_aarv_fig_casa_doppler}

Another property of Cas A is the strong bipolarity of the silicon/sulphur-rich
ejecta,
as indicated by optical \citep{fesen06}, X-ray \citep{vink04a,hwang04}, and
infrared observation \citep{hines04}. These observations show the presence
of Si/S-rich material outside the radius of the main shock 
in the northeast and southwest.
The  presence of ejecta outside the shock in the northeast was already
known for a long time, and its optical feature has long been
referred to as ``the jet''.
In X-rays the presence of two opposite jets is best shown by mapping the
ratio of Si XIII over Mg XI K$\alpha$ emission 
\citep[][Fig.~\ref{fig:orich}b]{vink04a,hwang04},
or the equivalent width of the Si XIII emission \citep{hwang00,yang08}.

The
hydrodynamical simulations by \citet{schure08} show that
these jets cannot have survived the interaction with a circumstellar shell.
Such a shell would be present if there has been 
a fast Wolf-Rayet star wind ploughing
through a red super giant wind (\sect~\ref{sec:csm}). 
This strongly suggests that the Cas A progenitor exploded while in the
red supergiant phase. This is consistent with a main sequence mass
below $\sim 25$~\msun, for which stars do not go through a Wolf-Rayet
star phase. 
Since there was likely no high mass loss Wolf-Rayet star phase,
the low ejecta mass is best explained with a binary star
scenario, in which a high mass loss is caused by a common envelope 
phase \citep{young06,vanveelen09}.
A similar scenario was proposed for the Type IIb supernova
1993J \citep{woosley94b},
whose optical spectrum closely resembles that of the
light echo spectrum of Cas A \citep{krause08}.
There is no candidate secondary star in Cas A, so the common envelope
phase may have led to a merger of the cores of the two stars.

The apparent bipolarity of the Cas A explosion may be related to the mass loss
history of the progenitor, which may have included a common envelope phase. 
It is not clear whether all core collapse
supernovae develop strong bipolarities either due to
the accretion process onto the collapsing core, or perhaps due to
rapid rotation (\sect~\ref{sec:supernovae}). In more massive progenitors,
these bipolarities may be smothered while the explosion ploughs through
the extended envelope of the star. Indeed, for supernovae the amount of
asphericity indicated by optical polarimetry seems to depend on how
much of the hydrogen-rich envelope was removed \citep{wheeler02}.
It may also be related with the common envelope merger, as it may have
increased the rotation of the stellar core.

A puzzling property of the jets in Cas A is their silicon richness.
If the jets really come from the core one expects the jets
to contain more iron-rich plasma.
Cas A does contain many iron-rich knots, but not so much in the jets, as 
well as at the outer edge of
the southeastern part of the main shell \citep{hughes00a,hwang03};
see the blueish colored regions in Fig.~\ref{fig:orich}a.
The projected radius of these iron knots, together with the inferred age of
Cas A $\sim 330$~yr \citep{thorstensen01}, indicate that the mean
velocity of these iron knots are higher than 7000~\kms.
This is very high indeed,
as core collapse simulations by \citet{kifonidis06} show iron-group elements
with velocities only up to 3300~\kms.

The morphology of Cas A as seen in Fe-line emission 
does not have an obvious axis of symmetry.
As Fig.~\ref{fig:orich}a. shows iron-rich regions
can be found in the southeast, the north, inside the silicon-rich shell,
and in the west. 
Although in the northern part the Fe-rich ejecta
seem to be on the inside of the Si-rich shell, Doppler measurements
using \xmm\ EPIC-MOS data show that in this region the velocity of
iron is higher, $\sim 2500$~\kms, than that of silicon,
$\sim 2000$~\kms\ \citep{willingale02}. It is therefore likely
that also in the northern part the Fe-rich
ejecta are outside the Si-rich shell.
The study by \citet{willingale02} also showed, by combining Doppler
information with the spatial structure, that the overall
morphology of the Si-rich resembles more closely a donut-like structure
than a spherical shell (Fig.~\ref{fig:casadoppler}), as already
suggested by Einstein Focal Plane Crystal Spectrometer data 
\citep{markert83} and ASCA data \citep{holt94}.

Finally, it is worth pointing out that Cas A is special in the sense that the detection of
$^{44}$Ti \gray\ and hard X-ray lines  from this SNR suggest a rather large production
of this element \citep[$1.6\times 10^{-4}$\msun, see also \sect~\ref{sec:radioactivity},][]{renaud06}.
This is much higher than model predictions, but consistent with what has been inferred for
SN\,1987A.  This higher than expected yield may be related to the fact that the Cas A explosion
was asymmetric \citep{nagataki98}. Something that may also apply to SN\,1987A. On the other hand,
\citet{prantzos11b} shows that the solar ratio of $^{44}$Ca/$^{56}$Fe, the stable products
of, respectively, $^{44}$Ti and $^{56}$Ni, is more consistent with the inferred yields for Cas A and
SN\,1987A than with model predictions. This suggests that asymmetric explosions
are more common than assumed, or that the production processes of these elements
is not yet well understood.

High resolution X-ray spectroscopy with the \chandra\ HETGS suggest that
the SMC SNR 1E\,0102.2-7219, has a similar, ``donut-like''
structure \citep{flanagan04}. This has been established using a special
property of transmission gratings, namely that the plus and minus
orders are each others mirrors in terms of wavelengths, whereas 
the projection of the object in a single wavelength is not mirrored
between the plus and minus orders. 
If for example 
one side of the object has  on average a redshift and the other side
a blueshift, this results in the blueshifted side to be shifted toward
the center (zeroth order) and the redshifted side to be shifted away from
the center. The monochromatic images in the 
plus or minus orders are, therefore,
squeezed in one order and a stretched in the opposite order.
1E\,0102.2-7219
shows exactly this 
behavior as shown in Fig.~\ref{fig:flanagan} for the Ne Ly-$\alpha$ line.

\input{jvink_aarv_fig_1e0102}

If we now turn the attention to another well studied oxygen-rich SNR
G292.0+1.8 (MSH 11-5{\it 4}) one also sees a non-spherical
distribution of elements \citep[][Fig.~\ref{fig:orich}c]{park02,park07}.
There is the enigmatic bar, running from east to west,
first seen in X-ray images taken with the Einstein satellite 
\citep{tuohy82}.  
The origin of this structure is not
so clear, but may be a pre-existing feature in the CSM
that was overtaken by the forward shock.
The idea that the bar is of CSM origin is supported
by the lack of evidence for
enhanced metal abundances,
as determined by  \cxo\ X-ray imaging spectroscopy
\citep{park02}.
The overall metal abundance distribution shows a striking asymmetry between
the northwest, which contains more Si-group elements, and the southwest,
which contains more oxygen, neon, magnesium. There is no evidence
that this asymmetry is caused by an asymmetry in the CSM.
 Thus an asymmetric explosion seems to have taken place 
\citep{park07}. \citet{gonzalez03} analyzed \chandra\ X-ray spectra
and came to the conclusion that the abundance pattern matches that
of a star with a main sequence mass of 30-40\msun. 

\input{jvink_aarv_fig_lopez}

If there is one thing that one that can be concluded from X-ray observations
of core-collapse SNRs, it is that their explosions are very irregular,
with their SNRs showing large
deviations from spherical symmetry and a lack of elemental stratification.
Material from the core (Fe-group elements)
can be ejected with speeds
in excess of those of layers higher up in the star. This behavior
is markedly different from that of the well stratified Type Ia SNRs.

Recently, \citet{lopez09,lopez11} made a quantitative analysis of the
morphological differences between core collapse and Type Ia SNRs, using a
a multipole statistical analysis of the X-ray
surface brightness distribution of SNRs.
Deviations from circular symmetry show up as large values for
$P_3/P_0$ (normalized third moment), whereas a mirror symmetry
gives large values for $P_2/P_0$. As Fig.~\ref{fig:lopez} shows,
Type Ia  SNRs are on average rather more spherical than core-collapse SNRs;
their mean values for both $P_2/P_0$ and $P_3/P_0$ are smaller.
In addition, there is the trend that if an individual Type Ia SNR 
shows deviations
from spherical symmetry (large $P_3/P_0$) it tends to have less mirror
symmetry. These difference makes that Type Ia and core collapse SNRs
occupy different areas in the  $P_3/P_0$ versus  $P_2/P_0$ diagram.

That there is a structural difference 
between core collapse and Type Ia SNRs is not too surprising,
since the energy of Type Ia SNRs
comes from explosive nucleosynthesis, and requires a regular
burning front to propagate through the white dwarf. Core collapse
supernovae are driven by a the gravitational energy from the collapsing
core, and much of that energy is deposited in the inner regions.
Irregularities will not impair the explosion of the rest of the star.

However, there are some signs of aspherical structures that may reveal
us something about the explosion process: there are the donut-like morphologies
of Cas A and 1E0102.2-7219, and, in particular, the bipolar
Si-rich jets in Cas A. Currently there is no theory to explain these features,
but it is tempting to relate the jets of Cas A to those of
long duration gamma-ray bursts. In fact, the
less energetic gamma-ray bursts, the X-ray flashes, have properties
rather similar to what has been inferred for Cas A.
For example, for the X-ray flash SN 2006aj \citep{mazzali07,laming06}
the inferred explosion energy is  $2\times 10^{51}$~erg, and the inferred
main sequence mass is $\sim 20$~\msun\ with an ejecta mass of $\sim 2$~\msun.

%% file: jvink_aarv_tab_orich.tex
\begin{table*}
\begin{center}
\caption{Oxygen-rich supernova remnants\label{tab:orich}}
\begin{tabular}{lclllll}\hline\hline\noalign{\smallskip}
Name                        & Location & Distance  & Radius   & High Res. Spec. & Other studies\\
                            &          & (kpc)  &   (pc) & \\
\noalign{\smallskip}\hline\noalign{\smallskip}
Cas A (G111.7-2.1)          & MW  &3.4 (1) 
& 2.6 (2) 
& (3,4) 
& (5,6,7,8,9)
\\

MSH 11-5{\it 4} (G292.0+1.8)& MW  & $\gtrsim 6$ (10) 
&  $\gtrsim 7.5$  &   (11) 
& (12,13) 
\\

Puppis A  (G260.4−3.4)      & MW  & 2.2 (14) 
& 17.6 & (15) 
& (16,17) 
\\

B0540-69.3                  & LMC  & 50 (18) 
& 8.4 & (19) 
& (20,21) 
\\

N132D                       & LMC & 50  & 11 & (22,23) 
& (24) 
\\

1E 0102.2-7219 (IKT 22)     & SMC & 60 (18) 
& 5.9  &
(25,26) 
& (27) 
\\

B0049-73.6 (IKT 6)          & SMC & 60  & 21.2 &  & (28,29)
\\

B0103-72.6 (IKT 23)         & SMC & 60  & 26.1 & (28) 
& (30) 
\\
\noalign{\smallskip}\hline\hline\noalign{\smallskip}
\end{tabular}
\end{center}
\scriptsize{MW=Milky Way, LMC=Large Magellanic Cloud, SMC=Small Small Magellanic
Cloud. \\
Selected references:
\\
1) \citet{reed95}; 2) \citet{gotthelf01a};
3)\citet{bleeker01}; 4) \citet{lazendic06}; 
5)  \citet{willingale02}; 6)  \citet{hwang03}; 7) \citet{hwang04};
8) \citet{maeda09}; 9)\citet{delaney10}; 
10) \citet{gaensler03b}; 11) \citet{vink04f}; 
12) \citet{gonzalez03}; 13) \citet{park07}; 
14) \citet{reynoso95};  15) \citet{winkler81}; 
16) \citet{hwang08}; 17) \citet{katsuda08b}; 
18) \citet{schaefer08};
19) \citet{vanderheyden01}; 20) \citet{hwang01}; 21) \citet{park10}; 
22) \citet{behar01}; 23) \citet{canizares01}; 24) \citet{borkowski07}; 
25) \citet{rasmussen01}; 26) \citet{flanagan04};
27) \citet{gaetz00}; 
28) \citet{vanderheyden04}; 29) \citet{hendrick05};
30) \citet{park03b}
}
\end{table*}

%% file: jvink_aarv_fig_orich.tex
\begin{figure*}
\centerline{
\includegraphics[trim=0 30 0 30,clip=true,width=\twofig]{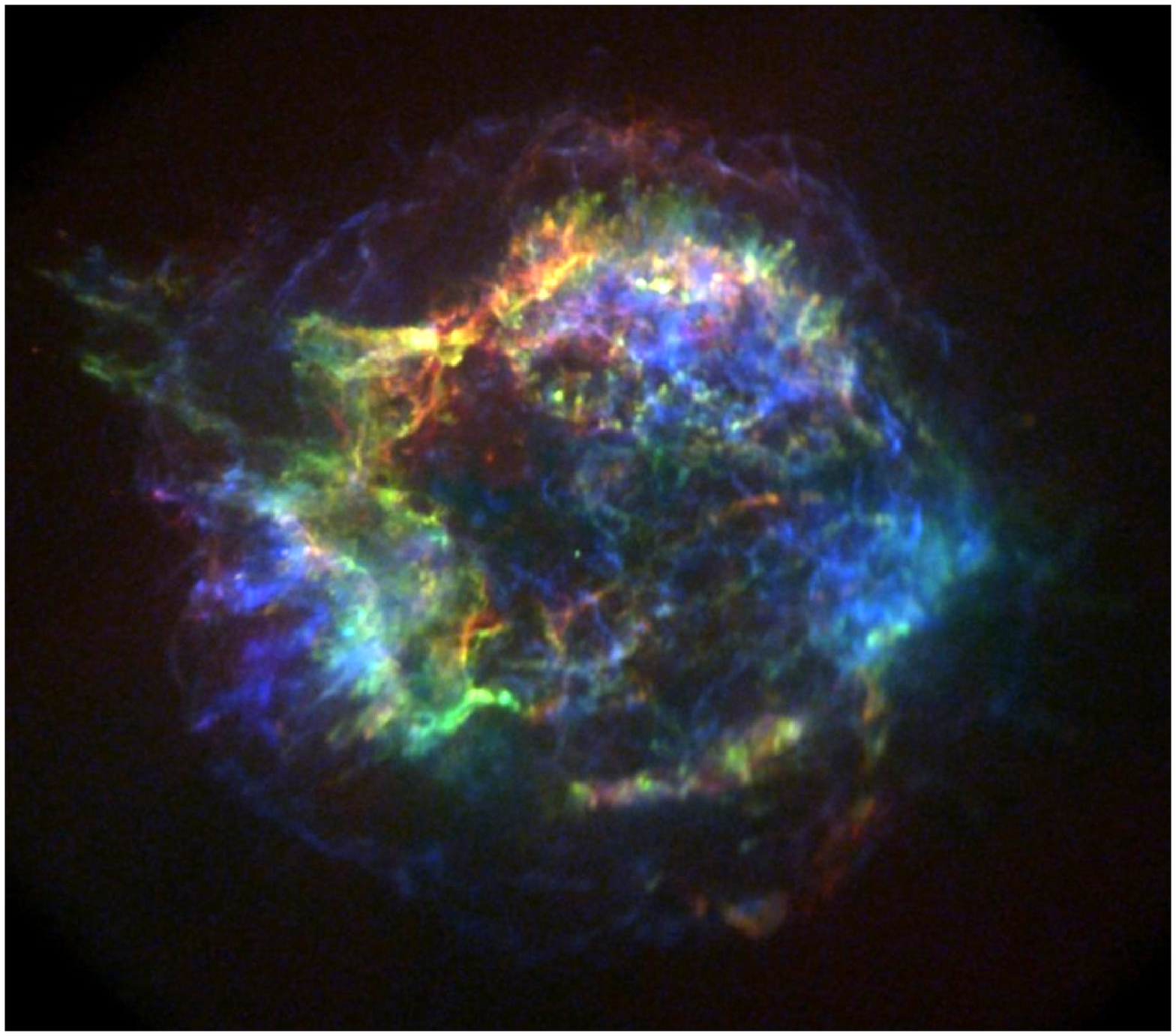}
\includegraphics[width=\twofig]{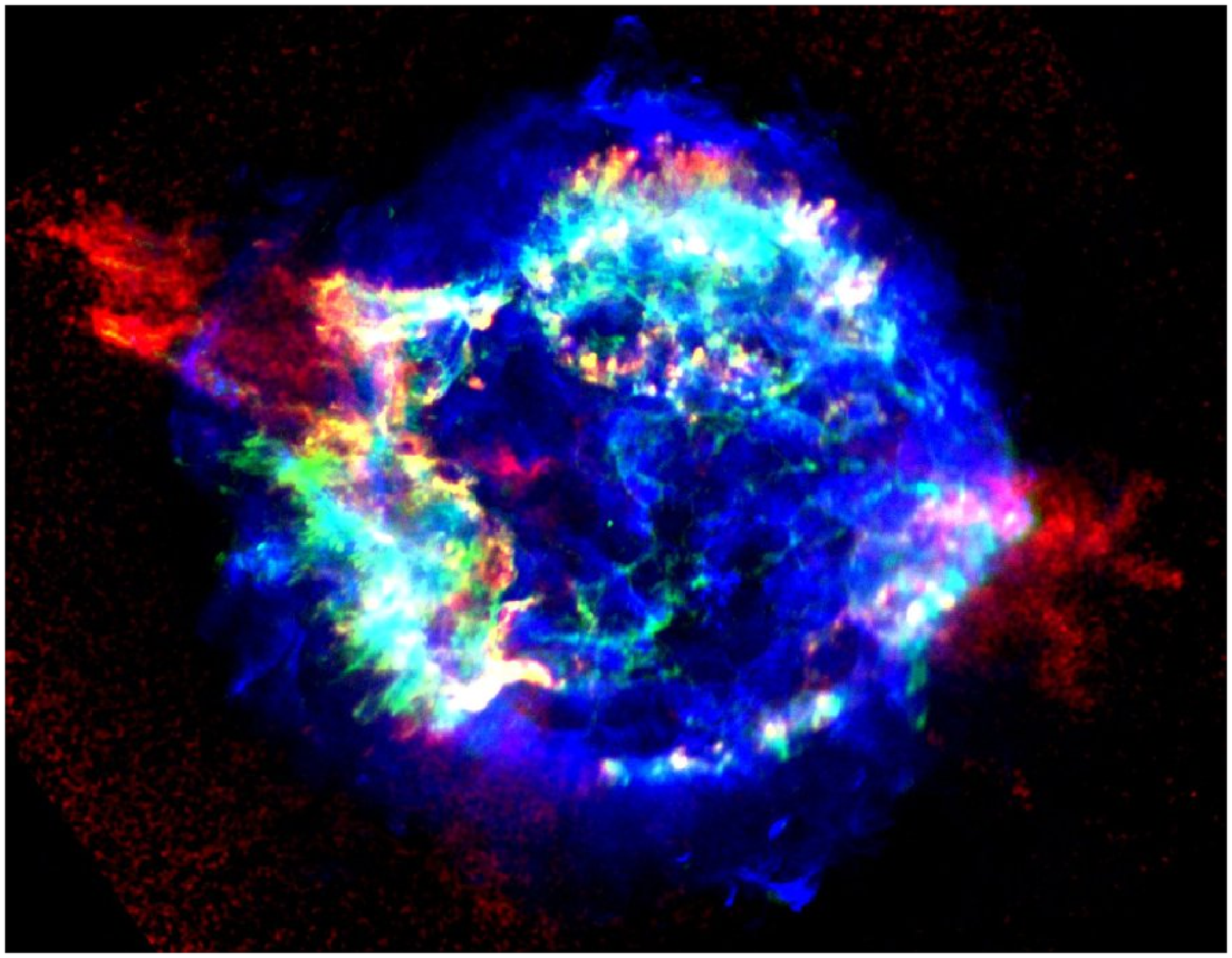}
}
\centerline{
\includegraphics[width=\twofig]{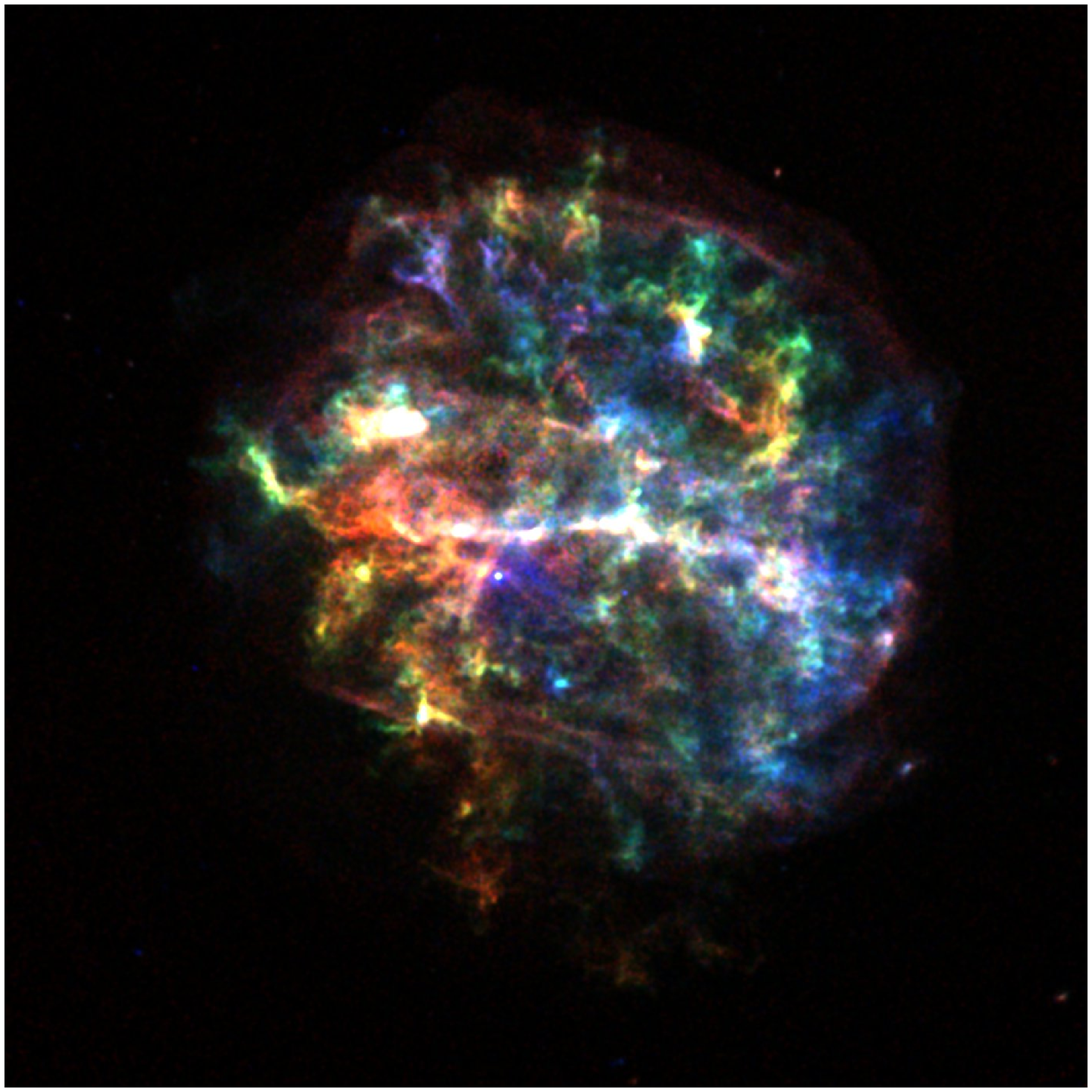}
\includegraphics[width=\twofig]{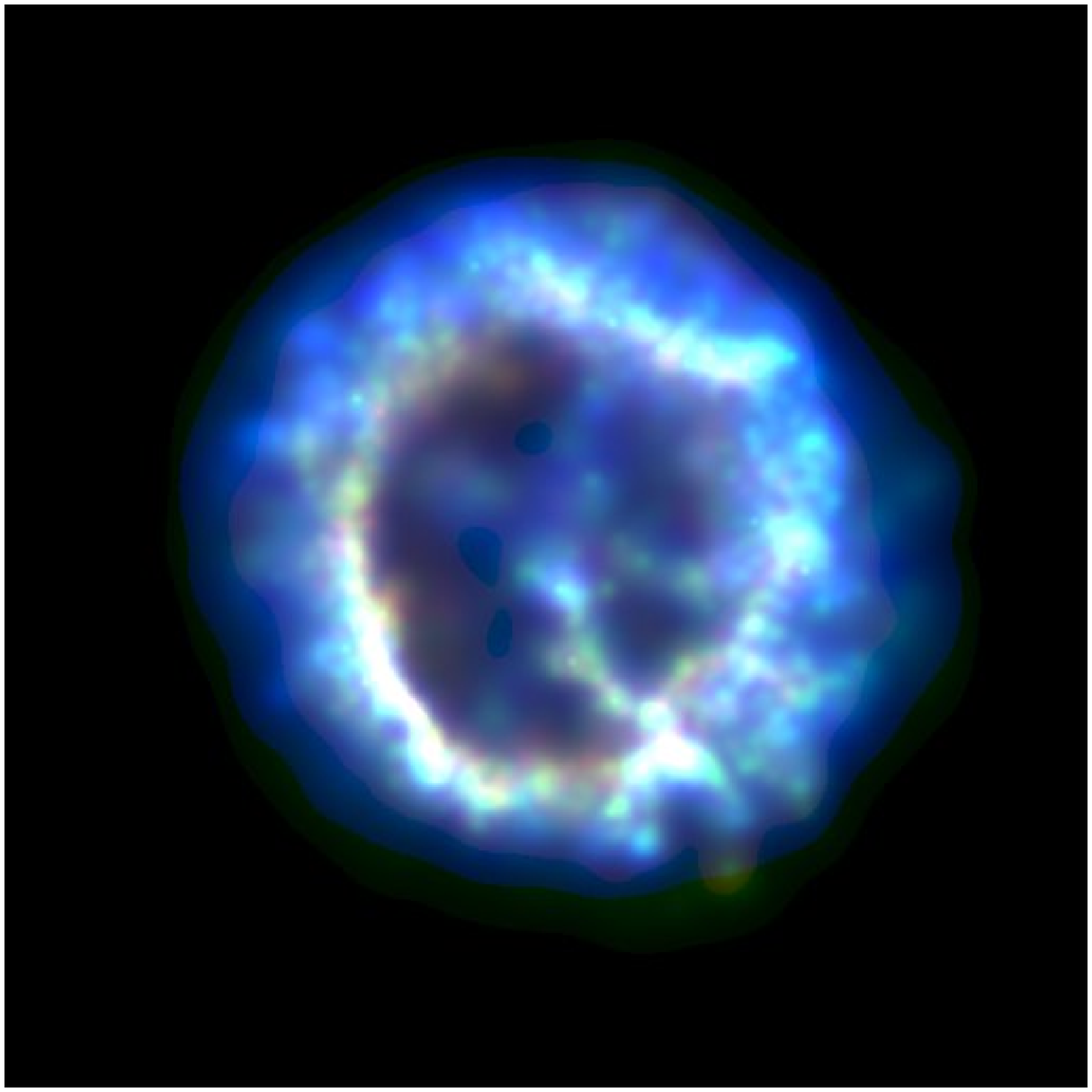}
}
\caption{
Multi-color \cxo\ X-ray images of four oxygen-rich SNRs 
(see Table~\ref{tab:orich} for the sizes).
Top left: Cas A, with red indicating O\,VIII Ly-$\beta$ emission,
green Si\,XIII He$\alpha$, and blue Fe XXVI\,He$\alpha$
\citep[based on the 1~Ms observation][]{hwang04}.
Top right: A different view of Cas A, based on 
a combination of radio (blue), X-ray Si XIII emission (green), and 
the ratio of Si\,XIII over Mg XI\,emission (red). The latter
brings out the jet-counter-jet system, which protrudes the radio/X-ray shell
in the northeast-southwest region \citep{vink04a,hwang04,schure08}.
Bottom left: G292.0+1.8, with red O\,VIII Ly-$\alpha$ emission, 
green is Ne X Ly$\alpha$, and
blue Si XIII HE$\alpha$ emission. Image based
on a  516~ks long observation \citep{park07}.
Bottom right:
The Small Magellanic Cloud SNR 1E 0102.2-7219, with red
O\,VII and O\,VIII emission, green Ne IX and Ne X emission, and
blue emission above 1.27 keV, which includes Mg XI, Mg XII, and Si XVIII
emission (see also Fig.~\ref{fig:mosrgs} and \ref{fig:flanagan}).
\label{fig:orich}
}
\end{figure*}

%% file: jvink_aarv_fig_casa_doppler.tex
\begin{figure*}
\includegraphics[width=\twofig]{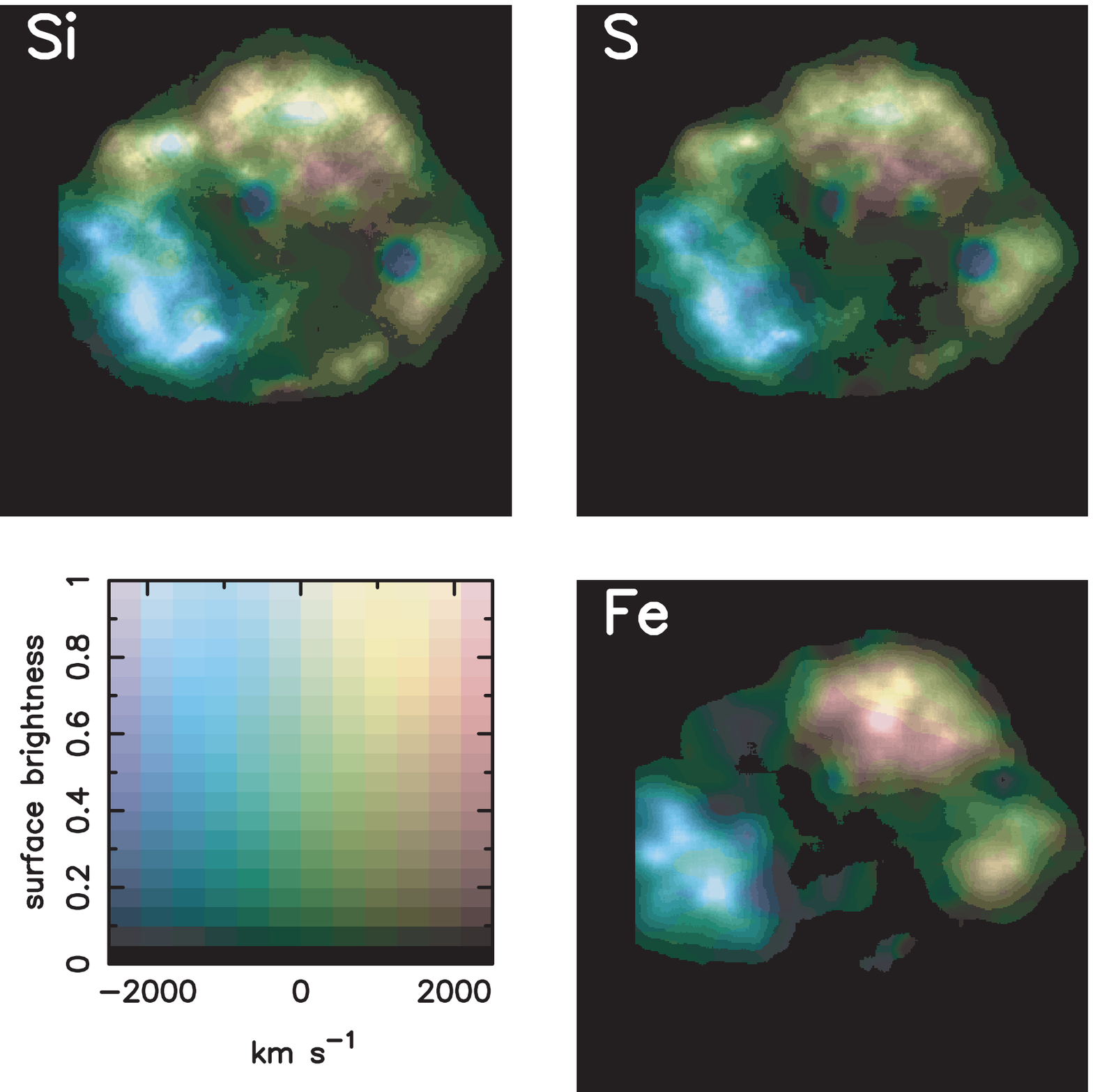}
\hskip 0.05\textwidth
\includegraphics[width=\twofig]{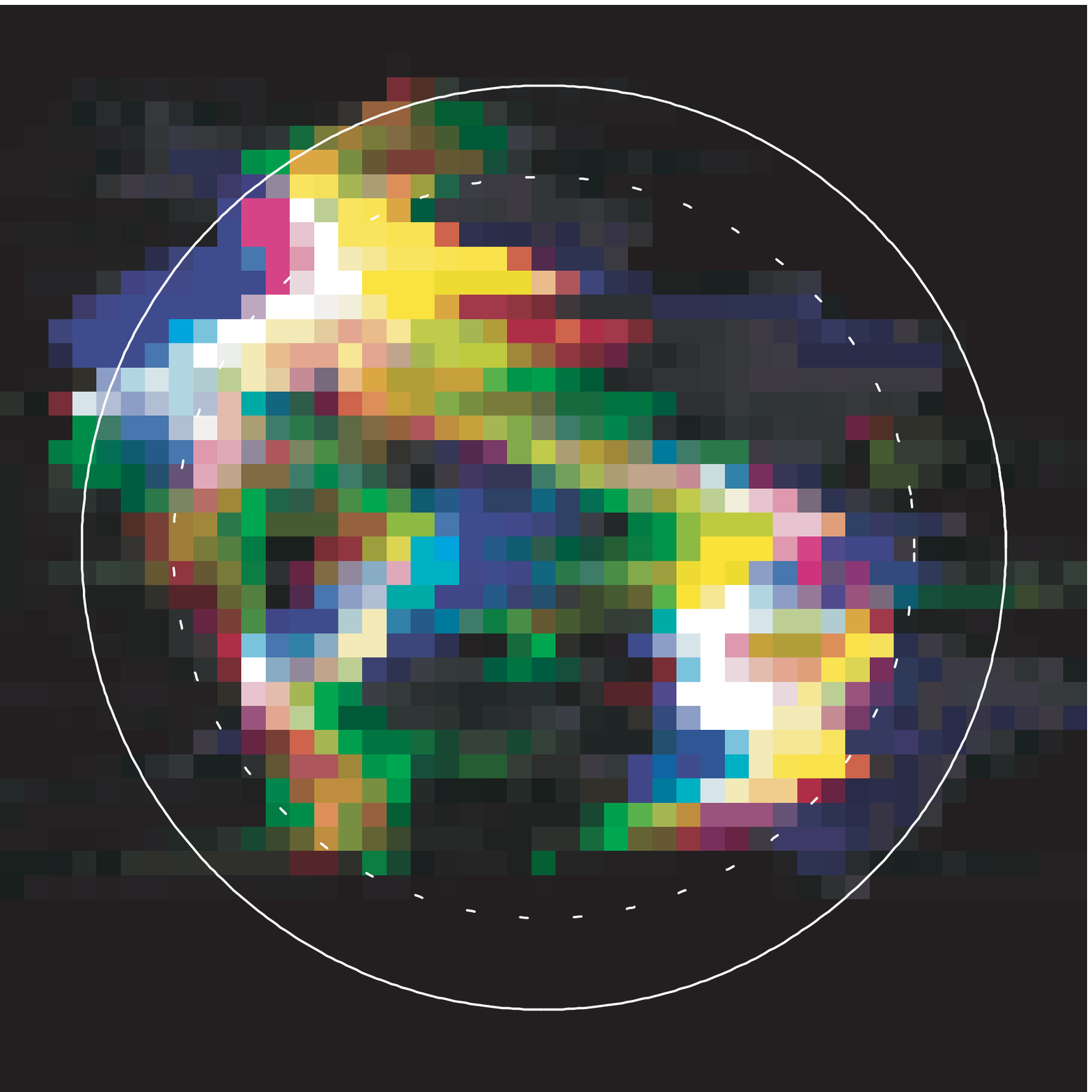}
\caption{
Left: Doppler maps of Cas A in the bands of He-like silicon, sulphur
and iron, based on \xmm-MOS data.
Right: Reconstructed ``side view'' of Cas A, based on the Doppler maps,
using a combination of north-south position and  Doppler velocity 
to reconstruct the third dimension. Red represent silicon, green silicon,
and blue iron. Note that in the North iron is in front of the
silicon-rich layer.
Both figures are based on figures from \citet{willingale03}.
\label{fig:casadoppler}
}
\end{figure*}

%% file: jvink_aarv_fig_1e0102.tex
\begin{figure}
\centerline{
\includegraphics[width=\medfig]{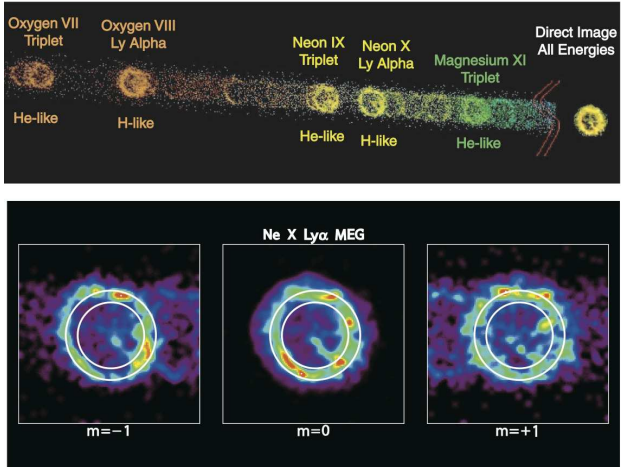}
}
\caption{
\cxo-HETGS spectrum of the oxygen-rich SNR 1E\,0102.2-7219.
The top panel shows part of the spectrum in the negative order
($m=-1$) as seen with the
Medium Energy Grating of the HETGS. The SNR in individual emission
lines shows up as images ordered by wavelength.
Bottom panel: a comparison between Ne X Ly$\alpha$ images in the negative,
zeroth order, and positive order. Differences in radial velocities 
over the face of the SNR result in distortions of the image with respect
to the zeroth order images. Deviations from spherical
symmetry show up as distortions of the images that are different
in the negative and positive orders,
as can be clearly seen here.
(Figure taken from \citet{flanagan04}.)
\label{fig:flanagan}
}
\end{figure}

%% file: jvink_aarv_fig_lopez.tex
\begin{figure}
\centerline{\includegraphics[width=\medfig]{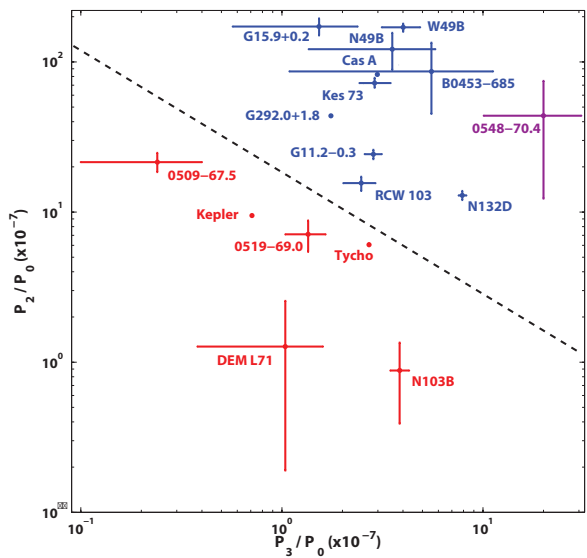}}
\caption{
Results from a statistical analysis of the Si XIII emission images (around 1.85 keV) as observed with the \cxo-ACIS instrument. The numbers $P_0,P_2,P_3$
are linked to different symmetry properties of the SNR emissivity
distribution, with $P_3$ indicating an overall deviation from circular
symmetry, whereas $P_2$ is sensitive to mirror-symmetry. Type Ia SNRs
(red) clearly occupy a different region of the diagram than core collapse
SNRs (blue). (Figure taken from \citet{lopez09}.)
\label{fig:lopez}}
\end{figure}

%% file: jvink_aarv_sn1987a.tex
The Type II supernova SN 1987A was discovered on February 23, 1987
in the LMC. Due its proximity, 
it remains, to date, the best studied supernova. 
The detection of $\sim 20$ neutrinos
during a short interval of $\sim 13$~s by Kamiokande II \citep{hirata87} and
IMB \citep{bionta87} confirmed beautifully the theory that Type II supernovae
are the result of the collapse of a stellar core into a neutron star,
and that most of the energy is released through a cooling population
of neutrinos. The detection of neutrinos also meant that 
the collapse of the core was very accurately timed, and that the core collapse
preceded the optical
detections by about 3 hr \citep[see][for a review]{arnett89}.

In many respects SN 1987A was a peculiar supernova.
Its maximum luminosity
was unusually low, corresponding to
$M_B=-15.5$, compared to $\sim -18$ for typical Type II 
supernovae. This low luminosity is the result of 
the compactness of the progenitor star at the moment of explosion: the 
progenitor was not
a red supergiant, but a B3 I supergiant, identified as the star
Sanduleak -69$^\circ$202. Such an evolved, blue giant is expected to have 
a fast, tenuous wind, creating a low density cavity within a red supergiant
wind of a previous mass loss phase. Instead, optical spectroscopy 
\citep{fransson89} provided evidence for dense 
($n_{\rm e} \sim 3\times 10^4$~cm$^{-3}$) 
circumstellar material enriched in nitrogen. 
Later optical imaging revealed the presence of
three rings \citep{crotts89,wampler90}, which was later confirmed
by a, now iconic, \hst\ image \citep{burrows95}.
The fact that the progenitor exploded as a blue supergiant, the
chemical abundances 
\citep{podsiadlowski92},
and the triple-ring system may be explained by the merging of a 15-20\msun\
star with a 5~\msun\ star, about 20,000 yr prior to explosion;
see the hydrodynamical simulations of \citet{morris07,morris09}.

\input{jvink_aarv_fig_sn87a_xray}

SN 1987A is of great importance, since it is the
first supernova for which one has been able to follow the evolution from supernova 
to SNR. Moreover, our knowledge about the supernova type and 
progenitor is much more detailed than for any other SNR.
One can say that the supernova stage
has given way to the SNR stage, when 
the emission is no longer dominated by radiation from
the cooling ejecta, but by radiation from shock-heated CSM/ejecta.\footnote{
Note that strictly speaking this means that radio supernovae \citep{weiler09}, 
should be considered SNRs, as their radio emission is due to interaction of the
ejecta with the dense CSM.}

A few months after the explosion SN 1987A was detected in hard X-rays
\citep{sunyaev87,dotani87}. The hard X-ray emission was probably the result
of multiple Compton scatterings of $\gamma$-rays generated by the decay
of radioactive material. Nuclear decay-line emission
from $^{56}$Co was detected suprisingly early on, indicating considerable
mixing of inner ejecta toward the surface of the supernova envelope
 \citep[e.g.][and references therein]{leising90}. Moreover,
the nuclear decay lines were redshifted \citep{sandie88,teegarden89,tueller90}.
This is an indication for an asymmetric explosion, bringing to mind the evidence
for an asymmetric explosion of the Cas A supernova in the form
of jets and high velocity
iron knots (\sect~\ref{sec:orich}).

The onset of the SNR stage of SN1987A, as far as the X-ray 
emission is concerned, 
was marked by the 
gradual increase of the soft X-ray emission as seen early on by
ROSAT, starting in February 1991 \citep{hasinger96}.
Since then all major X-ray observatories have observed SN 1987A on a regular
basis, following the X-ray
evolution of the SNR in considerable detail (Fig.~\ref{fig:sn1987a}).

The X-ray observations show that after an initial phase, in which the
outer shock wave was moving through the, relatively tenuous, medium 
associated with the progenitor's fast wind, between 1995-1999 
the shock wave started to interact with the
central dense ring seen in the optical. This ring has a size of 
1.7\arcsec\ by 1.2\arcsec, consistent with a ring of radius 0.19~pc
seen under an angle of $\sim 45$\deg. 
The dense ring itself has considerable
small-scale structure, with regions of different densities, and
fingers sticking inward \citep[][Fig.~\ref{fig:1987acartoon}]{mccray07}.
As a result the ring is not lighting up all at once, but several ``hot spots''
turned on; first in the northeast 
\citep{burrows00}, later
also in other regions. By now all along the ring one can find X-ray emitting
regions \citep[][see Fig.~\ref{fig:1987acartoon}]{racusin09}.

\input{jvink_aarv_fig_sn1987a}

The complexity of the interaction of the blast wave with the dense inner
ring do not fit into the simplified description of SNR hydrodynamics as
sketched in \sect~\ref{sec:hydro} \citep{borkowski97,michael02,mccray07}. 
The basic structure of a blast wave
heating the CSM, and a reverse shock heating of the ejecta
is still valid. But a range of shock velocities
must be considered, as the encounter of the blast wave
with the density enhancements in 
the ring and its protrusions lead to a system of
transmitted and reflected shocks. The transmitted shocks heat the material
of the ring, whereas the reflected shocks go back into the
plasma behind the blast wave. These reflected shocks
are low Mach number shocks, as they go through already shock-heated plasma,
and result in additional heating and compression of the already hot plasma.
The transmitted shocks will have a range of velocities, depending on the
density of the material and the obliquity with which the blast wave hits
the protrusions. The slowest transmitted shocks ($V_S \lesssim 200$~kms)
become radiative (\sect~\ref{sec:phases}), 
resulting in the bright optical radiation that
accounts for the bright knots that lie as beads all along the
optical ring as observed by the 
\hst\ \citep[][Fig.~\ref{fig:1987acartoon}]{kirshner07}. 

Because the blast wave reached the ring in about 5-10~yr, the average
blast wave velocity was initially $\sim 15,000-30,000$~\kms.
In contrast, recent expansion measurements indicate an expansion velocity
of $\sim 4000$~\kms\ in the radio \citep{ng08} and $1800\pm 600$~\kms\
in X-rays \citep{racusin09}. 
Until 2004 (6000 days after the supernova)
the expansion velocity in X-rays was similar
to that in the radio, indicating that the blast wave reached the denser
region of the inner rings around that time. The X-ray emission
also became brighter around that time \citep{park05}. Currently, the
light curve  has flattened again 
\citep[][see Fig.~\ref{fig:sn1987a}]{racusin09}.
The strong deceleration of the blast wave from 15,000-30,000 \kms\ to
1800-5000 \kms\ means
that a strong reverse shock must have developed with
shock velocities in the frame of the ejecta of $\sim 10,000$~\kms.
Note that the difference between the radio and X-ray expansion may
be partially due to the fact that the X-ray emission depends on
density as $\propto n_{\rm e}^2$ (\sect~\ref{sec:continuum}), 
whereas the radio synchrotron emission scales probably
as $\propto n_{\rm e}$ \citep[with some uncertainty, because synchrotron radiation
also depends on the electron acceleration properties of the shock and 
on the magnetic-field  strength, see also the discussion in][]{zhekov10}. 
The X-ray expansion may, therefore, be more skewed toward higher density
regions, than the radio expansion measurements.

From 2000 to 2010 SN 1987A was observed several times with both the
CCD  and grating spectrometers on board \chandra\ and \xmm.
The CCD spectra are well fitted with a two component model, a
soft component  with  $kT_{\rm e} =0.2-0.3$~keV, which is close to CIE, 
and a harder component with $kT_{\rm e} =2-3.5$~keV and 
\net$\approx 3\times 10^{11}$~\netunit\ \citep{park06a,zhekov10}. 
The soft CIE component is associated with the densest regions, heated by
the transmitted shock,
whereas the hotter, NEI component is generally
attributed to the plasma heated by reflected shocks \citep{park06a,zhekov10}.
However, it is not clear whether also reverse shock heated plasma
is responsible for some of the harder X-ray emission.

Note that even the reflected/reverse shock heated
plasma must be relatively dense as the ionization
parameter indicates $n_{\rm e}\approx 950\,t_{10}^{-1}$~\cc, 
with $t_{10}$ the time since the plasma was shocked in units of 10~yr.
As the SNR ages, the low temperature component is becoming
hotter ($kT_{\rm e}\approx 0.2$~keV in 1999 versus $kT_{\rm e}\approx 0.3$~keV
in 2005), whereas the high temperature component is becoming cooler
($kT_{\rm e}\approx 3.3$~keV and $kT_{\rm e}\approx 2.3$~keV, respectively),
see \citet{park06a,zhekov09,zhekov10}.

SN 1987A is an ideal target for the grating spectrometers of \chandra,
due to its small angular extent and its brightness
\citep[][Fig.~\ref{fig:sn87aspec}]{burrows00,michael02,dewey08,zhekov09},
which allows for resolving the 
He-like triplets (\sect~\ref{sec:linediag}).
The earliest high resolution spectrum, taken with
the \chandra\ HETGS, dates from 1999 \citep{burrows00}. It reveals
prominent emission lines from H-like and He-like transitions of
O, Ne, Mg, and Si, and  line widths consistent
with an expansion of $\sim 4000$~\kms.
Recent HETGS and LETGS spectra indicate a line broadening ranging from
1000-5000~\kms\ (FWHM), but with a 
bulk velocity that is surprisingly low, $\sim 100-200$~\kms, consistent with
the radial velocity of the LMC
\citep{dewey08,zhekov09}. This is a strong indication that the
hottest component is not due to the main blast wave, for which
the plasma velocity should move with $\frac{3}{4}V_S$,
but must result from either reflected shock \citep{zhekov09}
or from the reverse shock heated plasma.
This would reconcile the large line widths and hot temperatures,
indicating large shock velocities, with a relative low velocity
in the observers frame (see \sect~\ref{sec:hydro} and \sect~\ref{sec:kT}).

X-ray spectra taken with different instruments do not agree in all details. For example
the \chandra\ CCD spectra indicate a cool component consistent with
CIE \citep{park06a}, whereas the high spectral resolution
HETGS/LETGS spectra  indicate a plasma with a higher electron
temperature, 0.55~keV, and out of ionization
equilibrium, with \net $= 4\times 10^{11}$~\netunit\
 \citep{zhekov09}. The best fit RGS model gives a solution that
lies somewhere in between
 \net$\approx 8\times 10^{11}$~\netunit\ \citep{heng08}.

In reality a range of temperatures and \net\ values 
are expected, given the
complexity of the ring structure. Even the lowest \net\ values reported, 
imply densities in excess of $n_{\rm e}\approx 2000$~cm$^{-3}$.
Nevertheless,
these differences in fit parameters may also result in differences
in abundance determinations. In general
the abundance determinations from various studies are in good agreement
\citep{heng08,zhekov09}, indicating LMC abundances except for
oxygen, which seems underabundant, and nitrogen, which is mildy
overabundant. The RGS and LETGS/HETGS mainly disagree on the iron abundance,
which is underabundant according to fits to the RGS data \citep{heng08}.

SN 1987A will remain an important object for X-ray astronomy in the
coming decennia. It will brighten in the coming years as the transmitted
shock will penetrate deeper and deeper into the ring. The main blast wave,
which will pass around the ring, will at some point encounter
the shell that separates the tenuous wind of the progenitor from
the dense red supergiant wind. This shell is likely to have a
complicated shape, as indicated by the triple-ring system
seen in the optical, and the hydrodynamical simulations of
\citet{morris07}. The SNR will brighten as a result of encountering the shell,
first just outside the equatorial region, later also in the polar regions.
As the reverse shock penetrates deeper into the ejecta one will be able
to probe the composition of the inner ejecta of SN 1987A.

The evolution of  SN\,1987A has already shown us that the early evolution
of SNRs can be much more complicated  than described by
the standard evolutionary models described in \sect~\ref{sec:hydro}. 
SN\,1987A may be a special case,
with its triple-ring system and the possibility that its progenitor
was the result of a recent common envelope merging of a binary system.
However, such mergers are far from unique; even for the text-book shell-type 
SNR Cas A it has been argued that its progenitor was a merger product
(\sect~\ref{sec:orich}).

\input{jvink_aarv_fig_sn1987a_spec}

%% file: jvink_aarv_fig_sn87a_xray.tex
\begin{figure*}
\begin{center}
\includegraphics[width=\bigfig]{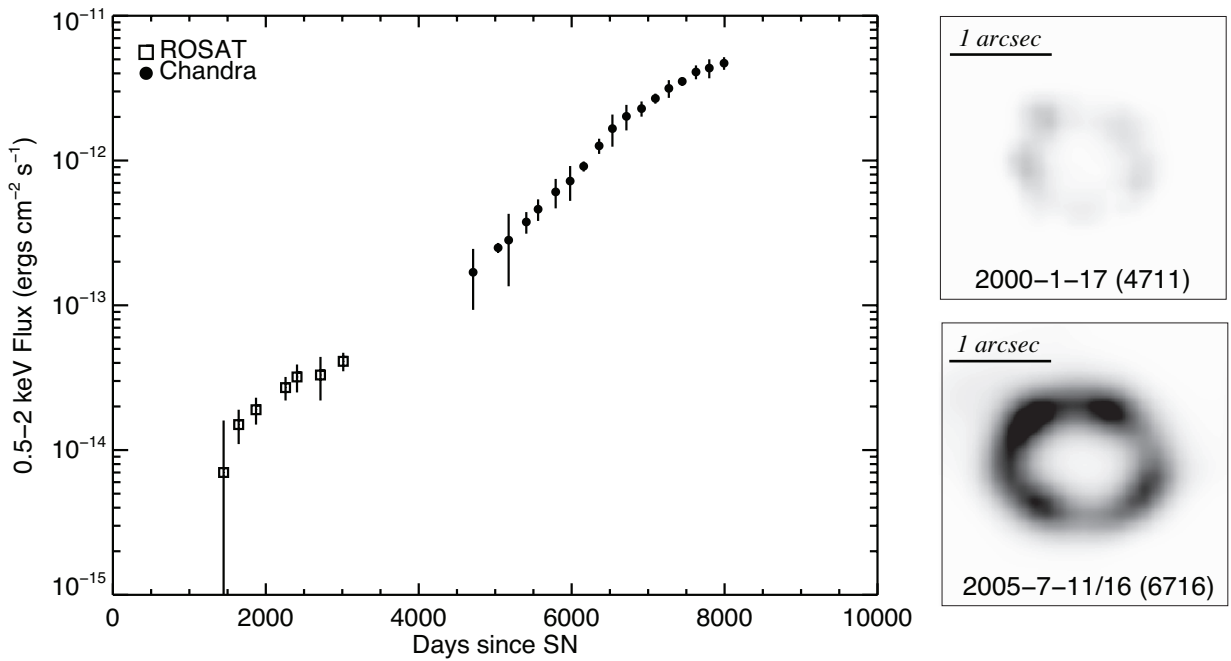}
\end{center}
\caption{
Left: The evolution of the X-ray flux of SN 1987A \citep{racusin09}.
Right: Two frames taken from Fig. 1 of \citet{park06a}, showing \chandra\
images of the SN1987A, which indicate that it did not only substantially 
brighten, but also changed
morphology as the shock heated more parts of the circumstellar ring.
The first regions to be heated were in the northeast, but by 2009
most of the ring is lit up in X-rays, although the brightest region is still
the northeast. In addition the radius of X-ray bright torus has increased
from 0.6\arcsec\ to 0.78\arcsec \citep{racusin09}.
\label{fig:sn1987a}
}
\end{figure*}

%% file: jvink_aarv_fig_sn1987a.tex
\begin{figure*}
\centerline{
\includegraphics[trim=170 140 150 141,clip=true,width=0.4\textwidth]{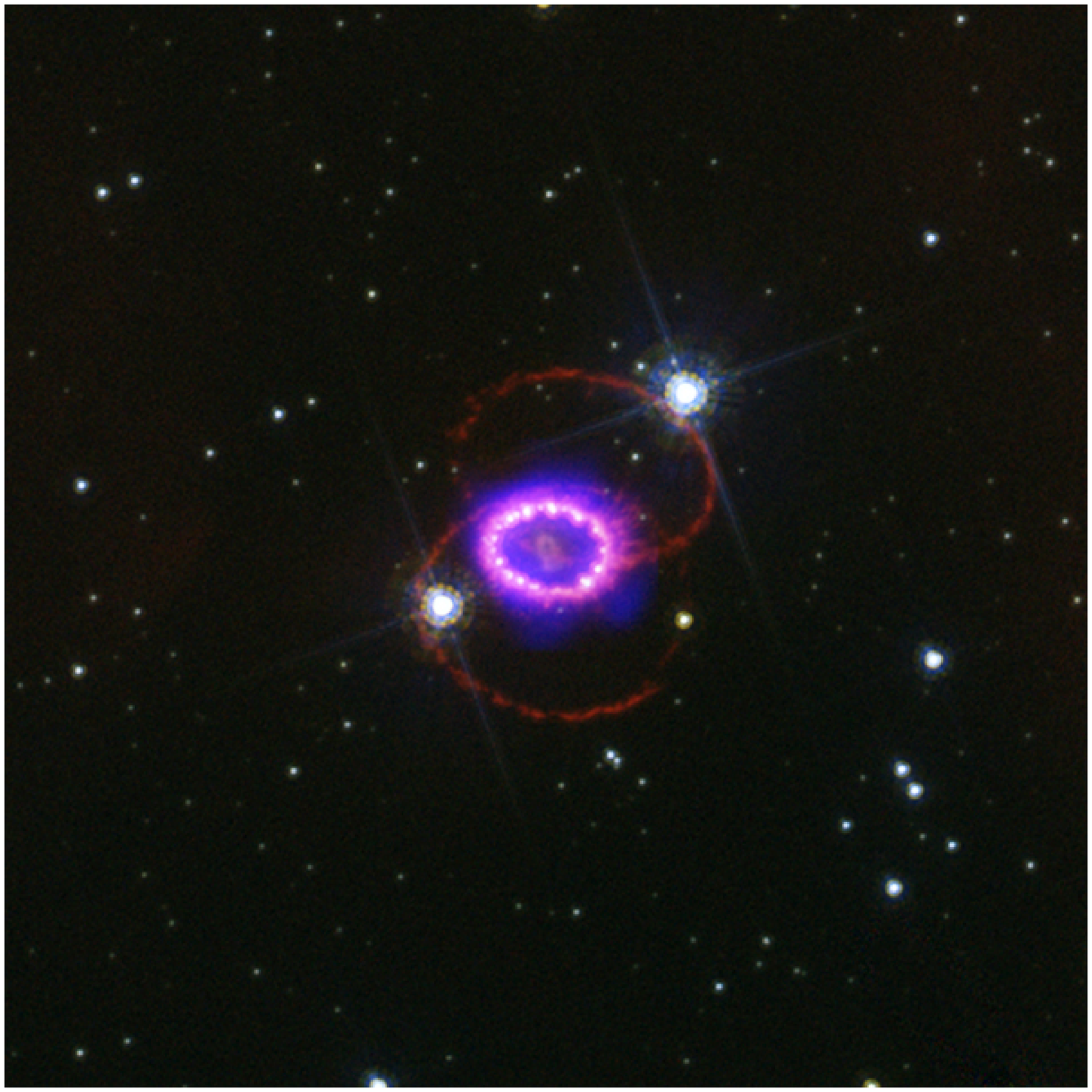}
\includegraphics[width=0.6\textwidth]{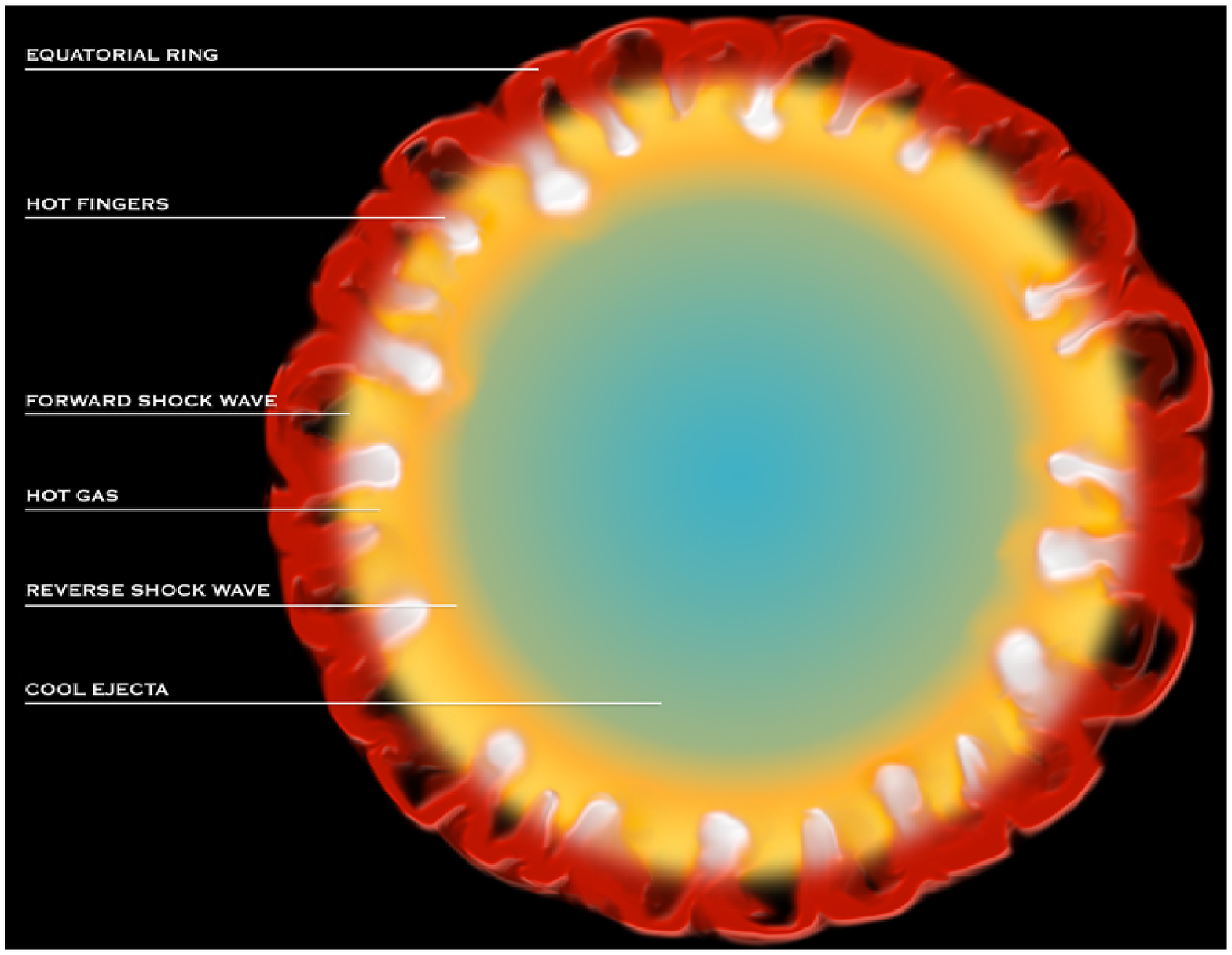}
}
\caption{
Left: Composite optical/X-ray image 
(Credit: X-ray: NASA/CXC/PSU/S.Park \& D.Burrows.; 
Optical: NASA/STScI/CfA/P.Challis).
Right:
Illustration of the interaction of the blast wave of SN1987A with the 
circumstellar ring. The blast wave shock heats the ``fingers'' on the inside
of the ring, and as a result both in X-rays and in the optical the 
emission comes from discrete bright spots rather than smooth structures.
(Source: NASA/CXO/M.Weiss)\label{fig:1987acartoon}
}
\end{figure*}

%% file: jvink_aarv_fig_sn1987a_spec.tex
\begin{figure*}
\centerline{
\includegraphics[width=\textwidth]{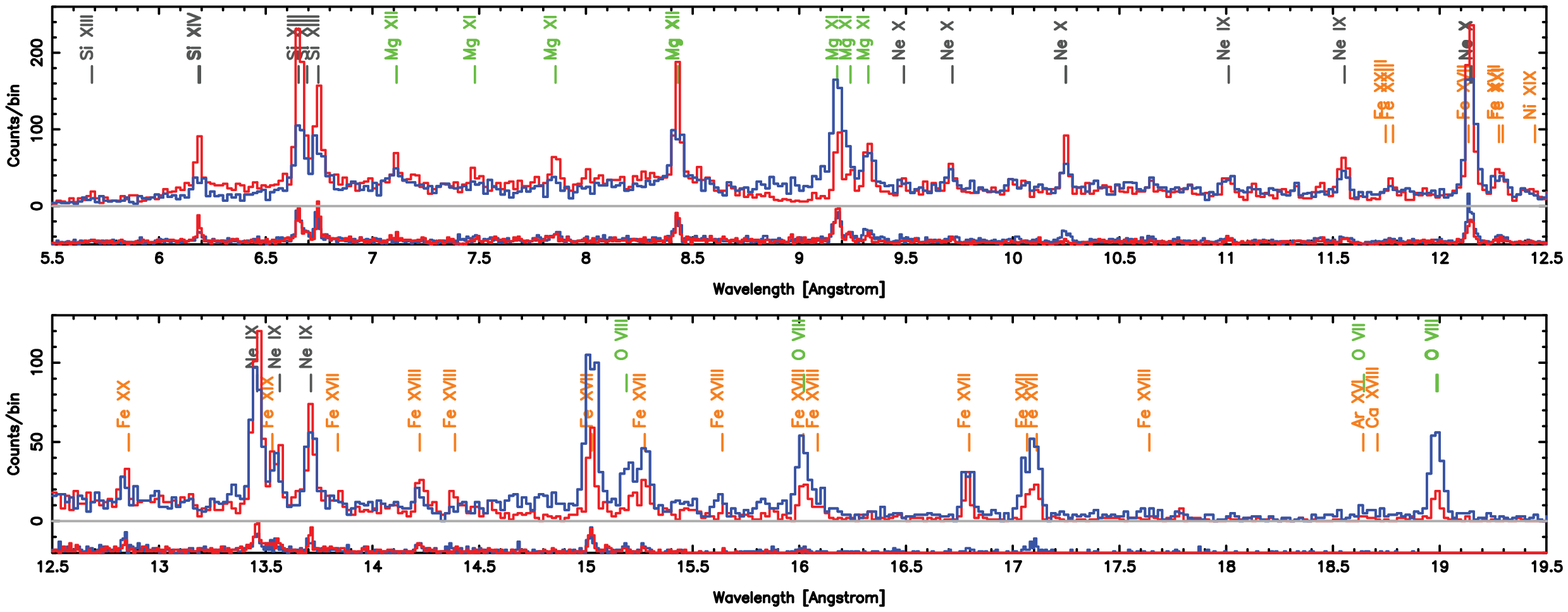}
}
\caption{
Combined \chandra\ HETGS count spectrum of SN 1987A, based on a total
exposure of 360~ks \citep{dewey08}. The positive grating order is colored
red, the negative order blue. (Credit: D. Dewey)
\label{fig:sn87aspec}
}
\end{figure*}

%% file: jvink_aarv_ns.tex
Core collapse SNRs are apart from their abundance patterns, also distinguished
by the presence of a neutron star. In the past it was assumed that this would
automatically imply the presence of a pulsar wind nebula \citep[PWN, see][for an early review on neutron stars in SNRs]{helfand84}, but over the last 20 years
it has become clear that not all neutron stars manifest themselves
through the creation of a PWN \citep[][for a review]{gaensler06}. 

There is good evidence that 5-10\% of
all neutron stars are born as magnetars, 
neutron stars with very high magnetic fields ($10^{14}-10^{15}$~G),
which emit predominantly in the soft and hard X-ray bands 
\citep[][for a review]{mereghetti08}, and whose emission is not powered
by pulsar spin-down, but by magnetic-field decay. These magnetars
rotate slowly ($P=2-12$~s), presumably due to a very rapid initial spin-down.
As a result, their spin-down power is currently too low to drive the
creation of a PWN.\footnote{There is evidence that the most rapidly
spinning magnetar (1E1547.0-5408  $P=2$), located inside a faint SNR 
\citep[G327.24-0.13][]{gelfand07} is
surrounded by a faint PWN \citep{vink09}, but this is disputed by \citet{olausen11}.}

Some neutron stars merely seem to manifest themselves as unresolved
X-ray sources. By lack of a better name, they are usually called
Compact Central Objects \citep[CCOs,][for a review]{deluca08}.
By definition they are associated with SNRs. It is not quite clear whether
CCOs form a real class on their own. The lack of associated PWNe around
them argues either for low surface magnetic fields, hence small period
derivates $\dot{P}$, or long rotational periods, as the
pulsar wind luminosity scales as $\dot{E}=4\pi^2 I \dot{P}/P^3$ (see
\sect~\ref{sec:snrclasses}).
It seems that at least some
of the CCOs are characterized by an unusual low magnetic field \citep[``anti-magnetars''][]{gotthelf08}. Examples are PSR J1210-5226 in SNR PKS 1209-51/52 
\citep[$P=0.42$~s, $B_{\rm p}<3.3\times 10^{11}$~G,][]{gotthelf07}, 
PSR J1852+0040 in SNR Kes 79
\citep[$P=0.10$~s, $B_{\rm p} < 1.5\times 10^{11}$~G,][]{gotthelf05,gotthelf08},
and PSR RX J0822-4300 in the oxygen-rich SNR Puppis A
\citep[$P=0.11$~s, $B_{\rm p}<9.8 \times 10^{11}$~G,][]{gotthelf09}.
Apart from their low magnetic fields these stand out because of their
relatively slow rotation period, which implies that they
are born slowly rotating.

No pulsation period has yet been found for the point source in Cas A,
but since it seems not surrounded by a PWN, it must either be a magnetar
or an anti-magnetar, and given its stable, but declining \citep{heinke10}, X-ray
emission, it is likely to be an anti-magnetar.

Another interesting CCO is 1E161348-5055 in RCW 103. 
This source stands out because
it is highly variable \citep{gotthelf99} and has an unusually large pulsational
period of 6.7 hr \citep[][]{garmire00,deluca06}. This long period, either 
suggests that its is not the neutron star spin period, but instead the orbital
period of the neutron star and an, as yet undetected, low luminosity companion,
or the neutron star is magnetar-like and has been slowed down due to magnetic 
braking in the presence of supernova fall-back material.


So what can SNRs tells us 
about the birth properties and
progenitor stars of the various types of neutron stars?
In particular, 
 a) what is the reason that neutron stars have a wide variety
of properties, and b)
how  SNRs can help to infer the initial birth properties of neutron stars?
These are questions for which observational
data is scarce, as we know little about neutron star progenitors,
and the birth of neutron stars
is hidden to us by the supernova event itself, with the exception of
the detection of neutrinos from SN 1987A,
which marked the birth of the neutron star
(\sect~\ref{sec:sn1987a}). Strangely enough, no neutron star has yet
been detected in SN\.1987A.

From a theoretical point it is not clear what kinds of progenitors
produce different types of neutron stars or black holes \citep{heger03}, 
and what the initial spin periods of neutron stars are \citep{ott06}. 
\citet{heger03} and \citet{nomoto10} have argued that less massive progenitors 
($M_{\rm MS} \lesssim  25$~\msun) produce neutron stars, whereas
more massive
progenitors result in the creation of a black hole, either directly
or due to late time supernova material fall back.
\citet{heger03} stress that this situation is modified by strong stellar winds,
which reduce the massive hydrogen envelope. As a result very massive
stars may create again neutron stars.
This also depends on the initial metallicity of the star, as wind loss
increases with metallicity. Indeed, there is evidence that
magnetars are associated with very massive stars, based on the young
ages of the stellar cluster in which they are found 
\citep{figer05,gaensler05c,muno06}. But the evidence
for this is not unambiguous \citep{davies09} and one should also keep in mind
that even if some magnetars are associated with the most massive 
stars, this does not necessarily imply that all magnetars are.

 The timing properties of pulsars
can be used to determine their characteristic age, 
\begin{equation}
\tau_{\rm c}=\frac{1}{2}\frac{P}{\dot{P}}, \label{eq:charage}
\end{equation}
which is, however, only a reliable age estimator if the initial
spin period $P_0\ll P$, and if the spectral braking index is $n=3$,\footnote{
$n$ is defined as $\dot{\Omega}= -k\Omega^n$} 
which is the value for braking due to  a rotating magnetic dipole.
The presence of neutron stars in SNRs helps to verify these assumptions
as $\tau_{\rm c}$ can be compared with age estimates of the SNR.
This helped establish that for the neutron star in Puppis A $P_0\approx P$
\citep{gotthelf09}. 

The presence of a SNR surrounding a PWN can also be used to constrain
the initial period in another way, namely by considering the pressure
confinement of the PWN by the shock heated SNR shell
\citep{vanderswaluw01,chevalier05}. The total pressure inside the PWN is
determined by the integrated energy loss of the pulsar, which in itself
is mostly determined by the initial spin period. A surprising result
of these studies is that there is a large range in initial spin periods,
which is theoretically not well understood \citep[e.g.][]{ott06}.
Moreover, and this brings us to the above formulated question a), there
is no clear relation between the initial period of a neutron star
and the type of supernova as determined from the SNR characteristics
\citep{chevalier05}.\footnote{See also \citet{chevalier10}.}

The same conclusion can be drawn when also considering the
other classes of neutron stars, CCOs and magnetars. First of all,
the discovery of these neutron stars owe much to the fact that unresolved
X-ray sources were discovered in SNRs. Most 
CCOs and magnetars that are more than a million year old are very difficult
to discover. The CCO will not be surrounded by a SNR anymore (and hence
no longer deserve their name), and will have cooled substantially.
Magnetic field decay makes that old magnetars
become radio-quiet dim X-ray sources.
Neutron stars in SNRs are therefore important to estimate the
the neutron star birthrate, especially of radio-quiet neutron stars,
and to estimate the initial spin-period and magnetic-field distributions.

As for the characteristics of the SNRs with CCOs and magnetars. At least
two CCOs are found in oxygen-rich SNRs (\sect~\ref{sec:orich}), 
 the remnants of the most massive stars. For one of them
(in Puppis A) we know that the neutron star has a low magnetic field.
For the other one (Cas A), no pulsation has yet been found, and its
nature is still unclear.
For magnetars, there is so far only circumstantial evidence linking
them to the core collapse of the most massive stars.
There is also evidence that a perfectly normal neutron star
can be created from  a very massive star:
The oxygen-rich SNR G292.0+1.8, has a pulsar with $P\approx 0.135$~s 
\citep{hughes03b}, and $B\approx 10^{13}$~G \citep{chevalier05}.
So the conclusion must be that the most massive
stars make neutron stars with any type of magnetic
field (low, normal, or magnetar-like).

This SNR/neutron star combination of G292.0+1.8 (Fig.~\ref{fig:orich},
\sect~\ref{sec:orich})
stands out for another reason as well.
Its X-ray spectrum has been used to determine an abundance
pattern that suggests that G292.0+1.8 is the result of
the explosion of a star with an initial mass of $M=30-40$~\msun\
\citep{gonzalez03}. This is the mass range for which 
theorists have argued that core collapse should lead
to the creation of a black hole, instead of a neutron star.
Clearly this is important for our understanding of neutron star and black
hole formation, but the evidence needs to be further studied.

In contrast, no neutron star has yet been found in the Cygnus Loop 
(Fig.~\ref{fig:morphology}, see \sect~\ref{sec:radiative}).
If this means that the event did not produce a neutron star but a black
hole, we have an example of a less massive core collapse event 
($M_{\rm MS}\approx 15$~\msun, \sect~\ref{sec:mature_metals}) that did
not produce a neutron star. But the Cygnus Loop is very big and perhaps
a neutron star is inside or in the vicinity of the SNR and has not
yet been identified.

As for  magnetars,
the origin of their magnetic fields is not clear. An interesting hypothesis
is that the magnetic fields result from a dynamo mechanism in
rapidly rotating proto-neutron stars \citep{duncan92}. This requires
rather short initial spin period of 
$P_0\approx 1-3$~ms. Since the rotational
energy of such a rapidly rotating neutron star is 
$E_{\rm rot} = \frac{1}{2} I \Omega^2\approx 3\times10^{52}
\bigl(\frac{P}{1\,\rm  ms}\bigr)^{-2}$~erg, one expects that a large
part of the rotational energy will power the supernova \citep{arons03,allen04}.
But an X-ray study of SNRs with magnetars by \citet{vink06c} shows that
the SNRs are consistent with typical explosion energies of 
$10^{51}$~erg. This is more consistent with the alternative
hypothesis concerning the origin of magnetar fields, namely that magnetars
are born from the core collapse of stars with very large magnetic
fields \citep[the fossile field hypothesis][]{ferrario06}.
There is another, albeit non-conclusive, 
argument against the idea that
magnetars are born rapidly rotating. As explained above, there is a large
spread in initial spin periods of normal neutron stars. In fact there
is very little evidence for pulsars born with periods $P_0 < 10$~ms
\citep{vanderswaluw01,chevalier05}. How likely is it then that
5-10\% of the neutron stars are born with extremely short periods, 
$P_0\sim 1-5$~ms?

The debate on the origin of magnetar magnetic fields 
has not yet settled,
but it is clear that SNRs play  an important role
in this debate \citep[e.g.][]{durant06,horvath11}.

%% file: jvink_aarv_cl_vela.tex
A couple of thousand years after the explosion
SNRs have swept up considerably more
mass than the supernova ejected.
SNRs in this phase of their evolution are often referred to
as ``mature SNRs'' (\sect~\ref{sec:hydro}). 
Once
their shocks have slowed down to below $V_s< 200$~\kms, the shocks
are becoming radiative, and the SNR evolution
can no longer be described by the Sedov-model (Eq.~\ref{eq:sedovr}). 
Slowing down
even more results in a post-shock region that 
is too cool to emit substantially in X-rays.
Nevertheless, some SNRs with slow shocks
have bright interior X-ray 
emission, as will be discussed in \sect~\ref{sec:mixed}.

The
prototypical ``mature'' SNRs are the
Cygnus Loop (Fig.~\ref{fig:morphology}a) and the Vela SNR,
which are on the border between phase II and phase III
(\sect~\ref{sec:hydro}). They are both considered to be core collapse SNRs,
as I will describe below.
They have earned their prototypical status predominantly by 
being {\em nearby examples} of mature SNRs. 
Their small distance also means that their X-ray emission can be studied
on very small physical length scales. And because the absorption column
is low, one can also study their soft X-ray and UV emission.

The Cygnus Loop is at a distance of
$540\pm100$~pc \citep{blair05}, and has an angular radius of
1.4\deg, corresponding to 13~pc, with a break-out region in the south.
Its age has been estimated to be 8,000 to 14,000~yr 
\citep{levenson98,levenson02,katsuda08}, and its shocks have velocities
in the range
of 150-400~\kms\ \citep{levenson98,levenson02,blair05,salvesen09}. This range in shock
velocities indicates  that the
Cygnus Loop has both radiative and non-radiative shocks. 
In fact, the Cygnus Loop derives part of its fame from
the beautiful optical nebula associated with
the radiative shocks, the so-called Veil Nebula.
The large contrasts in shock properties are best explained  by
the idea that the Cygnus Loop is evolving in a wind-blown cavity 
\citep{mccray79,charles85,levenson98,miyata99}. 
Earlier in its evolution the SNR  must have 
expanded rapidly within the tenuous medium inside the wind-blown bubble. But
currently most of the shock is interacting with the dense shell
swept up by the progenitor's wind (see also \sect~\ref{sec:hydro}).
In such a case parts of the shock may reach the radiative phase
at a relatively
young age. Another example is the probably even younger SNR RCW 86 
\citep[probably SN 185,][]{stephenson02}, which also shows a mixture of radiative and non-radiative shocks \citep[e.g.][]{smith97,ghavamian01}, and whose plasma 
ionization ages and emission measures testify of large density contrasts 
\citep{vink97,bamba00,bocchino00,rho02,vink06d}.
As a ``cavity SNR'' the Cygnus Loop may therefore not deserve its status
as a prototypical mature SNR. On the other hand, all SNRs when investigated
in detail appear to have some unique features.

X-ray spectroscopy of the Cygnus Loop has provided some peculiar results.
One example is that imaging spectroscopy
with \asca, \suz, \chandra\ and \xmm\ has revealed that the
abundances of the bright X-ray shell  are sub-solar, with typical
depletion factors
of $\sim 5$ \citep{miyata94,miyata07,katsuda08d, nemes08}.\footnote{ 
Although this is not confirmed by \citet{levenson02}, who only
analyzed a small portion of the SNR with \chandra.}
The low abundance of the inert
element Ne shows that this is  not due to
dust depletion. In contrast, \citet{katsuda08d} found that near
the rim of the bright shell there are abundance enhancements of about a factor 2.
The same team later attributed these enhanced abundances at the very
edge of the SNR to charge exchange reactions between unshocked neutral
hydrogen and shock ionized O VIII \citep{katsuda11}.  
However, as they point out, further investigation is clarify  the possible 
role of resonant-line scattering on the line equivalent
widths. 
As discussed in \sect~\ref{sec:linediag},
this process
results in scattering of photons with energies corresponding
to resonant lines out of the line of
sight. As the photons will eventually escape the SNR,
it will result in 
enhanced line emission at the edges of emission regions, and
an attenuation of resonant line emission from the regions projected
toward the interior. Indeed, there
is some evidence for resonant line scattering \citep{miyata08}, but
currently it is not clear whether charge exchange and/or
resonant line scattering can explain 
the X-ray line emission from the rim of the Cygnus Loop, or whether
the metallicity of the plasma is really depleted.
Future high resolution X-ray spectroscopy may be needed to
clarify this issue, as it is needed to resolve the
resonant lines from forbidden lines. Resonant line scattering will
boost the G-ratio toward the center of the SNR, and lower it at the
edge of the rim. Charge exchange will only enhance  the G-ratio
at the edge of the rim, but without affecting the G-ratio in the rest of
the shell.
In \sect~\ref{sec:mature_metals} I will discuss the X-ray spectroscopic results
of inner regions of the Cygnus Loop.

One striking feature of mature SNRs like the Cygnus Loop is the complexity
of the emission, which is caused by the interactions of the shocks
with density inhomogeneities (clouds, or cloudlets), and by 
a variety of shock velocities giving rise to both radiative and non-radiative
shocks. Moreover, these shock-cloud interactions lead to the creation
of reflected and transmitted shocks \citep{levenson96,patnaude02}, 
not unlike the situation described
in connection with SN 1987A (\sect~\ref{sec:sn1987a}).

\input{jvink_aarv_fig_vela}

A SNR that has been used quite extensively to investigate the interaction
of the blast with the inhomogeneous CSM is the Vela SNR.
The Vela SNR contains a pulsar with a rotational period of 89~ms, which is
 surrounded by a pulsar wind nebula
\citep{helfand01}. Its characteristic age is $\sim 11,000$~yr. 
The actual age of the pulsar and SNR may be older, $\sim 20,000$~yr 
\citep{aschenbach95}. VLBI parallax  measurements of the pulsar
show that its distance is $287\pm19$~pc \citep{dodson03}, at which distance the
angular size of the SNR, 4\deg, corresponds to a physical radius of 20~pc.
The SNR is located in a complex, starforming
region, which also contains the $\gamma^2$-Velorum massive stellar binary.
Both of them are embedded within the hot bubble/HII region
that is known as the Gum-nebula \citep{aschenbach95,sushch11}.

\rosat\ and
\xmm\ X-ray imaging spectroscopy of isolated regions, 
in combination with optical narrow filter
imaging show a complex structure \citep{bocchino00,lu00,miceli05} that is
best described in terms of three components: 1) a tenuous inter-cloud
medium, which is difficult to characterize in X-rays due to its diffuse
nature and low surface brightness, 2) soft X-ray emission from 
cloud cores, heated by slow transmitted shocks, and with temperatures
around $kT\approx 0.1$~keV, and 3) hotter regions surrounding
the clouds, consisting perhaps of plasma evaporated from the
clouds \citep[see also][]{kahn85}; this plasma could be heated
by thermal conduction from the hot intercloud medium, but may
otherwise cool by radiation and thermal conduction with the 
cool cloud cores.
The cloud cores have a moderately
high densities of $n\approx 5$~cm$^{-3}$.

%% file: jvink_aarv_fig_vela.tex
\begin{figure*}
\centerline{
\includegraphics[width=\medfig]{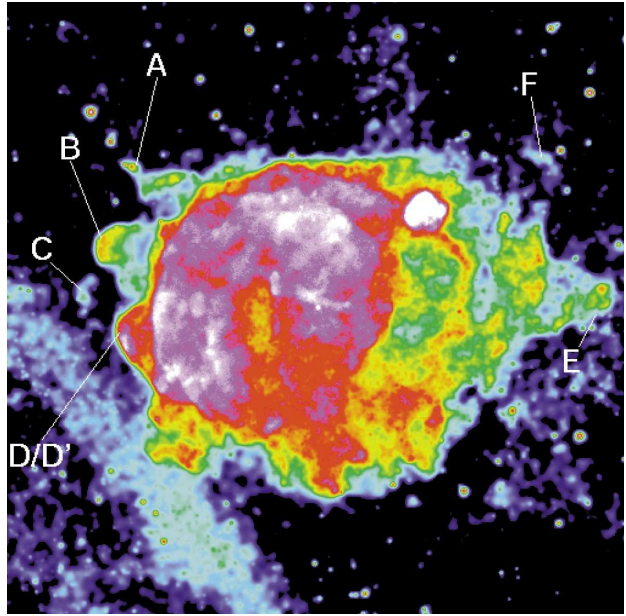}}
\caption{
False color map of the Vela SNR as observed in X-rays with
\rosat\ \citep{aschenbach95}. The figure points out the
location of the ``shrapnels'' and their designations.(Source:
Max-Planck-Institut für extraterrestrische Physik 
(\url{http://www.mpe.mpg.de}).)
\label{fig:vela}
}
\end{figure*}

%% file: jvink_aarv_mature_metals.tex
\input{jvink_aarv_fig_cygnusloop}

One striking result of imaging X-ray spectroscopy of mature
SNRs has been the identification
of metal-rich plasmas even within mature SNRs. 
In \sect~\ref{sec:typeia} and \sect~\ref{sec:orich} 
a few mature SNRs were mentioned that 
could be identified
as Type Ia (e.g. Dem L71) or core collapse/oxygen-rich SNRs 
(e.g. N132D and Puppis A) based on their abundances.
However, even older SNRs
such as Vela and the Cygnus Loop have regions
with high metal abundances.

Despite its large angular size, a significant fraction of
the Cygnus Loop has now been mapped with \suz\ and \xmm\ 
CCD-detectors \citep{uchida09a}. This has revealed that metal-rich plasma
is associated with the
low surface brightness interior of the SNR, which has a temperature
that is considerably hotter \citep[$kT\approx 0.6$~keV,][]{tsunemi07}
than the bright shell ($kT\approx 0.2$~keV). Even within this hotter
region there is some variation in abundance patterns, with O, Ne, and Mg,
more abundant in the outer regions, and Si, S, and Fe more abundant
in the central region
\citep[Fig.~\ref{fig:cygnusloop},][]{tsunemi07,uchida08}. 
Recently also the detection of Ar lines
was reported, indicating high Ar abundances \citep[8-9 times solar,][]{uchida11}.
This suggests a layered explosion, with the more massive elements
situated 
in the center. This is in contrast to Cas A, for which the Fe-rich ejecta
has (partially) overtaken the Si-rich ejecta (\sect~\ref{sec:orich}).

The overall abundance pattern of the inside of the Cygnus Loop is consistent
with that expected from a core collapse supernova with an initial progenitor
mass of $M \approx 15$~\msun, but with 5-10 times more
Fe than predicted by supernova explosion models \citep{tsunemi07}. 
As noted in \sect~\ref{sec:supernovae} 
the predicted Fe yield has considerable uncertainty. 
The idea that the Cygnus Loop progenitor was not a very massive star
($M>20$~\msun) is also consistent with the radius
of the stellar wind bubble of 13~pc, which is more in line
with an early type B-star than with an early type O-star 
\citep{chevalier90}.

It is tempting to speculate that the more layered metallicity
structure of the Cygnus Loop, as opposed to the more radially
mixed metallicity of Cas A, has to do with the different progenitor
types. The Cygnus Loop progenitor, with its relatively
low initial mass, probably exploded as a red supergiant, but may
still have had a significant fraction of the hydrogen envelope at the time
of explosion. The Cas A progenitor was almost completely stripped
of its hydrogen envelope at the time of explosion (\sect~\ref{sec:orich}),
which made it perhaps easier for convective motions inside the supernova
explosion to affect the layering of the ejecta. 
Indeed the simulations of \citet{kifonidis03} show a qualitative
difference in core ejecta velocities between Type II 
and Type Ib (i.e. stripped progenitor) supernovae.

The metal-rich plasma in the Vela SNR is not associated with the interior,
but rather with a  number of protruding plasma clouds, usually referred
to as the Vela ``bullets'' or ``shrapnels'' 
\citep[Fig.~\ref{fig:vela},][]{aschenbach95,strom95}. 
The name shrapnels is justified by their high abundances \citep{tsunemi99},
which indeed suggest that the clouds consist of supernova ejecta.
Although it is not clear whether this also means that the clouds
were ejected as such by the supernova, or whether they were formed due
to some hydrodynamical instabilities at the contact discontinuity
in the early SNR phase.
The morphology of the SNRs are suggestive of clouds surrounded by
bow-shocks, whose opening angles indicate 
Mach numbers $M\approx 3-4$. For the hot, tenuous
environment of the Vela SNR
\citep[$n\sim 0.01$~cm$^{-3}$][]{strom95},
this corresponds to velocities of $V\approx 300-700$~\kms.
The average shrapnel densities are $n_{\rm e}\approx 0.1-1~$cm$^{-3}$ with masses
for shrapnel A and D of, respectively, 0.005~\msun and 0.1~\msun
\citep{katsuda05,katsuda06}. 

Given their projected distance from the main shell of $\sim 1$\deg, 
corresponding to $l\approx5$~pc, the shrapnels must have survived passage through the
hot bubble for quite some time.
Even for a mean average velocity difference between shock and shrapnels
of $\Delta v\approx 1000$~\kms, the travel time is
$t_{\rm trav} \sim l/\Delta v \sim 5000$~yr.
This should be compared to the typical cloud-crushing time $t_{\rm cc}$
\citep{klein94}:
\begin{equation}
t_{\rm cc} = \frac{\chi^{1/2}R}{v_{\rm b}} = 980
\Bigl(\frac{\chi}{100}\Bigr)^{1/2} 
\Bigl(\frac{v_b}{1000~{\rm km\,s}^{-1}}\Bigr)^{-1}
\Bigl(
\frac{R}{1\ {\rm pc}}\Bigr) \ {\rm yr},\label{eq:cloud_crushing}
\end{equation}
with $\chi$ the density contrast between the shrapnel and its ambient medium.
Comparison of these time scales, assuming $v_{\rm b}\sim \Delta v$, 
shows that the apparent survival
of the shrapnels indicates that the density contrast between
the shrapnels and the ambient medium must be high $\chi\sim 1000$. 
A higher average velocity does not help, as it lowers both the travel time
and the destruction time.
A higher density contrast could either mean that  the initial density
of the knots was higher than it is now, or that the ambient density
is lower, $\sim 0.001$~cm$^{-3}$.
Either way, the Vela shrapnels may owe their survival to the low density
of the ambient medium and a fortuitous timing concerning the
phase in which we happen to observe the Vela SNR. There is
some morphological evidence for hydrodynamical instabilities
occurring in shrapnel D \citep{katsuda05}, which will 
lead to the destruction of this cloud.

As already indicated, imaging spectroscopy
with \asca\ \citep{tsunemi99},
\chandra\ \citep{miyata01}, and \xmm\ \citep{katsuda05,katsuda06}
have shown that the clouds represent
supernova ejecta.
These studies also indicate that the shrapnels originate from different
layers inside the supernova, with Shrapnel A being more rich in
the oxygen-burning product Si, and shrapnel D more rich in 
Ne and the carbon-burning products O, Ne, Mg (with enhanced abundances
by a factor 5-10 with respect to solar).
The shrapnels are apparent due to their projected distance from the Vela
SNR. However, more shrapnels may be projected unto the main SNR shell. 
Indeed,
\citet{miceli08} found various regions observed by \xmm\ that
have overabundances of O, Ne, Mg and Fe, which may be either
indicative of other shrapnels, or which may be regions containing
the remains of destroyed shrapnels.

%% file: jvink_aarv_fig_cygnusloop.tex
\begin{figure*}
\centerline{
\includegraphics[width=0.5\textwidth]{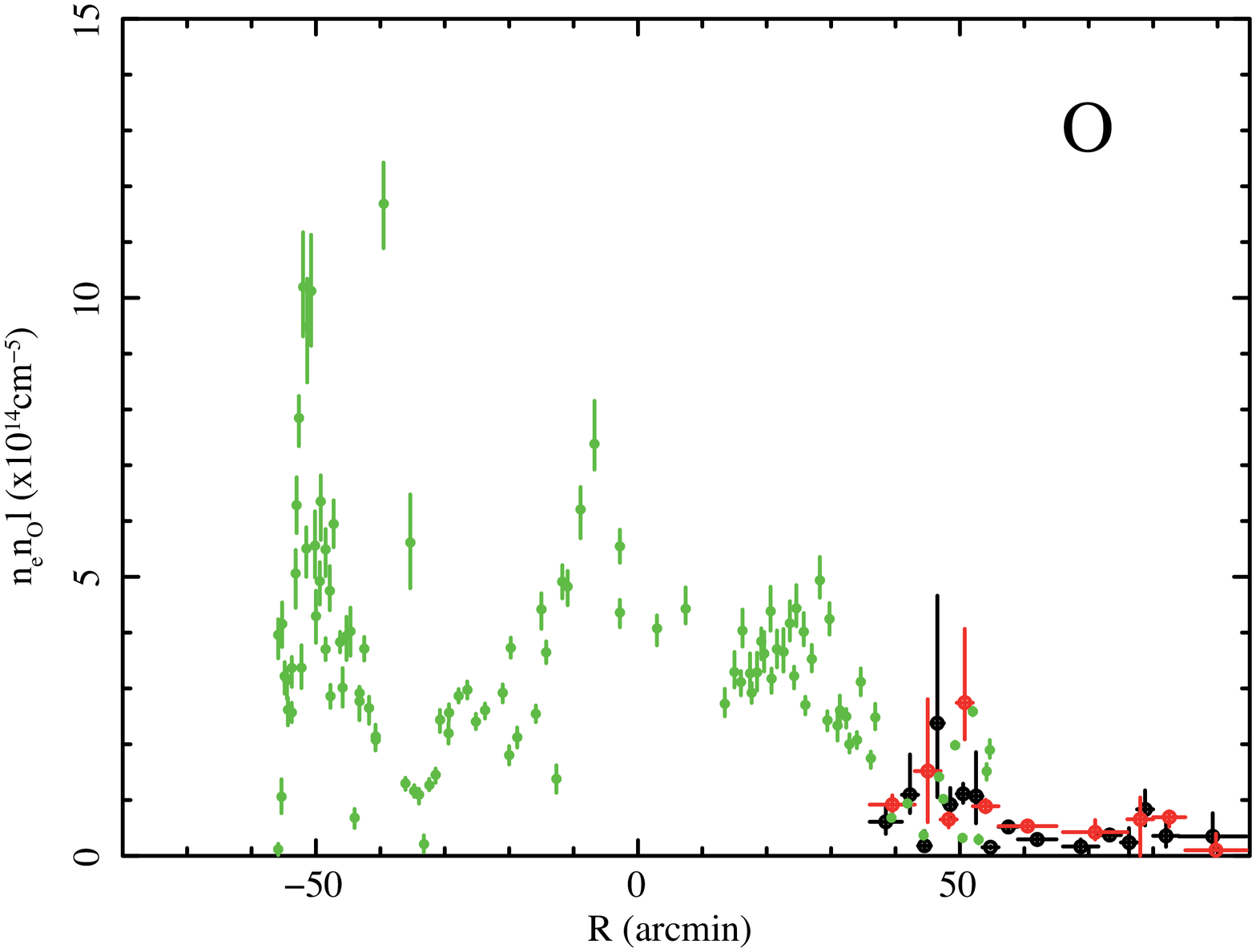}
\includegraphics[width=0.5\textwidth]{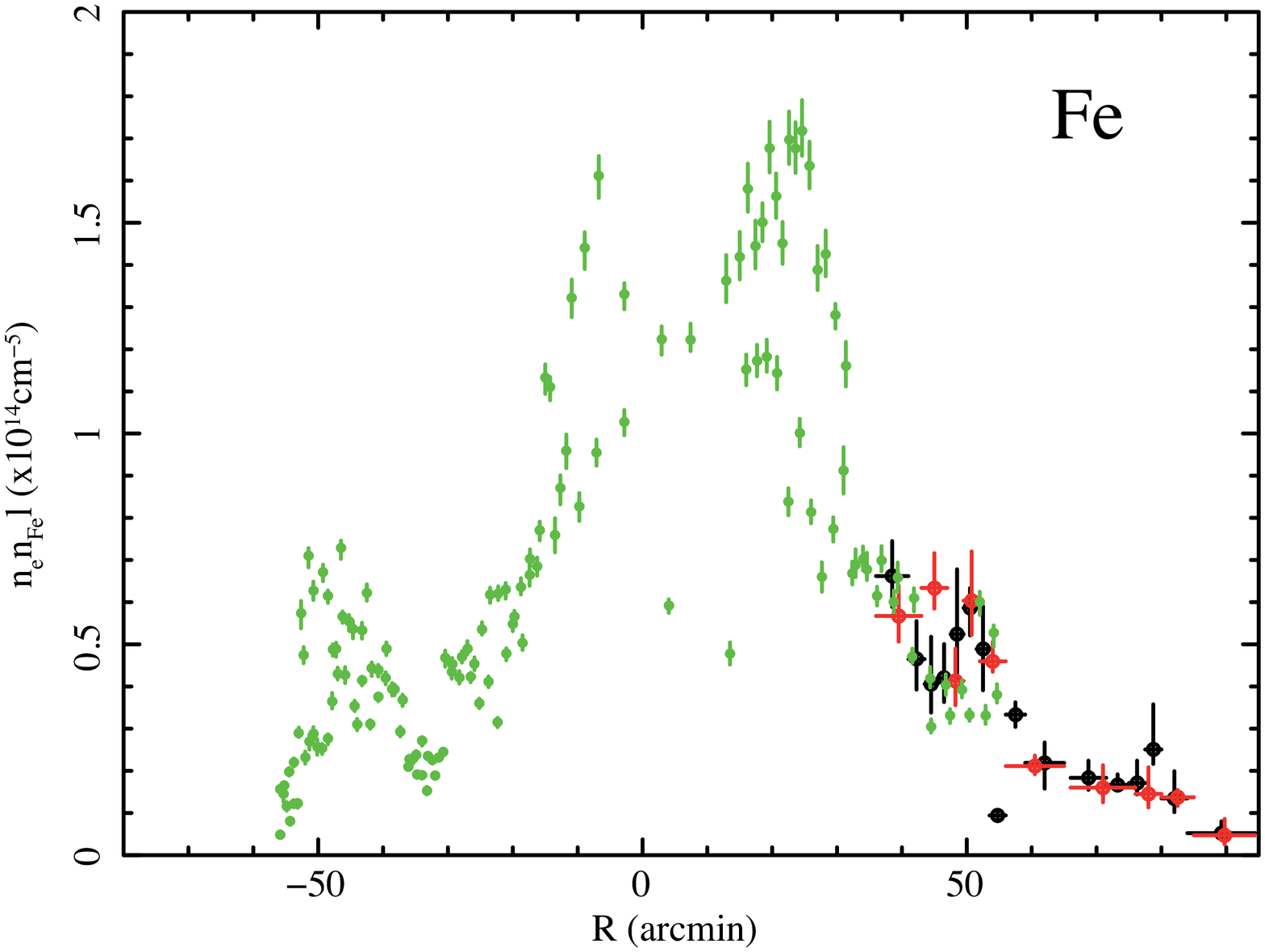}
}
\caption{
\label{fig:cygnusloop}
The emission pattern of oxygen (left) and iron line emission (right) from
the Cygnus Loop as a function of position within
the SNR, following a track from roughly northeast to southwest. 
The green symbols are taken from \citet{tsunemi07} and black and
red points were determined by \citet{uchida08}.
\citep[Figure from][]{uchida08}
}
\end{figure*}

%% file: jvink_aarv_mixed.tex
Among the mature SNRs, one class of objects sticks out:
the mixed-morphology (or thermal-composite) SNRs 
(\sect~\ref{sec:snrclasses}). They are characterized by centrally dominated, thermal X-ray emission, whereas
their radio morphology is shell-like. But apart from the curious
radio/X-ray morphology they 
have a number of other characteristics in common 
\citep[e.g.][]{cox99}. 
They tend to be older SNRs ($\gtrsim 20,000$~yr) 
and are associated with the denser parts
of the ISM.
Many of them are associated with OH masers, which is an indication
of interactions between the SNR shocks and molecular clouds
\citep[e.g.][Table~\ref{tab:mixed}]{claussen97,frail98}. 
Several  of them are also sources of GeV and even TeV 
\gray\ emission (Table~\ref{tab:mixed}),
which likely arises from
cosmic ray {\em nuclei}\footnote{As opposed to electrons. See also
the discussion in \sect~\ref{sec:synchrotron}} 
interacting with dense gas in the SNR shell, or from
escaping cosmic rays interacting with unshocked parts of  molecular clouds. 
These interactions produce pions, of
which the neutral pions decay into \gray\ photons. 
Finally, imaging  X-ray spectroscopy of mijxed-morphology SNRs shows that
their X-ray emitting plasmas have more or less homogeneous temperatures.

\input{jvink_aarv_tab_mixed_new}

The morphology of these SNRs is difficult to explain with
standard SNR evolution models (\sect~\ref{sec:hydro}). There have been
severals models to explain both the centrally enhanced X-ray emission,
which contrasts with their shell-type radio appearances 
(Fig.~\ref{fig:morphology}d).
The model of \citet{white91} assumes that the central thermal emission
is caused by dense cloudlets 
that have survived the passage of the forward shock,
but are slowly evaporating inside the hot medium due to saturated
thermal conduction. They found a self-similar
analytical model for the evolution of thermal-composite SNRs, which
in addition to the Sedov solutions (\sect~\ref{sec:hydro}) 
has two additional
parameters,  namely $C$, the ratio between the mass in the clouds
and the mass in the intercloud medium, and $\tau$, the ratio between the
evaporation time of the clouds and the SNR age.

An alternative model proposed by \citet{cox99} concentrates on
the contrast between the radio and X-ray morphology. According to their
model the SNRs evolve inside medium density CSM.
The forward shock has decelerated to velocities below $\sim 200$~\kms,
resulting in a strongly cooling shell that is too cool
to emit X-ray emission. 
Because the
shocks are radiative, the shock compression factors are large,
giving rise to strongly compressed magnetic fields,
and enhanced  cosmic-ray electron densities. 
As a result the radio synchrotron emission from the shells is strongly enhanced.
\footnote{This explanation for synchrotron emission from SNRs is called the 
Van der Laan mechanism \citep{vanderlaan62}.
\citet{uchiyama08} recently proposed that the radio emission may be additionally
enhanced by the presence of secondary electrons/positrons, i.e.
the electrons/positrons products left over from the decay of charged pions,
created due to cosmic ray nuclei colliding with the background plasma.
The presence of secondary electrons/positrons may also explain the flat 
spectral radio indices $\alpha_R$ in Table~\ref{tab:mixed}.
Finally, \citet{bykov00} discusses the possibility of electron
acceleration by magnetohydrodynamic turbulence behind high density
shocks.
}
The electrons are either accelerated by the SNR, or are simply ambient
relativistic electrons swept up by the radiative shock. 
The X-ray emission from the interior has a homogeneous 
temperature due to thermal conduction \citep{cui92} and
turbulent  mixing \citep{shelton04b}. 
Numerical hydrodynamical models based on the model
of \citet{cox99} indeed show  centrally enhanced X-ray morphologies
\citep{shelton99,velazquez04}.

Although the nature of mixed-morphology SNRs 
is still debated, there has been
considerable progress in our knowledge of them
over the last ten years. \citet{shelton04a} showed that
SNR G65.2+5.7, which is clearly in the radiative evolutionary stage
\footnote{See for example the well defined shell in optical [OIII] emission
\citep{boumis04}}, shows an X-ray filled morphology. Due to the low absorption
column toward this SNR, \rosat\ was also able to reveal
a shell-like outer part in X-ray emission below 0.3~keV.
These two characteristics are in qualitative agreement with
the explanation for mixed-morphology remnants by
\citet{cox99}. 

\input{jvink_aarv_fig_mixed}

Due to progress in X-ray imaging spectroscopy over the last 15 year,
several other clues about the nature
of mixed-morphology SNRs have appeared.
Many of the mixed-morphology SNRs show evidence for metal-rich
plasmas in the interior:
\citet{lazendic06b} list 10 out of a sample of 23 thermal-composite SNRs
with enhanced abundances.
Examples are W44 \citep{shelton04b},
HB 21, CTB 1 and HB 3 \citep{lazendic06b,pannuti10b}, IC 433 \citep{troja08} 
and Kes 27 \citep{chen08}.
\citet{kawasaki05} noted another feature
that may be generic for
this class of SNRs: their ionization
state is close to ionization equilibrium (\sect~\ref{sec:nei}),
unlike young SNRs and old SNRs like the Cygnus Loop.
Moreover, the thermal-composite SNRs IC 443 \citep{kawasaki02,yamaguchi09},
W49B \citep{ozawa09}\footnote{W49B is a young ejecta dominated SNR,
and it is not clear whether it belongs to the class of thermal-composite SNRs.
However, apart from its age it has three characteristics 
of thermal-composite SNRs: a high ambient density, centrally enhanced
X-ray emission \citep{rho98}, and \gray\ emission \citep{abdo10d}.
}, and G359.1-0.5 \citep{ohnishi11} 
show signs of overionization, in the form of radiative-recombination continua
(RRCs, \sect~\ref{sec:nei})
associated with metals like Si, S, and Fe (Fig.~\ref{fig:snr_rrc}).

It is not clear whether metal-richness is a generic feature of 
mixed-morphology SNRs, but it seems at least to be a common feature.
There are several ways in which metal-richness helps to explain
the characteristics of thermal-composite SNRs.
Firstly, the uniform temperatures in their interiors are often
attributed to thermal conduction \citep[e.g.][]{cox99}. A problem for
this explanation is that magnetic fields limit thermal conduction across
field lines (\sect~\ref{sec:conduction}).
However, if the interiors consist predominantly of supernova ejecta,
then the magnetic fields may be very low, simply because the
stellar magnetic field has been diluted by the expansion. 
Magnetic flux conservation gives $B = B_*(R_*/R)^2$,
with $R_*$ the stellar radius, and $R$ the radius of the SNR interior.
Taking typical values of $R_* \approx 10^{13}$~cm, 
$R\approx 10^{20}$~cm, $B_*= 1$~G, shows that the interior magnetic field
may be as low as $10^{-14}$~G, approaching the value
where conduction across field lines becomes
comparable to conduction along field lines. Hence, the conditions are optimal
for thermal conduction, but only up to the boundary between
ejecta dominated and swept-up matter,
where Rayleigh-Taylor
instabilities at the contact discontinuity may have enhanced the 
magnetic-field strengths
and may have made the magnetic fields more tangled, preventing strong thermal
conduction across the contact discontinuity.

Secondly, metal-rich plasma emits more X-ray emission, further increasing the
contrast in X-ray emission between metal-rich interior and the cool shell
of swept-up matter. An important additional aspect to consider here
is whether the interiors consist of enhanced metal abundances, i.e.
a mix of ejecta and swept-up matter, or of pure ejecta.
For SNR W49B, a rather young member of the mixed-morphology class, a pure metal abundance  seems likely, but for older
SNRs the plasma is probably best described by hydrogen/helium dominated plasma
with enhanced abundances.
One can illustrate the key features of the cool-shell/hot-interior 
scenario for mixed-morphology SNRs with
some simple numerical calculations, ignoring for the sake of argument the
more rigorous analytical treatment of the SNR evolution by \citet{cox99}.

Firstly, most mixed-morphology SNRs appear to be in the snowplough phase
of their evolution, with $V_s<200$~\kms, which results in cool, X-ray dim, but
optical/UV bright regions immediately behind the shock front.
The interior, presumably consisting of supernova ejecta, is more or less
homogeneous in temperature and, because of pressure equilibrium also reasonably
homogeneous in density. Given the age of the SNR and the relatively high
sound speed in the interior, one can assume approximate
equilibrium between the ram pressure
at the forward shock and the interior \citep[c.f.][]{cui92}. This gives:
\begin{equation}
n_{\rm interior} kT_{\rm interior}\approx \rho_0 V_s^2,\label{eq:pr_eq}
\end{equation}
with $n_{\rm interior}$ the total interior plasma density.
If we use the values for W44 listed by \citet{cox99}, 
$V_s=150$~\kms, $kT_{\rm interior}=0.6$~keV and $\rho_0=1.0\times 10^{-23}$~g\,cm$^{-3}$, we find a density of $n_{\rm interior}\approx 2.3$~cm$^{-3}$, which translates
into an electron density of $n_{\rm e}\approx 0.8$~\cc, in rough agreement with
X-ray measurements \citep{cox99}. W44 has a radius of $R\approx 10$~pc,
for an estimated distance of 2.5~kpc,
but the X-ray emitting interior has a radius of $R_{\rm interior}\approx 6$~pc,
corresponding to an interior volume of $V_{\rm interior}\approx 2.7\times 10^{58}$~cm$^{-3}$. This means that the total internal energy is 
$U=nkT V_{\rm interior}\approx 0.9\times 10^{50}$~erg 
(about 10\% of the explosion energy). The X-ray emitting mass is 
$M_{\rm interior} \approx 28$~\msun; somewhat higher than the expected
ejecta mass of a Type II supernova, but not too far off, given the crudeness
of the approximation. At least it suggests that a substantial fraction
of the interior plasma consists of ejecta, and that the interior
magnetic field may be low due to large expansion of the ejecta material,
or early shocked swept-up material.

Note that if we assume that the interior consists of metals only, this 
does not lead to a consistent result. For example, if we assume O VII to be the
dominant ion, then we have $n_{\rm OVII}= n/8 \approx 0.29$~cm$^{-3}$ and
$M_{\rm interior} \approx 100$~\msun. This is clearly much 
more oxygen than a massive star can produce. So pure metal abundances seem
unlikely, although perhaps there are some regions in the interior that have
unmixed ejecta. For a discussion on the  metal abundances in W44, see
\citet[][]{shelton04b}.

A SNR will reach the snowplough phase when its age is larger than
$t_{\rm rad}\approx 44,6000 (E_{51}/n_{\rm H})^{1/3}$~yr. 
For W44 ($n_H=6$~cm$^{-3}$), this corresponds to $\sim 25,000$~yr.
Hence, an ionization age of $4\times 10^{11}$~\netunit\ is expected
(\sect~\ref{sec:nei}).
This is relatively high, but below the ionization age for ionization equilibrium
($10^{12}$~\netunit). However, the situation is more complex,
as the interior density and temperature
may have been higher in the past, and, therefore, the degree of ionization may
be higher than indicated by the simple multiplication of the
present day age and electron density.
This could explain the observation by \citet{kawasaki05} that most
mixed-morphology SNRs are near ionization equilibrium. Note that the high
density in the interior of W44 and other mixed-morphology SNRs is a direct
consequence of the high ISM density (Eq.~\ref{eq:pr_eq}) and the
thermal conduction, which results in a more uniform interior density with
medium hot temperatures, rather than very high temperatures with very low
interior densities.

In the context of mixed-morphology
SNRs, it is of interest to show 
why some old SNRs are expected to
 have RRCs. As
noted by \citet{ohnishi11}, one may expect RRCs whenever the plasma
cools faster than the recombination time, for example due to
thermal conduction \citep{kawasaki02} or adiabatic expansion
\citep{itoh89,yamaguchi09}. \citet{yamaguchi09} calculated that the time
scale for cooling by conduction is too long, so adiabatic expansion
is probably the appropriate explanation. The time scale for adiabatic expansion
can be calculated from the thermodynamic relation $TV^{\gamma-1} = $ constant, which means that the adiabatic cooling time scale is given by
\citep{broersen11}:
\begin{equation}
       \tau_{\rm ad}= -\Bigl(\frac{T}{\dot{T}}\Bigl)_{\rm ad} =
-\frac{1}{1-\gamma}\frac{V}{\dot{V}} = -\frac{1}{3(1-\gamma)}\frac{R}{\dot{R}} =
\frac{1}{2}\frac{R}{\dot{R}}=
\frac{1}{2}\beta^{-1} t,
\end{equation}
with $\beta$ the expansion parameter (\sect~\ref{sec:hydro_analytical}), $\gamma=5/3$, and $t$ the age of the SNR. For SNRs in the snowplough phase we can use
$\beta=0.25$ (Eq~\ref{eq:snowplow}).  This suggests that $\tau_{\rm ad}\approx t/8$.
For overionization to occur it is first required
that the ionization age of the plasma
must be large enough (otherwise there could underionization). 
The second condition is that the timescale for recombination is
longer than the adiabatic cooling time scale:
\begin{equation}
\tau_{\rm ad} < \tau_{\rm rec} = \frac{1}{\alpha_{\rm rec} n_{\rm e}},
\end{equation}
with $\alpha_{\rm rec}$ the recombination rate.
The mixed-morphology SNRs show evidence for overionization
of Si XIII, for temperatures around $kT=0.2-0.7$~keV. For these
temperatures the recombination rate for Si XIV to Si XIII 
is approximately $\alpha_{\rm rec}= 5.9\times 10^{-14}$~cm$^{3}$s$^{-1}$
\citep{shull82}.
Using now that $n_{\rm e}=0.5n_{\rm interior}$ and Eq.~\ref{eq:pr_eq}
the condition for overionization in mixed-morphology SNR is:
\begin{align}\label{eq:rrccond}
t <  \frac{8}{\alpha_{\rm rec}n_{\rm e}} =
& \frac{8 kT_{\rm interior}}{0.5\alpha_{\rm rec}\rho_0V_s^2} = \\ \nonumber
&2\times 10^6 
n_{\rm H}^{-1} \Bigl(\frac{V_s}{150\, {\rm km\,s^{-1}}}\Bigl)^{-2}
\Bigl(\frac{kT_{\rm interior}}{0.5~{\rm keV}}\Bigl) \\ \nonumber 
& \times
\Bigl(\frac{\alpha_{\rm rec}}{6\times 10^{-14}~{\rm cm^3\,s^{-1}}}\Bigr)^{-1}
\ {\rm yr},
\end{align}
with $n_{\rm H}$ the pre-shock density. This equation shows that
overionization should be common, as most SNRs are much younger than
a few million year. More important is whether the average plasma
ever reached ionization equilibrium, i.e. whether $\int n_{\rm e}dt \gtrsim
10^{12}$~cm$^{-3}$s. The discovery of RRCs in mixed-morphology
SNRs has been  relatively recent. Given Eq.~\ref{eq:rrccond}, one 
expects that many more SNRs, with relatively dense interior plasmas,
should have RRCs in their X-ray spectra.

Finally, some word of caution considering the interpretation of
mixed-morphology SNRs. The explanation of their properties sketched
above is relatively simple, but nature is often not that simple.
For example, one may ask how it comes that SNRs with 
maser emission, and which are interacting with molecular clouds, have densities
of only $n_{\rm H}=5$~cm$^{-3}$. This is relatively high, but not nearly as
high as the dense molecular clouds usually associated with maser emission,
 $n_{\rm H}>1000$~cm$^{-3}$ \citep{frail98}.
But in fact, millimeter-wave and near infrared observations of W44 and W28 
by \citet{reach05} reveal the presence of dense molecular gas. 
However, this dense gas has only a small filling factor.
\citet{reach05} even note that some of the dense clumps may survive the shock passage,
which brings us back to the scenario proposed by \cite{white91}. The interclump
densities, which applies to 90\% of the volume are, however, consistent with
the estimates of \citet{cox99}, $n_{\rm H}\approx 5$~cm$^{-3}$. These
infrared observations show that old SNRs have a complex structure,
which,
for a proper understanding, requires multi-wavelength observations.

%% file: jvink_aarv_tab_mixed_new.tex
\begin{table*}
\begin{center}
\caption{Mixed morphology SNRs.\label{tab:mixed}}
{\small
\begin{tabular}{lllllllllll}\hline\hline\noalign{\smallskip}

SNR	&	Name	&	Size	&	$F_{\rm R}$	&	$\alpha_{\rm R}$	&	Dist	&Enh. 	&	RRC	&	GeV 	&	TeV 	& OH 		\\
	&		&		&		&		&		&	metals & 	&		&		&		Maser	\\
	&		&	arcmin	&	Jy	&		&kpc		&		&		&		&		&			\\
\noalign{\smallskip}\hline
\noalign{\smallskip}
G0.0+0.0	&	Sgr A East 	&	$3.5\times 2.5$	&	100?	&	0.8?	&	8	&	X	&		&	?	&	?	&	X (v)	\\
G6.4−0.1	&	W28	&	48	&	310	&	0.3	&	1.9	&		&		&	X (c)	&	X (g)   &	X (n)	\\
G31.9+0.0	&	3C391	&	$7\times 5$	&	24	&	0.49	&	8.5	&		&		&	X (k)  &		&	X (n)	\\
G33.6+0.1	&	Kes 79	&	10	&	22	&	0.5	&	7.8	&		&		&		&		&		\\
G34.7−0.4	&	W44	&	$35\times 27$	&	230	&	0.37	&		&	X	&		&	X (e)	&		&	X (n)	\\
G41.1−0.3	&	3C397	&	$4.5\times2.5$	&	22	&	0.48	&	7.5	&	X	&		&		&		&		\\
G43.3−0.2	&	W49B	&	4	&	38	&	0.48	&	10	&	X	&	X (s)	&	X (d)	&	X (j)	&		\\
G49.2−0.7	&	W51C	&	30	&	160	&	0.3	&	6	&		&		&	X (a)	&	X (y)	
&	X (o)	\\
G53.6−2.2	&	3C400.2	&	$33\times 28$	&	8	&	0.75	&	2.8	&		&		&		&		&		\\
G65.3+5.7	&	G65.3+5.7	&	$310\times 240$	&	52?	&	0.6?	&	0.8	&		&		&		&		&		\\
G82.2+5.0	&	W63	&	$95\times 65$	&	120	&	0.35	&		&	X	&		&		&		&		\\
G89.0+4.7	&	HB21	&	$120\times 90$	&	220	&	0.48	&	0.8	&	X	&		&		&		&		\\
G93.7−0.2	&	CTB 104A	&	80	&	65	&	0.65	&	1.5	&		&		&		&		&		\\
G116.9+0.2	&	CTB1	&	34	&	8	&	0.61	&		&	X	&		&		&		&		\\
G132.7+1.3	&	HB3	&	80	&	45	&	0.6	&	2.2	&	X	&		&	X (b)?	&		&		\\
G156.2+5.7	&	G156.2+5.7	&	110	&	5	&	0.5	&		&	X	&		&		&		&		\\
G160.9+2.6	&	HB9	&	$140 \times 120$	&	110	&	0.64	&	$<4$	&		&		&		&		&		\\
G166.0+4.3	&	VRO 42.05.01	&	$55\times 35$	&	7	&	0.37	&	4.5	& X (i)	&		&		&		&		\\
G189.1+3.0	&	IC 443	&	45	&	160	&	0.36	&	1	&	X (i)	&	X (x)	&	X (f)	&	X (h)	&	X (m)	\\
G272.2-3.2	&	G272.2-3.2	&	15?	&	0.4	&	0.6	&	2?	&	X (t)?	&		&		&		&		\\
G290.1-0.8	&	MSH 11-61A	&	19x14	&	42	&	0.4	&	7	&	X			&	&X (b)?	&		&		\\
G327.4+0.4	&	Kes 27	&	21	&	30	&	0.5	&	5	&	X (l)	&		&		&		&		\\
G357.1-0.1	&	Tornado	&	$8\times 3$	&	37	&	0.4	&	$>8$	&		&		&		&		&		\\
G359.1-0.5	&	G359.1-0.5	&	24	&	14	&	0.4	&		& X		& X (r)		&		&		&	X (q)	\\
\noalign{\smallskip}\hline
\end{tabular}
}
\end{center}
\scriptsize{
This list is based on the list by  \citet{lazendic06b}.
Sizes, distances, and radio properties (radio flux $F_{\rm R}$, and spectral index $\alpha_{\rm R}$)
were obtained from
Green D. A., 2009, ``A Catalogue of Galactic Supernova Remnants 
(2009 March version)'', Astrophysics Group, Cavendish Laboratory, Cambridge, United Kingdom (available at \url{http://www.mrao.cam.ac.uk/surveys/snrs/}.
The column names indicate: Enh. metals, the presence of plasma with overbaundances \citep[][unless indicated otherwise]{lazendic06b};
RRC, the presence of radiative recombination line continua; GeV, the detection in the GeV band (\fermi); TeV the detection of associated sources
in the TeV band with Cherenkov telescopes; OH Maser, the detection of associated OH masers.
The letters in parentheses refer to the following references:\\
a) \citet{abdo09}; 
b) \citet{abdo10e}; 
c) \citet{abdo10c}; 
d) \citet{abdo10d}; 
e) \citet{abdo10b};
f) \citet{abdo10f}; 
g) \citet{aharonian08a}; 
h) \citet{albert07b};
i) \citet{bocchino09};
j) \citet{brun11};
k) \citet{castro10}; 
l) \citet{chen08}
m) \citet{claussen97}; 
n) \citet{frail96} ; 
o) \citet{green97}
p) \citet{koralesky98b}; 
q) \citet{lazendic02};
r) \citet{ohnishi11};
s) \citet{ozawa09};
t) \citet{park09};
u) \citet{yusef-zadeh95};
v) \citet{yusef-zadeh96};
w) \citet{yusef-zadeh99};
x) \citet{yamaguchi09};
y) \citet{feinstein09}; \citet{krause11}.
}
\end{table*}

%% file: jvink_aarv_fig_mixed.tex
\begin{figure}
\centerline{\includegraphics[width=\medfigure]{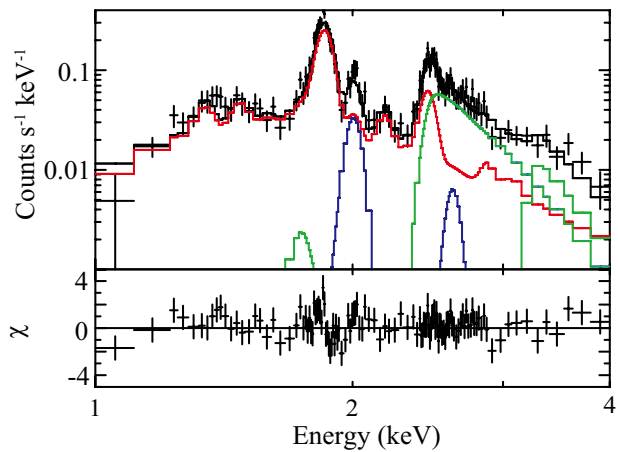}
}
\caption{
Suzaku SIS spectrum of G359.1-05 \citep{ohnishi11},
showing in green the Si and S Ly$\alpha$ line and in blue
the radiative-recombination continua.
\label{fig:snr_rrc}}
\end{figure}

%% file: jvink_aarv_thermaldoppler.tex
As explained in Sect.~\ref{sec:equilibration}, from a theoretical point of view
collisionless shocks do not naturally lead to equilibration of electron and ion 
temperatures in the immediate post-shock region. The degree of post-shock temperature 
equilibration by SNR shocks is therefore an important observational issue. 
Unfortunately, X-ray measurements of plasma temperatures usually
refer only to the {\em electron} temperature,
as it is the electron temperature that 
determines X-ray line ratios (\sect~\ref{sec:linediag}), 
and the shape of the thermal X-ray continuum (\sect~\ref{sec:continuum}).  

A first indication that the electron temperature is not equal to proton temperature, or at least that
Eq.~\ref{eq:kT_std} (\sect~\ref{sec:kT}) does not hold for young SNRs, 
is the fact that for young SNR no
X-ray temperatures above $kT_{\rm e} \approx 5$~keV have been reported.\footnote{
One of the highest temperatures reported is $kT_{\rm e} = 4.3$~keV 
for the SNR IC 131 in M33 \citep{tuellmann09}.} 
This is in contrast to the expected temperatures for some young SNRs
with high shock velocities.
For example, Cas A and Tycho have shock velocities of, respectively,
$V_s \approx 5100$~\kms \citep{vink98a,koralesky98,delaney03,patnaude09} 
and $V_s \approx 4000$~\kms\  \citep{reynoso97,katsuda10}.  
They should, according to Eq.~\ref{eq:kT_std},  have plasma components with
$kT_{\rm e} > 19$~keV. Even though part of the X-ray continuum of those two SNRs is due
to X-ray synchrotron emission (\sect~\ref{sec:synchrotron}), 
plasmas as hot as 20~keV should easily be detectable
as they give rise to relatively flat spectra with power-law indices around 1.5.
In contrast,
the observed spectral indices in the 4-6~keV continuum dominated band 
are typically $\approx 3$ \citep[e.g.][]{helder08}. 
Also the hard X-ray spectra do not show signs
of thermal X-ray emission from plasmas with $kT\gtrsim 5$~keV 
\citep{allen97,favata97,allen99,vink01a,renaud06}.

Apart from non-equilibration of electron-ion temperatures there may be another
reason for the low electron temperatures
in young SNRs: non-linear cosmic-ray acceleration
may be so efficient that the overall plasma temperatures are very low 
\citep[\sect~\ref{sec:kT},][]{drury09,helder09,vink10a}.

In order to estimate the {\em proton} or {\em ion} temperature, 
as opposed to the electron temperature,  one needs to measure the thermal Doppler
broadening. Thermal Doppler broadening gives rise to a Gaussian line profile with line width\footnote{
The lack of a factor 2 in the equation may be confusing. Note that \citet{rybicki} show that
the profile is proportional to $\exp[ -m_{\rm i} c^2 (E-E_0)^2/(2E_0 kT_{\rm i})]$. Comparing this
to a gaussian   $\exp[ -(x-x_0)^2/(2 \sigma^2)]$ shows the correctness of Eq.~\ref{eq:doppler}.}:
\begin{equation}
\sigma_{\rm E} = \Bigl(\frac{E_0}{c}\Bigr)
\sqrt{\Bigl(\frac{kT_{\rm i}}{m_{\rm i}}\Bigr)} .\label{eq:doppler}
\end{equation}
The line width is often given in terms of Doppler velocities, in which case
$\sigma_{\rm v} =\sqrt{kT_{\rm i}/m_{\rm i}}$.
In order to measure the thermal Doppler broadening one best uses a spectrum of a region
that is expected to be devoid of bulk line of sight motions. 
This is expected at the edges of SNRs,
where the expansion of the shell only results in
velocities in the plane of the sky.

Most measurements of thermal Doppler broadening are based on optical and UV observations
of Balmer and Lyman lines of hydrogen.
A comprehensive list of these measurements is given by \citet{ghavamian07}.
 Balmer and Lyman emission lines from SNR shocks require that the shocks move through
(partially) neutral gas. As the neutral hydrogen penetrates the shock heated region
the neutral atom may get excited, giving rise to narrow line emission
in the optical/UV.
In addition to excitations also charge transfer reactions may occur, 
i.e. a shock heated proton may pick up an electron from a neutral
particle. This process gives rise to thermally broadened Balmer or Lyman line 
emission \citep[e.g.][]{chevalier80, heng10} from which the hydrogen 
temperature immediately behind the shock can be measured 
\citep{vanadelsberg08}. 
The ratio between the narrow and broad line emission
is a function of shock velocity and the degree of
electron-proton temperature equilibration. 

\citet{ghavamian07} presented evidence 
that for low shock velocities the proton and electron temperatures
are equilibrated, whereas for shock speeds above $\sim 500$~\kms\, 
the temperatures become increasingly non-equilibrated, 
with a rough dependency of $T_{\rm e}/T_{\rm p} \propto 1/V_s^2$.
The work by \citet{vanadelsberg08} and \citet{helder11} 
does not corroborate the dependency, but
does confirm that slow shocks produce thermally equilibrated plasmas, 
whereas fast shocks do, in general, not. 
Note that the work of \citet{helder11}, and earlier work of \citet{rakowski03},
is based on electron temperatures measured using X-ray spectroscopy, whereas
proton temperatures were measured from broad H$\alpha$ line emission.
\citet{helder11} used the SNR RCW 86, which has large density and shock
velocity contrasts \citep{vink06b,helder09}, and may be efficient
in accelerating cosmic rays \citep{helder09,aharonian09}. 
\citet{rakowski03} measured temperature equilibration in the LMC
SNR DEM L71 with inferred shock velocities of 500-1000~\kms. For DEM L71
the $T_{\rm e}/T_{\rm p} \propto 1/V_s^2$ relation seems to be valid.

In addition to the hydrogen lines, also lines from other ions can be used to 
measure ion temperatures.
A noteworthy example is the UV line emission from the northwestern region of 
SN\,1006 as measured
by the {\em Hopkinson Ultraviolet Telescope} ({\em HUT}), which shows that the lines from
H I, He II, C IV, N V, and O VI all have the same width of $\sim 2300$~\kms 
\citep[FWHM][]{raymond95}. This indicates that
the ions are not equilibrated, since there is no dependency of the line width on the mass of
the ion (see Eq.~\ref{eq:doppler}).

\input{jvink_aarv_fig_sn1006}

Thermal Doppler broadening measurements in X-rays are rare, as one needs high resolution
spectroscopy of a region very close to the shock front, 
and the current grating spectrometers do
not easily allow to isolate particular regions of an extended object. 
But \citet{vink03b} 
measured the line broadening of the O VII He$\alpha$ line emission from  a bright knot 
 that stands out at the northeastern
rim of SN\,1006. 
The size of the knot is approximately 1\arcmin.
This measurement was possible for SN1006, because this
remnant is rather large for a young SNR (30\arcmin). As a result only a small amount of the
overall X-ray emission from the rest of the SNR is contaminating  the \xmm\ RGS spectrum. The
expected line profile for a narrow line consists of a sharp rise, as for a normal point source, but
it has a long tail, caused by emission from the interior of
SN\,1006 (Fig.~\ref{fig:sn1006a}). The shape of the short wavelength side of the
line profile is, therefore, sensitive to the thermal line broadening. \citet{vink03b} reported a line profile
with a width of $\sigma = 3.4 \pm 0.5$~eV, corresponding to $\sigma_v= 1777$~\kms.
Using Eq.~\ref{eq:doppler}, this translates into a temperature for O VII of 
$kT_{\rm O VII} = 528 \pm 150$~keV. This seems extremely hot, but  using Eq.~\ref{eq:kT_std}, with $m_{\rm i}=16m_{\rm p}$ instead of  $\mu m_{\rm p}$, 
shows that it corresponds to a shock velocity of $\sim 4000$~\kms, provided
that oxygen has not been equilibrated with protons.
The high O VII temperature should be compared to the electron temperature as measured with the
EPIC instruments on board \xmm, $kT_{\rm e} \approx 1.5$~keV. 
This is a clear indication that in this
electrons and oxygen ions are not equilibrated.

The nature of the X-ray emitting knot in SN\,1006 is not quite clear. It seems to have enhanced O and Si abundances,
and may therefore be an ejecta knot, in which case it was heated by the reverse shock. Another
important issue is whether non-thermal components, such as accelerated particles may have influenced
the thermal balance of the shock, as discussed in \sect.~\ref{sec:kT}.
The northeastern part of SN\,1006 seems, in that sense, a relatively good place to measure
equilibration processes in the absence of cosmic rays, as this region shows no evidence for
X-ray synchrotron emission (see also \sect~\ref{sec:synchrotron}). 

Although the RGS observations of SN\,1006 are perhaps the most direct measurements of
thermal Doppler broadening in X-rays, thermal Doppler broadening is important to take into
account when using the overall line broadening of SNRs. An example is the Doppler line
broadening of LMC SNRs like 0509-67.5 \citep{kosenko08} and 0519-69.0 \citep{kosenko10}
as measured with the \xmm\ RGS. Thermal line broadening may even affect 
the line width as measured with CCD instruments. \citet{furuzawa09} reported that Fe-K line
broadening of X-ray line emission from Tycho's SNR, and its spatial distribution, cannot be easily
modeled with bulk velocities alone, but that thermal Doppler broadening is needed
as well. From this 
they infer an iron temperature of $(1-3) 10^{10}$~K (860-2600 keV)!
So there are several pieces of evidence that electrons and ions 
may indeed have different temperatures.
In X-rays, however, the evidence seems at the moment to pertain to plasma heated by
the {\em reverse} shock. 
The evidence for non-equilibration of temperatures of plasma heated by the
forward shock is mostly based on 
H$\alpha$ measurements. For these measurements the amount of
equilibration is based on the narrow to broad line flux ratio,
but the interpretation is model dependent and does not
always give consistent results \citep{vanadelsberg08}.
Moreover, current models do not take into account the
effects of cosmic-ray precursor physics, which is likely important
for the correct interpretation of the line ratios
\citep{vanadelsberg08,rakowski08,helder11,raymond11}.

It could be that the dichotomy between equlibrated shocks below
500~\kms, and unequilibrated shocks above 500~\kms, is somehow
related to cosmic-ray acceleration, possibly due to an increased
role of plasma waves \citep[e.g.][]{rakowski08}. 
However, this different equilibration behavior between slow and fast shocks
may also have another origin:
as pointed out by \citet{bykov04}, electrons 
have thermal velocities larger than the
shock velocity for Mach numbers $M<\sqrt{m_{\rm p}/m_{\rm e}}=43$ . 
The velocity with which the electrons are therefore
scattered behind  the shock
 is largely determined by the pre-shock temperature rather than
by the shock velocity. This may lead to additional electron heating mechanisms.
A Mach number of $\sim 43$ corresponds, for a typical sound speed
of 10~\kms, to a shock velocity of 400~\kms. Interestingly, low Mach numbers
may also occur for high shocks speeds in the presence of cosmic
rays. This happens when the cosmic-ray precursor significantly preheats the
incoming plasma. The different equilibration properties of the TeV source
RCW 86 \citep{helder11} and DEM  L71 \citep{rakowski03} may be explained
by the more prominent role of cosmic-ray acceleration in RCW 86.
This is suggestive of an intimate relation between
cosmic-ray acceleration and temperature equilibration at the shock.

In the future, with the availability of microcalorimeter spectrometers on 
\astroh\ and \ixo, it will be possible to
routinely obtain
line broadening measurements from regions close to SNR shock fronts. 
The advantage
that X-ray measurements have over $H\alpha$ measurements is that no neutral hydrogen is needed
upstream of the shock. Right now 
the sample of young SNRs with thermal Doppler broadening 
measurements is skewed toward Type Ia SNRs, as they seem to be more likely to have
H$\alpha$ emission from the shock region (\sect~\ref{sec:typeia}).
The problem with this is that the presence of a large fraction of
neutral atoms may dampen Alfv\'en waves. This
may result in different shock-heating properties and may also quench
efficient cosmic-ray acceleration \citep{drury96}. X-ray measurements of ion
temperature, therefore, provides a more unbiased probe of shock-heating 
properties. The disadvantage is that on average the 
X-ray emission comes from further behind the shock front 
than H$\alpha$ emission.

High resolution X-ray spectroscopy may also be used to gather information on the post-shock
equilibration process itself, because different plasma ionization ages 
should also correspond to different degrees 
of equilibration (i.e. full/partial non-equilibration
and equilibration, see \sect~\ref{sec:collisionless}).

%% file: jvink_aarv_fig_sn1006.tex
\begin{figure*}
\centerline{
\parbox{0.35\textwidth}{
\includegraphics[width=0.33\textwidth]{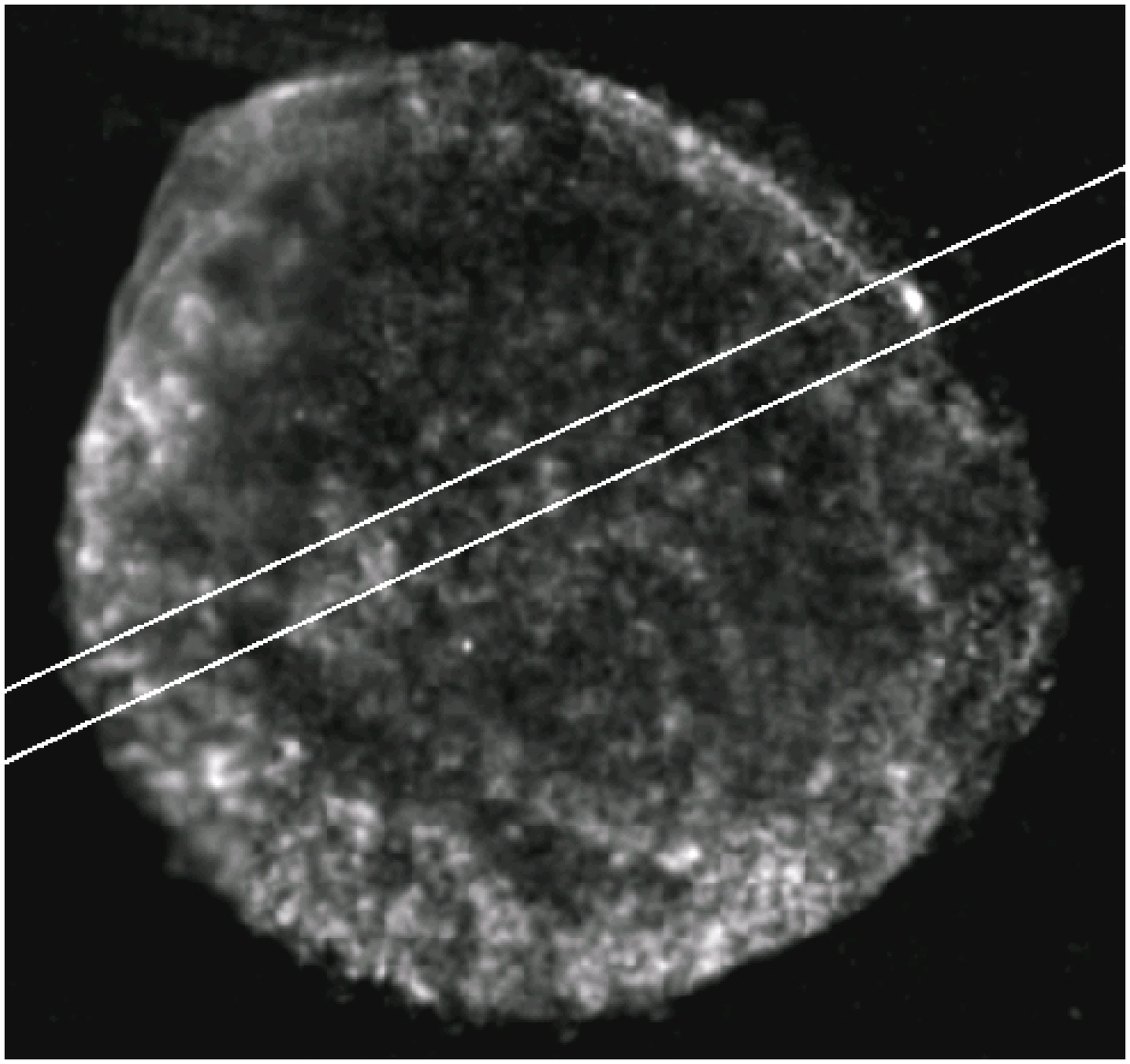}}
\parbox{0.45\textwidth}{
\includegraphics[angle=-90, width=0.47\textwidth]{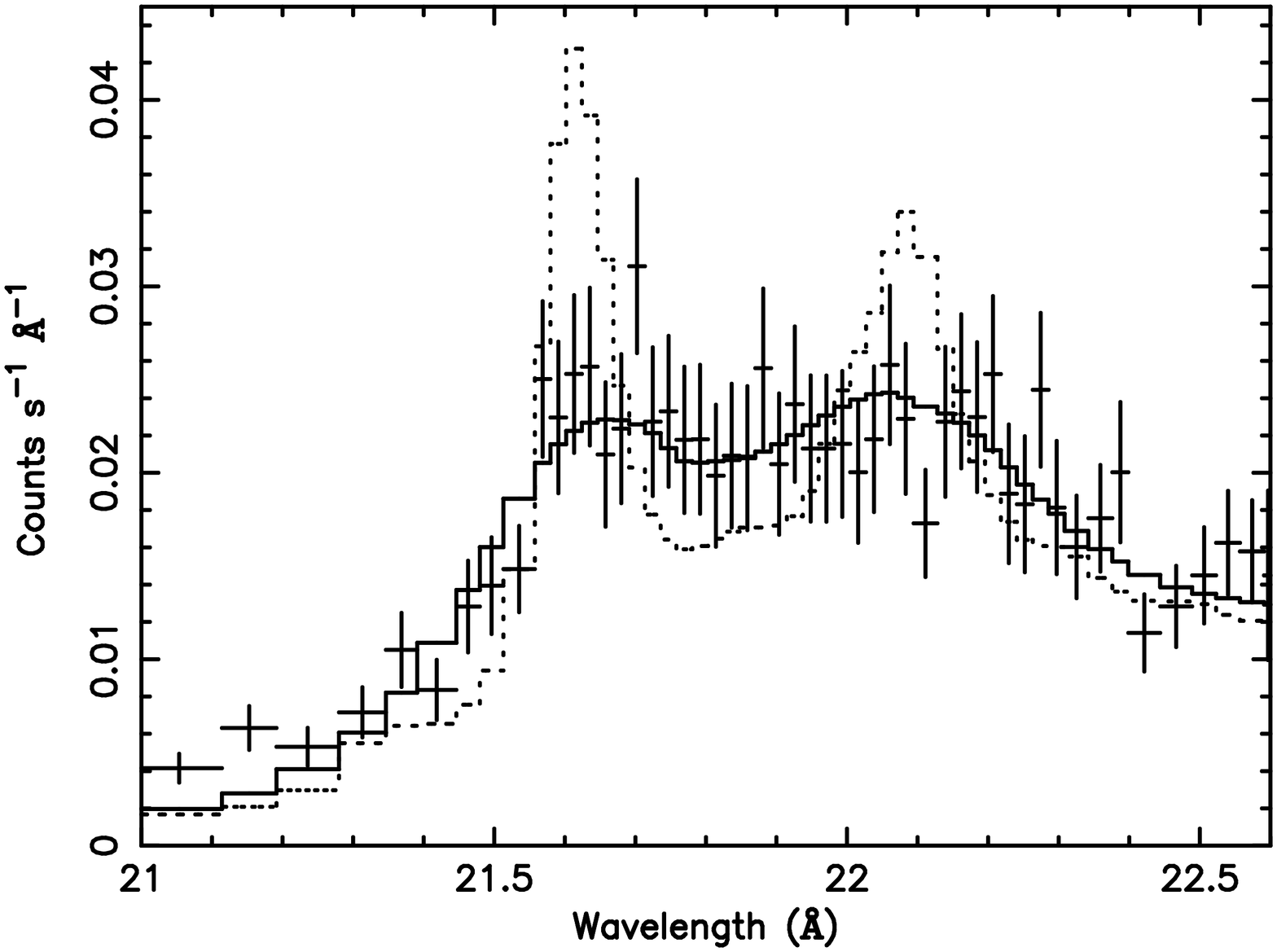}}
}
\caption{Left: Map of O\,VII emission made from
several \chandra-ACIS observations (see also Fig.~\ref{fig:sn1006b}). The lines indicate
the region observed by the \xmm\ RGS instrument.
The target was the bright knot in the northeast inside the RGS region.
Right: Detail of the RGS1 spectrum of the northeastern knot, showing
O\,VII He$\alpha$ line emission.
The dotted line is the best fit model without line broadening,
whereas the solid line shows the model including thermal line 
broadening \citep[adapted from][]{vink03b}.
\label{fig:sn1006a}}
\end{figure*}

%% file: jvink_aarv_synchrotron.tex
\input{jvink_aarv_fig_sn1006b}
Although SNRs are thought to be the main sources of Galactic cosmic rays
(\sect~\ref{sec:cr}), until 1995 the observational 
evidence for this consisted mostly of the detection of 
radio synchrotron radiation from SNRs. The electrons responsible
for radio emission 
have energies in the GeV range. So radio synchrotron radiation
informs us that SNRs can accelerate electrons, but
does not proof that they can accelerate particles to energies comparable
to the ``knee'' in the cosmic ray spectrum ($\sim 3\times 10^{15}$~eV).

In 1995 it was for the first time convincingly 
demonstrated that a SNR, SN\,1006, emitted {\em X-ray} synchrotron radiation 
\citep{koyama95}. That the X-ray spectrum of this source
was unusual was already clear from observations with the
solid-state spectrometer on board the \einstein\ X-ray observatory
\citep{becker80}. The spectrum appeared featureless, and  was
completely different
from the line-rich X-ray spectra of a young SNRs like Tycho or Cas A.
This was interpreted by \citet{reynolds81} as the result
of synchrotron emission from shock accelerated electrons.
However, \citet{hamilton86b} proposed an ingenious interpretation in
which most of the emission was coming from carbon-rich, shocked ejecta,
which would provide a large thermal bremsstrahlung emissivity (Eq.~\ref{eq:bremss}) 
without much line radiation in the 0.5-7 keV range.
The case for X-ray synchrotron radiation by \citet{koyama95}
was settled thanks to the spectral-imaging capabilities of \asca, which
established that thermal radiation was being emitted from all over 
the face of SN\,1006, but that the thermal 
emission was overwhelmed by non-thermal  continuum
emission coming from two regions close to the forward
shock (Fig.~\ref{fig:sn1006b}).
This had important consequences for the idea that SNRs can accelerate particles,
as X-ray synchrotron emission implies that electrons are accelerated up
to energies of $10-100$~TeV. This requires rather small
diffusion coefficients (see Eq.~\ref{eq:syn_max} for the loss-limited case).
If indeed maximum electron energies are limited by radiative losses,
then ions can in principle be accelerated to even higher energies, possibly
even beyond $10^{15}$~eV.
But the X-ray synchrotron radiation does not proof that this is the
case.

Around the same time that X-ray synchrotron radiation from SN\,1006 was
established non-thermal hard X-ray emission was detected from
the young SNR Cas A by the
\cgro-OSSE \citep{the96}, \rxte-HEXTE \citep{allen97} and 
\sax-PDF \citep{favata97} instruments. This hard X-ray emission is likely
caused by synchrotron radiation as well, although other explanations have been
proposed
\citep[e.g.][]{allen97,vink99x,laming01a,vink08a,helder08}. 
Moreover, \asca\ observations also
established that two newly discovered SNRs RX J1713.7-3946 
\citep[G347.3-0.5,][]{pfeffermann96} and RX J0852.0-4622 
\citep[G266.2-1.2,][]{aschenbach98} were similar to SN\,1006 in that
their X-ray emission was completely dominated by synchrotron emission
\citep{koyama97,slane99,slane01ax}. 
In fact, no thermal X-ray emission from these SNRs has 
yet been detected \citep[e.g.][]{hiraga05,acero09,slane01ax,pannuti10}.

The spatial resolution of \chandra\ helped to establish
that not only rather large SNRs are source of X-ray synchrotron radiation
(SN\,1006, RX J1713.7-3946, and  RX J0852.0-4622 are all larger
than 30\arcmin).
Also Cas A \citep{gotthelf01a},
Tycho \citep{hwang02} and Kepler \citep{reynolds07} have regions close
to the shock front that are continuum dominated. These regions are, however,
very thin (1-2\arcsec). It was soon shown that the narrowness of these filaments
must be the result of relatively large magnetic fields \citep[$\sim 100-600~\mu$G,][]{vink03a,berezhko03c,bamba04,bamba05,voelk05,ballet06,parizot06}.

\citet{vink03a} determined a high magnetic field for Cas A by assuming the
width of the filament is determined by the radiative loss time of
X-ray synchrotron emitting electrons. As the plasma is advected away from
the shock front, a radiative loss time also corresponds to a physical length
scale. The plasma velocity with respect to the shock front
is given by $v_2= V_{\rm s}/\chi$ (c.f. Eq.~\ref{eq:massflux}), whereas the
synchrotron loss time is given by Eq.~\ref{eq:tau_syn}. The synchrotron
loss time, therefore, corresponds to an advection length scale of
\begin{equation}
l_{\rm adv}=v_2\tau_{\rm syn}=\frac{V_{\rm s}\tau_{\rm syn}}{\chi}.\label{eq:ladv}
\end{equation}
Since $\tau_{\rm syn}$ depends on both the magnetic field and electron
energy ($\tau_{\rm syn}\propto B^{-2}E^{-1}$), one needs additional information
to determine $B$ and $E$ separately. For that one can use the dependence
of the typical photon energy on the magnetic-field strength and electron
energy ($\propto E^2B$, Eq.~\ref{eq:nu_char}).
This is 
graphically depicted in Fig.~\ref{fig:casaslope}, showing that for Cas A
$B \sim 100~\mu$G near the shock front.
This method is not the only way to determine the magnetic field.
One can also assume that the width of the synchrotron emitting 
filaments are equal to
the diffusion length scale ($l_{\rm diff}\propto B^{-1}E$, Eq.~\ref{eq:ldiff})
appropriate for the energies of the
synchrotron emitting electrons \citep{berezhko03c}.

 \input{jvink_aarv_fig_casa_nonthermal}

Both methods give roughly the same magnetic-field estimates \citep{ballet06}. 
In principle, the X-ray synchrotron rims should be always wider than, or
comparable to, the diffusion length scale, as this is the width of
the region from which electrons are actively being accelerated (indicated
by the shaded region in Fig.~\ref{fig:casaslope}b). But
the diffusion length
scale is always comparable to the advection length scale 
for electrons with energies close to the maximum energy
\citep{parizot06,vink06d}.
This is easily demonstrated using the basic condition for
determining the maximum electron energy, i.e.
the acceleration time equals the synchrotron cooling time,
$\tau_{\rm acc}\approx \tau_{\rm syn}$ (Eq.~\ref{eq:emax}). 
The acceleration time scale is within an order of magnitude given by
$\tau_{\rm acc} \approx D/v_2^2$ (c.f. Eq.~\ref{eq:tau_cr}).
Using Eq.~\ref{eq:ladv} one can rewrite the condition for reaching the maximum
electron energy,
$\tau_{\rm acc}\approx \tau_{\rm syn}$, in $\tau_{\rm acc} \approx l_{\rm adv}/v_2$.
Eq.~\ref{eq:ldiff} shows that 
$l_{\rm diff} \approx D/v_2 \approx \tau_{\rm acc}v_2$.
Hence, the condition $\tau_{\rm acc}\approx \tau_{\rm syn}$ also
implies $l_{\rm diff}\approx l_{\rm adv}$. 

By combining
Eq.~\ref{eq:tau_syn} and Eq.~\ref{eq:emax} one obtains a
relation between the typical width of the X-ray synchrotron
emitting region, assumed to be $l_{\rm adv}$,
and the magnetic-field
strength, under the assumption that we observe synchrotron
emission from electrons near the cut-off energy
\citep[c.f.][]{parizot06,vink06d}: 
\begin{eqnarray}
B_2 \approx  
26 \Bigl(\frac{l_{\rm adv}}{1.0\times10^{18} {\rm cm}}\Bigr)^{-2/3}\eta^{1/3}\Bigl(\chi_4-\frac{1}{4}\Bigr)^{-1/3}
\ 
{\rm \mu G},\label{eq:ladv2}
\end{eqnarray}
with $B_2$ the average magnetic field in the shocked plasma, $\chi_4$ the
total shock compression ratio in units of 4, and $\eta$ as defined in Eq.~\ref{eq:diff_const}. Note that this magnetic field estimate is independent of 
the shock velocity and depends only weakly on the shock compression ratio
and $\eta$. 
Since the actual width is determined by a combination of diffusion
and advection the actual observed width is larger than either 
$l_{\rm adv}$ or $l_{\rm diff}$, and one can take $l_{\rm obs}\approx \sqrt{2}l_{\rm adv}$.

It is worth mentioning that not all X-ray synchrotron radiation may
come from loss-limited spectra. For large SNRs like RX J1713.7-3946,
with a rather broad X-ray synchrotron emitting shell,
one may wonder if the magnetic field is not so large that the
X-ray synchrotron spectrum is age limited. Also for the
youngest known Galactic SNR,  G1.9+0.3 \citep{reynolds08b,
borkowski10}, it is not clear whether
the X-ray synchrotron emission is age or loss limited.
One possible way to test this is too monitor the non-thermal X-ray emission
for some time. For example, recently it has been shown that the X-ray synchrotron
brightness of Cas A has declined between 2000 and 2010. At the same time 
the spectrum has softened \citep{patnaude11}. 
This can be understood assuming a loss-limited spectrum, as the exponential
cut-off energy depends solely on $V_{\rm s}^2$ (Eq.~\ref{eq:syn_max}). 
Since the shock is decelerating
the cut-off energy will decrease, thereby making the spectral index larger
(corresponding to a softer spectrum) and the
brightness smaller.  
For an age-limited synchrotron spectrum the cut-off photon
energy depend on the age of the SNR. 
Thus the cut-off energy will increase with time. Depending on
the magnetic-field evolution in the shock region this could in principle
make the X-ray synchrotron spectrum harder \citep{carlton11}. 
The secular evolution of the X-ray synchrotron emission does, however, also depend on other
variables, 
such as the number of electrons accelerated as a function of age, 
but the evolution of the exponential cut-off is likely to be 
the dominant
contribution to the spectral evolution \citep{patnaude11,katsuda10b}.

%% file: jvink_aarv_fig_sn1006b.tex
\begin{figure}
\centerline{
\includegraphics[trim=0 80 0 0,clip=true,width=\medfig]{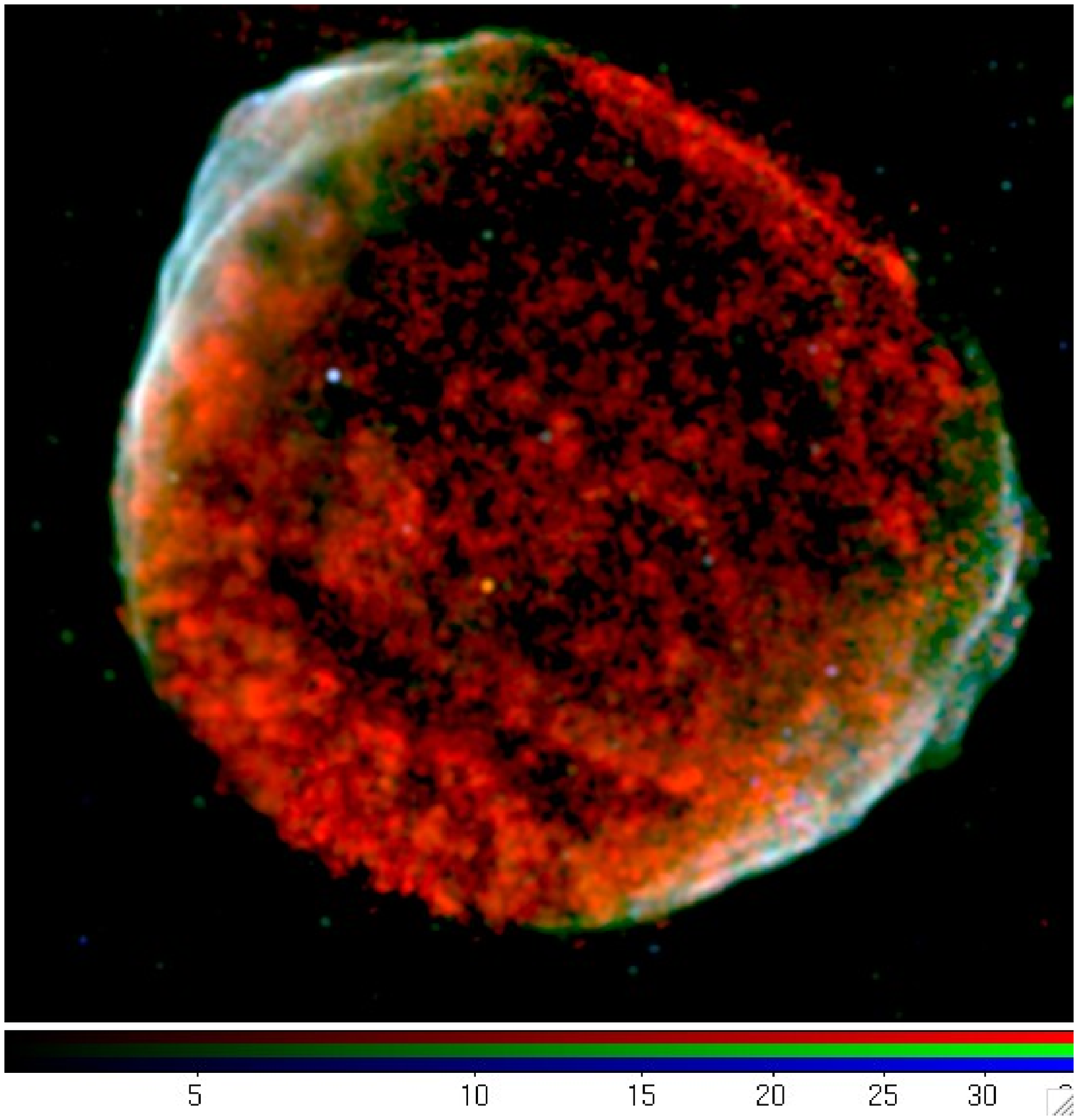}
}
\caption{
\chandra\ image of SN 1006 based on a mosaic of
several observations. The
three colors correspond to the 0.5-0.6 keV band (O VII He-$\alpha$, red),
1-1.7 keV (continuum with some Mg XI emission, green), and 2-5 keV (continuum
dominated, blue). 
SN 1006 was the first SNR for which imaging spectroscopy
with the  \asca\
showed a clear spatial separation of continuum emission, identified
as synchrotron emission mostly from near the rims of the SNR, 
and thermal emission (dominated by O VII), which dominates the emission
from the interior \citep{koyama95}.}
\label{fig:sn1006b}
\end{figure}

%% file: jvink_aarv_fig_casa_nonthermal.tex
\begin{figure*}
\centerline{
\includegraphics[trim=0 0 0 0,clip=true,width=0.45\textwidth]{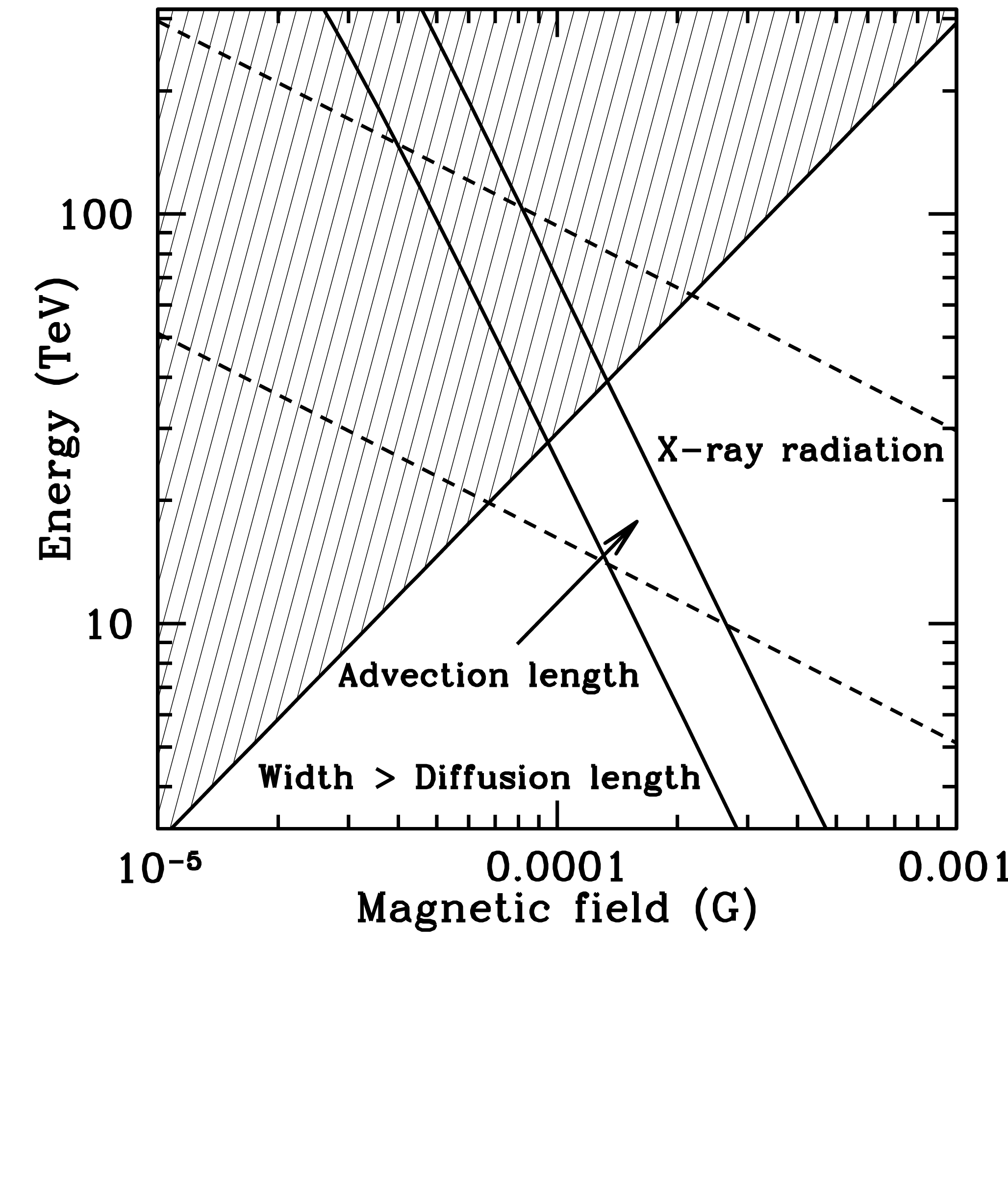}
\includegraphics[trim=0 0 250 0,clip=true,width=0.55\textwidth]{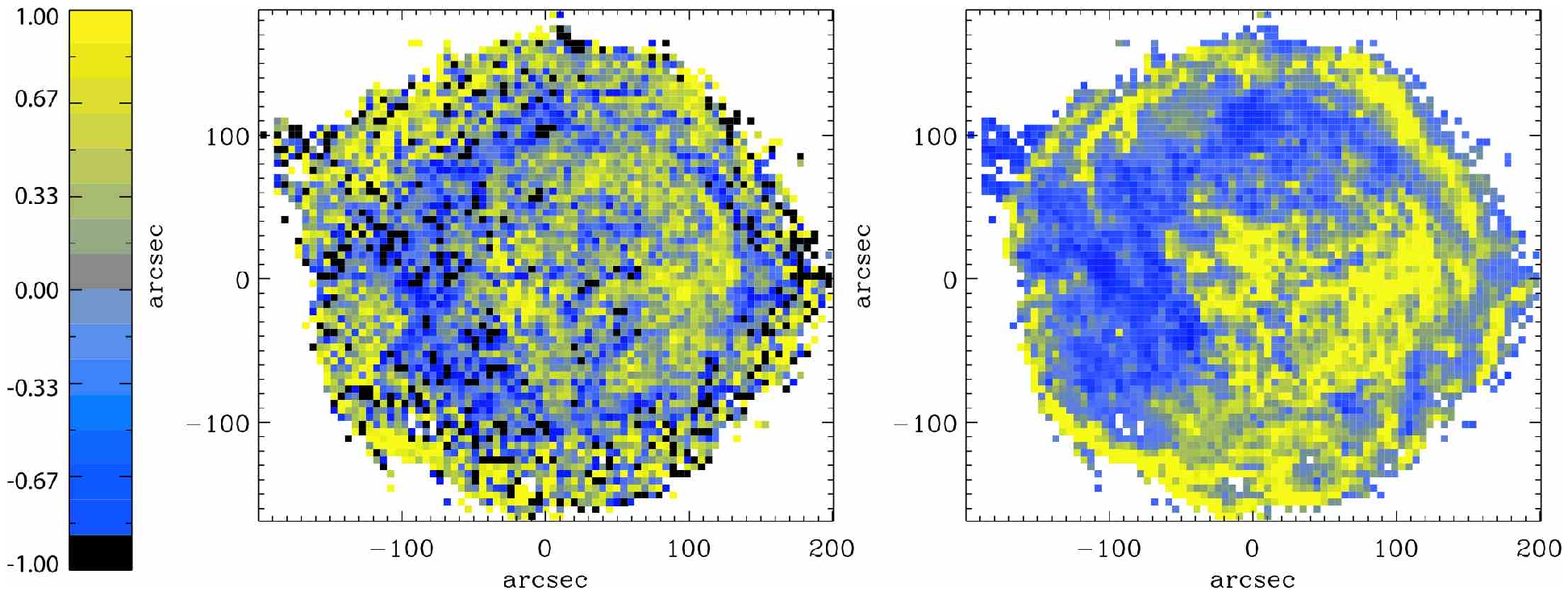}
}
\caption{
Left:
The maximum cosmic-ray electron
energy versus magnetic-field strength
for the region just downstream of Cas A's shock front, as determined
from the thickness of the filaments \citep{vink03a}. The shaded area
is excluded, because the filament width cannot be smaller than the minimum
possible diffusion length \citep[c.f.][]{vink03a}.
(These figures were published before in \citep{vink06b}.)
Right:
The power slope of the X-ray continuum in the 4-6 keV band based
on the 1 MS \chandra\ observation of Cas A \citep{helder08}. The color
scale indicates the slope difference with respect to the mean slope
of $\Gamma= 3.2$. Brighter color mean harder spectra, and signifiy a
larger contribution of synchrotron radiation. Note that there seem
to be two general regions of synchrotron radiation, the outer edge and
the inner region near the reverse shock.
A high resolution \chandra\ X-ray image of Cas A is shown in Fig.~\ref{fig:orich}, where the X-ray synchrotron can be seen as a faint blue wisp at the
edge of the SNR.
\label{fig:casaslope}
}
\end{figure*}

%% file: jvink_aarv_efficient_crs.tex
One of the surprises that came out of the magnetic-field determination
was that the magnetic fields were typically $100-600~\mu$G,
larger than might be expected, if they were merely caused by
the compression of the interstellar magnetic field 
$(B_{\rm ISM}\approx 5~\mu$G). 
Additional evidence for these high magnetic fields is that the
X-ray synchrotron emitting regions seem to vary on time scales of
several months to years, which may be indicative of the 
synchrotron cooling time scales, and, hence high magnetic fields
\citep{patnaude07,uchiyama07,uchiyama08}.\footnote{However, an alternative
explanation may be varying magnetic fields due to passages of large
scale plasma waves \citep{bykov08}.}
The high magnetic fields suggest
that some magnetic-field amplification mechanism is
operating near the shocks of young SNRs. 

There are several theories that link magnetic-field amplification to
efficient cosmic-ray acceleration.
A particular promising mechanism is that discovered by
\citet{bell04}, in which cosmic rays streaming outward result in an
electron return current. These electrons
are then deflected by the magnetic field,
thereby creating a very turbulent, amplified magnetic field. This mechanism
is expected to give $u_B\approx V_s u_{\rm cr} \propto \rho V_S^3$,
with $u_B$ and $u_{cr}$ respectively 
the magnetic field and cosmic-ray energy density,  and $\rho_0$ the
ambient plasma density. 

Combining the magnetic-field estimates of several young SNRs shows that
there is indeed a clear trend that SNRs evolving in low density environments
seem to have lower magnetic fields than those evolving
in high density media
\citep{bamba05,voelk05,ballet06,vink08d}.
This also explains the trend that the wider X-ray synchrotron emitting
regions, hence lower magnetic fields, are found in larger SNRs:
these SNRs have a combination of high shock velocity and large radius,
because they evolve in a low density region
\citep{ueno06}.

\input{jvink_aarv_fig_bfield}

The relation between magnetic-field energy density and shock velocity is less
clear. The reason is that for  X-ray synchrotron emitting SNRs the dynamic range
in velocity is relatively small ($\sim 2000-6000$~\kms).
\citet{voelk05} find that $B^2 \propto \rho_0V_S^2$, but they
used an outdated measurement of the shock velocity of Cas A ($\sim 2500$~\kms).
\citet{vink08d} instead favors a dependency
of  $B^2/(8\pi) \propto \rho_0V_S^3$, but noted that the dynamic
range makes the dependency on $V_s$ uncertain. 
Including the magnetic-field determination for SN 1993J
\citep[$V_S\approx 20,000$~\kms, and $B\approx 64$~G,][]{fransson98,tatischeff09} indicates that 
$B^2\propto V_S^3$ seems more likely (Fig.~\ref{fig:bfield}).
But it should be noted that
it is not quite clear whether the magnetic-field amplification mechanism
in SN1993J has a similar origin as in young SNRs.

There has been some concern that amplified, turbulent magnetic fields dampen
rapidly, even to the point that the width of the narrow X-ray synchrotron
filaments in young SNRs may not be determined by the advection or diffusion length
scale, but by the magnetic-field damping time scale \citep{pohl05}. However,
this would affect the radio and X-ray synchrotron profile similarly, and
this does not appear to be the case \citep{cassam07,gotthelf01a}.

The X-ray synchrotron emission from SN\,1006 has another characteristic that
may be of importance for a better understanding of cosmic-ray acceleration:
it appears that the synchrotron radiation is coming from just two regions
of the SNR, i.e. there seems to be very little synchrotron emission from
the center of the SNR \citep{willingale96,rothenflug04,cassam08}. 
A likely explanation
is that this is due to initial the magnetic field configuration. Given the
inferred polar geometry, the implication is that X-ray synchrotron radiation
is favored by a magnetic field that is parallel to the shock normal.
On the other hand, the presence of X-ray synchrotron radiation requires
a turbulent magnetic field. This seems a contradiction, but the
two conditions can be combined by assuming that efficient 
cosmic-ray acceleration favors a parallel magnetic field to start, 
but once it has started it creates its own turbulent magnetic field
\citep{voelk03}. SN\,1006 may be special in that its background
magnetic field was much more regular than that of other X-ray synchrotron
emitting SNRs.

One question that may be raised is whether the magnetic fields in young SNRs
are caused by amplification, or that even creation of magnetic
fields is possible. 
There is at least some indication for magnetic-field creation, because
\citet{helder08} showed, on the basis of  a spectral analysis of \chandra\ X-ray data
(Fig.~\ref{fig:casaslope}b), that most of the X-ray synchrotron
emitting filaments are not associated with the forward shock, but
with the reverse shock \citep[see also][]{uchiyama08}. 
Since these filaments are also narrow, the
local magnetic fields must be of the order of $100-500~\mu$G. A problem
with this is that the ejecta, due to their large expansion must have very
low magnetic fields. So this suggests that Bell's mechanism, or any
other amplification mechanism, even operates for very low seed magnetic
fields, or perhaps does not even require a seed magnetic field.
Alternatively, cosmic rays diffusing from the forward shock to
the reverse shock may have provided a seed population
of cosmic rays that resulted in magnetic-field amplification through
Bell's mechanism \citep{schure10}. 
Note that there are also suggestions that the reverse shock regions
of RCW 86 \citep{rho02} and  RX J1713.7-3946 \citep{zirakashvili10}
are sources of X-ray synchrotron radiation.

Irrespective of the exact magnetic-field 
dependence on shock velocity, the evidence for magnetic-field
amplification in young SNRs has important consequences.
The presence of amplified magnetic fields imply  large
cosmic-ray energy densities. This makes it likely
that cosmic rays are indeed being efficiently accelerated, i.e.
a large fraction of the shock energy may be transferred to cosmic
rays.  This in turn suggests that
the post-shock temperatures are expected to be low
and the shock 
compression ratios should be larger than four (\sect~\ref{sec:kT}). 

As argued in \sect~\ref{sec:thermaldoppler},
the lack of very hot electron temperatures in young SNRs may
be caused either by a lack of electron-proton temperature equilibration,
or by efficient acceleration. Proton temperatures are less
affected by equilibration effects, and can be determined
from H$\alpha$ line widths. Using this technique 
\citet{helder09} showed that in the northeastern part of RCW 86 the 
proton temperature is $kT\approx 2$~keV, much lower than the expected
$kT> 10$~keV. They derived from this a post-shock cosmic-ray
pressure contribution of $>50$~\%. Note that RCW 86 is 
a TeV source \citep{aharonian09} and that the northeastern part 
emits X-ray synchrotron radiation \citep{bamba00,borkowski01,vink06d}.
A similar study of the rapidly expanding LMC SNR B0509-67.5 
(\sect~\ref{sec:typeia})
indicates that there in the southwest the post-shock pressure is for $>20$\% provided by
cosmic rays \citep{helder10}.
This method can also be employed using X-ray spectroscopy once high resolution, non-dispersive, 
spectrometers will be available .

\input{jvink_aarv_fig_curved}

Another effect of efficient cosmic-ray acceleration is
high shock compression ratios (\sect~\ref{sec:kT}).
There is indeed evidence for high shock 
compression ratios in  Tycho \citep{warren05,cassam07}
and SN\,1006 \citep{cassam08}. Although the compression ratio cannot
be measured directly, X-ray imaging spectroscopy of
both SNRs shows that the ejecta that are located very close
to the forward shock, which is an observational consequence
of a high compression ratio \citep[][]{decourchelle00}.

Efficient acceleration should also lead to a curvature
of the relativistic electron spectrum. The reason is
that 
particles of different energies
experience different compression ratios during the acceleration process,
depending on whether they experience the overall shock compression ratio,
the gas-shock (sub-shock) compression ratio, or something in between 
(see \sect~\ref{sec:cr}).
Radio spectra of a few SNRs do indeed show evidence for spectral
hardening with increasing frequencies \citep{reynolds92}.
For X-ray spectroscopy the consequence of spectral curvature is that the
X-ray synchrotron brightness cannot be simply estimated from
an extrapolation of the radio synchrotron spectrum, but it should be brighter
than expected.
This has indeed been demonstrated to
be the case for the northeastern rim of RCW 86 
\citep[][see Fig.~\ref{fig:curved}]{vink06d}
and SN\,1006 \citep{allen08}.

%% file: jvink_aarv_fig_bfield.tex
\begin{figure}
\centerline{
\includegraphics[angle=-90,width=\medfig]{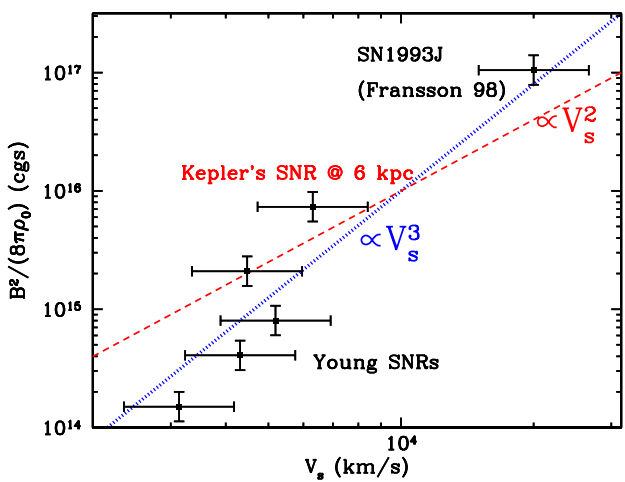}
}
\caption{
The dependence of the 
post-shock magnetic field energy density $B^2/(8\pi)$
on the shock velocity $V_s$. The dashed line shows a $V_s^2$ dependency
and the dotted line a $V_s^3$ dependency.
The input values can be found in \citep{vink06b}, 
except for SN1993J \citep{fransson98} and Kepler's SNR, 
for which the shock velocity in the southwest
was taken from \citep{vink08b} and the density from \citep{cassam04}.
\label{fig:bfield}
}
\end{figure}

%% file: jvink_aarv_fig_curved.tex
\begin{figure}
\centerline{\includegraphics[angle=-90,width=\medfig]{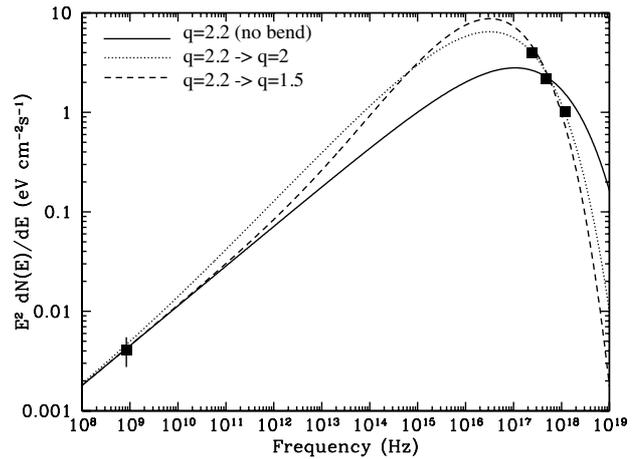}}
\caption{
Radio to X-ray synchrotron spectrum of the northeastern part
of RCW 86 \citep{vink06d}. The models show that a curved synchrotron
spectrum is needed. A simple synchrotron spectrum with a spectral radio
index $\alpha=0.6$ (corresponding to an electron particle
index of $q=2.2$, \sect~\ref{sec:synchrotron}) 
and an exponential cut-off in electron
energy cannot simultaneuously fit both the radio and X-ray data points.
\label{fig:curved}
}
\end{figure}

%% file: jvink_aarv_xray_gamma.tex
The field of GeV and TeV \gray\ astronomy has made tremendous progress  over the last decade,
thanks to technical progress in building sensitive TeV  Cherenkov telescopes \citep[see][for a review]{hinton09}
and, very recently the launch of the NASA \fermi\ satellite, which is sensitive
to $\sim 0.1-100$~GeV \grays.
Together these telescopes have greatly expanded the number of  \gray\ sources,
and, over the last decade have resulted in the first firm detections of continuum \gray\
emission from shell-type SNRs. The first such detection in the TeV band was Cas A with
\hegra\ \citep{aharonian01}, but the list of TeV detected telescopes has greatly
expanded since then.\footnote{At the moment of writing 13 shell-type
SNRs are now claimed to be TeV sources, see \url{http://tevcat.uchicago.edu}.}

The $0.1-100$~GeV band was sampled by \gray\  satellites since the 1970ies, but the first
evidence for associations between GeV \gray\ sources and SNRs were found with
\egret, and concerned mature SNRs interacting with molecular clouds \citep[][]{esposito96}. \fermi\ has confirmed these early associations, and greatly enhanced
the number of \gray\ sources among mature SNRs interacting with molecular clouds (\sect~\ref{sec:mixed}). In addition, also a number of younger SNRs have now been detected.
The first catalog of \fermi\ listed 41 sources as possibly associated with SNRs \citep{abdo10e}
and this list is still expanding.\footnote{Note that the \gray\ may not necessarily come
from the SNR shell, but in some cases from an embedded pulsar wind nebula.}

The great importance of GeV and TeV \gray\ astronomy is that it can give us a direct view
of the accelerated cosmic-ray nuclei\footnote{Protons and other ions. These
are often named "hadronic cosmic rays", as opposed to leptonic cosmic
rays, which are mostly electrons and positrons.}, which make up 99\% of the cosmic
rays observed on earth. These nuclei produce \gray\ emission if they
collide with atomic nuclei,
thereby creating, among others, neutral pions,
which decay in two \gray\ photons. 
However,  two other important \gray\ radiation processes
originate from relativistic electrons: interactions with background
photons result in inverse Compton up-scattering, whereas interactions
with ions in the SNR result in bremsstrahlung.
X-ray observations are crucial for disentangling all these contributions.
The thermal X-ray emission is important for determining the local 
plasma density, which determines the \gray\
non-thermal bremsstrahlung and pion-decay contributions. 
The X-ray synchrotron emission
can be used to determine the local magnetic field. The combination of magnetic
field and radio and X-ray synchrotron flux can then be used to
infer the inverse Compton contribution to the \gray\ emission.

Although X-ray observations seem to indicate that
cosmic-ray acceleration is efficient in young SNRs,
the recent observations of SNRs in GeV and TeV \grays\
provide a more ambiguous story.
Although many SNRs have been detected in TeV \grays\
with TeV telescopes like \hegra\ \citep{aharonian01}, 
\hess\ \citep[e.g.][]{aharonian04}, \magic\ \citep{albert07}
and \veritas\ \citep[e.g.][]{acciari11} and the \fermi\ satellite,
 the \gray\ observations do not provide clear evidence
that cosmic rays are being accelerated {\em efficiently}, nor that they
SNRs accelerate ions to beyond "the knee".

\input{jvink_aarv_fig_rxj1713}

There has been a strong debate
about the nature of TeV emission X-ray synchrotron radiation dominated SNRs.
A central role in this debate concerns the brightest TeV emitting SNR,
RX J1713.7-3946.
For example  \citet{berezhko10} advocates a dominant pion-decay contribution
to the TeV \gray\ emission, whereas \citet{katz08} and \citet{ellison10}
argue that the \gray\ emission is dominated by inverse Compton scattering.
The lack of thermal X-ray emission is an important ingredient in this
debate, as to some it indicates low densities, and hence low pion-decay
luminosities, whereas others \citep[notably][]{drury09} have argued that
it could be the result of extremely low temperatures caused by
efficient cosmic-ray acceleration (see \sect~\ref{sec:kT}).
An argument in favor of an electron origin for the \gray\ emission
of RX J1713.7-3946 is the close resemblance between the TeV 
\gray\  and X-ray brightness distribution, which suggests that both
originate from the same population of particles: electrons
 \citep[Fig.~\ref{fig:rxj1713},][]{acero09}.
 
However, there are two arguments in favor of pion decay. Namely the
relatively hard spectrum in TeV, which extends up to 100~TeV, and the
inferred high magnetic field in the SNR, $>$~mG \citep{uchiyama07}.
Higher magnetic fields imply lower
relativistic electron densities for a given X-ray synchrotron flux. Hence, the
electron contribution to the TeV flux is expected to be low, and pion
decay becomes
the most likely dominant source of TeV \gray s.
The high magnetic-field estimates are based on the the observation of rapid
X-ray flux changes and comparing them to the synchrotron loss time scales
(see Eq.~\ref{eq:tau_syn}). But the fluctuations may also originate from
fluctuations of due to long wavelength Alfv\'en waves \citep{bykov08}.
Another argument in favor of relatively high magnetic
fields  are the small widths of the
X-ray synchrotron rims (\sect~\ref{sec:synchrotron}).
But for this measurement the  problem is  what region one should measure
the width of: the whole shell
or just a bright filament?
For example, one can take the overall
shell size of RX J1713.7-3946 to be the size of the filament (since for this SNR
the whole shell seems to emit synchrotron emission), or use some filamentary
sub-structures. This problem is illustrated
in (Fig.~\ref{fig:rxj1713prof}).
If one takes the whole shell region, at best one comes up with
a length scale of 100\arcsec\  corresponding to $25-30~\mu$G.

\input{jvink_aarv_fig_rxj1713_Xray}

The most serious challenge to the pion-decay model for RX J1713.7-3946
comes from recent observations by \fermi\ \citep{abdo11}, which show that the
shape of the broad band \gray\ spectrum is more consistent
with inverse Compton scattering than pion decay.
The implication of this result is surprising,
because RX J1713.7-3946 was often used as a
key object in the argument that SNRs can accelerate ions to very high energies.

An important result of \fermi\ \citep{abdo10} is the detection of GeV \gray\ emission from
Cas A, which indicates 
 that the total cosmic-ray energy content 
is less than 4\% of the total explosion energy. This is disappointing
if one thinks that young SNRs should be able to
accelerate cosmic rays efficiently, and convert $>10$\% 
of the explosion energy to cosmic rays (\sect~\ref{sec:cr}).
The case for pion decay as origin for the \gray\ emission is stronger for Tycho's SNR,
given the recent detection with the \veritas\ TeV telescope \citep{acciari11}.

The apparent lack of evidence for a high cosmic ray acceleration efficiency for Cas A
is also somewhat surprising, given
the high magnetic field of Cas A, as argued in \sect~\ref{sec:synchrotron}, which is 
suggestive of 
efficient cosmic-ray acceleration. It could be that a large part of the energy
in cosmic rays has already escaped the SNR. Since Cas A evolves inside
the wind of its progenitor, the background plasma that the escaped cosmic
rays are in likely has a lower density than the SNR shell.

Indeed, several theories about
non-linear cosmic ray acceleration point toward the importance of
cosmic ray escape \citep{blasi05,vladimirov08,reville09,vink10a}.
Future very sensitive TeV telescopes
like the Cherenkov Telescope Array \citep[CTA,][]{cta10} could in principle
be used to search for ``cosmic-ray haloes'' around young SNRs.
Hopefully, the improvement in sensitivity in TeV astronomy can be met
with a similar improvement in sensitivity and spectral resolution in
X-ray spectroscopy, which will help to detect thermal X-ray emission
and provide temperatures for X-ray synchrotron emitting plasmas.

Interestingly enough, the case for a pion-decay origin of the \gray\ emission
is most unambiguous for some of the mature SNRs that are interacting with or are near
molecular clouds. An example is  W44 \citep[][see also \sect~\ref{sec:mixed}]{abdo10b}, for which
it is argued that inverse Compton scattering, or bremsstrahlung do not provide an adequate description
of the spectrum. But the \gray\ spectrum of this source steepens around 10~GeV, which is consistent
with the idea that cosmic rays with energies above $10^{15}$~eV are present. This break in the spectrum could be evidence that cosmic rays with larger energies have already escaped
the SNR in the past. This is also consistent with several SNRs that have \gray\ sources nearby, which
could be caused by cosmic rays that escaped  from the SNR  and are now interacting with the high
density medium of molecular clouds. The GeV/TeV sources associated with SNR W28
is an example for such an association \citep{aharonian08a,abdo10c}.\footnote{\url{http://tevcat.uchicago.edu/} lists 6 SNRs/Molecular Cloud associations as TeV sources.}

%% file: jvink_aarv_fig_rxj1713.tex
\begin{figure*}
\centerline{
\includegraphics[width=0.5\textwidth]{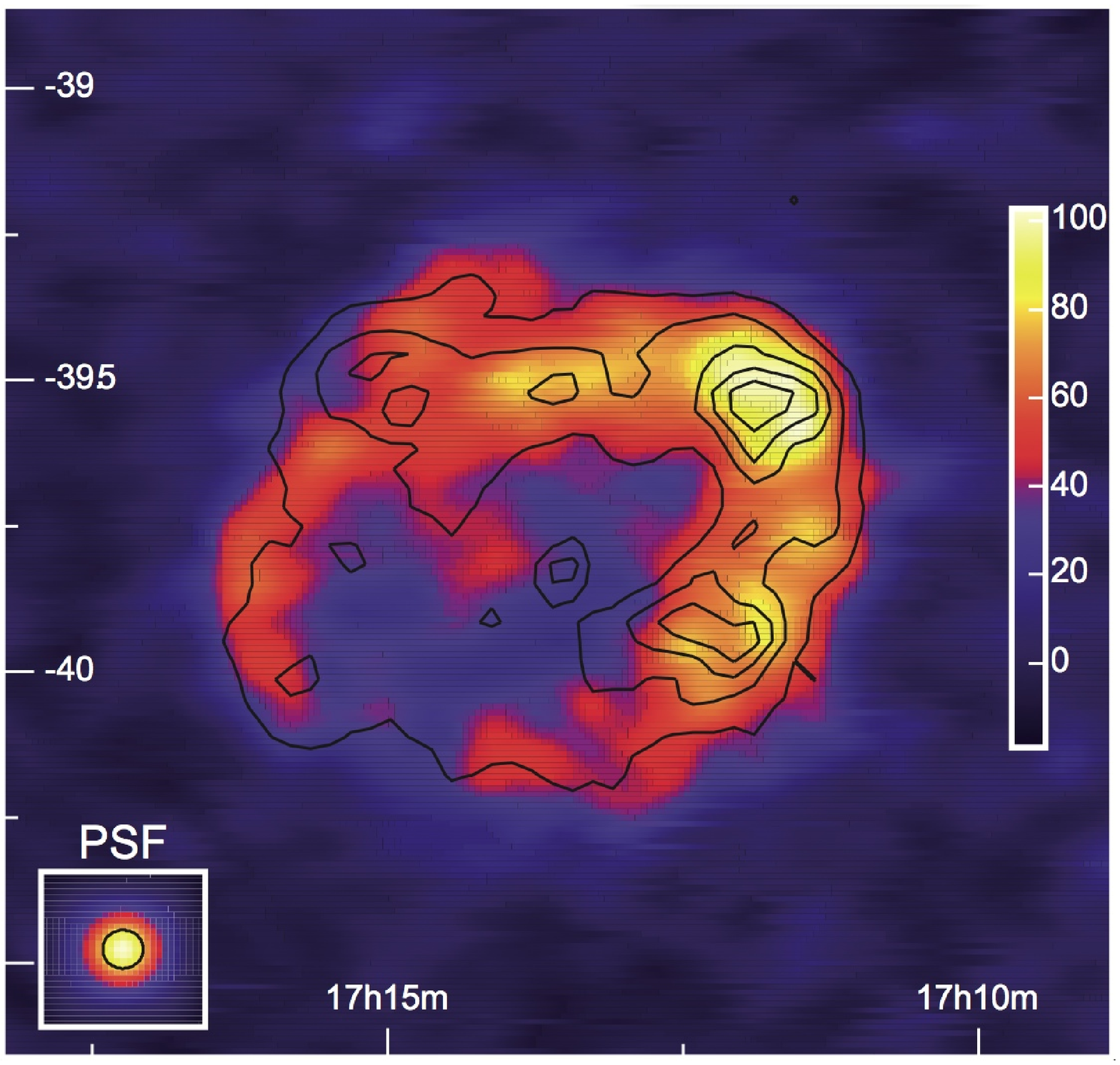}
\includegraphics[trim=30 10 90 10,clip=true,width=0.5\textwidth]{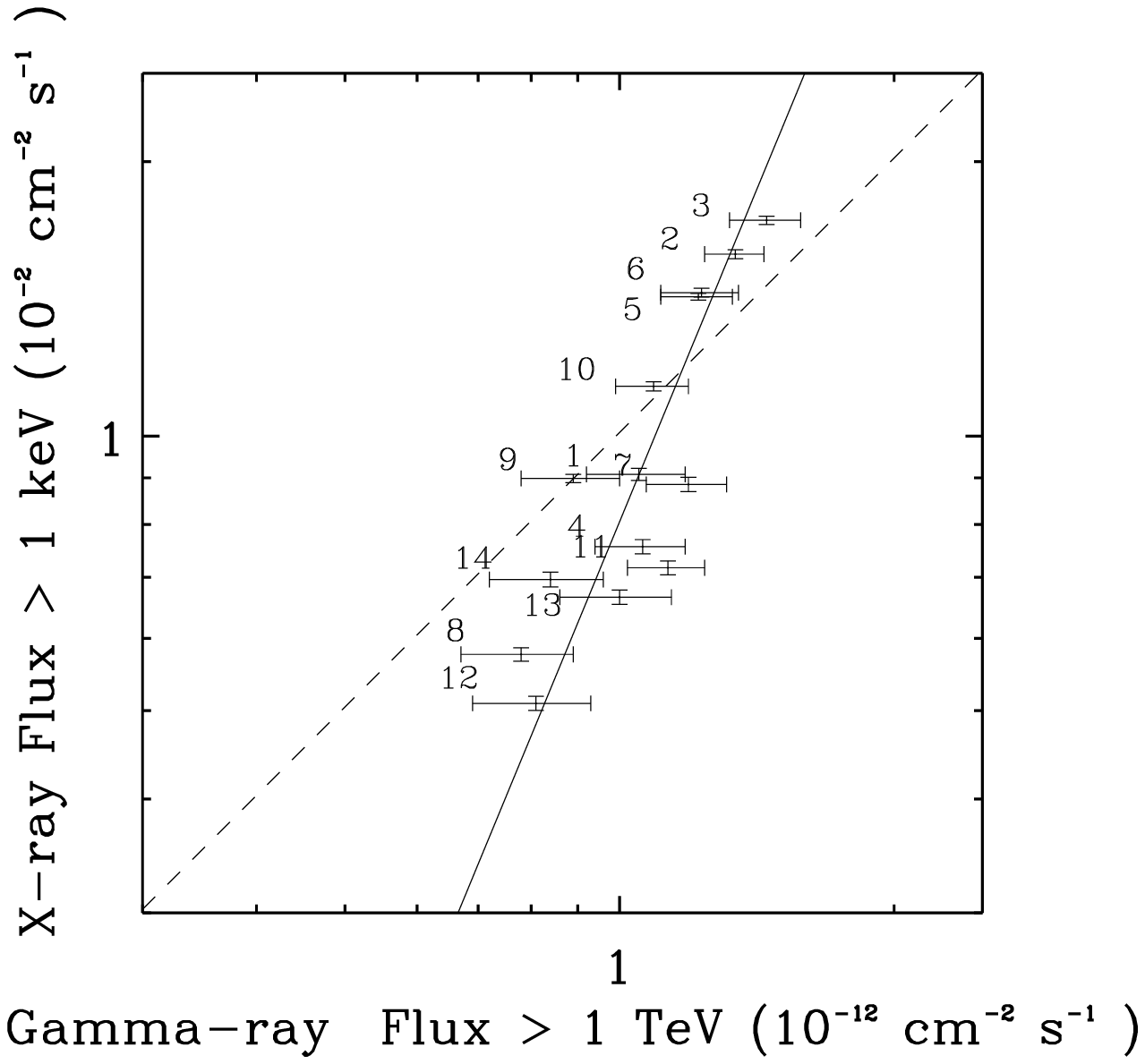}
}
\caption{
Left: \gray\ excess count map of RX J1713.7-3946 as observed by \hess .
Overlayed are 1-3 keV X-ray (\asca) intensity contours
\citep{aharonian07}.
Right: A comparison of the X-ray flux (1-10 keV,\xmm) and
HESS \gray\ integrated flux \citep[1-10 TeV band, Figure taken from][]{acero09}.
\label{fig:rxj1713}}
\end{figure*}

%% file: jvink_aarv_fig_rxj1713_Xray.tex
\begin{figure}
\centerline{
\includegraphics[angle=-90,width=\medfig]{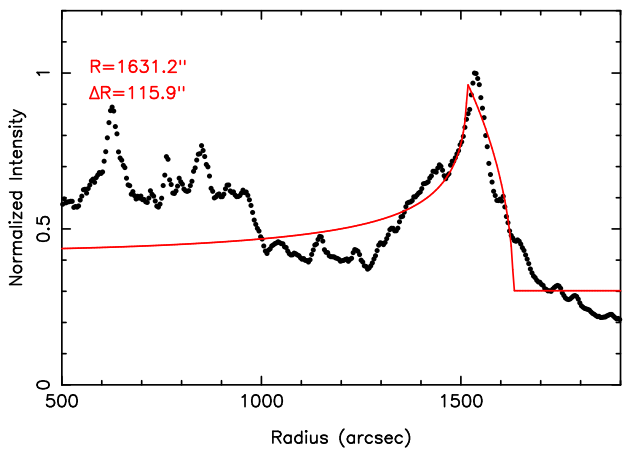}
}
\caption{
X-ray emission profile of the southeastern part of RX J1713.7-3846 \citep[extracted by the author from the \xmm\ image reduced by][]{acero09}.
It illustrates the difficulty
in determining for this SNR what the width of the X-ray synchrotron emitting
region is. The red line indicates a simple model of a projected sphere
with a uniform emissivity with a shell width of 116\arcsec. For
a distance of 1~kpc this corresponds, according to Eq.~\ref{eq:ladv2},
to 25~$\mu$G.\label{fig:rxj1713prof}
}
\end{figure}

%% file: jvink_aarv_conclusion.tex
Our understanding of SNRs has greatly increased over the last decade
due to the X-ray observatories \chandra, \xmm, and \suz. The second half of this
review essentially summarized the advances in SNR research made with these
observatories.  I think that a number of these results deserve more recognition
outside the field of SNR research. I will summarize some of them here.

X-ray spectroscopy has proven to 
be a promising tool to separate young Type Ia 
(\sect~\ref{sec:typeia}) 
from core collapse
SNRs (\sect~\ref{sec:orich}). X-ray emission for Type Ia SNRs has given us important clues toward
the nature of the progenitor systems: the progenitors did not produce
fast winds, and delayed detonation models seem to explain the abundance
patterns well. Type Ia SNRs are also well stratified, unlike core collapse
SNRs. 

An important question concerning core collapse supernovae is what drives
the explosion (neutrino absorption, or other mechanisms?) and what types
of massive stars produce neutron stars and what types make
black holes. X-ray observations show that the core collapse process
is chaotic, with pure metal plasma obtaining high velocities in seemingly
random directions. But for one SNR, Cas A, there
is evidence for an ejecta jet/counter jet. 
These are in Si-group material, not in Fe-rich material. 
With future
observations we may learn more about rare elements, which may indicate the
neutron content of the inner most ejecta. This may help us to
identify the role of neutrino absorption in the explosion.

As for the question what determines the creation of a black hole. There is
at least one SNR, G292.0+1.8, for which there is an indication that it produced
a neutron star, but had nevertheless a progenitor mass above 30~\msun\
(\sect~\ref{sec:snrs_ns}).
It is important to confirm this results with further X-ray studies,
as a popular hypothesis is that stars above 25~\msun\ leave behind a black hole.

Another uncertainty in the theory of core collapse supernovae is how much
iron they produce. The iron yield is greatly affected by the details of
the explosion, such as explosion asymmetries and fall back on the neutron star.
There is evidence that the Cygnus Loop (\sect~\ref{sec:mature_metals}) 
contains more iron than predicted by explosion models. Whereas for
Cas A, iron seems to have obtained a higher velocity than predicted by
explosion models.

One surprising result of X-ray imaging spectroscopy is that even mature
SNRs still have regions with enhanced metal abundances. This is true
for the Cygnus Loop (\sect~\ref{sec:mature_metals})  and many
mixed-morphology SNRs (\sect~\ref{sec:mixed}). For the Vela SNR
X-ray imaging spectroscopy confirmed that the bullets outside the main
shock are indeed ejecta ``shrapnels'' that have enhanced abundances.

Finally, X-ray imaging spectroscopy has played an important role
in increasing our understanding of cosmic-ray acceleration. A role it shared with
the rapidly evolving
fields of TeV astronomy with Cherenkov telescopes, and GeV astronomy
with the \fermi\ satellite.
In particular, X-ray detections
of narrow width synchrotron filaments
have provided evidence for magnetic-field amplification near SNR shocks
(\sect~\ref{sec:synchrotron}).
This implies that SNRs are efficiently accelerating cosmic rays. However,
the overall interpretation of both X-ray and \gray\ observations is still
ambiguous, in particular concerning the total energy in cosmic-rays contained
in young SNRs (\sect~\ref{sec:xray_gammarays}).

As is clear from the above, 
our understanding of SNRs has greatly benefitted from the 
coming of age of X-ray imaging spectroscopy. The impact of high
resolution X-ray spectroscopy has been more modest, as \chandra\
and \xmm\ have grating spectrometers, which are not
ideal for observations of extended sources as SNRs. Nevertheless,
for bright sources of small angular extent important results have been obtained.

In the near future high resolution X-ray spectroscopy of SNRs will
likely become much more important as microcalorimeters will become
operational. These will have a spectral resolution of 
$\Delta E \approx 5$~eV at 6~keV ($R\approx 1200$) and also have some
limited spatial resolution.
The first opportunity to use them will start after
the launch of \astroh\footnote{Formerly known as \next\ \citep{takahashi08}.}.
After \astroh\ the next step forward will hopefully be
\athena, a mission concept for an 
European X-ray observatory that will offer a large
effective area ($>1$~m$^2$ at 1 keV) and a spatial resolution of $<10$\arcsec, 
 which should be compared
to 1\arcmin\ for \astroh. 

The tantalizing glimpse of high resolution X-ray spectroscopy 
of SNRs offered by \chandra\ and \xmm, shows that SNR research beyond
2014 will enter a new era.